\documentclass[twocolumn]{aastex631}

\shorttitle{SOMA IV - Isolated Protostars}
\shortauthors{Fedriani et al.}
\graphicspath{{./}{figures/}}
\usepackage{multirow}

\begin{document}

\title{The SOFIA Massive (SOMA) Star Formation Survey. IV.\\
Isolated Protostars}

\author[0000-0003-4040-4934]{Rub\'{e}n Fedriani}
\affil{Department of Space, Earth \& Environment, Chalmers University of Technology, 412 93 Gothenburg, Sweden}


\author[0000-0002-3389-9142]{Jonathan C. Tan}
\affiliation{Department of Space, Earth \& Environment, Chalmers University of Technology, 412 93 Gothenburg, Sweden}
\affiliation{Department of Astronomy, University of Virginia, Charlottesville, Virginia 22904, USA}
\author[0000-0001-6465-9590]{Zoie Telkamp}
\affiliation{Department of Astronomy, University of Virginia, Charlottesville, Virginia 22904, USA}
\author[0000-0001-7511-0034]{Yichen Zhang}
\affiliation{Department of Astronomy, University of Virginia, Charlottesville, Virginia 22904, USA}
\affiliation{RIKEN Cluster for Pioneering Research, Wako-shi, Saitama, 351-0198, Japan}
\author[0000-0001-8227-2816]{Yao-Lun Yang}
\affiliation{RIKEN Cluster for Pioneering Research, Wako-shi, Saitama, 351-0198, Japan}
\affiliation{Department of Astronomy, University of Virginia, Charlottesville, Virginia 22904, USA}
\author[0000-0001-6159-2394]{Mengyao Liu}
\affiliation{Department of Astronomy, University of Virginia, Charlottesville, Virginia 22904, USA}
\author[0000-0001-7378-4430]{James M. De Buizer}
\affiliation{SOFIA-USRA, NASA Ames Research Center, MS 232-12, Moffett Field, CA 94035, USA}
\author[0000-0003-1964-970X]{Chi-Yan Law}
\affiliation{Department of Space, Earth \& Environment, Chalmers University of Technology, 412 93 Gothenburg, Sweden}
\author[0000-0003-3315-5626]{Maria T. Beltran}
\affiliation{INAF-Osservatorio Astrofisico di Arcetri, Largo E. Fermi 5, I-50125 Firenze, Italy}
\author[0000-0001-8596-1756]{Viviana Rosero}
\affiliation{National Radio Astronomy Observatory, 1003 L\'opezville Road, Socorro, NM 87801, USA}
\author[0000-0002-6907-0926]{Kei E. I. Tanaka}
\affiliation{Center for Astrophysics and Space Astronomy, University of Colorado Boulder, Boulder, CO 80309, USA}
\affiliation{National Astronomical Observatory of Japan, National Institutes of Natural Sciences, 2-21-1 Osawa, Mitaka, Tokyo 181-8588, Japan}
\author[0000-0001-5551-9502]{Giuliana Cosentino}
\affiliation{Department of Space, Earth \& Environment, Chalmers University of Technology, 412 93 Gothenburg, Sweden}
\author[0000-0003-1602-6849]{Prasanta Gorai}
\affiliation{Department of Space, Earth \& Environment, Chalmers University of Technology, 412 93 Gothenburg, Sweden}
\author[0000-0002-5851-2602]{Juan Farias}
\affiliation{Department of Space, Earth \& Environment, Chalmers University of Technology, 412 93 Gothenburg, Sweden}
\author[0000-0001-9040-8525]{Jan E. Staff}
\affiliation{College of Science and Math University of the Virgin Islands St Thomas, VI 00802 USA}
\author{Barbara Whitney}
\affiliation{Space Science Institute, 4765 Walnut St, Suite B, Boulder, CO 80301, USA}



\begin{abstract}
We present $\sim10-40\,\mu$m SOFIA-FORCAST images of 11 \textit{isolated} protostars as part of the SOFIA Massive (SOMA) Star Formation Survey, with this morphological classification based on 37\,$\mu$m imaging. We develop an automated method to define source aperture size using the gradient of its background-subtracted enclosed flux and apply this to build spectral energy distributions (SEDs). We fit the SEDs with radiative transfer models, developed within the framework of turbulent core accretion (TCA) theory, to estimate key protostellar properties. Here, we release the sedcreator python package that carries out these methods. The SEDs are generally well fitted by the TCA models, from which we infer initial core masses $M_c$ ranging from $20-430\:M_\odot$, clump mass surface densities $\Sigma_{\rm cl}\sim0.3-1.7\:{\rm{g\:cm}}^{-2}$ and current protostellar masses $m_*\sim3-50\:M_\odot$. From a uniform analysis of the 40 sources in the full SOMA survey to date, we find that massive protostars form across a wide range of clump mass surface density environments, placing constraints on theories that predict a minimum threshold $\Sigma_{\rm cl}$ for massive star formation. However, the upper end of the $m_*-\Sigma_{\rm cl}$ distribution follows trends predicted by models of internal protostellar feedback that find greater star formation efficiency in higher $\Sigma_{\rm cl}$ conditions.
We also investigate protostellar far-IR variability by comparison with IRAS data, finding no significant variation over an $\sim$40 year baseline. 
\end{abstract}


\section{Introduction} \label{sect:introduction}

Massive stars are the engines that drive the evolution of galaxies. Their energetic radiation, winds, and supernovae also impact their surrounding environments, including protoplanetary disks around lower-mass stars that are forming in the same protocluster. In spite of their importance, many fundamental questions remain unanswered about the origins of massive stars, including the basic nature of their formation mechanism, e.g., whether it is via an extension of standard core accretion theory \citep[e.g.,][]{mckee2003} or whether it requires chaotic, competitive accretion in the center of a dense protocluster of low-mass protostars \citep[e.g.,][]{bonnell1998,wang2010,grudic2022}.

The SOFIA Massive (SOMA) Star Formation Survey (PI: Tan) aims to characterize a sample of $\ga$ 50 high- and intermediate-mass protostars over a range of evolutionary stages and environments with their $\sim$ 10-40\,$\mu$m emission observed with the SOFIA-Faint Object infraRed CAmera for the SOFIA Telescope (FORCAST) instrument \citep{herter2018}. In Paper I of the survey \citep{debuizer2017}, the first eight sources were presented, which were mostly massive protostars. In Paper II \citep{liu2019}, seven additional high-luminous sources were presented, corresponding to some of the most massive protostars in the survey. In Paper III \citep{liu2020}, 14 intermediate-mass sources were presented and analyzed. Here, in Paper IV in the series, we present 10 regions that harbor a total of 11 sources, selected based on the nature of their environment, i.e., appearing to be relatively \textit{isolated} in $37\,\mu$m imaging. We note that another set of eight regions that are relatively crowded in their $37\,\mu$m morphology, i.e., \textit{clustered} sources, will be presented in Paper V in this series (Telkamp et al., in prep.). Thus, we consider these samples will help to probe the environmental dependence of star formation, e.g., being of particular interest for testing the prediction of competitive accretion models that massive protostars should be surrounded by clusters of lower-mass protostars.

Our approach follows the same general methods developed in Papers I-III to build the spectral energy distributions (SEDs) of the protostars, measuring fluxes from infrared (IR) images, especially from Spitzer, SOFIA and Herschel facilities. We then fit these SEDs with the \citet[][hereafter ZT18]{zhang2018} protostellar radiative transfer (RT) models to estimate intrinsic source properties. However, here we introduce a number of new improvements to the SOMA analysis methodology, including an algorithmic way of choosing the aperture size and a new python module that updates the SED-fitting tool, including a revised method of assessing uncertainties in background-subtracted fluxes. We measure fluxes and fit SEDs for all the SOMA Papers I-IV sources with the new methods to produce a sample of 40 massive protostars that have been analyzed in a uniform way. We are thus able to more reliably examine trends in source properties among these sources.

Similar studies based on the SED fitting with the ZT18 models have been carried out by \citet{towner2019} on 12 extended green objects (EGOs); \citet{lim2019} on 41 massive protostar candidates in W51A; \citet{moser2020} on about 30 sources in the Infrared Dark Cloud (IRDC) G28.37+00.07. Furthermore, the methods developed here have also been applied by \citet{costa2022} in IRAS~18264-1152, \citet{law2022} in G28.20-0.05, and \citet{taniguchi2022} in G24.78+0.08.

The relatively isolated nature of the sources of this paper also enables a search for mid-IR (MIR) to far-IR (FIR) variability by comparing our SOFIA and Herschel-fitted SEDs with the flux measurements of IRAS made $\sim$40 yr earlier. Variability at MIR to FIR wavelengths has been reported in a few massive protostars \citep[e.g.,][]{caratti2017nature,hunter2017,stecklum2021,chen2021} and interpreted as being caused by accretion bursts. However, it is very uncertain what fraction of massive protostars undergo such bursts and what fraction of mass is accreted in such events.

The observations and data utilized in this paper are described in \S\ref{sect:observations}. The analysis methods are summarized in \S\ref{sect:methods}. We present the MIR to FIR imaging and SED-fitting results in \S\ref{sect:results}. We discuss our results in \S\ref{sect:discussion} and give a summary in \S\ref{sect:conclusions}.

\section{Observations} \label{sect:observations}

We used the Stratospheric Observatory for Infrared Astronomy (SOFIA\footnote{SOFIA is jointly operated by the Universities Space Research Association, Inc. (USRA), under NASA contract NAS2-97001, and the Deutsches SOFIA Institute (DSI) under DLR contract 50 OK 0901 to the University of Stuttgart.}) together with FORCAST \citep{herter2018} instrument to observe ten regions of massive star formation that harbor 11 protostars. Four filters were used that are centered at 7.7, 19.7, 31.5, and 37.1\,$\mu$m (see Table\,\ref{tab:sofia_obs} for details). Source selection for the SOMA survey has mainly utilized the CORNISH survey \citep{hoare2012}, complemented by radio-quiet MIR sources in IRDCs \citep{butler2012}. 

The photometric and astrometric calibration methods are the same as those used in Papers I-III. For SOFIA observations the photometric calibration error is estimated to be in the range of $\sim3\% - 7\%$. The astrometric precision is about 0\farcs1 for the SOFIA 7.7\,$\mu$m image, and 0\farcs4 for longer wavelength in SOFIA images. See Paper I for further details. Pipeline-reduced and calibrated data from the SOFIA archive were used.

In addition to SOFIA observations, when available, we also retrieved publicly-available images of Spitzer/IRAC \citep{fazio2004,werner2004} at 3.6, 4.5, 5.8, and 8.0\,$\mu$m from the Spitzer Heritage Archive, and Herschel/PACS and SPIRE \citep{griffin2010} at 70, 160, 250, 350, and 500\,$\mu$m from the ESA Herschel Science Archive, and IRAS \citep{neugebauer1984} at 12, 25, 60, and 100\,$\mu$m from the NASA/IPAC Infrared Science Archive. We use the HIRES results of the IRAS data to achieve a resolution of $\sim1\arcmin$. The astrometric precision is about 20\arcsec--30\arcsec. Flux measured from HIRES agrees with those of the IRAS Point Source Catalog (PSC2) to within 20\% and a ring of low flux that may appear around a point source can contain up to another 10\% uncertainty of the flux of the point source.

\begin{deluxetable*}{lllcccccc}
\tablecaption{SOFIA FORCAST Observations: Observation Dates \& Exposure Times (seconds)\label{tab:sofia_obs}}
\tablehead{
\colhead{Source} & \colhead{R.A.(J2000)} & \colhead{Decl.(J2000)} & \colhead{$d$ (kpc)} & \colhead{Obs. Date} & \colhead{7.7$\,{\rm \mu m}$} & \colhead{19.7$\,{\rm \mu m}$} & \colhead{31.5$\,{\rm \mu m}$} & \colhead{37.1$\,{\rm \mu m}$}
}
\startdata
AFGL~2591 & 20$^h$29$^m$24$\fs$8916 & $+$40$\arcdeg$11$\arcmin$19$\farcs$388 & 3.3 & 2016 Sep 20 & 404 & 779 & 642 & 1504  \\
G25.40-0.14 & 18$^h$38$^m$08$\fs$2700 & $-$06$\arcdeg$45$\arcmin$57$\farcs$820 & 5.7 & 2015 Jun 05 & 278 & 701 & 482 & 743 \\
G30.59-0.04 & 18$^h$47$^m$18$\fs$9000 & $-$02$\arcdeg$06$\arcmin$17$\farcs$600 & 11.8 & 2018 Sep 08 & 492 & 1319 & 825 & 2020 \\
G32.03+0.05 & 18$^h$49$^m$37$\fs$0520 & $-$00$\arcdeg$46$\arcmin$50$\farcs$150 & 5.5 & 2015 Nov 04 & 281 & 899 & 818 & 281 \\
G33.92+0.11 & 18$^h$52$^m$50$\fs$2730 & $+$00$\arcdeg$55$\arcmin$29$\farcs$594 & 7.1 & 2015 Nov 20 & 116 & 308 & 162 & 630 \\
G40.62-0.14 & 19$^h$06$^m$01$\fs$6000 & $+$06$\arcdeg$46$\arcmin$36$\farcs$200 & 2.2 & 2015 Jun 03 & 337 & 664 & 386 & 466 \\
IRAS~00259+5625 & 00$^h$28$^m$42$\fs$6000 & $+$56$\arcdeg$42$\arcmin$01$\farcs$110 & 2.5 & 2015 Nov 20 & 116 & 308 & 162 & 630 \\
IRAS~00420+5530 & 00$^h$44$^m$58$\fs$5842 & $+$55$\arcdeg$46$\arcmin$45$\farcs$675 & 2.2 & 2015 Jun 03 & 337 & 664 & 386 & 466 \\
IRAS~23385+6053 & 23$^h$40$^m$54$\fs$5171 & $+$61$\arcdeg$10$\arcmin$27$\farcs$768 & 4.9 & 2015 Jun 03 & 337 & 664 & 386 & 466 \\
HH 288 & 00$^h$37$^m$13$\fs$2580 & $+$64$\arcdeg$04$\arcmin$15$\farcs$020 & 2.0 & 2015 Nov 06 & 334 & 806 & 488 & 1512 \\
\enddata
\tablecomments{
The source positions listed here are the same as the positions of the
black crosses denoting the radio continuum peak for each of the sources shown in Figures \ref{fig:AFGL2591}-\ref{fig:HH288}. Source distances are from the literature are discussed below (see Sect.\,\ref{sect:sources}).}
\end{deluxetable*}

\section{Methods}\label{sect:methods}

In this paper, we introduce a number of new and updated analysis methods. The main update is the release of \verb+sedcreator+, which is an open-source python package hosted in both GitHub\footnote{\url{https://github.com/fedriani/sedcreator}} and PyPi\footnote{\url{https://pypi.org/project/sedcreator/}} (the documentation can be accessed at this URL \url{https://sedcreator.readthedocs.io/}). The main two sets of tools of \verb+sedcreator+ are encapsulated into SedFluxer and SedFitter. SedFluxer helps one construct an SED by providing tools to measure fluxes on a given image. \textit{SedFitter} fits an SED with massive star formation radiation transfer model grid by \citet{zhang2018}. It updates and replaces the earlier version of the fitting tool written in IDL \citep{zhang2018IDL}.

\subsection{SedFluxer}\label{sect:sedfluxer}

We follow methods similar to those of Papers I-III to construct SEDs. In brief, this involves defining an aperture for each source that is based on consideration of the Herschel $70\,\mu$m or SOFIA $37\,\mu$m (when Herschel $70\,\mu$m is not available) images with the goal to include most of the source flux within a relatively compact scale. Conceptually, we need to define an aperture to measure fluxes within the core scale, which is embedded inside the larger clump. The preferred method is to use a fixed aperture size at all wavelengths, i.e., the one defined by the $70\,\mu$m and/or $37\,\mu$m images (see \S\ref{sect:opt_rad}).
Background subtraction is carried out by estimating the average background intensity in an adjacent annulus, i.e., from one to two aperture radii, but excluding any regions that are within the aperture of another defined source (this is especially relevant for the \textit{clustered} sources to be studied in Paper V, but also for some sources previously studied in Papers I to III). The estimator used is the median value in the annulus. Then this value is multiplied by the area of the main aperture to account for the background in the entire aperture.

SedFluxer uses a number of functions from the Photutils \citep{photutils2020} and astropy \citep{astropy2013,astropy2018} python packages to measure fluxes on any image, including unit transformation (provided correct units are present in the header). These tools set the workflow for the parameters of SOMA studies as described above, but complete freedom is given to the user to change the inputs to their needs. SedFluxer can be used either with or without the use of SedFitter (see \S\ref{sect:sedfitter}).

\subsubsection{Optimal aperture algorithm}\label{sect:opt_rad}

With the release of \verb+sedcreator+, we have also developed a new algorithm to choose the aperture size in an unbiased and reproducible way for extended sources. The algorithm selects the optimal aperture radius for a given source through the following process. First, it samples a range of aperture radii within a user-defined lower and upper boundary and with a given step size. Next, it calculates the background-subtracted flux enclosed by each sampled aperture. Starting from the innermost scale, the algorithm searches for the condition when a 30\% increase in the radius results in a smaller than 10\% increase in the background subtracted flux. The first time this condition is met defines the radius of the optimal aperture. However, if the condition is not met in the search range, then the radius is set to the location where the smallest fractional increase in background-subtracted enclosed flux occurs. To assess the performance of this algorithm, we made the experiment of retrieving the optimal aperture for an artificial 2D Gaussian profile with an idealized background set at zero counts. For this case, an aperture is set at a radius of $\sim2.2\times$\,FWHM, which retrieves $\sim90\%$ of the total flux.

\begin{figure}[!htb]
\includegraphics[width=0.5\textwidth]{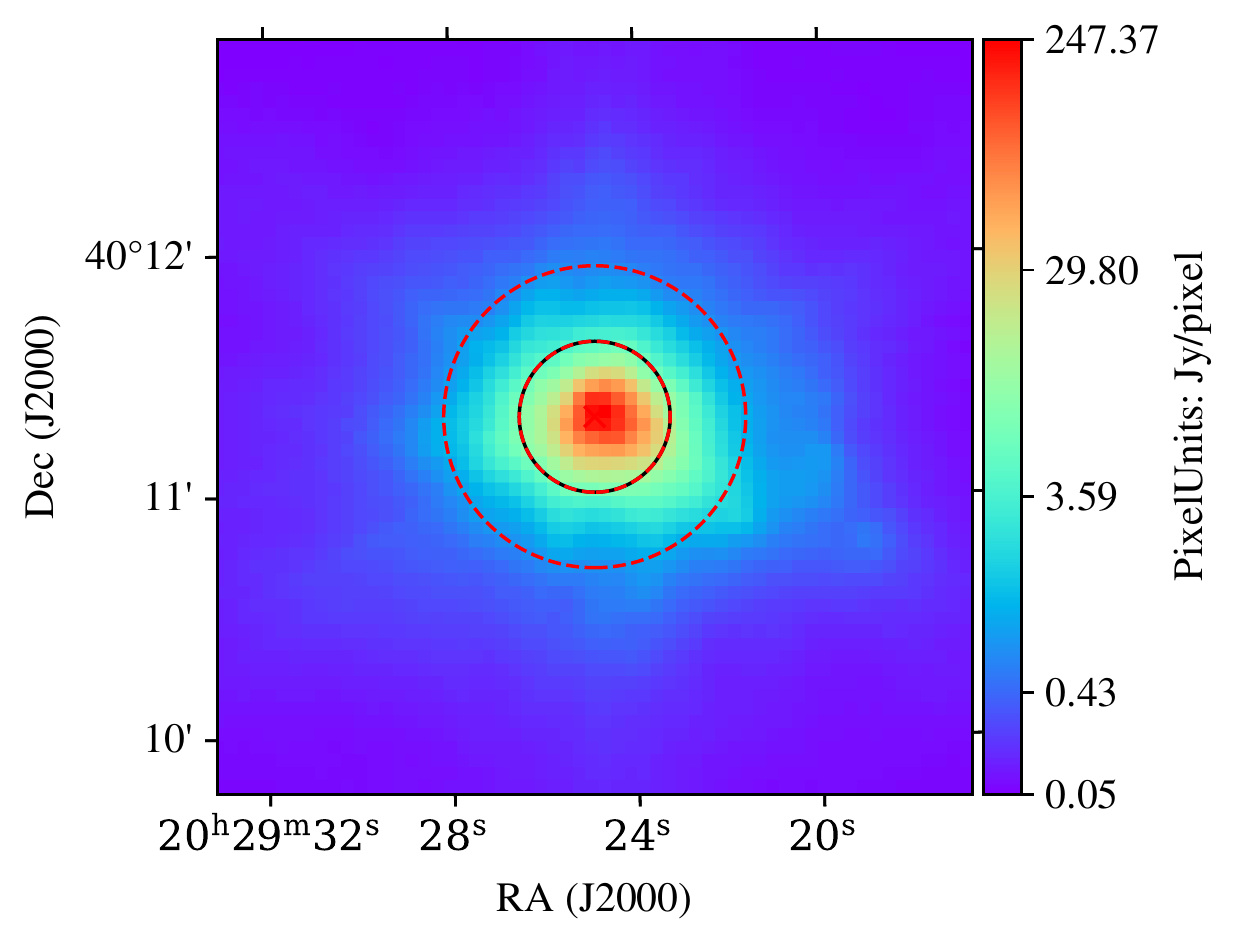}
\includegraphics[width=0.5\textwidth]{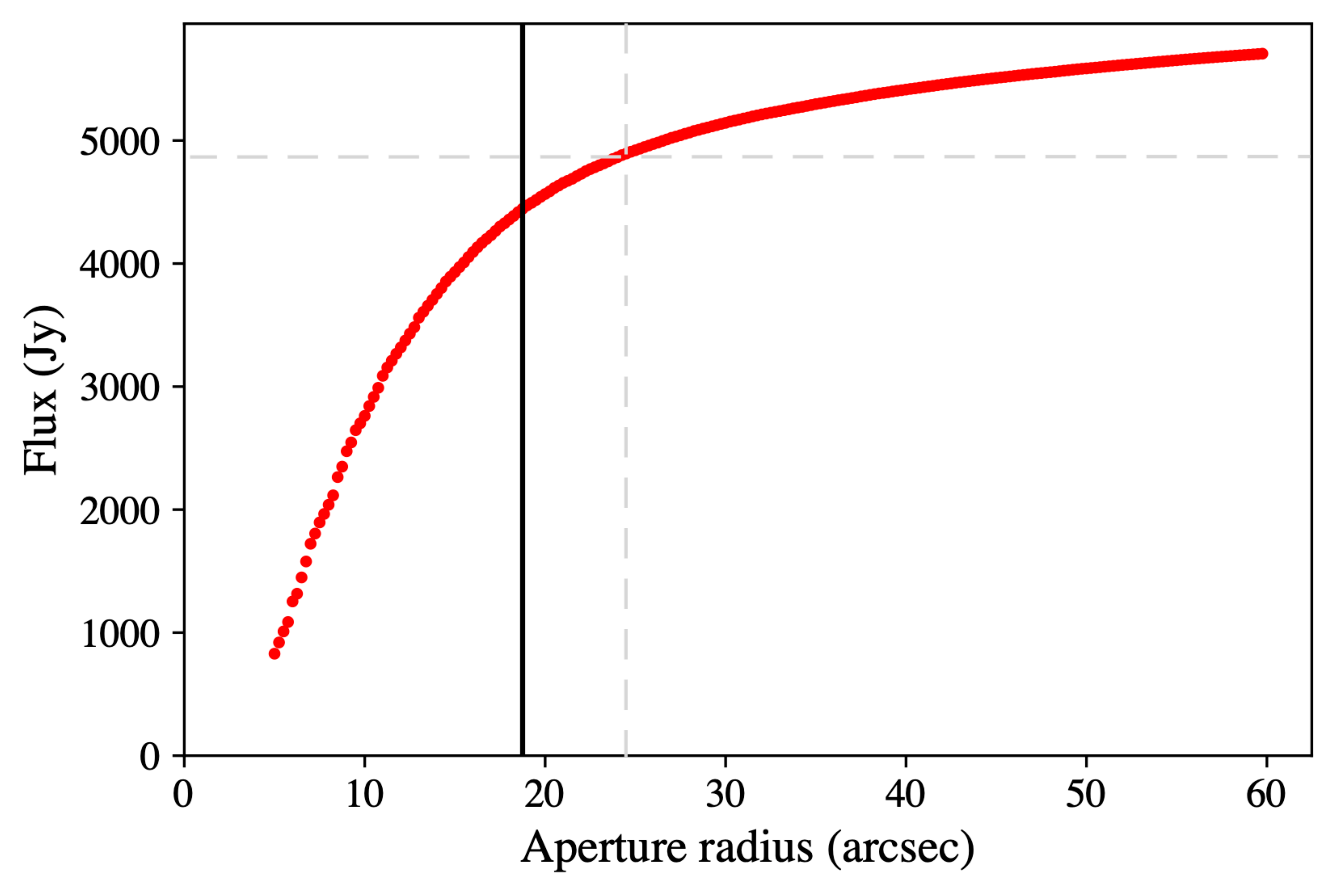}
\caption{Demonstration of the optimal aperture algorithm for the AFGL~2591 source.
\textit{Top panel:} Herschel $70\,\mu$m image of AFGL~2591 with logarithmic intensity stretch. The inner circle is the optimal aperture chosen by the algorithm and the region from the inner to outer circle defines the annulus over which the background emission is estimated.
\textit{Bottom panel:} Background-subtracted flux versus aperture radius. The vertical black line shows the radius of the optimal aperture,  while the horizontal and vertical gray dashed lines indicate the 10\% increase in flux and 30\% increase in radius, respectively, with respect to the optimal aperture for reference (see the text). \label{fig:opt_aper}}
\end{figure}

Figure\,\ref{fig:opt_aper} shows an example for the source AFGL~2591 The top panel shows the $70\,\mu$m image with the optimal aperture represented by the inner circle and the annular region used for background estimation for this aperture being from the inner to outer circles. The bottom panel shows the dependence of background-subtracted flux on the aperture radius, with the vertical black line indicating the radius of the optimal aperture. For all the sources analyzed in this paper, we have used the optimal aperture calculated using the Herschel $70\,\mu$m image, or if this was not available, then the SOFIA $37\,\mu$m image.

\subsubsection{Error estimation}

In previous SOMA papers, we have assumed as the error estimator for the fluxes, the background measured in the annulus region (see \S\ref{sect:sedfluxer}). In this paper, we revise the method for estimating the errors for those wavelengths that are not affected (or negligibly affected) by cold clump contamination, i.e., $\lambda<100\,\mu$m. For longer wavelengths ($\lambda\geq100\mu$m), we still use the background error as the emission at Herschel bands, especially those at $160,250,350,$ and $500\mu$m. are contaminated by the emission of the cold clump.

For the new error estimator, the fluctuations on the flux in a region from the annulus equivalent to the main aperture is evaluated. Therefore, the annulus region is divided into three sectors with areas equal to the circle of the main aperture with radius $r$. Recall that the fiducial case for the annulus definition is to take $r_\mathrm{inner}=r$ and $r_\mathrm{outer}=2r$. Therefore, the area of each of the three sectors ($\pi(r_\mathrm{outer}^2-r_\mathrm{inner}^2)/3=\pi r^2$) is equal to the area of the main aperture ($\pi r^2$). To simplify, each sector is approximated by four circles with radius $r/2$ (which sum an area of $4\pi(r/2)^2=\pi r^2$). We then estimate the fluctuation of the three sectors by calculating the standard deviation in their measurements. In order to avoid bias and missing regions within the annulus, the three sectors are aliased 6 times, from 0 to 75 degrees in steps of 15 degrees, to cover the full annulus. For each aliased position, the standard deviation of the measurement of the three sectors is calculated. Finally, the mean value of the six standard deviations is calculated and considered as the fluctuation error.

For the final error considered in the fits, a systematic error of $10\%$ of the background-subtracted flux is added in quadrature to each error estimator, i.e. fluctuation error for $\lambda<100\,\mu$m and background error for $\lambda\geq100\,\mu$m.

\subsection{SedFitter}\label{sect:sedfitter}

SedFitter is an update of the IDL code that fits a given SED with the RT model grid of \citet[][]{zhang2018}, see \S\ref{sect:ZT_model_grid} for more details about the model grid. The main changes in the code include: (i) the method of convolution of instrument filter profiles and foreground extinction and (ii) the methods of fitting models to data. Regarding (i), in the IDL code unextincted model SEDs were convolved by the filter responses to obtain model fluxes of each band, which were then extincted with a given level of foreground extinction evaluated at a reference wavelength for each filter. We have updated this aspect to first apply foreground extinction to the models, i.e., with a finite grid of foreground extinction values, and then convolve these SEDs with the filter response functions. It is worth noting that the model flux computed with these two methods does not change dramatically, but this still can introduce modest differences in the properties of the best-fitted physical model results. 

Regarding (ii), the previous IDL version did a grid search for every physical model (i.e., 8640 models including different inclinations of viewing angle). For every value of the visual extinction that was considered in the array, the $\chi^2$ was calculated and the best one kept. Finally, all models were ordered based on this $\chi^2$ value to yield a final result of 8640 models, each with their best visual extinction. To improve efficiency, for each of the 8640 models, we now minimize the $\chi^2$ function \citep[see Equation 1 in][]{debuizer2017} over $A_V$ using the python package scipy.optimize (in particular the routine minimize). We now calculate the $\chi^2$ function with fluxes and errors in linear space. This allows us to better constrain the model grid as the bad models will have a very large $\chi^2$ value. The new version includes an ``\verb+idl+'' method for backward compatibility and to be able to reproduce the results of Papers I-III using the exact method of the older IDL code.

\subsubsection{Zhang and Tan (ZT) Model Grid}\label{sect:ZT_model_grid}

Here, we provide basic information on the RT model from the Zhang and Tan (ZT) series, but more details can be found in Section 3.2 in \citet{debuizer2017}, \citet{liu2019} and \citet{zhang2018}.

Based on the turbulent core model from \citet{mckee2003}, the evolution of high- and intermediate-mass protostars has been developed in a series of papers \citep{zhang2011,zhang2013model,zhang2014,zhang2018}. For massive star formation, the initial conditions are pressurized, dense, massive cores embedded in high-mass surface density clump environments. Massive stars are assumed to form from preassembled massive prestellar cores supported by internal pressure from a combination of magnetic fields and turbulence. In these models, there are two main parameters that set the initial conditions: the initial mass of the core ($M_c$) and the mean mass surface density of the clump environment ($\Sigma_\mathrm{cl}$). A third parameter is the protostellar mass ($m_*$), which defines the location along an evolutionary track from a given initial condition. The properties of protostellar cores, including the protostar, disk, infall envelope, outflow, and their evolution, are calculated self-consistently from the given initial conditions. There are also two secondary parameters, which are the inclination angle of the line of sight to the outflow axis ($\theta_\mathrm{view}$), and the level of foreground extinction ($A_V$). Thus, there are five parameters ($M_c-\Sigma_\mathrm{cl}-m_*-\theta_\mathrm{view}-A_V$) that determine a protostellar SED, i.e., that are constrained by fitting the model grid to a given observed SED. We note that other properties, such as accretion rate, infall envelope mass, outflow cavity opening angle, and disk size, are prescribed for a given set of values of $M_c-\Sigma_\mathrm{cl}-m_*$.

In the current model grid \citep{zhang2018} the main parameters are sampled as follows: $M_c$ is sampled at 10, 20, 30, 40, 50, 60, 80, 100, 120, 160, 200, 240, 320, 400, and 480\,$M_\odot$ and $\Sigma_\mathrm{cl}$ is sampled at 0.10, 0.316, 1.0, and 3.16\,$\mathrm{g\,cm^{-2}}$, which makes a total of 60 evolutionary tracks. For each track, $m_*$ is sampled at 0.5, 1, 2, 4, 8, 12, 16, 24, 32, 48, 64, 96, 128, and 160\,$M_\odot$ (although not all these masses are feasible from a given initial core, which typically has a formation efficiency of about 50\%). In the end, this yields a total of 432 physical models that have different combinations of $M_c-\Sigma_\mathrm{cl}-m_*$. Then, for each physical model, there are 20 viewing angles sampled uniformly at $\cos\theta_\mathrm{view}=0.975,0.925,\cdots,0.025$, i.e., equally distributed from 1 (face-on) to 0 (edge-on). Therefore, we have $432\times20=8640$ model SEDs. Finally, the visual extinction is constrained from $A_{V,{\rm min}}$ to $A_{V,{\rm max}}$ which are user-defined values (with the default set to $0-1000$\,mag).

\subsubsection{Average ``Good'' models}\label{sect:average_models}

It is known that SED fitting is subject to degeneracies, i.e., a range of protostellar properties can yield SEDs that are consistent with a given observed SED. Thus, rather than only consider the best-fitting model, we also evaluate 
an average of  models, with these defined via set criteria on the reduced $\chi^2$ of the fit. 
We adopt the following method for this averaging. We consider all physical models, including different viewing angles, i.e., 8640 models, and evaluate the value of $\chi^2$ of the best model, i.e., $\chi^2_\mathrm{min}$. If $\chi^2_\mathrm{min}<1$, then we average over all models that satisfy $\chi^2<2$. If $\chi^2_\mathrm{min}>1$, then we average over all models that satisfy $\chi^2<2\chi^2_\mathrm{min}$. As our fiducial method, we also require models to satisfy the condition that $R_\mathrm{core}<2 R_\mathrm{ap}$, i.e., we only consider models where the radius of the core is within the chosen aperture radius (within a factor of 2). Note, we first apply the cut in aperture radius to define the best-allowed model and then apply the $\chi^2$ cut. Model properties are averaged in log space, i.e., geometric means, except for $A_V$, $\theta_\mathrm{view}$, and $\theta_\mathrm{w,esc}$ which are averaged in linear space, i.e., arithmetic means.

\subsection{2D Gaussian Fitting on IRAS-HIRES Images}\label{sect:iras}

Since our sources are not resolved in IRAS-HIRES images, we perform two-dimensional Gaussian fitting to derive the source flux, in order to reduce potential contamination from nearby sources. We extract a square area containing the source on the HIRES image for the 2D Gaussian fitting and explore how the choice of fitting area affects the result, i.e., by varying the size of the side of the square from 1.5-2.0 times the IRAS beam major axis.

The resolution in the HIRES image is not the same across an entire image\footnote{https://irsa.ipac.caltech.edu/applications/Hires/docs/instructions.html}. HIRES provides beam sample maps, which consist of a field of point-source spikes superimposed on a highly smoothed version of the image, as well as the corresponding beam FWHM and position angle (P.A.) at each location. We tried two ways to perform the 2D Gaussian fitting: (1) treating all the parameters as free parameters and (2) deriving the major and minor FWHM and the P.A. from the beam profiles provided by HIRES and keeping them fixed, and then fitting for the other parameters of the 2D Gaussian. Most of the time we used the beam profile located at the center of the map, where the source is located. But occasionally the beam profiles are significantly impacted by source strength, and in these cases, we find the nearest beam profile that is not contaminated to make sure the beam profile is consistent with the appearance of the unresolved source in the image. We note that this method of Gaussian fitting also includes a constant background term, so the derived flux can be considered to be background subtracted.

\section{Results}\label{sect:results}

\subsection{Individual sources} \label{sect:sources}

Below we describe the main properties of each source, along with a description of the new SOFIA-FORCAST images and relevant ancillary data.

\subsubsection{AFGL 2591}

\begin{figure*}[!htb]
\includegraphics[width=1.0\textwidth]{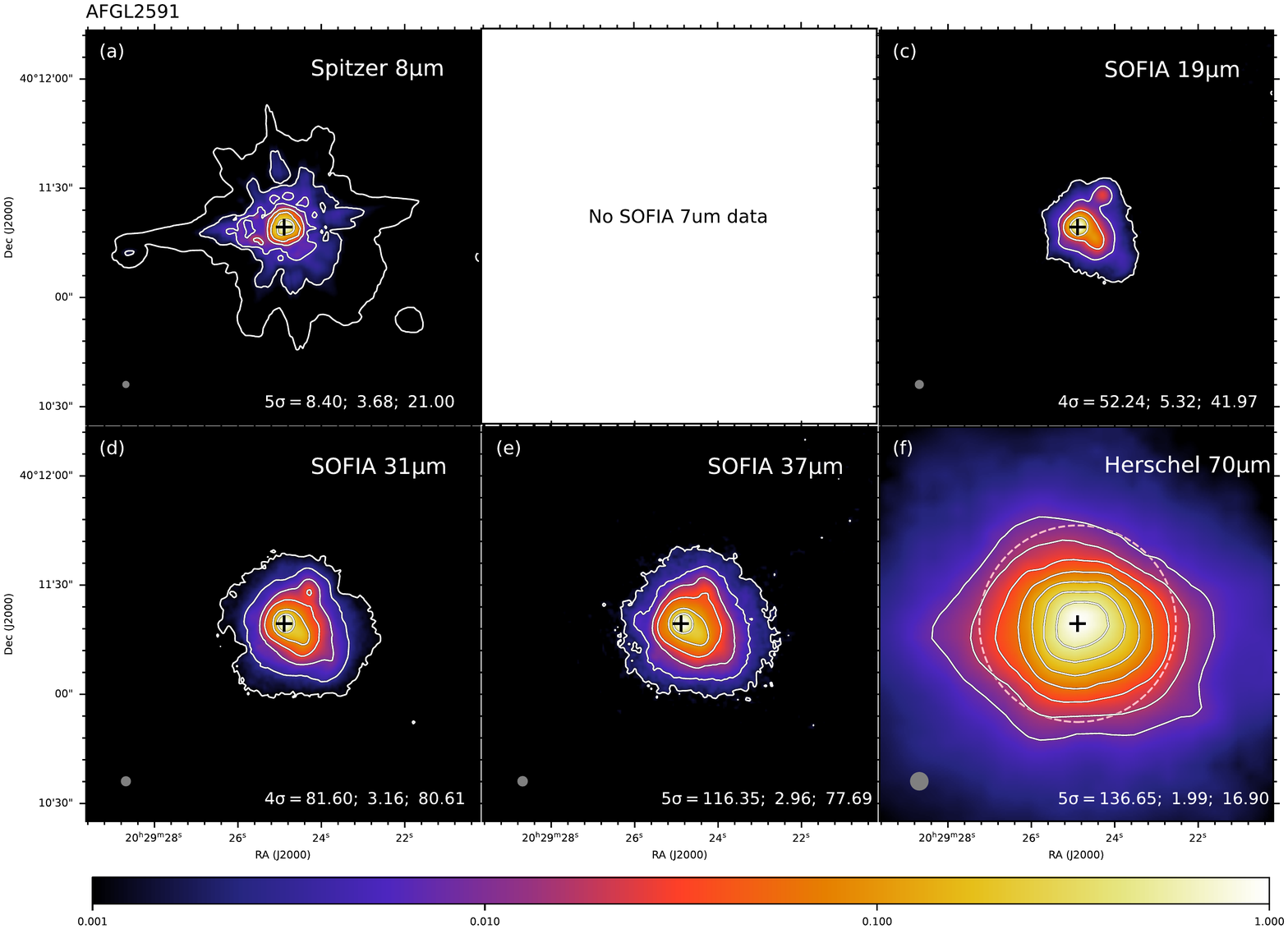}
\caption{Multiwavelength images of AFGL~2591 with the facility and wavelength given in the upper right of each panel. Contour level information is given in lower right: lowest contour level in the number of sigma above the background noise and corresponding value in mJy per square arcsec; then the step size between each contour in $\log_{10}$ mJy per square arcsec, then peak flux in Jy per square arcsec. The color map indicates the relative flux intensity compared to that of the peak flux in each image panel. The pink dashed circle shown in (f) denotes the aperture used for the fiducial photometry. Gray circles in the lower left show the resolution of each image. The black cross in all panels denotes the position of the 3.6\,cm radio source VLA3 in \citet{trinidad2003} at  R.A.(J2000) = $20^h29^m24\fs8916$, Decl.(J2000) =  $+40\arcdeg11\arcmin19\farcs388$. The data used to create this figure and the SED are available.\label{fig:AFGL2591}}
\end{figure*}

AFGL~2591, also known as IRAS~20275+4001, was first reported in the literature by \citet{rieke1973}, although they stated that it was discovered in a survey by Walker \& Prince in 1972. It was then observed in the IR from 2.8-14\,$\mu$m by \citet{merrill1974}, where they discussed the remarkable similarity of AFGL~2591 with the BN object in Orion. \citet{wynn1977} reported radio and IR observations to confirm that the IR source and the ultracompact (UC) HII region, initially thought to coincide, are separated by $7^{\prime\prime}$. They also observed an H$_2$O maser coinciding with the IR source. This maser was later used to determine a distance of $3.33\pm0.11$\,kpc from its trigonometric parallax with the Very Long Baseline Array \citep{rygl2012}. \citet{simon1981} observed this region with the VLA at 6 and 2~cm. They confirmed the findings of \citet{wynn1977} and tentatively associated a weak radio source with the IR source. Subsequently, many radio sources have been detected in this star-forming region. VLA 1, 2, 4, and 5 have been classified as HII regions \citep{trinidad2003,johnston2013}, whereas VLA 3 has been identified as the main high-mass protostar \citep{trinidad2003,gieser2019}.
\citet{indriolo2015} used SOFIA/EXES to observe the $\nu_2$ rovibrational band of water in the MIR $6.086-6.135\,\mu$m range and suggested that the background source is only partially covered by the absorbing gas or that the absorption arises within the $6\,\mu$m emitting photosphere. AFGL~2591 was also observed as part of the CORE survey at 1.37mm with NOEMA where the authors concluded that this source contains three cores that have a mean separation of about 15\,000\,au \citep{beuther2018b}.

AFGL~2591 drives a powerful molecular outflow as revealed by CO and HCO$^+$ observations \citep{bally1983,lada1984,hasegawa1995}. This outflow is oriented toward the east-west direction with an approximate P.A. of $260^\circ{}$, an extent of $\sim1.5$\,pc, and a dynamical age of $\sim2\times10^4$\,yr \citep{poetzel1992,preibisch2003}. The presence of the outflow is hinted at in Figure\,\ref{fig:AFGL2591} as all SOFIA images show some extension in the east-west direction. The NIR K band image at $\sim2\,\mu$m shows a well-defined conical structure delineating the outflow cavity walls \citep[e.g.,][]{hodapp1994}. \citet{poetzel1992} found several Herbig-Haro (HH) objects toward the main IR source. They found [N II] and [S II] emission in the optical and H$_2$ in the NIR and measured radial velocities in the range of $200-500\,\mathrm{km\,s^{-1}}$. The blue-shifted lobe of the outflow is located toward the west and the red-shifted lobe is toward the east \citep[e.g.,][]{poetzel1992,preibisch2003,olguin2020}. \citet{hasegawa1995} estimated an opening angle of $<90^\circ{}$ and an inclination angle to the line of sight of $<45^\circ{}$ from CO emission. The inclination angle was later constrained to a range between $26^\circ{}$ and $38^\circ{}$ \citep{vandertak2006}, although other values have been found by fitting the SED \citep[e.g.,][]{johnston2013,simpson2013}. See \citet{olguin2020} for a comprehensive and extensive review of this object. On the large scale, multiple studies have argued that the HII region is part of a larger star cluster that is revealed in NIR imaging, but in the MIR images, this source appears to be isolated.

Figure\,\ref{fig:AFGL2591} shows the multiwavelength images of AFGL 2591. The peak emission at IR wavelengths coincides with the position of the VLA3 source at 3.6\,cm \citep{trinidad2003}. In the SOFIA images, there is an elongation in the SW-NE direction, which suggests the presence of an outflow consistent with previous studies \citep{poetzel1992,preibisch2003}. In fact, this elongation is brighter to the west of the centimeter peak, further suggesting that the blue-shifted outflow is located toward this direction and the red-shifted outflow toward the east \citep[see also Figure\,5 of][]{preibisch2003}.

\subsubsection{G25.40-0.14}

\begin{figure*}[!htb]
\includegraphics[width=1.0\textwidth]{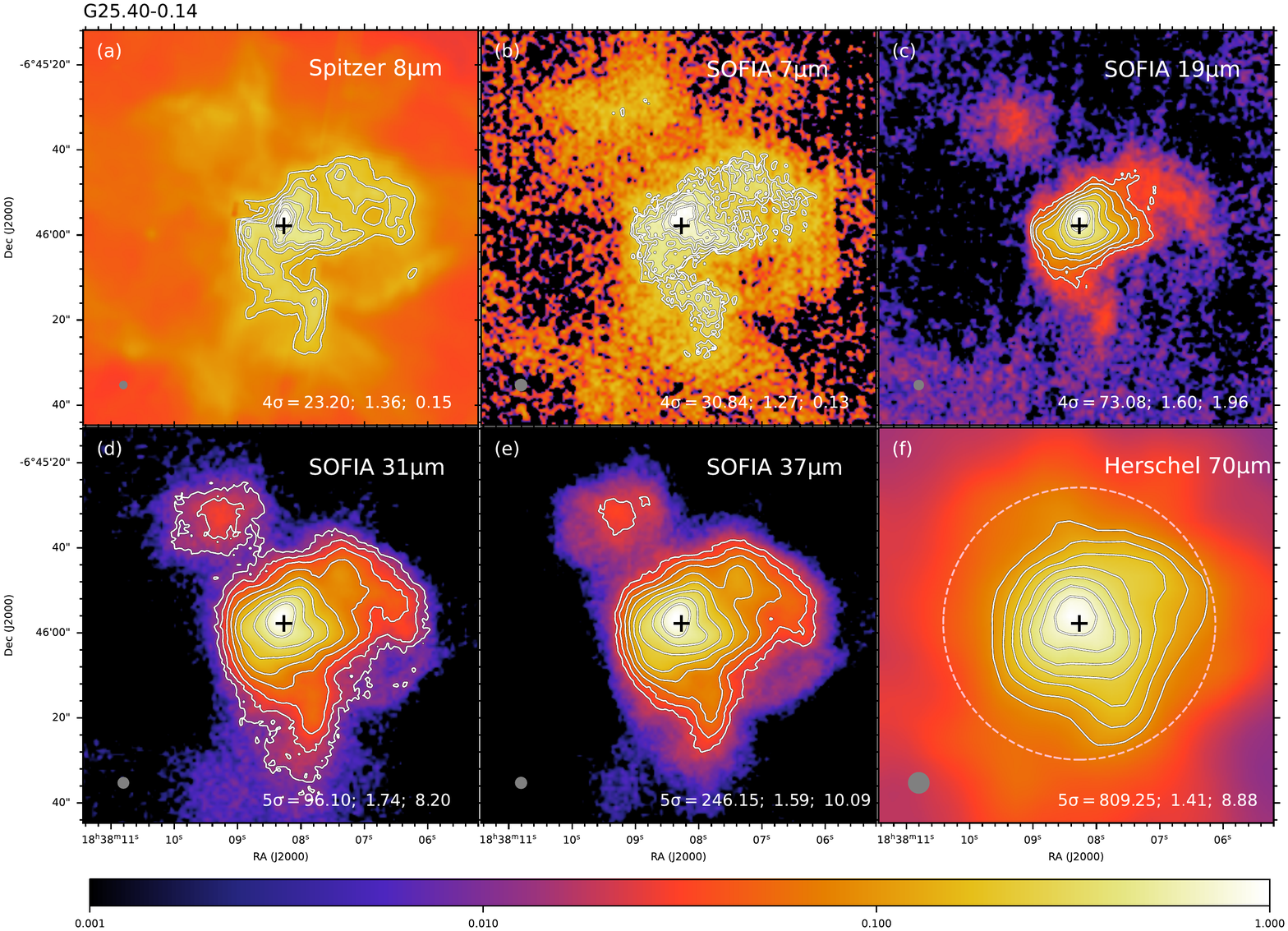}
\caption{Multiwavelength images of G25.40-0.14, following the format of Figure\,\ref{fig:AFGL2591}. The black cross in all panels denotes the peak position of the 6\,cm continuum emission from \citet{giveon2005} at R.A.(J2000) = 18$^h$38$^m$08$\fs$270, Decl.(J2000) = $-$06$\arcdeg$45$\arcmin$57$\farcs$82. The data used to create this figure and the SED are available.\label{fig:G25.40}}
\end{figure*}

Located at $5.7\,$kpc \citep{zhu2011,ai2013}, G25.40-0.14 is a UC HII region with a core halo structure \citep{garay1993}, also referred as G25.4NW in \citet{dewangan2015b}. Even though G25.40-0.14 lies in the direction of W42, it is not thought to be associated with the W42 complex due to its very different $^{13}$CO velocity components ($58–69\,\mathrm{km\,s^{-1}}$ for W42 and $88–109\,\mathrm{km\,s^{-1}}$ for G25.40-0.14) as discussed by \citet{ai2013} and \citet{dewangan2015b}. \citet{ai2013} estimated a bolometric luminosity of $10^{5.6}\,L_\odot$ for this region, which corresponds to an 06 zero-age main-sequence star (ZAMS). The source also sits in the direction of the bipolar nebula N39 \citep{churchwell2006,beaumont2010,deharveng2010}. \citet{deharveng2010} found a filament or sheet-like structure along the bipolar nebula, which is also seen in the $8\mu$m image.

The SOFIA images show clearly the core halo structure (Figure\,\ref{fig:G25.40}). The brightest IR emission is located at the centimeter peak \citep{giveon2005} and an arc-like structure is seen toward the south. An extension to the NW is also seen, as well as a potential secondary source to the NE.

\subsubsection{G30.59-0.04}

\begin{figure*}[!htb]
\includegraphics[width=1.0\textwidth]{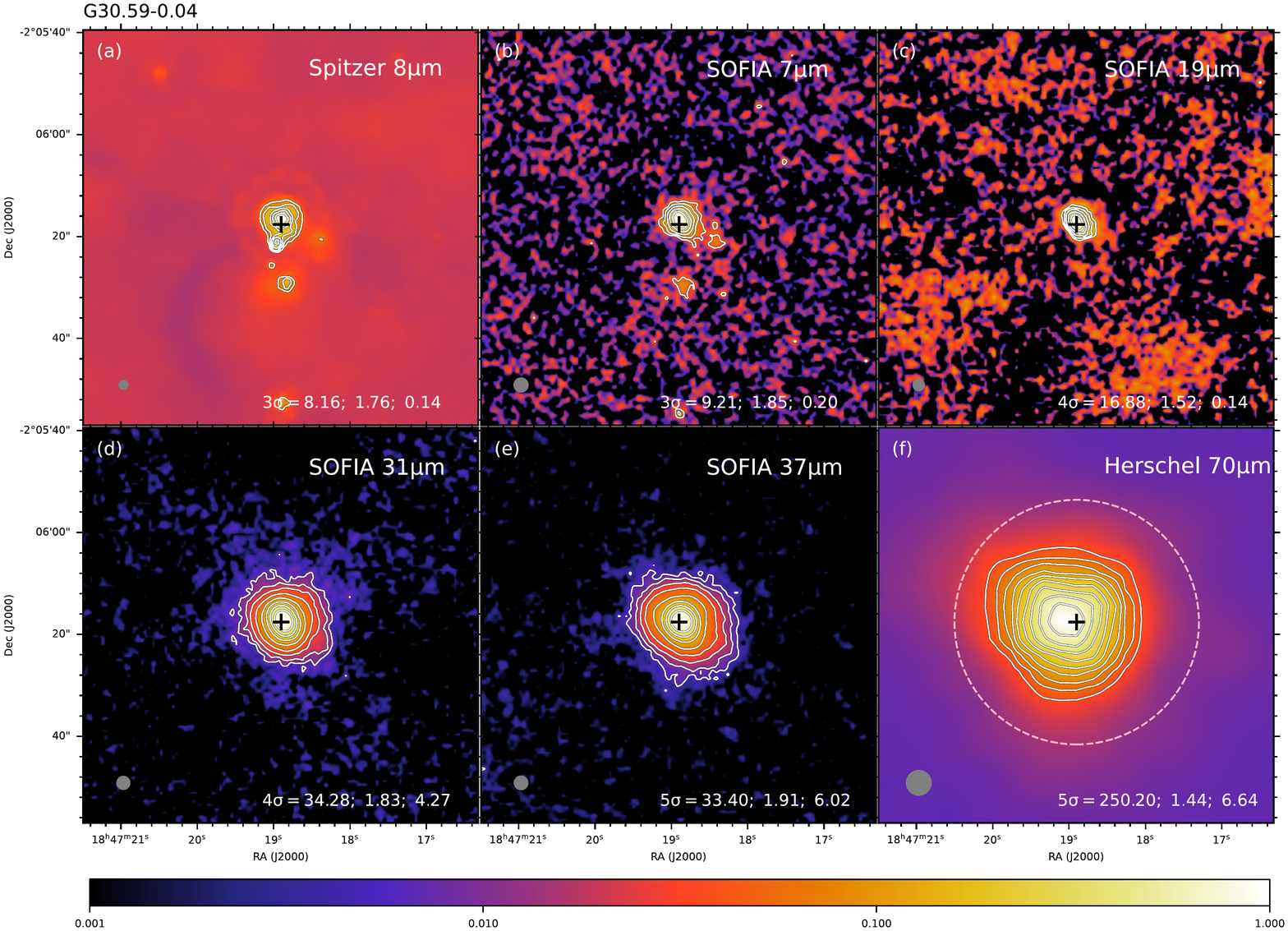}
\caption{Multiwavelength images of G30.59-0.04, following the format of Figure\,\ref{fig:AFGL2591}. The black cross in all panels denotes the peak position of the 6\,cm continuum emission from \citet{giveon2008} at R.A.(J2000) = $18^h47^m18\fs9$, Decl.(J2000) = $-02\arcdeg06\arcmin17\farcs6$. The data used to create this figure and the SED are available.\label{fig:G30.59}}
\end{figure*}

G30.59-0.04 was first identified in a VLA survey by \citet{fish2003}. Some studies \citep[e.g.,][]{hill2009} adopt a near 3~kpc distance, while others \citep[e.g.,][]{fish2003} adopt a far distance of 11.8 kpc. The studies using the near distance adopted such a distance for consistency among samples \citep[e.g.,][]{hill2009}. In the same study, the authors checked and argued that the near-distance assumption has no strong influence on the conclusion in their study. On the other hand, studies such as that of \citep{purcell2006}, which attempted to resolve the ambiguity, favor the far distance of $11.8~$kpc using the technique presented in \citet{solomon1987}, see also Table 2 in \citet{purcell2006} assuming the Galactic rotation curve of \citet{brand1993}. Recently, the source distance was reevaluated by \citet{mege2021} as part of the Hi-GAL survey. They found a distance of $11.7$~kpc, consistent with that found by \citet{urquhart2018} in the ATLASGAL survey ($11.5$\,kpc). Hence, we also adopt the far distance in this work. \citet{hill2009} performed SED analysis and estimated that the clump has upper limits of mass and luminosity of $1200\:M_{\odot}$ and $2.4\times10^4\:L_{\odot}$, respectively. We recomputed the luminosity given by \citet{hill2009} using the far distance as assumed in our work and obtained a value of $3.5\times10^5\,L_\odot$.

Figure\,\ref{fig:G30.59} shows multiwavelength images of G30.59-0.04, where one main source is identified. In the SOFIA $7\,\mu$m image there are two tentative detections of secondary sources toward the south and SW of the centimeter peak, which are also seen in the Spitzer $8\,\mu$m image. There is a hint of elongation in the NE to the SW direction, especially clear in the SOFIA 19 and $37\,\mu$m images.

\subsubsection{G32.03+0.05}

\begin{figure*}[!htb]
\includegraphics[width=1.0\textwidth]{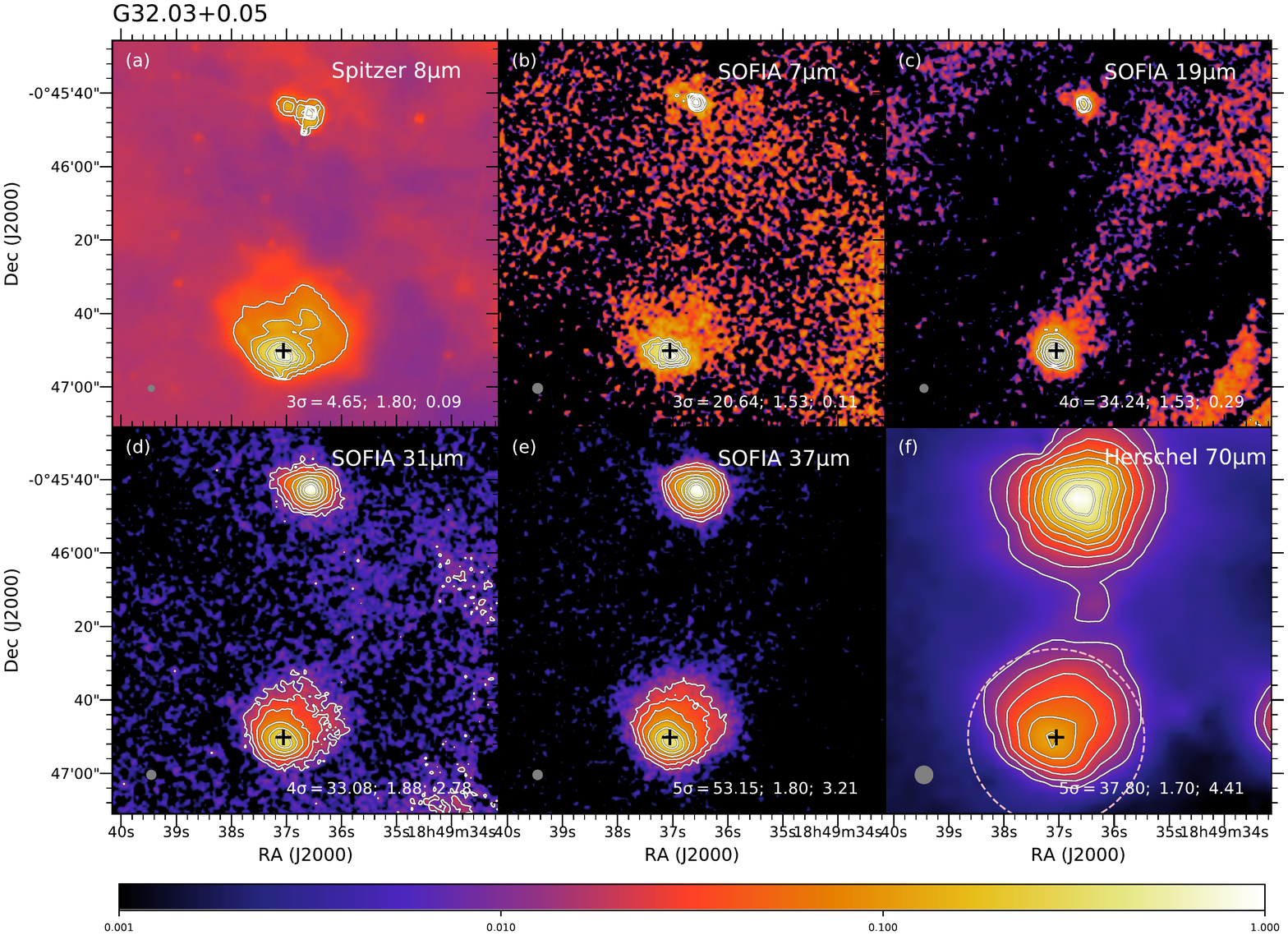}
\caption{Multiwavelength images of G32.03+0.05, following the format of Figure\,\ref{fig:AFGL2591}. The black cross in all panels denotes the peak position of the 6\,cm continuum emission from \citet{white2005} at R.A.(J2000) = 18$^h$49$^m$37$\fs$052, Decl.(J2000) = $-$00$\arcdeg$46$\arcmin$50$\farcs$15. The secondary source towards the north, denoted G32.03+0.05N, has central coordiantes R.A.(J2000) = 18$^h$49$^m$36$\fs$55, Decl.(J2000) = $-$00$\arcdeg$45$\arcmin$42$\farcs$40. The data used to create this figure and the SED are available.\label{fig:G32.03}}
\end{figure*}

G32.03+0.05 was first reported as part of the 5\,GHz VLA Survey of the Galactic plane carried out by \citet{becker1994}. The authors associated the source at 6\,cm with IRAS~18470-0050 as their radio and the IR coordinates differed by less than $2^{\prime\prime}$. Later, this source was also part of the deeper VLA survey by \citet{white2005}, where they confirmed the detection at 6\,cm of \citet{becker1994}, but reported non-detection at 20\,cm. The coordinates reported in Figure\,\ref{fig:G32.03} are the ones from \citet{white2005}. The near (5.5\,kpc) and far (8.7\,kpc) kinematic distances were reported for this region by \citet{anderson2012}. In this work, we assume the near distance, following \citet{battersby2014}.

\citet{battersby2014} observed the IRDC\,G32.02+0.06 region in NH$_3$(1,1), (2,2), and (4,4) with the VLA. They identified two main clumps, which they denoted active and quiescent. Our source is located within the active clump, which they find has a mass of $\sim5,000-10,000\,M_\odot$ and displays clear signs of star formation. This includes the presence of 6.7\,GHz methanol maser \citep{pestalozzi2005}, emission at 8 and 24\,$\mu$m, and radio continuum emission \citep{white2005}. \citet{battersby2014} identified our main source (which corresponds to the bright source toward the south in Figure\,\ref{fig:G32.03}) as a young UC HII region and no NH$_3$ cores were observed here. 

The source toward the north shown in our Figure\,\ref{fig:G32.03}, is identified in \citet{battersby2014} as a warm core complex and includes three NH$_3$ cores (cores 2, 6 and 8), which all show clear signatures of NH$_3$(1,1), (2,2) and (4,4) emission. From these data, core masses of $\sim100\,M_\odot$ and temperatures of 25\,K were estimated \citep[see Table 2 of][for the properties derived]{battersby2014}.

Figure\,\ref{fig:G32.03} shows the multiwavelength images of G32.03+0.05. In the IR images, two sources are clearly visible, consistent with previous studies. We treat the southern source that coincides with the centimeter peak as the primary. We carried out an SED analysis for both sources. So this region yields two protostars for our sample.

\subsubsection{G33.92+0.11}

\begin{figure*}[!htb]
\includegraphics[width=1.0\textwidth]{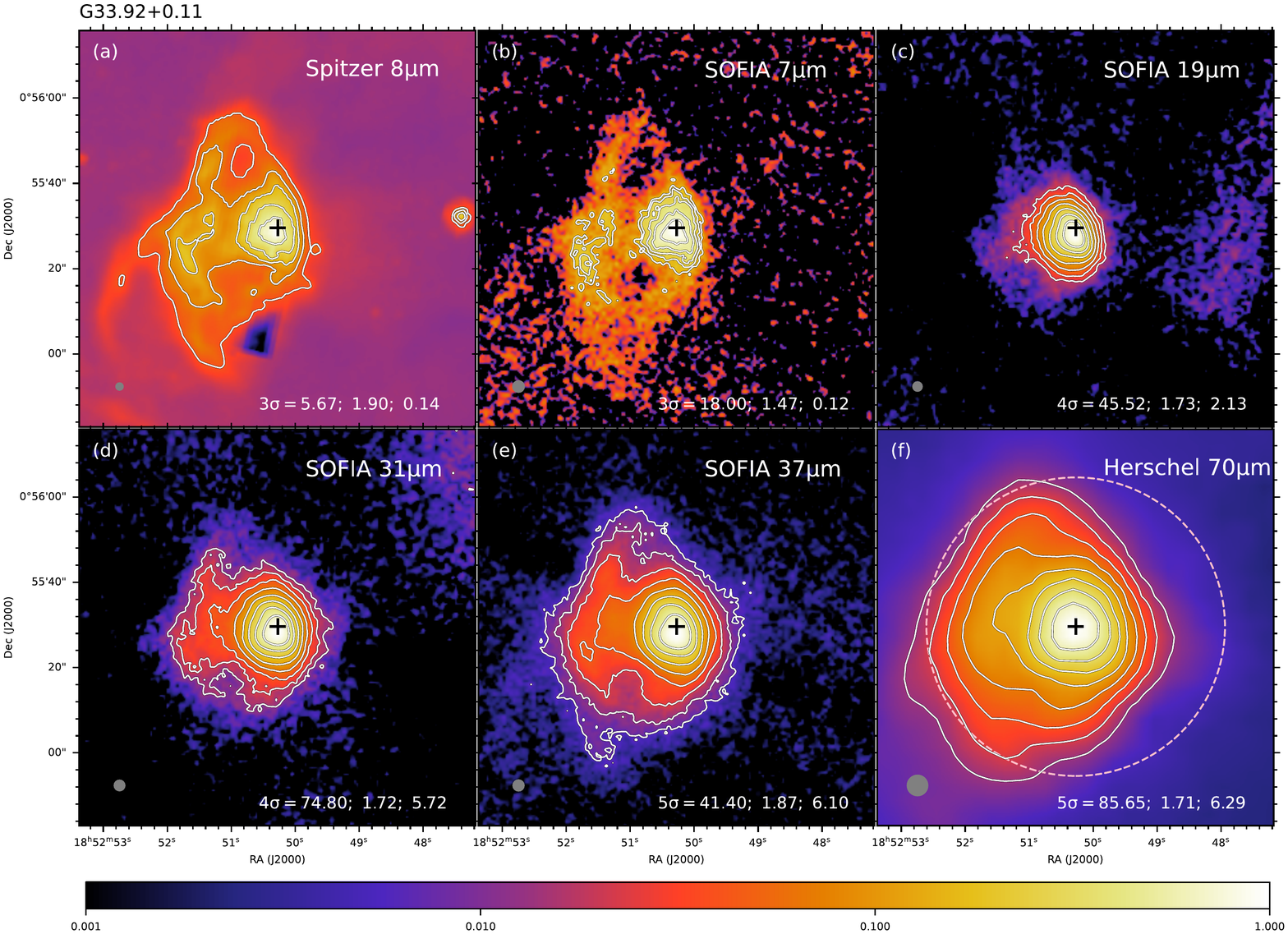}
\caption{Multiwavelength images of G33.92+0.11, following the format of Figure\,\ref{fig:AFGL2591}. The black cross in all panels denotes the peak position of the 2.7\,mm, 2\,cm, and 3.6\,cm continuum emission from \citet{watt1999} at R.A.(J2000) = $18^h52^m50\fs273$, Decl.(J2000) = $+00\arcdeg55\arcmin29\farcs594$. The data used to create this figure and the SED are available.\label{fig:G33.92}}
\end{figure*}

G33.92+0.11 has been classified as a cometary UC HII region, with an estimated distance of $7.1 \pm 1.3$\,kpc and a systemic velocity of $107.6\,\mathrm{km\,s^{-1}}$ \citep{fish2003,liu2015}. \citet{liu2012} described the dense gas in the region as having a hub-filament structure, with multiple parsec-long and spiral-like filaments converging to the central massive ($\sim 3\times 10^{3}\:M_{\odot}$) hub \citep{liu2015}. Atacama Large Millimeter/submillimeter Array observations were performed toward the inner region of the filaments \citep{liuHB2019}. The authors argued that the filaments are feeding a massive OB protocluster by gravitationally driven inflow and resolved the inner structures down to 1000\,au scales. Five outflow sources were identified in the study by Liu et al. (2012) with $^{12}$CO(2-1) and shock tracers such as SiO(5-4), SO, and OCS. The authors identified high-velocity molecular gas toward the five regions as high as $31.4~$km~s$^{-1}$. Assuming an excitation temperature of $50~$K and X(CO) = $10^{-4}$, the authors estimated the outflow mass, momentum, and energy of the outflows. The upper limits are $2.3\:M_{\odot}$, $12\:M_{\odot}\:{\rm km\:s}^{-1}$ and $20\times 10^{44}\:$erg, respectively.
The source also presented high chemical complexity with detection of various hot core lines \citep{minh2016,minh2018}.

The multiwavelength images of G33.92+0.11 are shown in Figure\,\ref{fig:G33.92}. The SOFIA $7\,\mu$m morphology resembles that seen in the Spitzer $8\,\mu$m image and shows the arc-like structure towards the east of the centimeter peak. Longer wavelengths also show stronger emission to the east, including the Herschel $70\,\mu$m image. There is a source to the west of the main protostar that is detected in the Spitzer $8\,\mu$m, but which is not seen in the SOFIA $7\,\mu$m or other images. This may indicate that this source is variable, potentially being a protostar that has evolved from an outbursting active state to a quiescent one during the last $\sim 10$~yr. However, the observations for this source were also affected by the so-called \textit{window contamination} problem, which can prevent the observation of faint sources in the field.

\subsubsection{G40.62-0.14}

\begin{figure*}[!htb]
\includegraphics[width=1.0\textwidth]{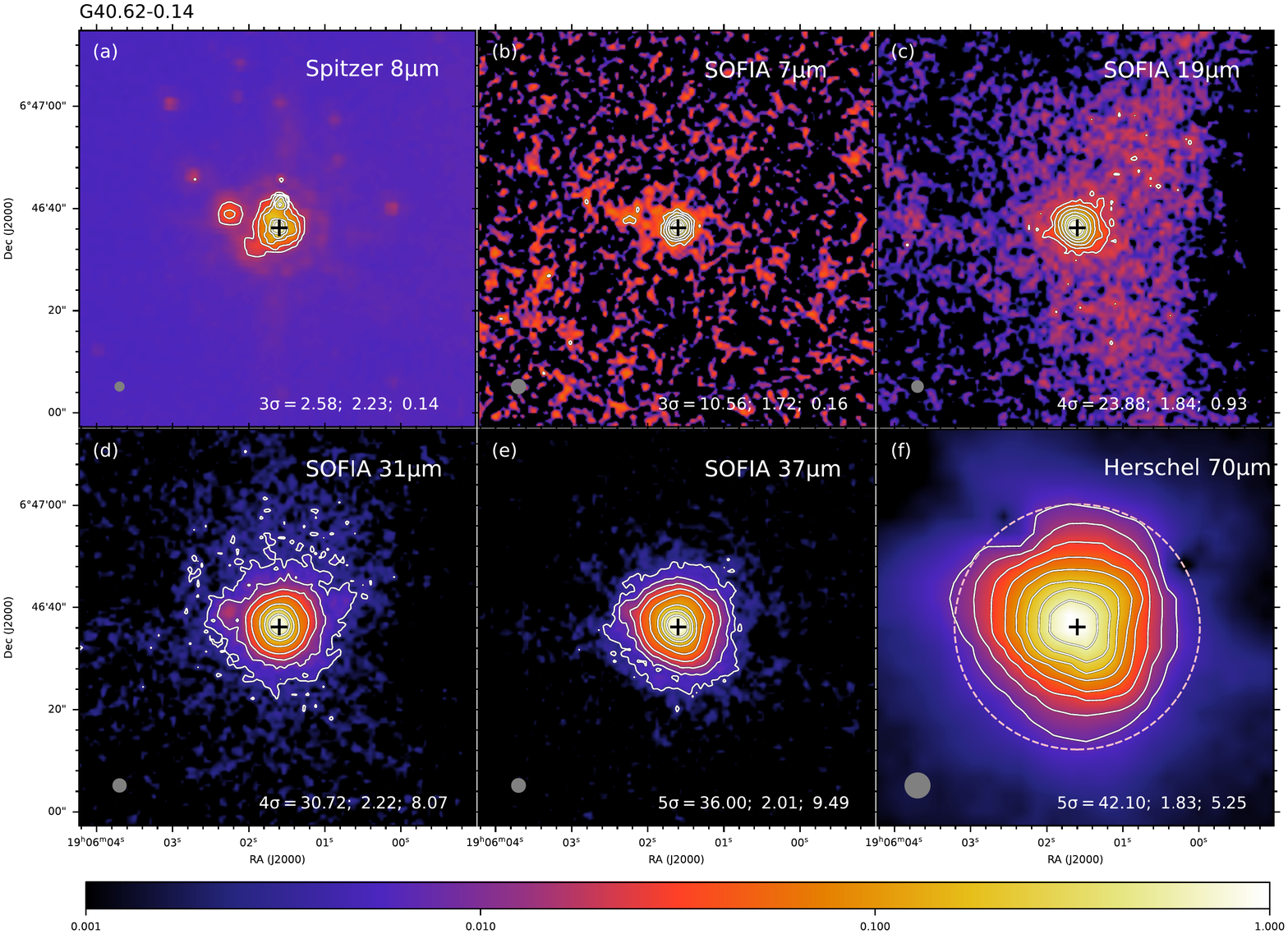}
\caption{Multiwavelength images of G40.62-0.14, following the format of
Figure\,\ref{fig:AFGL2591}. The black cross in all panels denotes the peak position of the 1.3\,cm and 6\,cm continuum emission of I19035-VLA2 in \citet{rosero2016} at R.A.(J2000) =
19$^h$06$^m$01$\fs$60, Decl.(J2000) =
$+$06$\arcdeg$46$\arcmin$36$\farcs$2. The data used to create this figure and the SED are available.\label{fig:G40.62}}
\end{figure*}

G40.62$-$0.14, also known as IRAS~19035+0641, is a massive star formation region located at 2.2 kpc and with a bolometric luminosity of $\sim 8 \times 10^{3}\:L_{\odot}$ \citep{sridharan2002}. Single-dish data toward G40.62$-$0.14 shows the presence of CO and HCO$^{+}$ molecular outflows \citep{beuther2002b,lopez-sepulcre2010} oriented in the NW-SE direction. \citet{sanchez-monge2011thesis} reported two radio continuum emission sources in this region: I19035-VLA1, a small cometary UC HII region and I19035-VLA2, a very faint ($4\sigma$) detection. Furthermore, \citet{rosero2016} confirmed the detection of I19035-VLA2 (their source 19035+0641 A) and reported that the source has a jet-like morphology oriented in the NE-SW direction.  Based on the spectral index, the centimeter emission morphology and associations with outflow tracers (e.g., maser emission and molecular emission), \citet{rosero2019b} concluded that I19035-VLA2 is an ionized jet. Also, they show that this source is associated with NIR emission seen by UKIDSS, which is elongated in the same direction as the ionized jet and is most likely tracing scattered light from the central young stellar object (YSO) that is escaping through an outflow cavity. I19035-VLA2 is reported to have very weak or no H$_{2}$ emission \citep{rosero2019b}. I19035-VLA2 is located at the peak of the main ammonia clump detected toward this region and the ammonia emission has a velocity gradient in the same direction as the molecular outflow detected in the region \citep{sanchez-monge2011}. I19035-VLA2 is associated with 6.7 GHz CH$_{3}$OH, H$_{2}$O \citep{beuther2002c,rosero2019b} and OH maser emission, with the latter two distributed linearly in the same direction as the ionized jet and interpreted by \citet{debuizer2005} as tracing the outflow into the cavity. I19035-VLA2 is spatially associated with MIR emission seen by Spitzer/IRAC \citep{sanchez-monge2011}, likely tracing dust emission from a deeply embedded YSO. I19035-VLA2 is the source that coincides with our MIR peak at $39\,\mu$m. I19035-VLA1 is offset from the peak of the MIR and NIR emission and the millimeter clump. \citet{rosero2019b} suggest that based on its low emission measure and electron density, I19035-VLA1 is a UC HII region ionized by a B1 ZAMS star that has formed near the edge of the dust clump where the density of the surrounding medium is much lower than in the center.  With this picture, G40.62$-$0.14 is composed of I19035-VLA2, a YSO that is deeply embedded in dust and it is the driving source of at least one of the molecular outflows detected in the region, and I19035-VLA1, which is a slightly more evolved stellar source. \citet{rosero2019b} suggest that the observed misalignment between the centimeter continuum emission and the dominating molecular outflow in the region could be explained either by the existence of multiple overlapping outflows or by precession, where the outflow axis changes from the small to the large scale. 

Figure\,\ref{fig:G40.62} presents the multiwavelength images of G40.62-0.14. At wavelengths shorter than 19$\mu$m, the MIR data shows a weak source located slightly NE of the central emission. This source is not associated with any ionized emission or millimeter dust clump. Also, our SOFIA data at 31 and 37$\mu$m and the Herschel 70$\mu$m show a slight NE-SW elongation, which is in the same direction as the NIR emission and the ionized jet associated with I19035-VLA2. This suggests that the MIR/FIR data is also tracing the outflow cavity driven by the YSO in I19035-VLA2.

\subsubsection{IRAS 00259+5625}

\begin{figure*}
\includegraphics[width=1.0\textwidth]{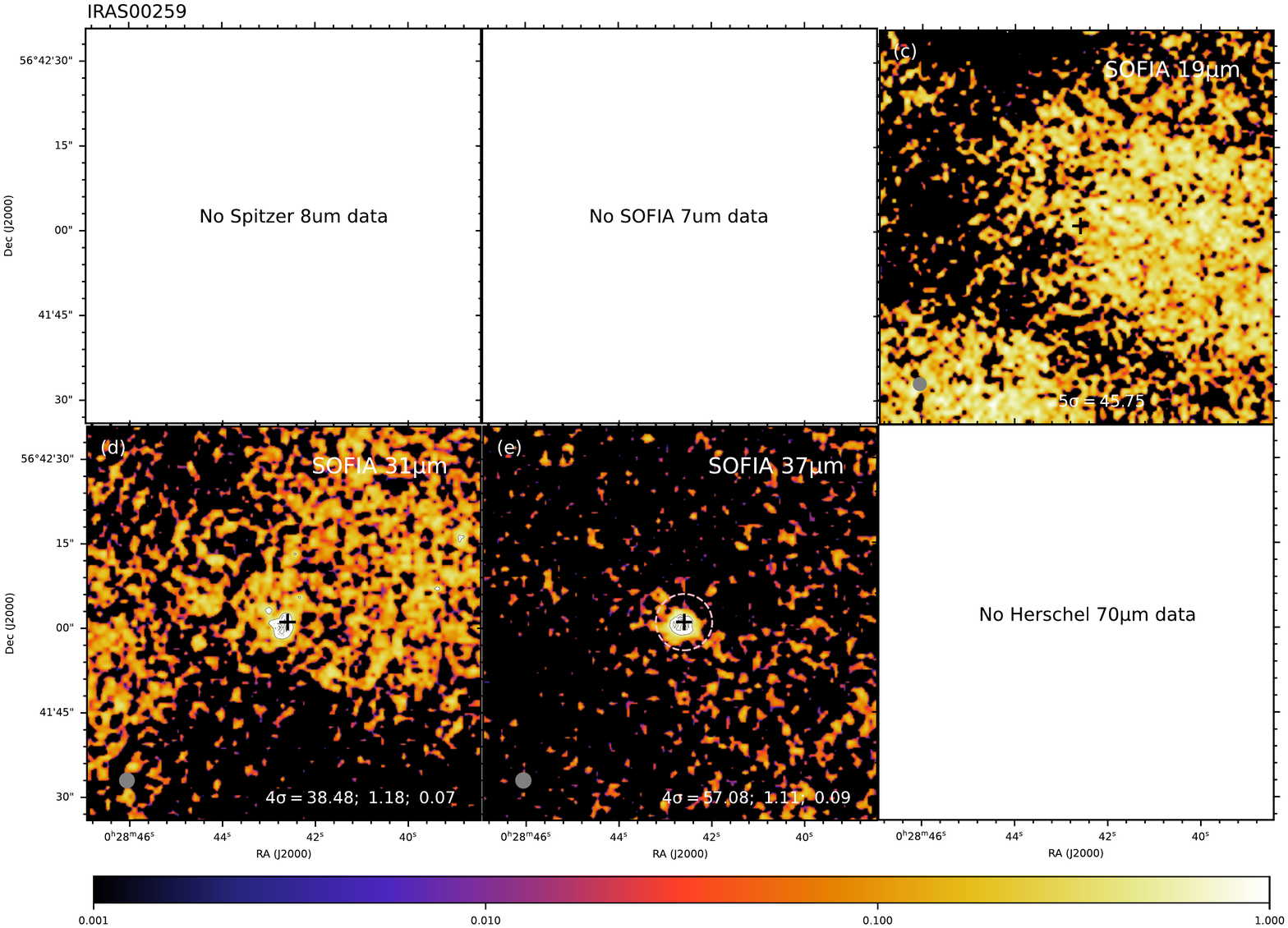}
\caption{Multiwavelength images of IRAS~00259+5625, following the format of Figure\,\ref{fig:AFGL2591}. The black cross in all panels denotes the position of the 1.3\,mm and 3.3\,mm source CB3-1 from \citet{fuente2007} at R.A.(J2000) = 0$^h$28$^m$42$\fs$60, Decl.(J2000) =
$+$56$\arcdeg$42$\arcmin$01$\farcs$11. The data used to create this figure and the SED are available.\label{fig:IRAS00259}}
\end{figure*}

IRAS~00259+5625, also known as CB3 \citep{clemens1988}, is identified as a Bok globule. The large distance (2.5 kpc) and high luminosity make CB3 stand out from typical Bok globules, which are usually low-mass star-forming cores, suggesting that CB3 is an intermediate-mass star-forming region \citep{launhardt1997a}. \citet{launhardt1997b} estimated a kinematic distance of 2.4 kpc (which is the one adopted in this work), while \citet{yun1994} obtained a CO velocity distance of 2.1 kpc using the rotation curve in \citet{clemens1985}. The water maser's proper motion suggests a distance of 2.6$^{+1.0}_{-0.6}$ kpc, consistent with previous estimates \citep{sakai2014}.  \citet{launhardt1997b} compiled the 12--1300 \micron\ flux of the continuum source, CB3-mm, and derived a luminosity of 930 $L_{\odot}$ and a gas mass of 72 $M_{\odot}$.  Using PdBI, \citet{fuente2007} resolved the continuum source into a binary source at 1.3 mm and 3 mm, CB3-1 (00$^h$28$^m$42\fs{6}/56\arcdeg42\arcmin01\farcs{11}) and CB3-2 (00$^h$28$^m$42\fs{2}/56\arcdeg42\arcmin05\farcs{11}), separated by 0.06 pc ($\sim$5\arcsec\ at a distance of 2.5 kpc).  The mass of CB3-1 and CB3-2 derived from the 1.3 mm emission is 0.62 and 0.24 $M_{\odot}$, respectively.  \citet{yun1995} identified a source in \textit{J}, \textit{H}, and \textit{K} bands at $\sim$32\arcsec\ (00$^h$28$^m$46\fs{3}/56\arcdeg41\arcmin38\arcsec) to the east of the millimeter source; however, no corresponding continuum source has been found in submillimeter and millimeter wavelengths.  Clear outflow signatures in the N-S direction have been detected in CO, CS, SO, SO$_2$, SiO, and CH$_3$OH, where the SiO and CH$_3$OH shows collimated jet-like morphology \citep{yun1994,launhardt1998,codella1999}.  In particular, the SiO emission is directly associated with the binary source, showing two episodes of mass loss corresponding to a time scale of 10$^4$ and 10$^5$ years \citep{codella1999}.  At a scale of $\sim$5\arcmin$\times$5\arcmin, \citet{lundquist2014} identified four candidate YSOs using Wide-field Infrared Survey Explorer (WISE) and Two Micron All Sky Survey (2MASS) photometry, including three intermediate-mass YSOs (4.0, 2.8, and 3.0 $M_{\odot}$) and a low-mass YSO (0.8 $M_{\odot}$).  All of this evidence suggests that the IRAS~00259+5625 is a Class 0 intermediate-mass star-forming cluster.

Figure\,\ref{fig:IRAS00259} shows the SOFIA images of IRAS~00259+5625. Neither Spitzer nor Herschel data are available for this region. The SOFIA image at 19~$\rm \mu m$ does not show any clear emission for this source, in part because of extended diffuse emission, which is attributed to a window contamination problem affecting this particular FORCAST observation. The source is detected in the longer-wavelength SOFIA images, which allow a flux measurement for the SED.

\subsubsection{IRAS 00420+5530}

\begin{figure*}[!htb]
\includegraphics[width=1.0\textwidth]{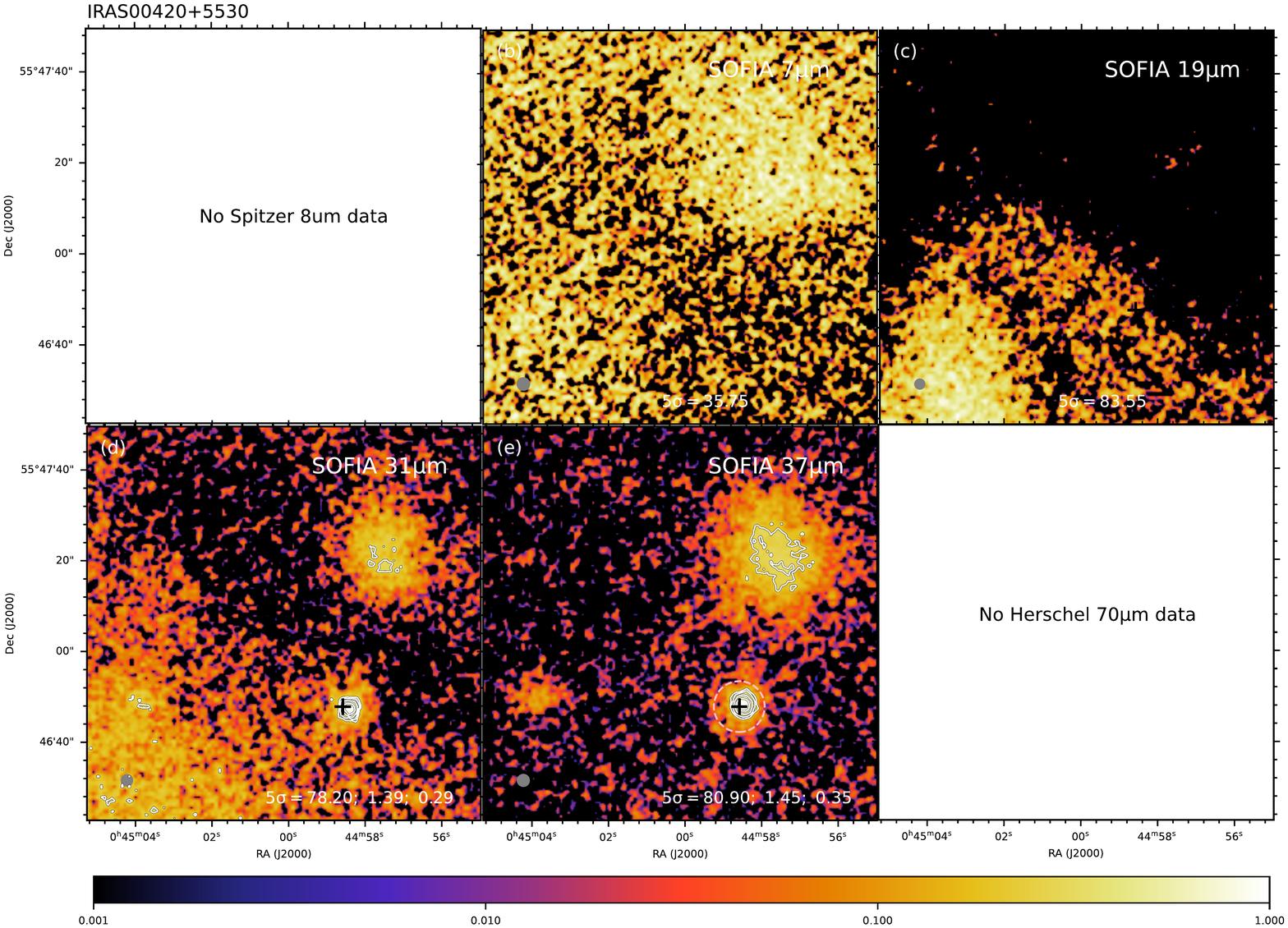}
\caption{Multiwavelength images of IRAS~00420+5530, following the format of Figure\,\ref{fig:AFGL2591}. The black cross in all panels denotes the position of the 3.6\,cm source VLA3 from \citet{molinari2002} at R.A.(J2000) = 0$^h$44$^m$58$\fs$5842, Decl.(J2000) = $+$55$\arcdeg$46$\arcmin$45$\farcs$675. The data used to create this figure and the SED are available.\label{fig:IRAS00420}}
\end{figure*}

IRAS~00420+5530, also known as Mol 3, is a high-mass protostellar candidate first identified by \citep{molinari1996,molinari1998a}.  The distance to IRAS~00420+5530 is 2.17$\pm$0.05\,kpc, measured from the parallax of water masers \citep{moellenbrock2009} (while \citet{molinari2002} and \citet{zhang2005} derived kinematic distances of 5.0 and 7.72\,kpc, respectively). In this work, we adopt a distance of 2.17\,kpc.  \citet{kumar2006} suggested the presence of a cluster of 380 $M_{\odot}$ associated with IRAS~00420+5530. Using high-resolution interferometry, \citet{molinari2002} detected two 3.4 mm continuum sources, MM1 (00$^h$44$^m$58\fs{2}/55\arcdeg46\arcmin46\arcsec) and MM2 (00$^h$44$^m$57\fs{3}/55\arcdeg46\arcmin57\arcsec), with an equal flux of $\sim$0.5 mJy along with HCO$^+$ $J=1-0$ emission coincided with MM1. The separation between MM1 and MM2 is $\sim$10\arcsec\ with MM 2 located to the NW. These authors also found two 3.6 cm sources, i.e., VLA 3, which is centered on MM1 and VLA 5, which is located $\sim$20\arcsec\ toward the NE.  
IRAS~00420+5530 also shows high-velocity CO emission with a complex morphology consisting of multiple peaks \citep{zhang2005}.  The red-shifted CO emission extends toward the south, while the blue-shifted emission concentrates on the millimeter sources.  Using WISE photometry and H$_2$ 2.12 \micron\ emission, \citet{wolf-chase2017} identified three Class I candidates and one Class 0 candidate in $\sim$2\arcmin$\times$2\arcmin\ field around IRAS~00420+5530.  The binaries, MM1 and MM2, is classified as Class I candidates, and they are associated with a chain of H$_2$ knots.  Most notably, they discovered an arc-like structure of strong H$_2$ emission to the north of the millimeter sources.  A Class I candidate (00$^h$44$^m$57\fs{3}/55\arcdeg47\arcmin18\farcs{1}) is identified at the apex of the arc.  The other two sources are more than 30\arcsec\ away from the mm sources.  At a larger $\sim$12\arcmin$\times$12\arcmin\ scale, \citet{lundquist2014} identified nine YSOs using WISE and 2MASS photometry, including eight intermediate-mass YSOs and a low-mass YSO.  Most of the YSOs are located at least $\sim$5\arcmin\ away to the west of the millimeter sources, while two YSOs are located at a separation of $\sim$0\farcm{5}--1\arcmin.

Figure\,\ref{fig:IRAS00420} shows the multiwavelength images for IRAS~00420+5530. Neither Spitzer nor Herschel data are available for this region. The shorter wavelength SOFIA images show high noise levels and extended emission that is expected to be caused by window contamination and no emission could be retrieved from these images. Longer-wavelength SOFIA images show concentrated emission around the centimeter peak as well as extended emission toward the NW.

\subsubsection{IRAS 23385+6053}

\begin{figure*}[!htb]
\includegraphics[width=1.0\textwidth]{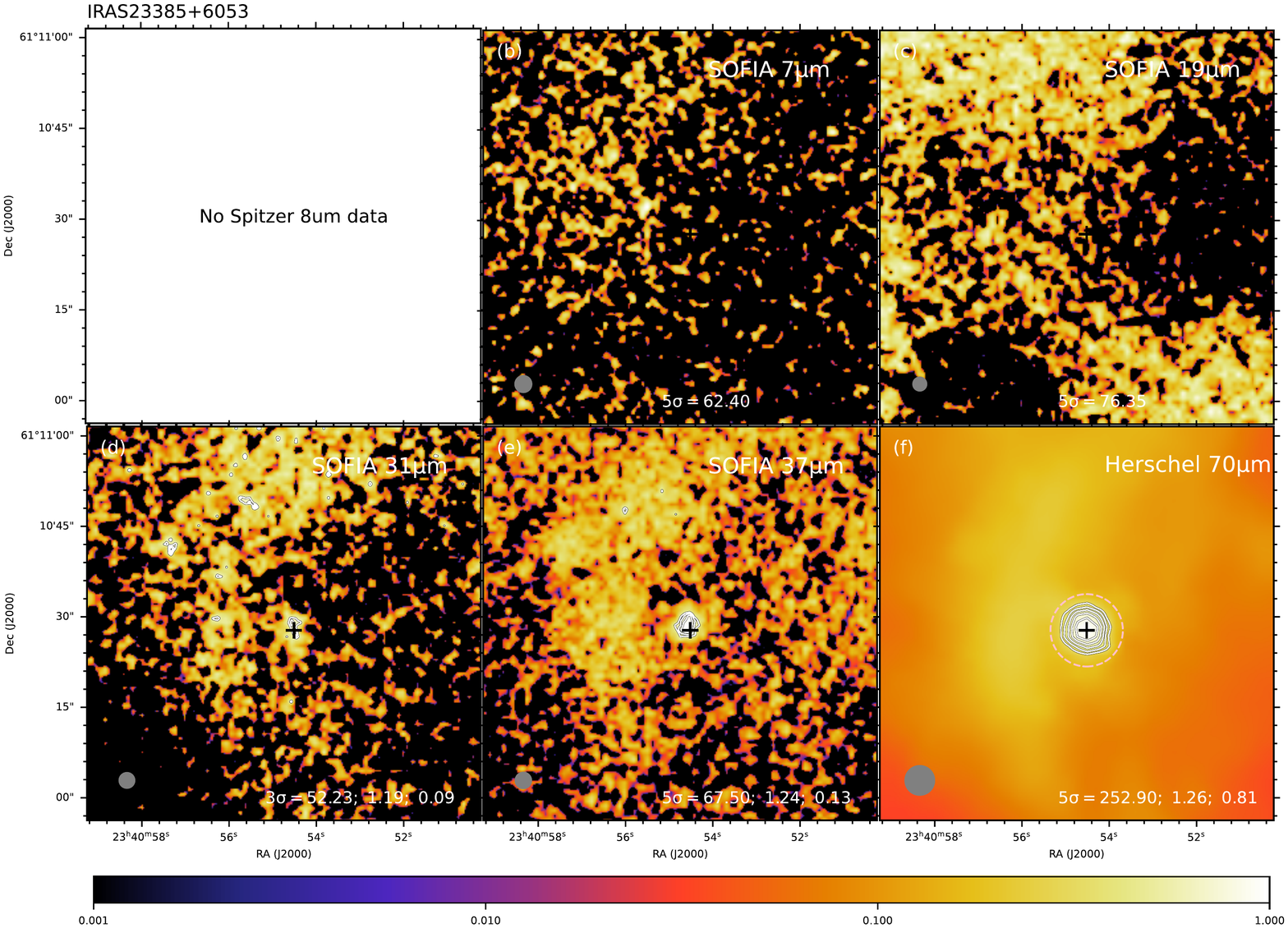}
\caption{Multiwavelength images of IRAS~23385+6053, following the format of Figure\,\ref{fig:AFGL2591}. The black cross in all panels denotes the position of the 3.4\,mm source Mol 160 from \citet{molinari2002} at R.A.(J2000) = 23$^h$40$^m$54$\fs$5171, Decl.(J2000) = $+$61$\arcdeg$10$\arcmin$27$\farcs$768. The data used to create this figure and the SED are available.\label{fig:IRAS23385}}
\end{figure*}

IRAS~23385+6053, also known as Mol 160, was identified as a possible precursor of a UC HII region, located at 4.9\,kpc \citep{molinari1996,molinari1998a,molinari1998b}.  Further observations from NIR to centimeter wavelengths revealed an embedded massive protostellar cluster, along with two HII regions at $\sim$30\arcsec\ to the east and $\sim$20\arcsec\ to the west \citep{molinari2002,cesaroni2019}. The powering sources of the two HII regions remain unclear.  \citet{molinari2008b} found bright 24 \micron\ emission surrounding the dominant continuum source.  The 24 \micron\ emission shows an arc-like structure to the east of the dominant continuum source and another extended structure peaks at $\sim$30\arcsec\ to the west, which has been suggested as having originated from a photodissociation region due to the two HII regions.  A massive core with a luminosity of $\sim$3000 $L_{\odot}$ dominates the cluster, which is the point-like source detected by SOFIA-FORCAST \citep{fontani2004,wolf-chase2012}.  Observations of molecular emission, such as SiO and HCO$^+$, show broad line widths, suggesting the presence of outflows.  The HCO$^+$ emission extends in the NE-SW direction with a velocity gradient, indicating a NE-SW outflow \citep{wolf-chase2012}. Similar outflow signatures are detected in H$_2$ 2.12\micron\ emission \citep{wolf-chase2012,wolf-chase2017}.  \citet{cesaroni2019} identified six cores in the cluster. The previously known massive core becomes a binary source, A1 and A2, each moving at a different velocity. The A2 source is half as massive as the A1 source and/or in an earlier evolutionary phase. Other cores are less massive and colder, appearing to be on the verge of collapse. By fitting a Keplerian rotating disk to the emission of CH$_3$CN, they further estimate a stellar mass of $\sim$9 $M_{\odot}$ in the A1 source. \citet{molinari2008a} constructed a SED model grid assuming either an embedded ZAMS or a graybody envelope if the embedded ZAMS model fails. In the end, IRAS~23385+6053 was fitted with a graybody envelope model that has a $L_\mathrm{bol}$ of 1.75$\times10^4 L_{\odot}$ and $M_\mathrm{env}$ of $222\:M_{\odot}$.

Figure\,\ref{fig:IRAS23385} shows the IR images for IRAS~23385+6053. No Spitzer data are available for this source. The SOFIA images are relatively noisy and emission from the main source is only discernible at 31 and $37\,\mu$m. However, extended emission from an eastern arc, which is visible in the Herschel~70$\rm \mu m$ image, is also visible at 31 and $37\,\mu$m.

\subsubsection{HH 288}

\begin{figure*}[!htb]
\includegraphics[width=1.0\textwidth]{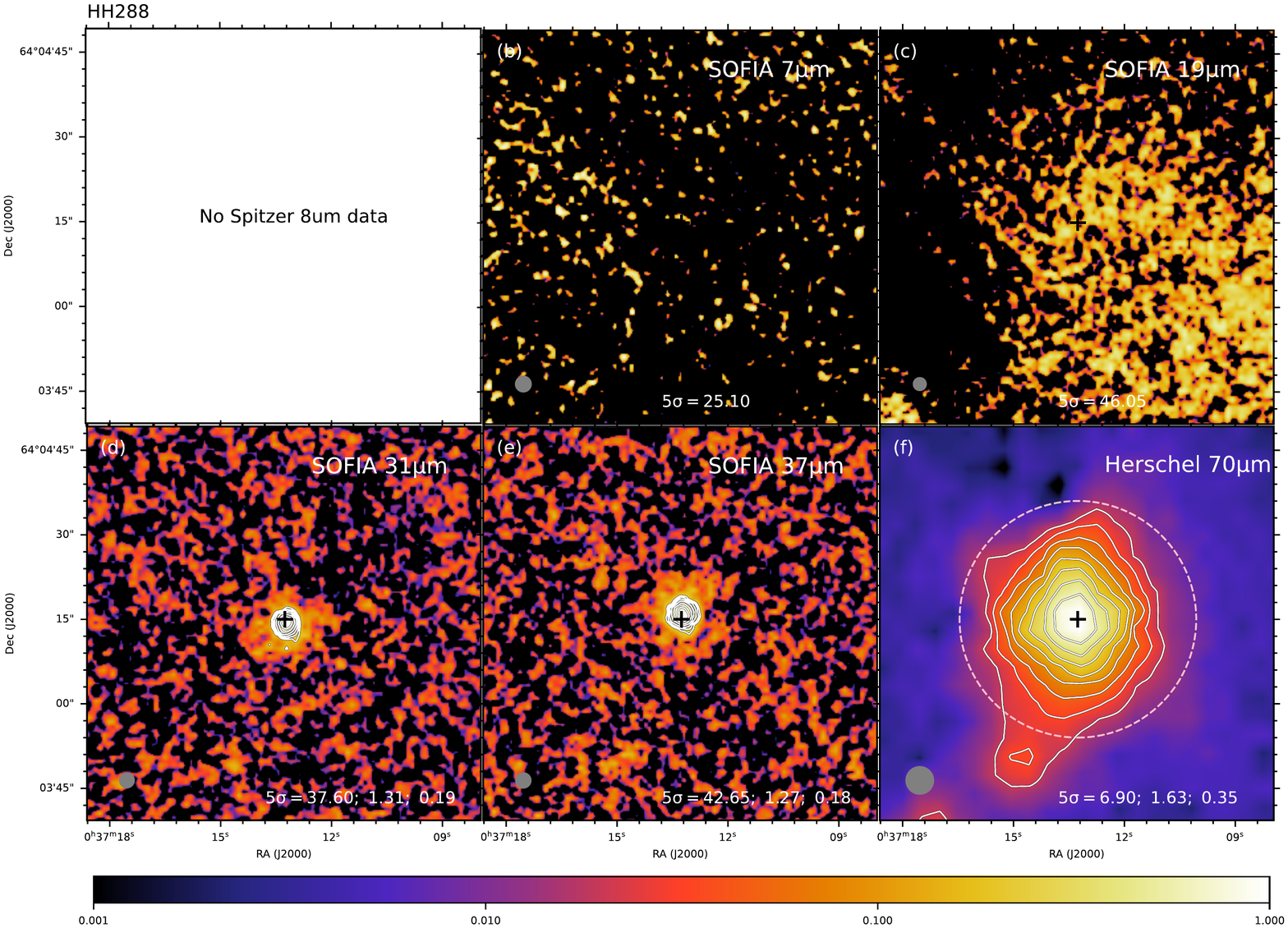}
\caption{Multiwavelength images of HH 288, following the format of Figure\,\ref{fig:AFGL2591}. The black cross in all panels denotes the position of the 3.5\,cm source 2a from \citet{franco2003} at R.A.(J2000) = 0$^h$37$^m$13$\fs$258, Decl.(J2000) = $+$64$\arcdeg$04$\arcmin$15$\farcs$02. The data used to create this figure and the SED are available.\label{fig:HH288}}
\end{figure*}

The HH object HH288 is located in the Cepheus region and it is associated with the exciting source IRAS~00342+6347 \citep{dent1998,gueth2001}. \citet{wouterloot1993} reported H$_2$O masers in the surroundings of this source. Two bipolar outflows have been observed in molecular tracers, including CO $J=1-0$, $J=2-1$ in the millimeter regime and H$_2$ in the NIR \citep{gueth2001}. The systemic velocity of the associated molecular emission is $-29\mathrm{\,km\,s^{-1}}$, which implies a kinematic distance of $\sim2.0$\,kpc \citep{gueth2001}. The large molecular outflow extends $\sim2$\,pc in the north-south direction with the CO blue-shifted emission located toward the south and the red-shifted toward the north, whereas the other smaller outflow is aligned toward east-west direction with the CO blue-shifted emission being more prominent toward the east and the red-shifted toward the west. The P.A.'s of these two outflows differ by $\sim65^\circ{}$ as estimated from Figure\,3 of \citet{gueth2001}: Even though these authors do not explicitly give the P.A.'s of the two outflows, we have estimated a P.A. of $\sim145^\circ{}$ and $\sim80^\circ{}$ for the north-south and the east-west flows, respectively, from their Figure 3.

The confirmation of HH288 having two independent outflows was later given by \citet{franco2003} as they revealed two centimeter sources associated with IRAS~00342+6347, called VLA2a and VLA2b. While they confirmed the association of VLA2a with the north-south outflow, they could not unambiguously associate VLA2b with the east-west outflow as this source was unresolved in their high-resolution ($\sim0.3^{\prime\prime}$) VLA observations.

Figure\,\ref{fig:HH288} shows the multiwavelength images for HH288. No Spitzer images are available for this region and SOFIA shorter wavelengths do not show significant emission. At 31 and $37\,\mu$m the central region shows some emission. We note that the extended 19~$\rm \mu m$ emission is likely due to a known ``window contamination'' problem that affected this observational dataset.

\subsection{Three-color images}

\begin{figure*}
\includegraphics[width=1.0\textwidth]{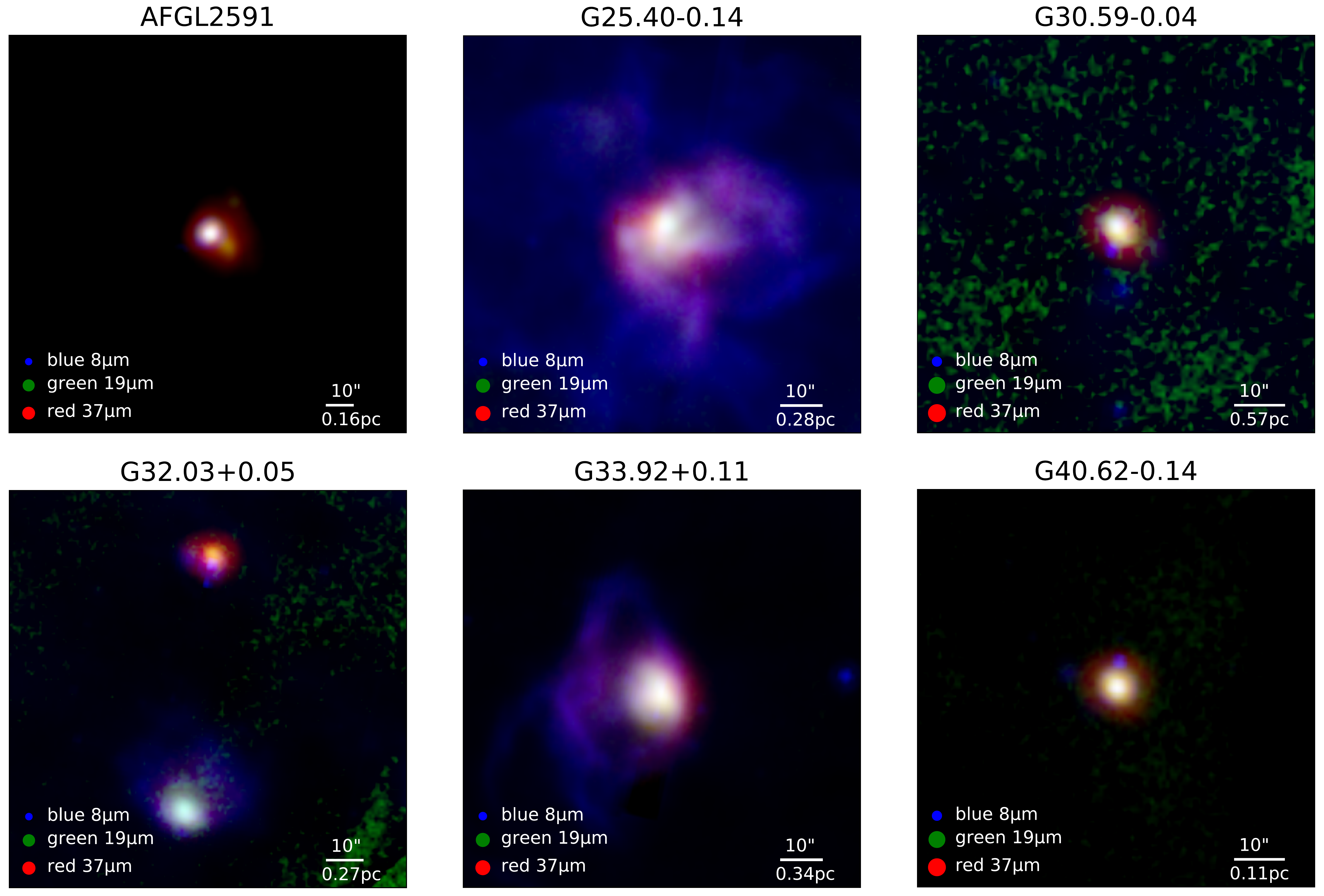}
\caption{Gallery of RGB images of the six protostellar sources, as labeled. Red denotes SOFIA-FORCAST 37\,$\mu$m, green denotes SOFIA-FORCAST 19\,$\mu$m, and blue is Spitzer IRAC 8\,$\mu$m. The color intensity scales are stretched as arcsinh and show a dynamic range of 100 from the peak emission at each wavelength. The bottom-left corner shows the wavelengths used and the beam sizes. The rest of the sources, i.e., IRAS~00259+5625, IRAS~00420+5530, IRAS~23385+6053, and HH288 are not shown due to the lack of Spitzer IRAC 8\,$\mu$m images and relatively poor signal-to-noise ratio in the SOFIA images. \label{fig:SOMA_rgb}}
\end{figure*}

Figure\,\ref{fig:SOMA_rgb} shows ``RGB'' images (i.e., based on 8, 19, and 37~$\rm \mu m$ MIR to FIR images) for the six regions analyzed in this paper that have the highest quality FORCAST imaging, i.e., AFGL~2591, G25.40-0.14, G30.59-0.04, G32.03+0.05, G33.92+0.11, and G40.62-0.14. These images allow for the inspection of color gradients in the regions. In particular, one prediction of TCA models is that near-facing outflow cavities will appear relatively brighter at shorter wavelengths. More generally, sources and regions of greater extinction will appear redder in these images. This is exactly what we see in the RGB images shown in Figure\,\ref{fig:SOMA_rgb}, redder colors toward the massive protostars with greater extinction and bluer color in the surroundings probably cleared by the outflow activity. In the case of AFGL~2591, the SW elongation is evident suggesting the location of the near-facing outflow. More diffuse blue emission is present in the G25.40-0.14 region with concentrated redder emission toward the center, whereas in the regions G30.59-0.04 and G40.62-0.14, less extended emission is revealed. It is interesting to note that in the case of the G32.03+0.05 region the northern source appears redder than the southern one. This may indicate that the northern source is younger as it is more extincted and has not had enough time to clear the surroundings. Finally, G33.92+0.11 shows the arc-like shape at shorter wavelengths and more concentrated and redder emission toward the location of the massive protostar.

\subsection{Results of SED Model Fitting}\label{sect:sed_results}

Here, we present the results of the measured fluxes and constructed SEDs of the 11 new sources analyzed in this paper, and their fits using the \citet{zhang2018} model grid. In the Appendix\,\ref{sect:appendix_soma} all SOMA sources published so far, i.e., from Papers I-III, are revisited with the new fitting pipeline.

\subsubsection{Measured SEDs}\label{sect:measured_sed_results}

For SOMA IV sources, we measure fluxes at 3.6, 4.5, 5.8, and 8.0$\,\mu$m from Spitzer-IRAC images; 7.7, 19.1, 31.5, and 37.1$\,\mu$m from SOFIA-FORCAST; and 70, 160, 250, 350, and 500$\,\mu$m from Herschel-PACS/SPIRE. Not all fluxes were measured for all sources due to either missing data or because the source was saturated at that wavelength. Additionally, when available, we measure the fluxes from IRAS-HIRES data to check for variability (\S\ref{sect:iras}). Note, these fluxes were not used to fit the SED model grid, partly because the sources are poorly resolved in the IRAS images, but also because the fitted SED is used to make a prediction of the IRAS flux for the variability analysis. Table\,\ref{tab:soma_iv_fluxes} summarizes the measured fluxes in this study using the aperture radius indicated in the second column. As in SOMA I-III, the fiducial method for measuring fluxes across all wavelengths is to set a fixed aperture for all images and perform background subtraction (this is what is reported throughout the paper). This aperture was chosen based on the Herschel $70\,\mu$m image or the SOFIA $37\,\mu$m when the former was not available (\S\ref{sect:opt_rad}). 

In Figure\,\ref{fig:sed_1D_results_soma_iv} we show the measured fluxes as red squares with their error bars. The error bars are set to be the larger of either 10\% of the background-subtracted flux density or the value of the estimated background flux density. It should be recalled that everything at $8\,\mu$m and shorter wavelengths is treated as upper limits due to the RT models not including PAH emission and single photon transient heating effects on small dust grains  \citep[see][]{debuizer2017,zhang2018}. Most of the sources show the complete set of Spitzer and Herschel data, with the latter constraining very well, together with SOFIA data, the peak of the SED. However, for sources IRAS~00420+5530 and IRAS~00259+5625, only 19.1, 31.5, and 37.1$\,\mu$m images could be used to build their SEDs.

\begin{longrotatetable}
\begin{deluxetable}{lcccccccccccccccccc}
\tabletypesize{\tiny}
\tablecaption{Integrated Flux Densities\label{tab:soma_iv_fluxes}}
\tablehead{
\colhead{Source} & \colhead{radius} & \colhead{$F_{3.6}$} & \colhead{$F_{4.5}$} & \colhead{$F_{5.8}$} & \colhead{$F_{7.7}$} & \colhead{$F_{8.0}$} & \colhead{$F_{12}$} & \colhead{$F_{19.1}$} & \colhead{$F_{25}$} & \colhead{$F_{31.5}$} & \colhead{$F_{37.1}$}  & \colhead{$F_{60}$} & \colhead{$F_{70}$} & \colhead{$F_{100}$} & \colhead{$F_{160}$} & \colhead{$F_{250}$}  & \colhead{$F_{350}$} & \colhead{$F_{500}$} \\
\colhead{} & \colhead{(\arcsec/pc)} & \colhead{(Jy)} & \colhead{(Jy)} & \colhead{(Jy)} & \colhead{(Jy)} & \colhead{(Jy)} & \colhead{(Jy)} & \colhead{(Jy)} & \colhead{(Jy)} & \colhead{(Jy)} & \colhead{(Jy)} & \colhead{(Jy)} & \colhead{(Jy)} & \colhead{(Jy)} & \colhead{(Jy)} & \colhead{(Jy)} & \colhead{(Jy)} & \colhead{(Jy)}
}
\startdata
\multirow{3}{*}{AFGL2591} & \multirow{2}{*}{18.75} & 74.45 & 173.42 & 342.88 & $\cdots$ & 308.61 & 522.29 & 681.70 & 1486.42 & 2437.74 & 3196.32 & 6988.93 & 4445.46 & 7132.77 & 2172.18 & 535.94 & 151.56 & 25.50 \\
& \multirow{2}{*}{0.30} & (74.83) & (174.28) & (348.25) & $\cdots$ & (313.06) & $\cdots$ & (679.86) & $\cdots$ & (2463.83) & (3223.91) & $\cdots$ & (4618.42) & $\cdots$ & (2416.41) & (660.42) & (201.93) & (45.07) \\
&  & [10.53] & [24.53] & [48.50] & $\cdots$ & [43.65] & $\cdots$ & [96.43] & $\cdots$ & [344.94] & [452.75] & $\cdots$ & [632.37] & $\cdots$ & [392.45] & [145.74] & [54.74] & [19.90] \\
\hline
\multirow{3}{*}{G25.40-0.14} & \multirow{2}{*}{48.5} & $\cdots$ & $\cdots$ & $\cdots$ & 87.83 & $\cdots$ & 58.36 & 155.98 & 618.83 & 1641.73 & 2513.93 & 5244.28 & 5549.39 & 10150.80 & 3535.18 & 1146.80 & 405.21 & 115.15 \\
& \multirow{2}{*}{1.34} & $\cdots$ & $\cdots$ & $\cdots$ & (84.73) & $\cdots$ & $\cdots$ & (202.50) & $\cdots$ & (1447.83) & (2000.58) & $\cdots$ & (6383.78) & $\cdots$ & (4498.69) & (1739.51) & (650.73) & (206.07) \\
&  & $\cdots$ & $\cdots$ & $\cdots$ & [20.41] & $\cdots$ & $\cdots$ & [50.81] & $\cdots$ & [289.94] & [380.30] & $\cdots$ & [834.98] & $\cdots$ & [1085.50] & [614.49] & [252.12] & [92.37] \\
\hline
\multirow{3}{*}{G30.59-0.04} & \multirow{2}{*}{13.5} & $\cdots$ & $\cdots$ & 1.68 & 3.39 & 2.13 & 1.12 & 3.57 & 16.02 & 113.51 & 182.21 & 631.11 & 875.04 & 863.71 & 616.98 & 170.58 & 43.76 & 3.32 \\
& \multirow{2}{*}{0.77} & $\cdots$ & $\cdots$ & (2.28) & (2.87) & (4.22) & $\cdots$ & (5.46) & $\cdots$ & (114.10) & (182.64) & $\cdots$ & (932.24) & $\cdots$ & (798.99) & (277.49) & (93.98) & (23.59) \\
&  & $\cdots$ & $\cdots$ & [0.24] & [0.78] & [0.30] & $\cdots$ & [1.30] & $\cdots$ & [16.10] & [25.77] & $\cdots$ & [124.64] & $\cdots$ & [201.84] & [109.60] & [50.60] & [20.28] \\
\hline
\multirow{3}{*}{G32.03+0.05} & \multirow{2}{*}{22.75} & 0.32 & $\cdots$ & 2.43 & 8.10 & $\cdots$ & $\cdots$ & 12.38 & $\cdots$ & 69.87 & 114.25 & $\cdots$ & 171.12 & $\cdots$ & 129.50 & 62.29 & 23.75 & 8.91 \\
& \multirow{2}{*}{0.61} & (0.42) & $\cdots$ & (3.04) & (4.89) & $\cdots$ & $\cdots$ & (5.43) & $\cdots$ & (79.01) & (103.49) & $\cdots$ & (184.66) & $\cdots$ & (203.10) & (145.75) & (73.65) & (28.54) \\
&  & [0.06] & $\cdots$ & [0.36] & [2.35] & $\cdots$ & $\cdots$ & [4.55] & $\cdots$ & [11.50] & [16.44] & $\cdots$ & [26.46] & $\cdots$ & [75.84] & [83.93] & [50.01] & [19.66] \\
\hline
\multirow{3}{*}{G32.03+0.05N} & \multirow{2}{*}{13.5} & 0.09 & $\cdots$ & 1.31 & 2.23 & $\cdots$ & $\cdots$ & 1.76 & $\cdots$ & 65.35 & 105.86 & $\cdots$ & 543.84 & $\cdots$ & 600.94 & $\cdots$ & 61.97 & 11.06 \\
& \multirow{2}{*}{0.36} & (0.13) & $\cdots$ & (1.60) & (1.92) & $\cdots$ & $\cdots$ & $\cdots$ & $\cdots$ & (67.69) & (103.36) & $\cdots$ & (567.51) & $\cdots$ & (716.45) & $\cdots$ & (107.73) & (26.67) \\
&  & [0.02] & $\cdots$ & [0.19] & [0.90] & $\cdots$ & $\cdots$ & [0.92] & $\cdots$ & [9.93] & [15.65] & $\cdots$ & [77.79] & $\cdots$ & [143.41] & $\cdots$ & [46.60] & [15.70] \\
\hline
\multirow{3}{*}{G33.92+0.11} & \multirow{2}{*}{25.25} & 0.78 & $\cdots$ & $\cdots$ & 22.56 & 17.71 & 35.29 & 138.75 & 340.07 & 542.56 & 699.50 & 2089.71 & 1723.21 & 2612.35 & 1612.85 & $\cdots$ & 217.43 & 51.50 \\
& \multirow{2}{*}{0.87} & (0.90) & $\cdots$ & $\cdots$ & (17.00) & (21.33) & $\cdots$ & (128.00) & $\cdots$ & (535.73) & (709.91) & $\cdots$ & (1784.91) & $\cdots$ & (1862.11) & $\cdots$ & (301.87) & (85.89) \\
&  & [0.40] & $\cdots$ & $\cdots$ & [4.39] & [2.65] & $\cdots$ & [20.78] & $\cdots$ & [76.91] & [99.33] & $\cdots$ & [249.72] & $\cdots$ & [337.87] & $\cdots$ & [89.87] & [35.15] \\
\hline
\multirow{3}{*}{G40.62-0.14} & \multirow{2}{*}{13.5} & 0.09 & $\cdots$ & $\cdots$ & 2.47 & 1.81 & 1.69 & 15.81 & 84.14 & 182.25 & 269.50 & 800.55 & 694.32 & 1110.80 & 417.82 & 125.57 & 26.85 & 4.20 \\
& \multirow{2}{*}{0.14} & (0.12) & $\cdots$ & $\cdots$ & (1.97) & (2.32) & $\cdots$ & (19.42) & $\cdots$ & (189.54) & (273.43) & $\cdots$ & (716.24) & $\cdots$ & (487.06) & (171.61) & (56.62) & (13.59) \\
&  & [0.03] & $\cdots$ & $\cdots$ & [0.38] & [0.26] & $\cdots$ & [2.51] & $\cdots$ & [25.82] & [38.11] & $\cdots$ & [98.60] & $\cdots$ & [91.02] & [49.35] & [30.01] & [9.41] \\
\hline
\multirow{3}{*}{IRAS00259} & \multirow{2}{*}{7.0} & $\cdots$ & $\cdots$ & $\cdots$ & $\cdots$ & $\cdots$ & 0.33 & 0.18 & 0.57 & 1.51 & 2.21 & 25.34 & $\cdots$ & 100.53 & $\cdots$ & $\cdots$ & $\cdots$ & $\cdots$ \\
& \multirow{2}{*}{0.08} & $\cdots$ & $\cdots$ & $\cdots$ & $\cdots$ & $\cdots$ & $\cdots$ & (0.92) & $\cdots$ & (1.77) & (0.85) & $\cdots$ & $\cdots$ & $\cdots$ & $\cdots$ & $\cdots$ & $\cdots$ & $\cdots$ \\
&  & $\cdots$ & $\cdots$ & $\cdots$ & $\cdots$ & $\cdots$ & $\cdots$ & [0.51] & $\cdots$ & [0.45] & [0.45] & $\cdots$ & $\cdots$ & $\cdots$ & $\cdots$ & $\cdots$ & $\cdots$ & $\cdots$ \\
\hline
\multirow{3}{*}{IRAS00420} & \multirow{2}{*}{7.0} & $\cdots$ & $\cdots$ & $\cdots$ & $\cdots$ & $\cdots$ & 5.23 & 0.45 & 11.43 & 5.81 & 8.09 & 188.40 & $\cdots$ & 218.18 & $\cdots$ & $\cdots$ & $\cdots$ & $\cdots$ \\
& \multirow{2}{*}{0.07} & $\cdots$ & $\cdots$ & $\cdots$ & $\cdots$ & $\cdots$ & $\cdots$ & (0.20) & $\cdots$ & (6.69) & (8.01) & $\cdots$ & $\cdots$ & $\cdots$ & $\cdots$ & $\cdots$ & $\cdots$ & $\cdots$ \\
&  & $\cdots$ & $\cdots$ & $\cdots$ & $\cdots$ & $\cdots$ & $\cdots$ & [0.61] & $\cdots$ & [1.02] & [1.28] & $\cdots$ & $\cdots$ & $\cdots$ & $\cdots$ & $\cdots$ & $\cdots$ & $\cdots$ \\
\hline
\multirow{3}{*}{IRAS23385} & \multirow{2}{*}{5.75} & $\cdots$ & $\cdots$ & $\cdots$ & 0.10 & $\cdots$ & $\cdots$ & -$\cdots$ & $\cdots$ & 0.96 & 1.78 & $\cdots$ & 22.38 & $\cdots$ & 37.96 & 22.82 & 2.89 & 0.75 \\
& \multirow{2}{*}{0.14} & $\cdots$ & $\cdots$ & $\cdots$ & $\cdots$ & $\cdots$ & $\cdots$ & $\cdots$ & $\cdots$ & (0.80) & (2.57) & $\cdots$ & (34.68) & $\cdots$ & (57.99) & (42.11) & (14.31) & (5.17) \\
&  & $\cdots$ & $\cdots$ & $\cdots$ & [0.34] & $\cdots$ & $\cdots$ & $\cdots$ & $\cdots$ & [0.44] & [0.55] & $\cdots$ & [4.17] & $\cdots$ & [20.73] & [19.56] & [11.43] & [4.43] \\
\hline
\multirow{3}{*}{HH288} & \multirow{2}{*}{7.5} & $\cdots$ & $\cdots$ & $\cdots$ & $\cdots$ & $\cdots$ & 0.02 & 0.23 & 0.89 & 3.84 & 4.33 & 30.41 & 27.20 & 104.26 & 37.57 & 6.88 & 1.52 & 0.05 \\
& \multirow{2}{*}{0.07} & $\cdots$ & $\cdots$ & $\cdots$ & $\cdots$ & $\cdots$ & $\cdots$ & (0.38) & $\cdots$ & (5.84) & (4.63) & $\cdots$ & (30.40) & $\cdots$ & (48.53) & (17.45) & (6.32) & (1.50) \\
&  & $\cdots$ & $\cdots$ & $\cdots$ & $\cdots$ & $\cdots$ & $\cdots$ & [0.18] & $\cdots$ & [0.58] & [0.64] & $\cdots$ & [3.90] & $\cdots$ & [12.18] & [10.61] & [4.80] & [1.44] \\
\enddata
\tablecomments{
{\tiny $F_{3.6}$, $F_{4.5}$, $F_{5.8}$, and $F_{8.0}$ refer to fluxes from Spitzer-IRAC at 3.6, 4.5, 5.8, and 8.0\,$\mu$m, respectively. $F_{7.7}$, $F_{19.1}$, $F_{31.5}$, and $F_{37.1}$ refer to fluxes from SOFIA-FORCAST at 7.7, 19.1, 31.5, and 37.5\,$\mu$m, respectively. $F_{70}$, $F_{160}$, $F_{250}$, $F_{350}$ and $F_{500}$ refer to fluxes from Herschel-PACS/SPIRE at 70, 160, 250, 350, and 500\,$\mu$m, respectively.  $F_{12.0}$, $F_{25.0}$, $F_{60.0}$, and $F_{100.0}$ refer to fluxes from IRAS-HIRES at 12.0, 25.0, 60.0, and 100.0\,$\mu$m, respectively. The three dots refer to either data not found or saturated at that wavelength. The first row for each source refers to background subtracted flux. The second row for each source with fluxes in parenthesis refers to non-background-subtracted flux (in many FORCAST $7.7\:{\rm \mu m}$ images we find that the estimated background has a slightly negative value, which results from a calibration offset problem; hence, background subtracted values appear larger in this table; however, we still expect these background subtracted values to be accurate). The third row for each source with square brackets refers to the associated error to the background-subtracted fluxes. $\dagger$ No Herschel-PACS 70\,$\mu$m data is available and SOFIA-FORCAST 37\,$\mu$m was used to find the optimal aperture (\S\ref{sect:opt_rad}). For the radius column for each source, the first row refers to \arcsec and the second row to pc.}}
\end{deluxetable}
\end{longrotatetable}


\subsubsection{SED model fitting}\label{sect:sed_fit}

Figure~\ref{fig:sed_1D_results_soma_iv} shows all good (see \S\ref{sect:average_models}) SED models with distinct values of $M_c$, $\Sigma_{\rm cl}$, $m_*$, and $\theta_\mathrm{view}$ that fit the SED data of each of the SOMA IV massive protostars. The properties of the best five of these models, along with the average and dispersion of all the good models, are shown in Table~\ref{tab:best_models_soma_iv} (note, these averages are equally weighted over all good models, including with different viewing angles for the same physical model, but only counting one best value of $A_V$ in each case). It should be noted that the errors used as weights in the evaluation of $\chi^2$ may not have Gaussian distributions, so care should be taken in the interpretation of these numerical values.

Figure\,\ref{fig:sed_2D_results_soma_iv} presents the 2D distribution of the three main physical parameters of the model grid, i.e., in $\Sigma_{\rm cl}$ - $M_{c}$ space (left), $m_{*}$ - $M_{c}$ space (center) and $m_{*}$ - $\Sigma_{\rm  cl}$ space (right), color coded by $\chi^{2}$ (that of the best model at each location in the 2D parameter space is shown). These plots help illustrate the extent of degeneracies that are present in the SED fitting.

In general, the observed SEDs are reasonably well fitted by the models and thus consistent with being massive protostars forming via TCA. In most cases, the best model for each source has $\chi^2_{\rm min}\lesssim 1$, with the exceptions being AFGL~2591 (with $\chi^2_{\rm min}=1.7$), G25.40-0.14 (with $\chi^2_{\rm min}=3.2$), G32.03+0.05N (with $\chi^2_{\rm min}=1.6$) and G33.92+0.11 (with $\chi^2_{\rm min}=4.7$). These relatively poor fits are typically caused by discrepancies at the longest wavelengths indicating that there is additional relatively cold dusty material present around the protostar. This may result from imperfect clump background subtraction, especially if the adopted aperture radius is significantly larger than that of the model core. It should also be noted that values $\chi^2<1$ are either due to the small number of data points being fitted, and therefore the small degree of freedom, or the relatively large errors in the fluxes greater than $100\,\mu$m.

Considering the average properties of the protostellar models, we see that initial core masses range from $M_c\sim20$ to $\sim400\,M_\odot$, while clump mass surface densities range from $\Sigma_{\rm cl}\sim 0.2$ to $2\:{\rm g\:cm^{-2}}$. The values of current protostellar mass range from $m_*\sim 2$ to $40\:M_\odot$.

We now describe the results of the SED fitting for each source in the following:

\textit{AFGL~2591:} This source is the most massive protostar in SOMA IV with $m_*=51^{+23}_{-16}\,M_\odot$ accreting at $6.9^{+9.3}_{-4.0}\times10^{-4}\,M_\odot\,\mathrm{yr^{-1}}$ from a $313^{+159}_{-105}\,M_\odot$ core in a $\Sigma_\mathrm{cl}=0.699^{+1.879}_{-0.509}\,\mathrm{g\,cm^{-2}}$. This source has an average bolometric luminosity of $5.6^{+4.3}_{-2.4}\times10^5\,L_\odot$ being one of the most luminous sources in the SOMA survey to date. The SED fitting for this source predicts a highly extincted region of $67\pm42$\,mag. This case has a viewing angle ($38\pm16^\circ{}$) close to the cavity opening angle ($37\pm8^\circ{}$), so that there are high levels of shorter wavelength emission as can be seen in both the observed and modeled SEDs.

\textit{G25.40-0.14:} The predicted protostellar mass for this source is $42^{+30}_{-18}\,M_\odot$ making it the second most massive protostar in SOMA IV. It is forming in a core with mass $436^{+70}_{-61}\,M_\odot$ with a surface density $\Sigma_\mathrm{cl}=1.678^{+1.318}_{-0.738}\,\mathrm{g\,cm^{-2}}$, viewed at an angle $44\pm13^\circ{}$. This source is also at the upper end of bolometric luminosity in our survey with $5.4^{+4.6}_{-2.5}\times10^5\,L_\odot$ and it is accreting material at high rates $1.5^{+0.4}_{-0.3}\times10^{-3}\,M_\odot\,\mathrm{yr^{-1}}$ In this case, the predicted visual extinction is low and on average is 7\,mag.

\textit{G30.59-0.04:} This source is also among the most massive protostars in the SOMA sample with $m_*=31^{+13}_{-9}\,M_\odot$ forming in a massive core of $M_\mathrm{c}=409^{+96}_{-78}\,M_\odot$ in a high-mass surface density clump of $\Sigma_\mathrm{cl}=1.584^{+1.204}_{-0.684}\,\mathrm{g\,cm^{-2}}$ accreting at high rates $1.2^{+0.5}_{-0.4}\times10^{-3}\,M_\odot\,\mathrm{yr^{-1}}$. The SED fitting predicts a viewing angle of $63\pm18^\circ{}$ and a narrow cavity opening angle of $19\pm6^\circ{}$. This source also sits among the most luminous sources in our sample with $L_\mathrm{bol}=3.5^{+1.7}_{-1.1}\times10^5\,L_\odot$.

\textit{G32.03+0.05:} This source continues with the trend of high-mass protostars with a predicted current mass of $m_*=20^{+15}_{-8}\,M_\odot$ forming in a relatively high-mass core of $M_\mathrm{c}=148^{+139}_{-72}\,M_\odot$ but relatively low-$\Sigma$ clump with $0.222^{+0.475}_{-0.151}\,\mathrm{g\,cm^{-2}}$, which implies it is accreting material at a relatively low rate of $1.4^{+1.3}_{-0.7}\times10^{-4}\,M_\odot\,\mathrm{yr^{-1}}$. Its bolometric luminosity is $6.7^{+13.2}_{-4.4}\times10^4\,L_\odot$. The low value of $\Sigma_{\rm cl}$ implies it takes a relatively long time to form the protostar, i.e., the current stellar age (not shown in the main tables) for this source is $\sim2\times10^5$\,yr, which is older than G32.03+0.05N (see below). The low value of $\Sigma_{\rm cl}$ also implies a low level of internal extinction, which is also reflected in its bluer IR colors.

\textit{G32.03+0.05N:} This is the second source found in the G32.03+0.05 region  and it is somewhat more massive ($31^{+36}_{-17}\,M_\odot$) forming in a higher-mass core ($245^{+187}_{-106}\,M_\odot$) with a high mass surface density clump ($0.465^{+1.063}_{-0.324}\,\mathrm{g\,cm^{-2}}$) and accreting material at a higher rate ($3.7^{+4.1}_{-1.9}\times10^{-4}\,M_\odot\,\mathrm{yr^{-1}}$). This source is also more luminous ($2.1^{+5.2}_{-1.5}\times10^5\,L_\odot$) than the southern source present in the region. The SED-fitting outputs a current stellar age of $\sim1.4\times10^5$\,yr, which is younger than G32.03+0.05 (see above). Its higher $\Sigma_{\rm cl}$ implies higher internal extinction, which is reflected in its redder IR colors.

\textit{G33.92+0.11:} The SED-fitting results of this source also involve a relatively massive core of $339^{+155}_{-106}\,M_\odot$ in density environment with $\Sigma_\mathrm{cl}=0.934^{+1.639}_{-0.595}\,\mathrm{g\,cm^{-2}}$ that harbors a current protostellar mass of $52^{+32}_{-20}\,M_\odot$. The protostar is accreting at high rates $9.0^{+7.6}_{-4.1}\times10^{-3}\,M_\odot\,\mathrm{yr^{-1}}$ with a narrow cavity opening angle of $35\pm11^\circ{}$ and viewed at $54\pm17^\circ{}$. This source also is among the most luminous sources with $L_\mathrm{bol}=6.3^{+5.2}_{-2.8}\times10^5\,L_\odot$.

\textit{G40.62-0.14:} This source is still in the high-mass regime with $m_*=15^{+8}_{-5}\,M_\odot$, but forming in a relatively low-mass core of $64^{+34}_{-22}\,M_\odot$ and also low mass surface density clump of $\Sigma_\mathrm{cl}=0.615^{+1.011}_{-0.382}\,\mathrm{g\,cm^{-2}}$. It is accreting material at rates of $2.1^{+1.9}_{-1.0}\times10^{-4}\,M_\odot\,\mathrm{yr^{-1}}$ and has a bolometric luminosity of $4.9^{+7.0}_{-2.9}\times10^4\,L_\odot$. 

\textit{IRAS~00259+5625:} Note that for this source, which lacks Spitzer and Herschel data, there are only three effective data points constraining the models. The values of $\chi^2$ are small, i.e., about 0.01, for the best-fit case. When considering the good models, they indicate $3^{+6}_{-2}\,M_\odot$ protostars in relatively low-$\Sigma$ cores $0.517^{+1.012}_{-0.342}\,\mathrm{g\,cm^{-2}}$ with low masses of $45^{+57}_{-25}\:M_\odot$ viewed at $55\pm22^\circ$. It is accreting at low rates of $8.6^{+15.0}_{-5.4}\times10^{-5}\,M_\odot\,\mathrm{yr^{-1}}$. This source represents one of smaller bolometric luminosity ($L_\mathrm{bol}=2.5^{+12.5}_{-2.0}\times10^3\,L_\odot$) in the SOMA IV sample. Longer wavelength data would obviously be helpful here to break some of these degeneracies.

\textit{IRAS~00420+5530:} Note that also for this source, which also lacks Spitzer and Herschel data, there are only three effective data points constraining the models. The values of $\chi^2$ are small, i.e., about 0.33, for the best-fit case. When considering the good models, they indicate $3^{+3}_{-2}\,M_\odot$ protostars in $\Sigma$ cores $0.365^{+0.619}_{-0.230}\,\mathrm{g\,cm^{-2}}$ viewed at $49\pm23^\circ{}$ and accreting at low rates $6.1^{+5.5}_{-2.9}\times10^{-5}\,M_\odot\,\mathrm{yr^{-1}}$. This source is on the lower end of bolometric luminosities with $1.5^{+3.5}_{-1.0}\times10^3\,L_\odot$ framing it into the intermediate-mass protostars. Longer wavelength data would obviously be helpful here to break some of these degeneracies.

\textit{IRAS~23385:} The SED fitting for this source predicts $m_*=5^{+10}_{-3}\,M_\odot$ forming from a core of $52^{+54}_{-26}\,M_\odot$ at relatively low-$\Sigma$ clump $0.517^{+1.067}_{-0.384}\,\mathrm{g\,cm^{-2}}$ and accreting material at a rate of $1.1^{+1.1}_{-0.5}\times10^{-4}\,M_\odot\,\mathrm{yr^{-1}}$. Its luminosity is $L_\mathrm{bol}=5.6^{+27.4}_{-4.7}\times10^3\,L_\odot$ and it is on the edge of being considered a massive protostar given the current protostellar mass. However, it should be noted that good models cover a large range of parameter space, i.e., source properties are not well constrained by the current data.

\textit{HH288:} The SED fitting for this source predicts a current protostellar mass that is in the intermediate-mass regime with $m_*=3^{+3}_{-1}\,M_\odot$, forming from a core of initial mass $18^{+9}_{-6}\,M_\odot$ within a clump of mass surface density of $0.449^{+1.046}_{-0.314}\,\mathrm{g\,cm^{-2}}$. This source is also accreting at low rates ($5.5^{+5.9}_{-2.8}\times10^{-5}\,M_\odot\,\mathrm{yr^{-1}}$) and has the lowest bolometric luminosity ($1.1^{+1.7}_{-0.7}\times10^3\,L_\odot$) in the SOMA IV sample.

As has been noted in previous SOMA papers, MIR to FIR SED fitting can be subject to significant degeneracies. Follow-up analysis of source MIR and FIR images \citep[e.g.,][]{zhang2013}, centimeter continuum emission from ionized gas \citep[e.g.,][]{rosero2019a} and protostellar outflow properties are some of the methods that can be used to help break such degeneracies.

\begin{figure*}[!htb]
\includegraphics[width=0.5\textwidth]{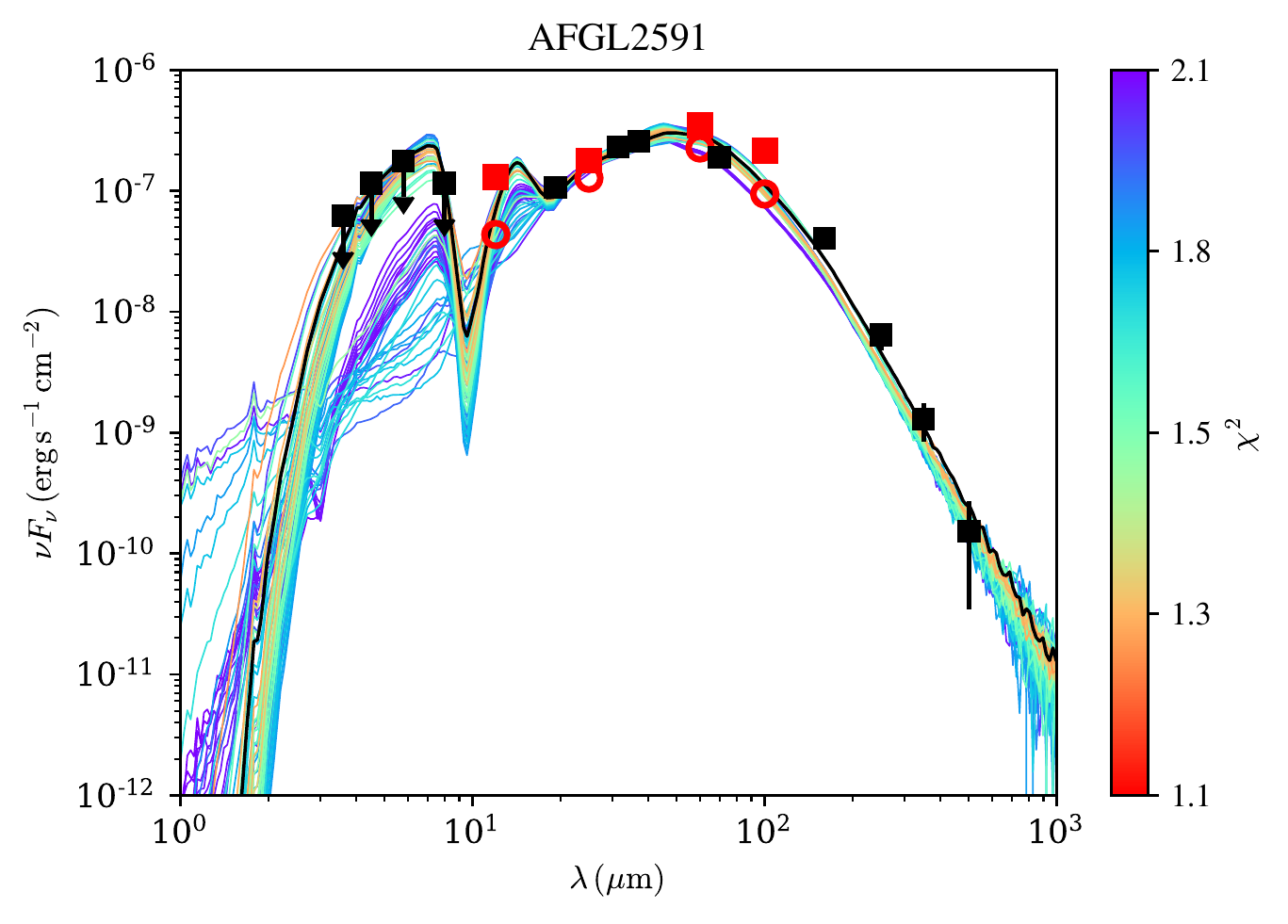}
\includegraphics[width=0.5\textwidth]{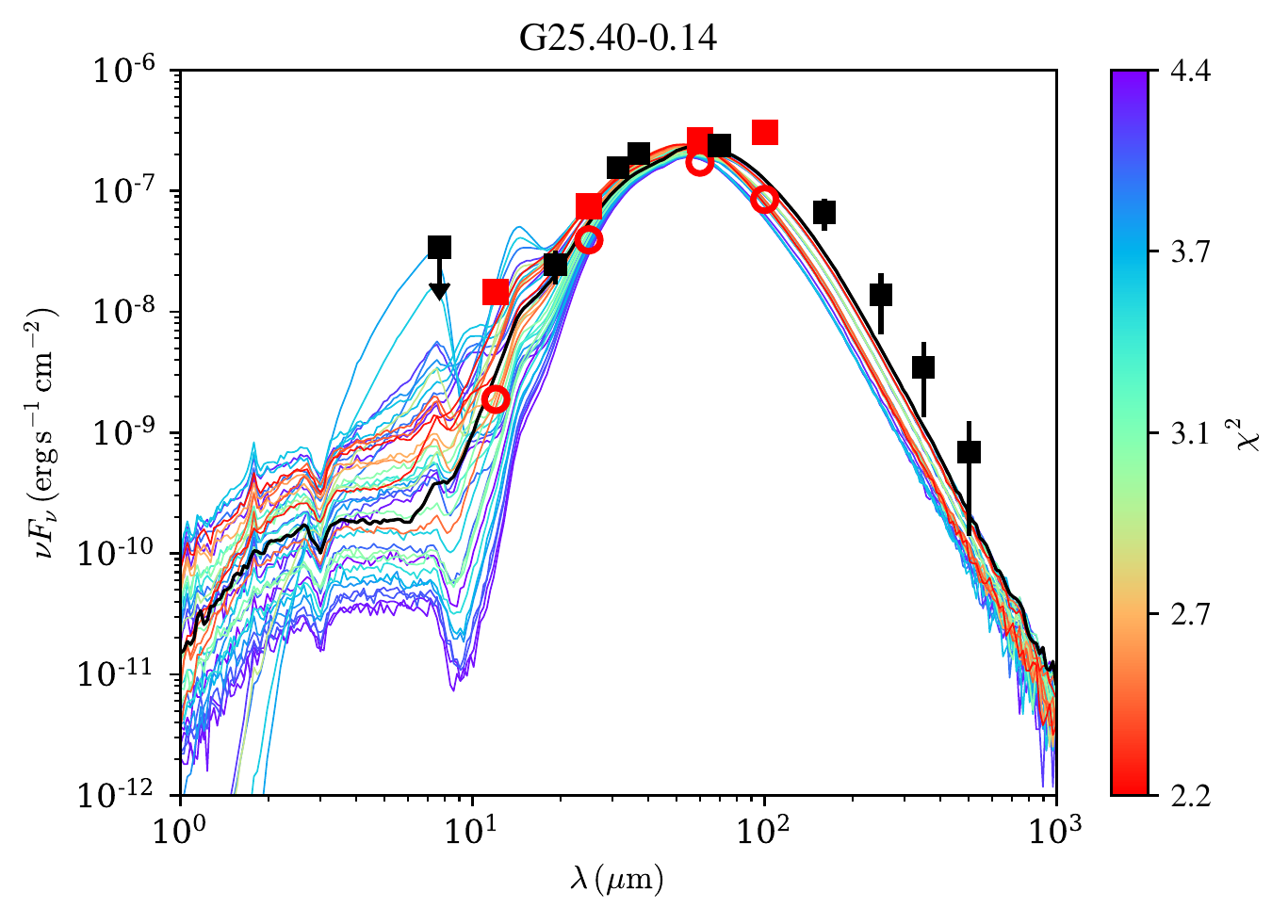}
\includegraphics[width=0.5\textwidth]{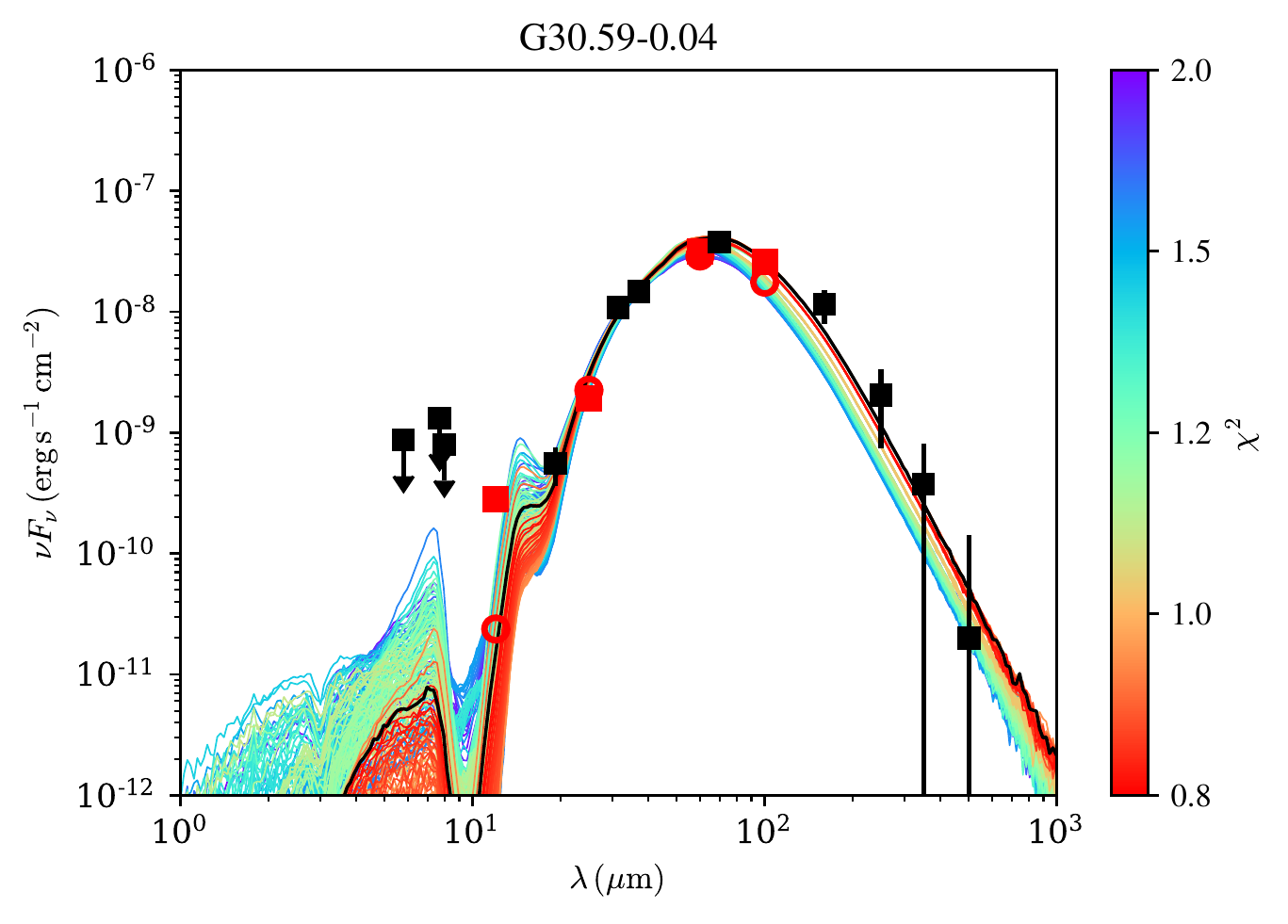}
\includegraphics[width=0.5\textwidth]{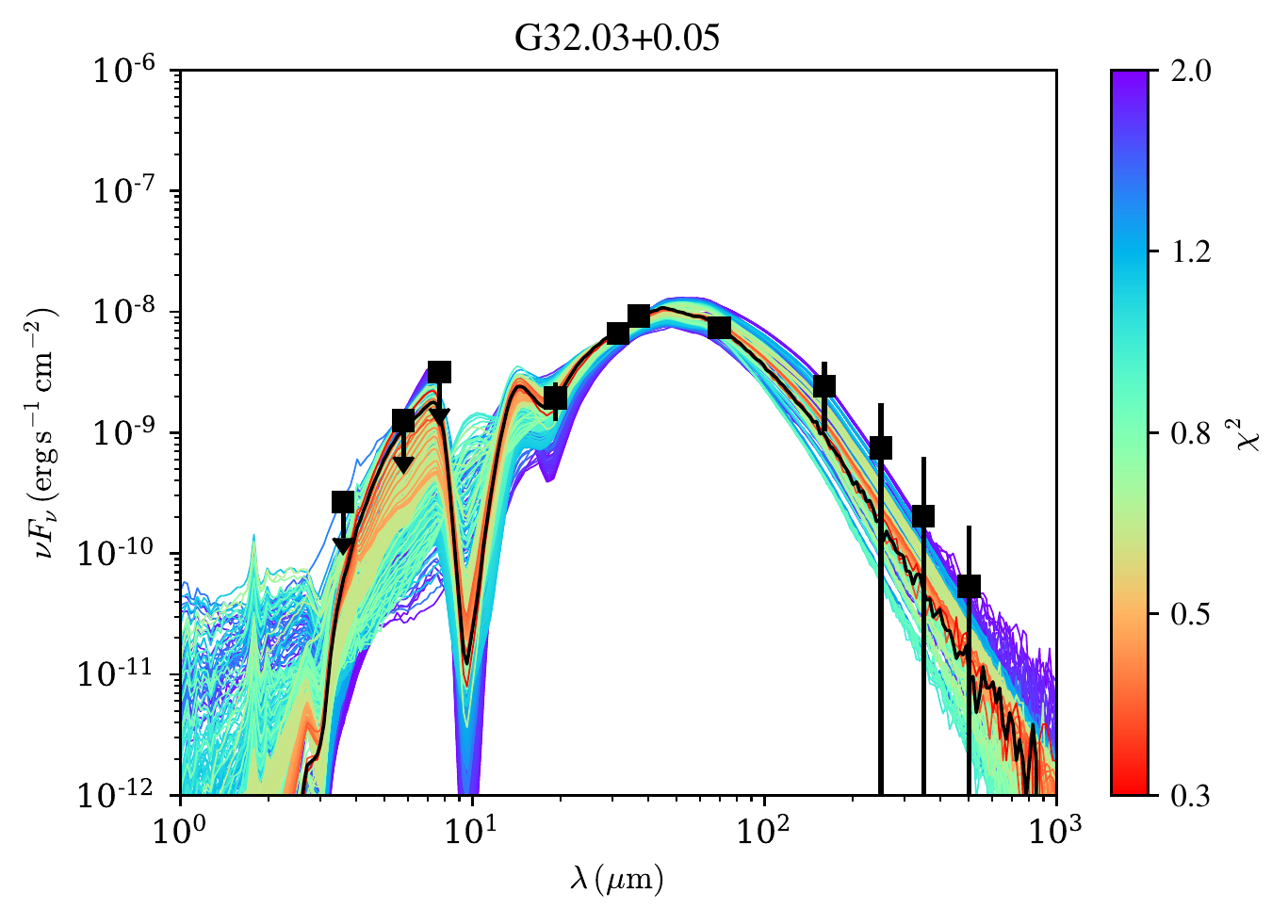}
\includegraphics[width=0.5\textwidth]{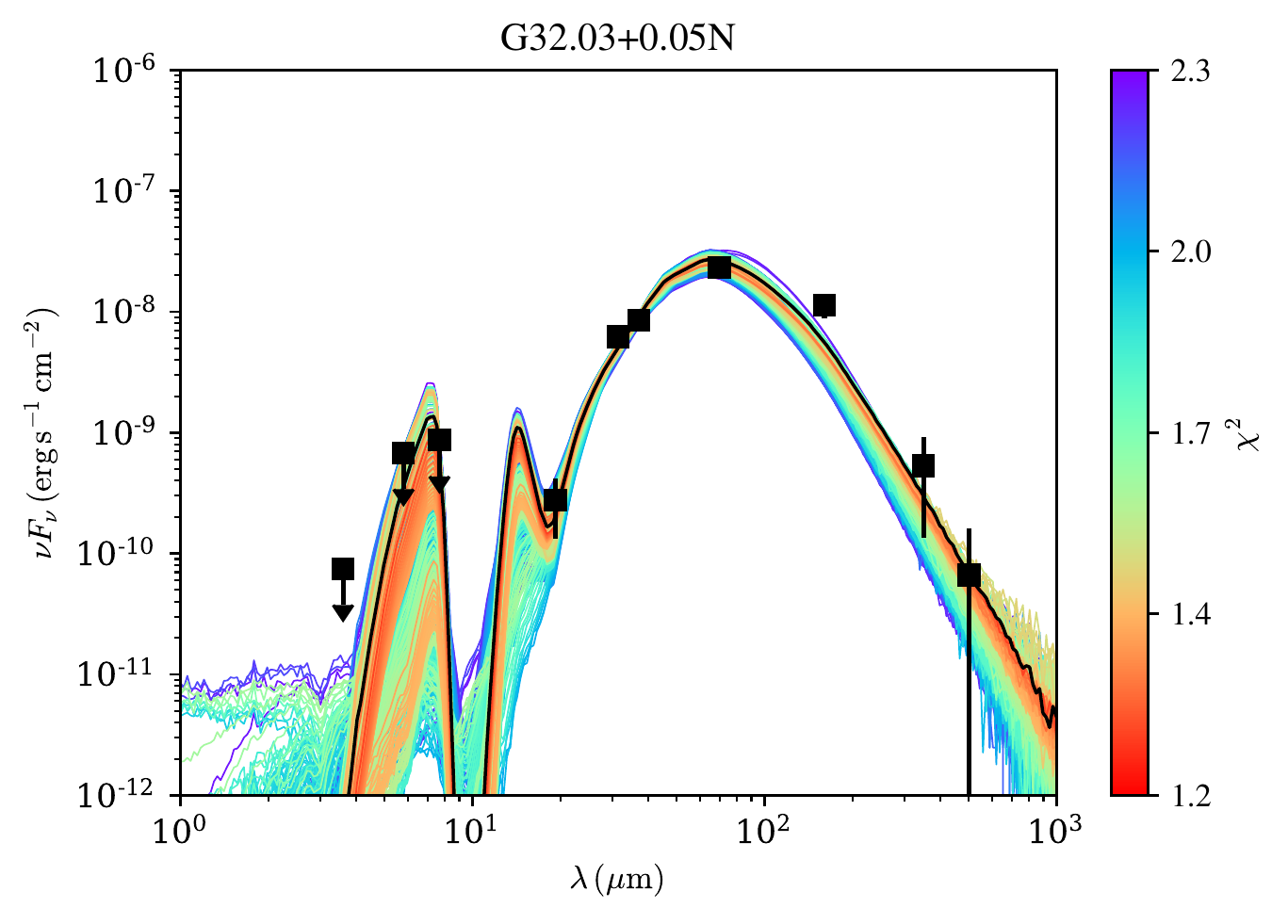}
\includegraphics[width=0.5\textwidth]{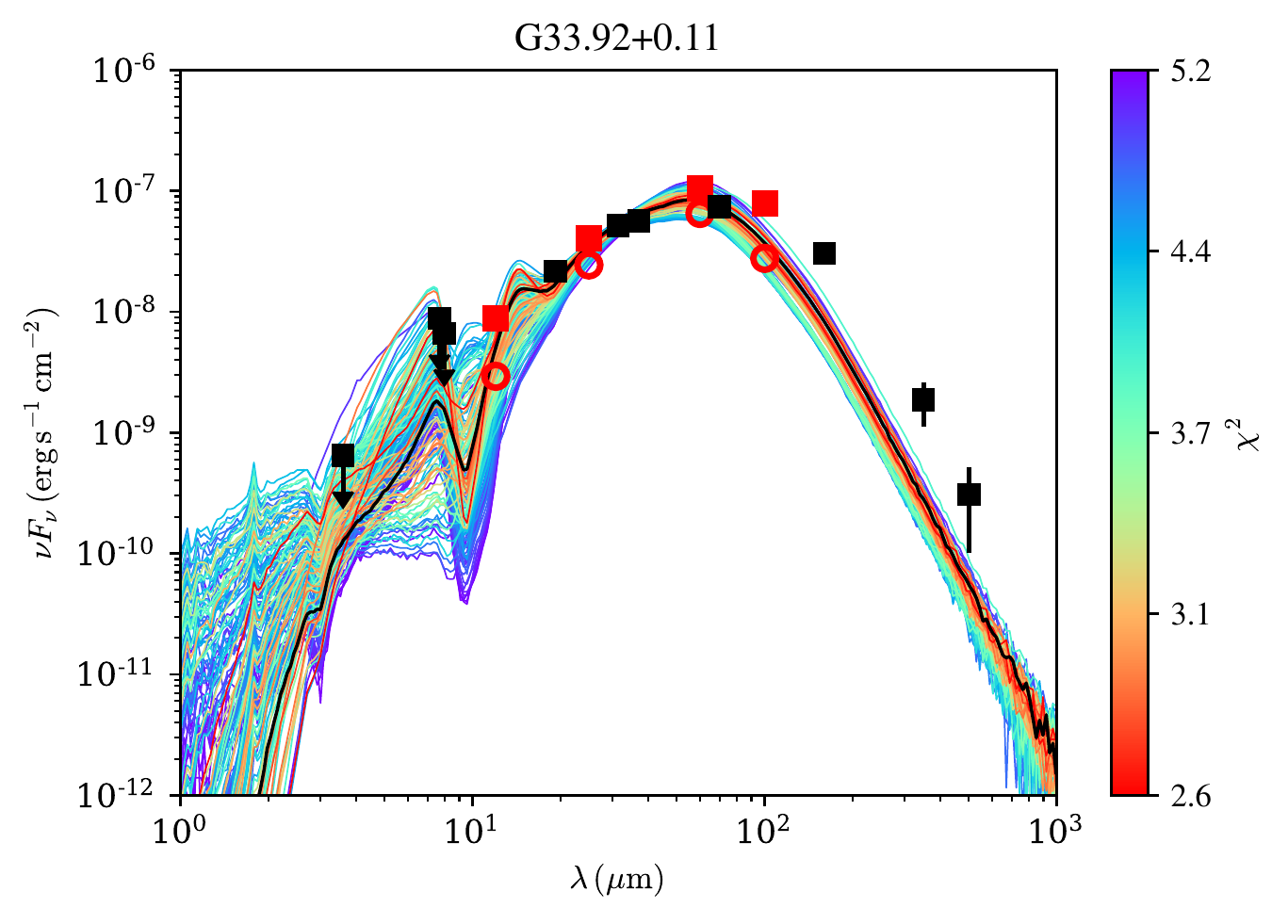}
\caption{
Protostellar SEDs with fluxes evaluated via the fixed aperture method and shown after background subtraction  for each source noted on top of each plot. Flux values are those from Table\,\ref{tab:soma_iv_fluxes} and are represented as black squares. Solid black squares are used in the model fitting, while red squares (i.e., IRAS fluxes) are not. Note that the data at $\lesssim8\,{\rm \mu m}$ are treated as upper limits for SED model fitting (see the text).
The best-fitting protostar model is shown with a black line, while all other good model fits (see the text) are shown with colored lines (red to blue with increasing $\chi^2$). Red empty circles denote the geometric mean prediction for good models at IRAS wavelengths to test for source variability (see the text). \label{fig:sed_1D_results_soma_iv}}
\end{figure*}

\renewcommand{\thefigure}{\arabic{figure}}
\addtocounter{figure}{-1}
\begin{figure*}[!htb]
\includegraphics[width=0.5\textwidth]{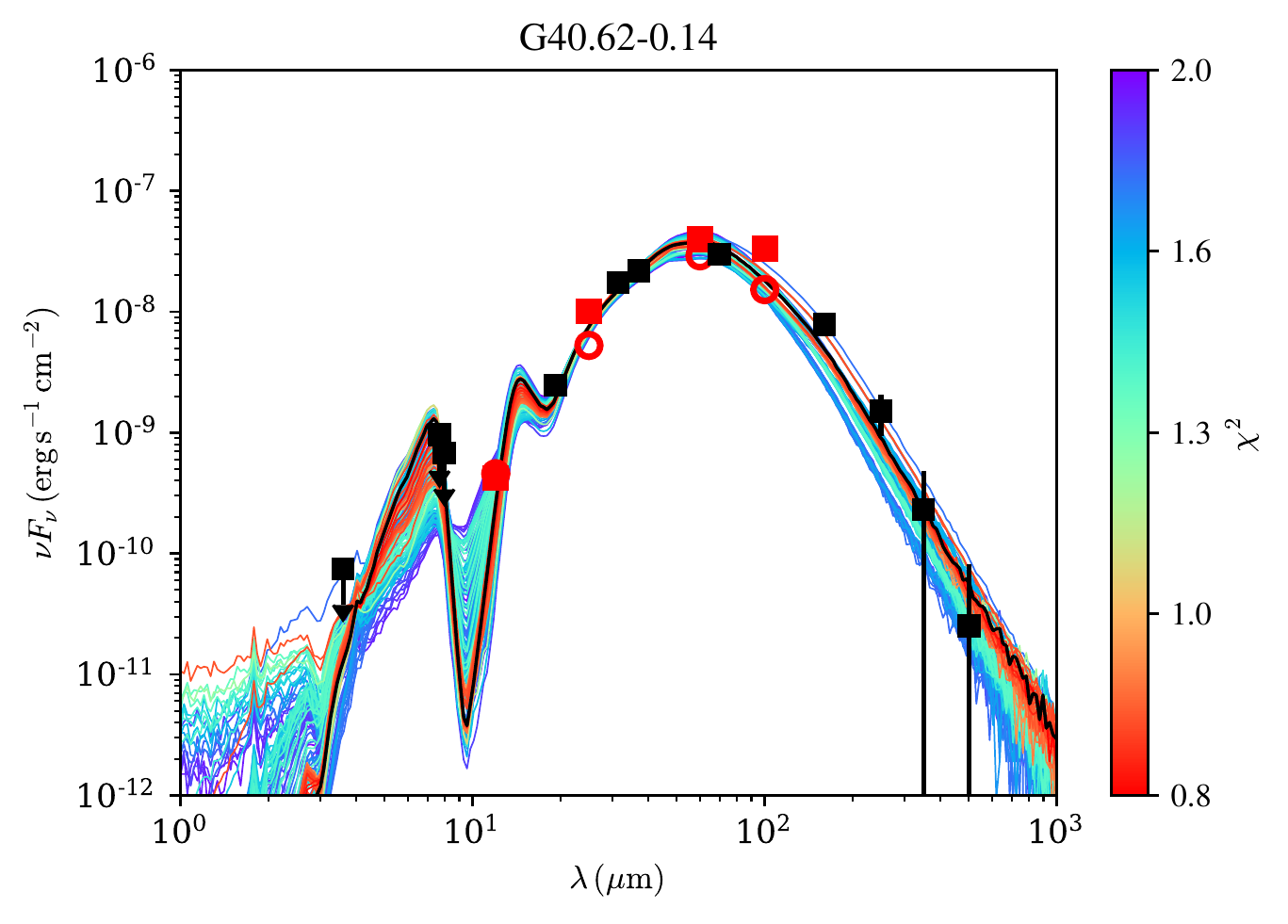}
\includegraphics[width=0.5\textwidth]{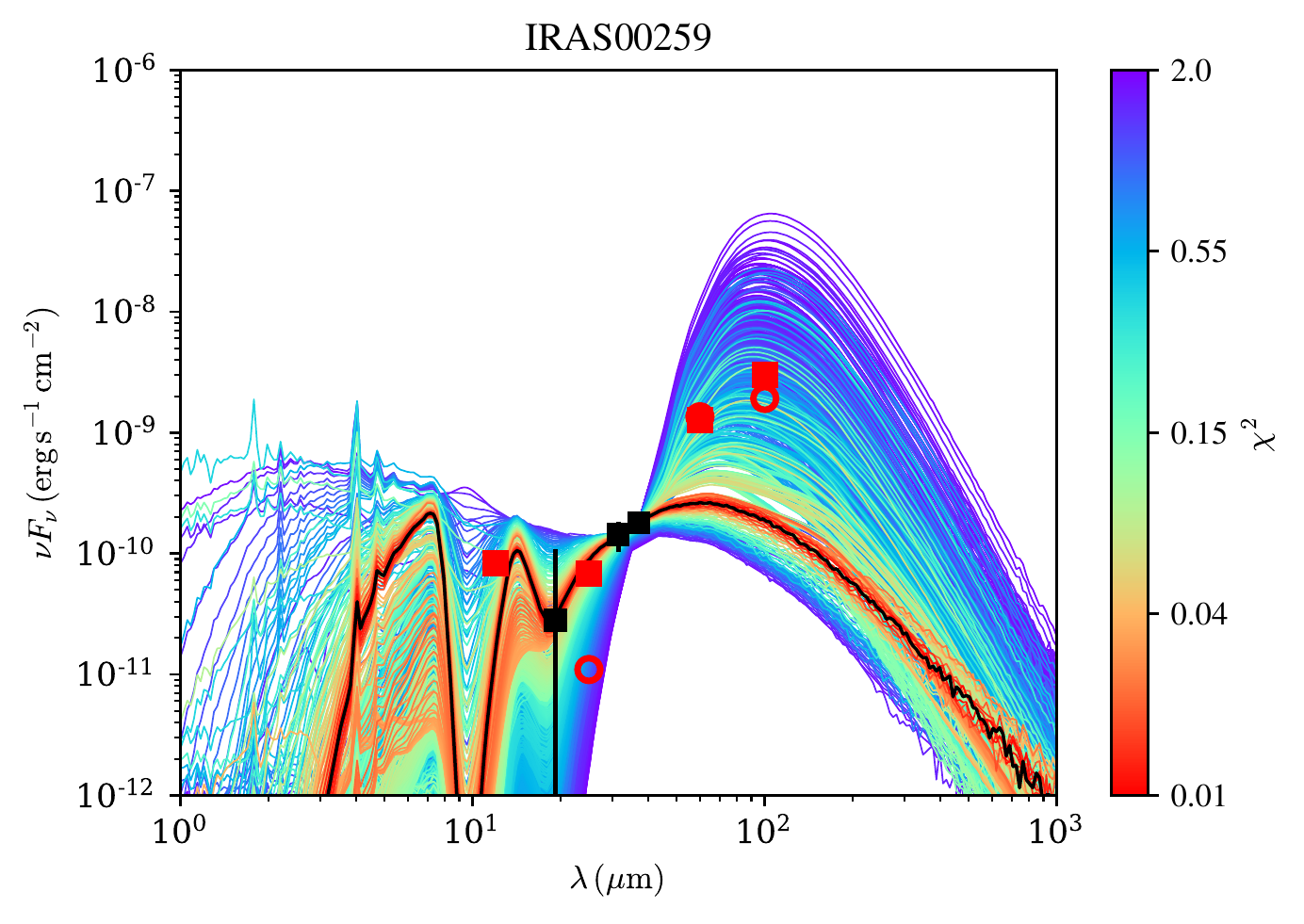}
\includegraphics[width=0.5\textwidth]{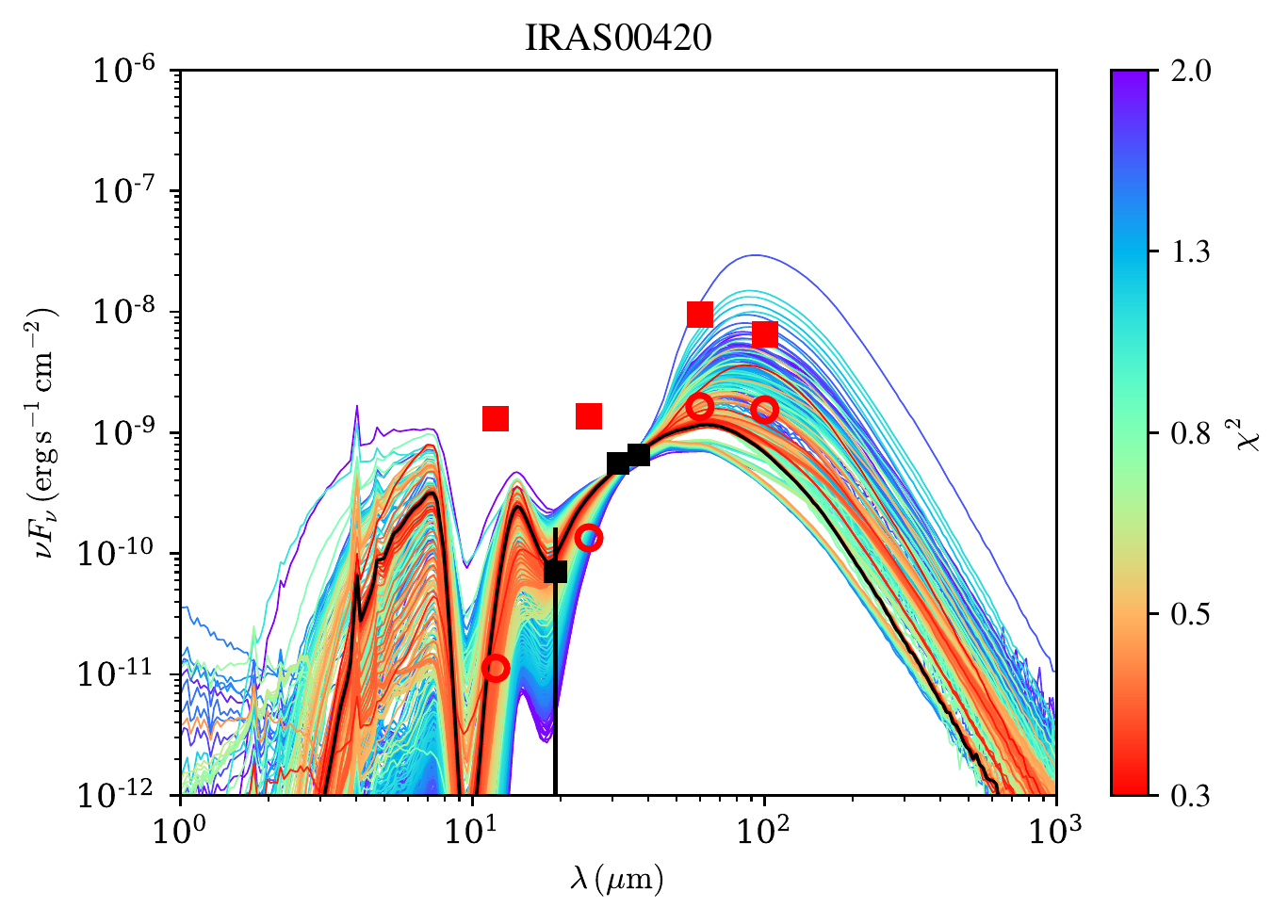}
\includegraphics[width=0.5\textwidth]{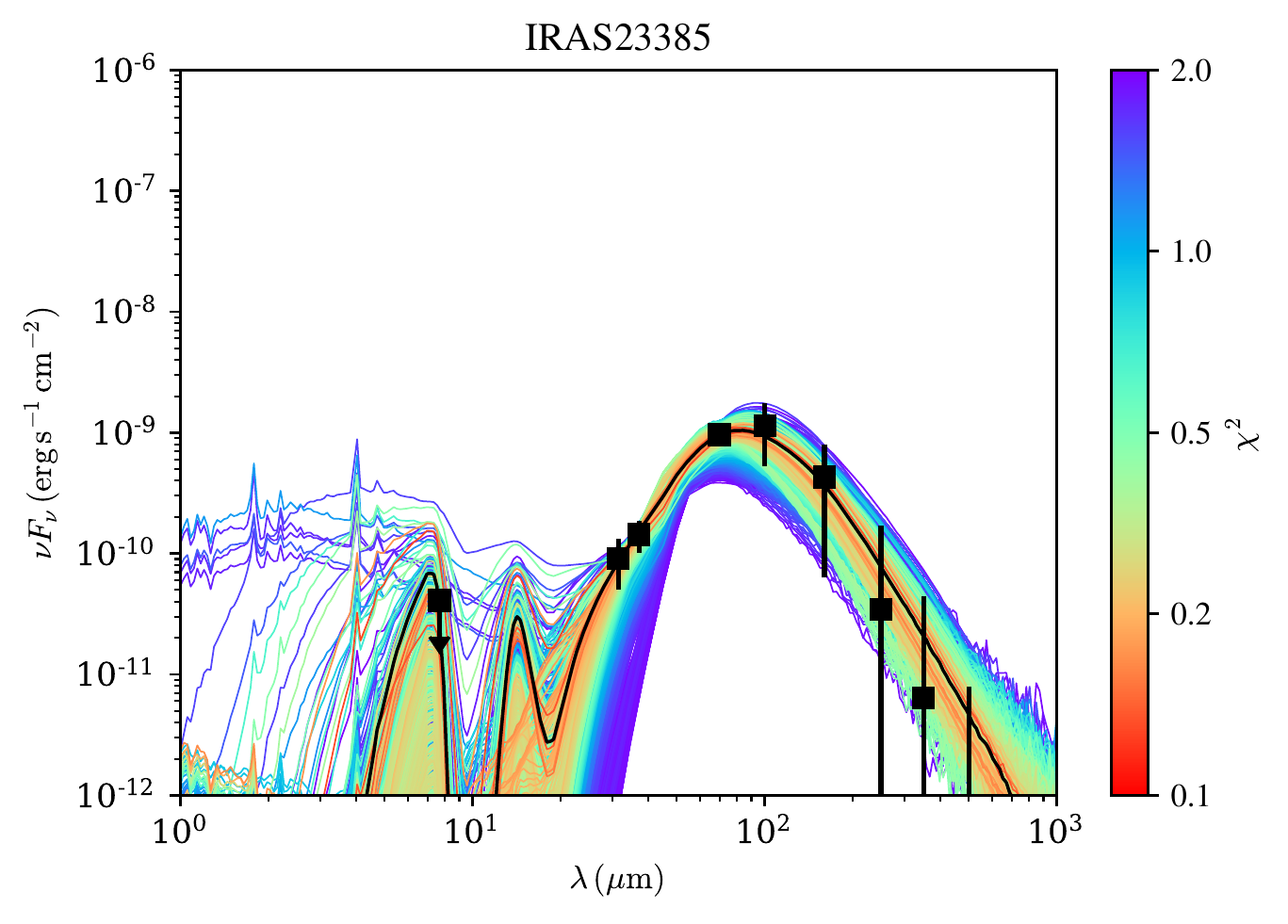}
\includegraphics[width=0.5\textwidth]{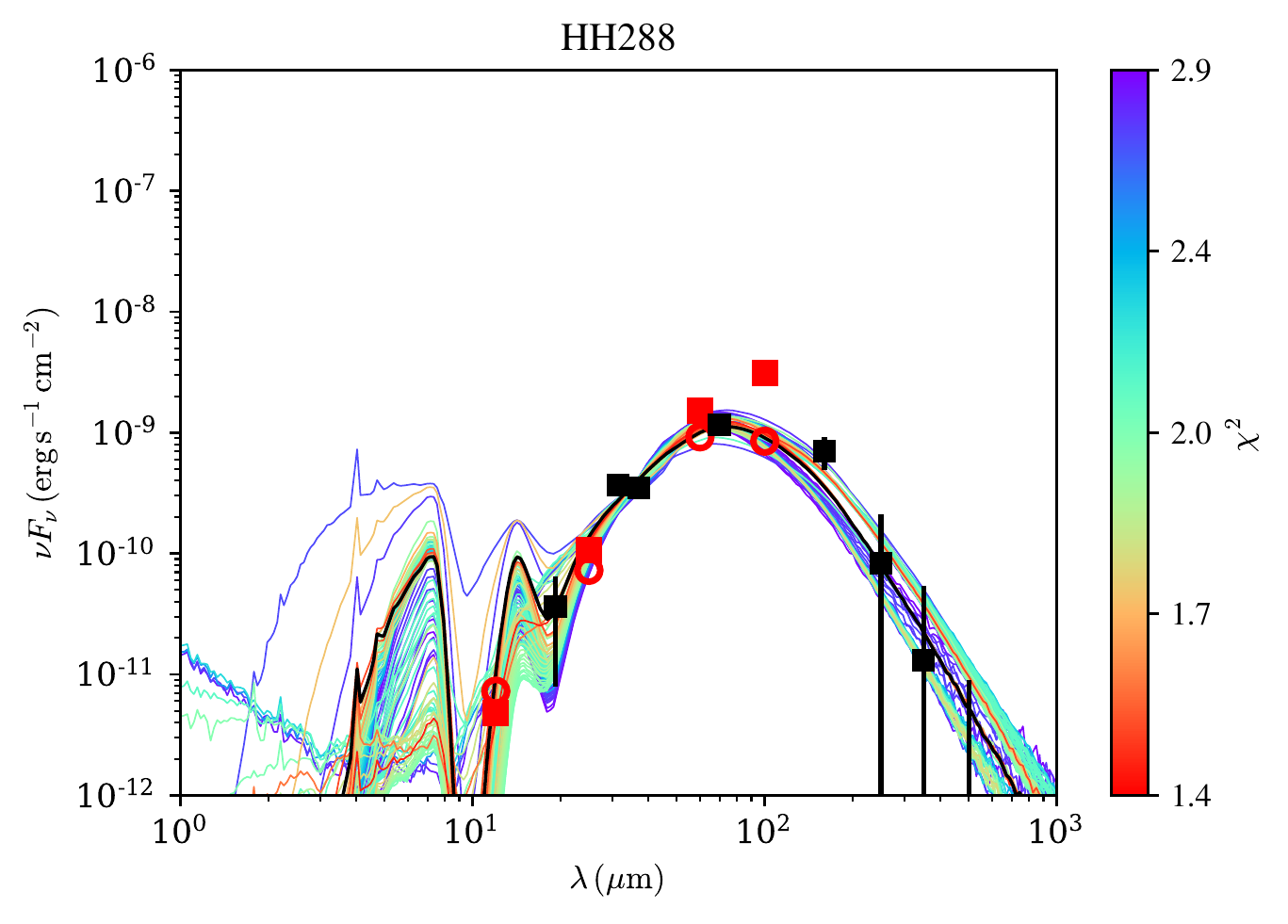}
    \caption{(Continued.)}
\end{figure*}

\begin{figure*}[!htb]
\includegraphics[width=1.0\textwidth]{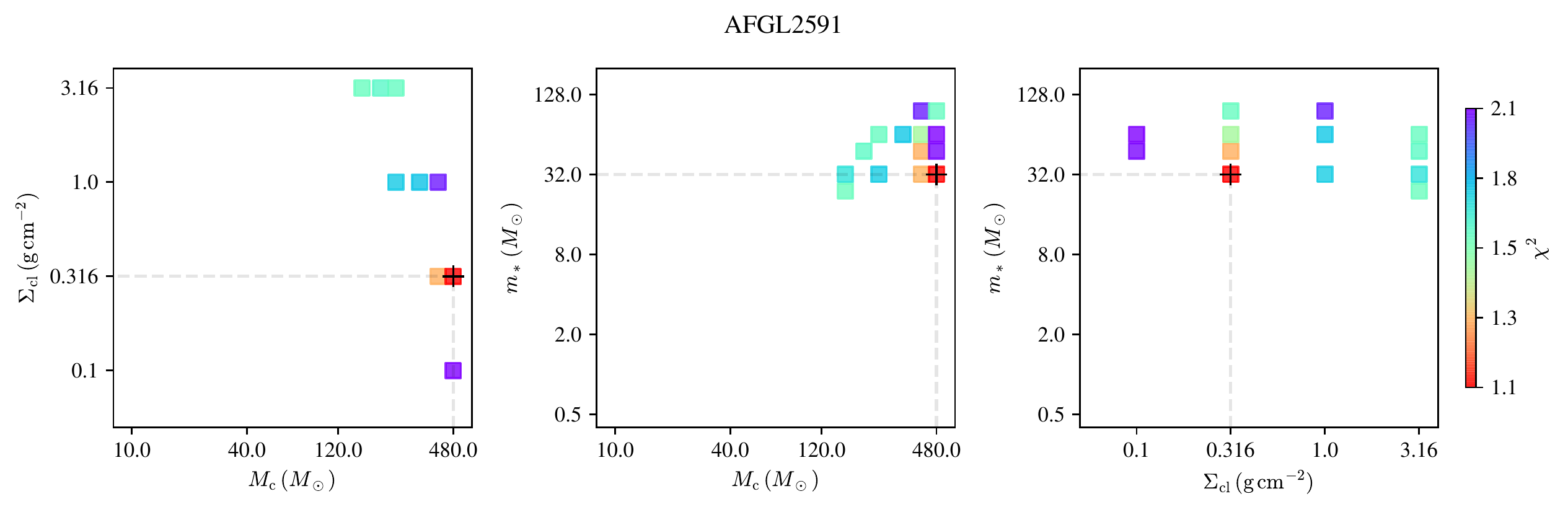}
\includegraphics[width=1.0\textwidth]{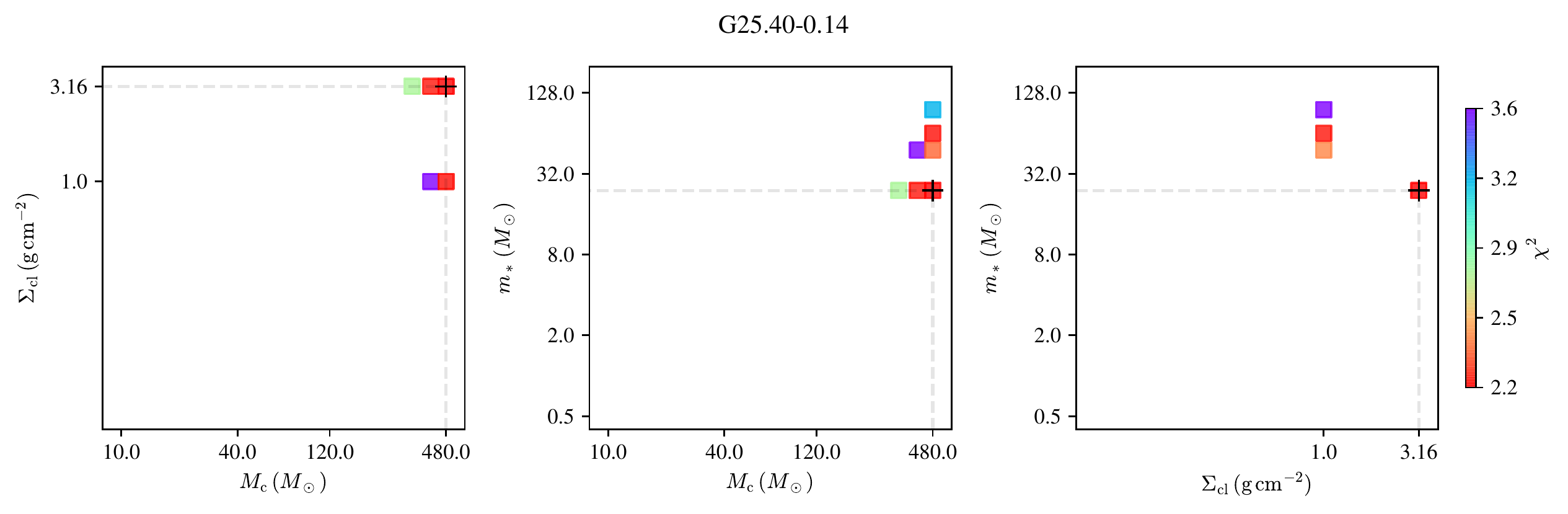}
\includegraphics[width=1.0\textwidth]{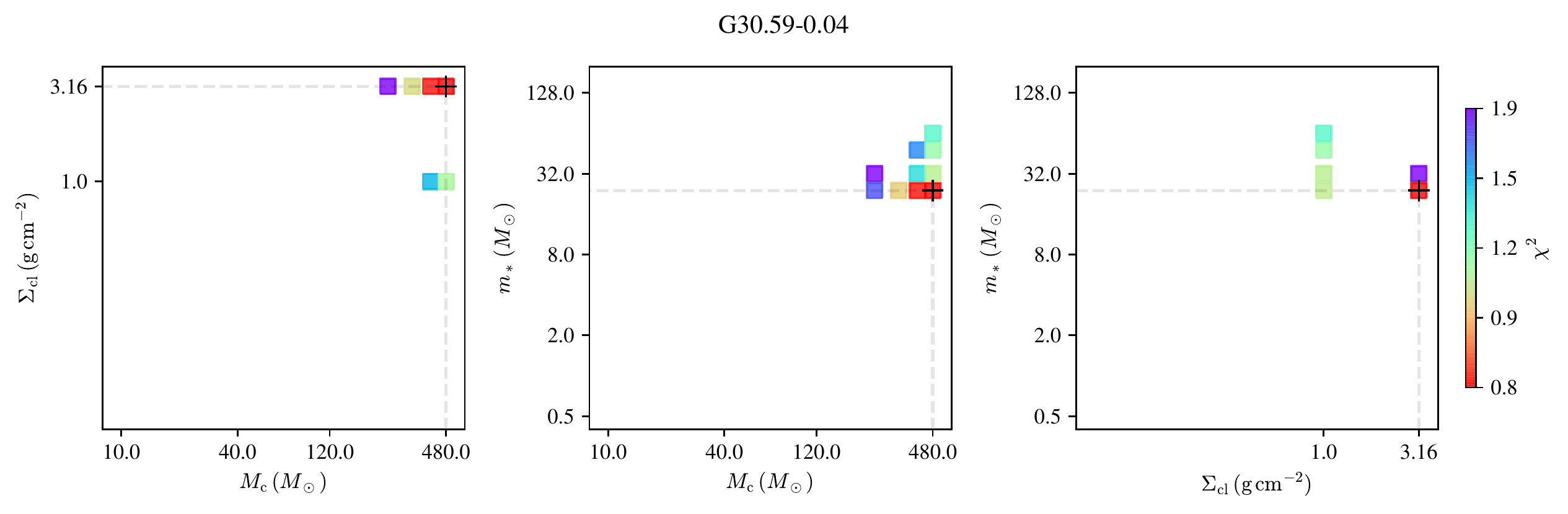}
\caption{Diagrams of the $\chi^{2}$ distribution in the $\Sigma_{\rm cl}$ - $M_{c}$ space (left column), $m_{*}$ - $M_{\rm c}$ space (center column) and $m_{*}$ - $\Sigma_{\rm  cl}$ space (right column) for each source noted on top of each plot. The black cross denotes the best model.
\label{fig:sed_2D_results_soma_iv}}
\end{figure*}

\renewcommand{\thefigure}{\arabic{figure}}
\addtocounter{figure}{-1}
\begin{figure*}[!htb]
\includegraphics[width=1.0\textwidth]{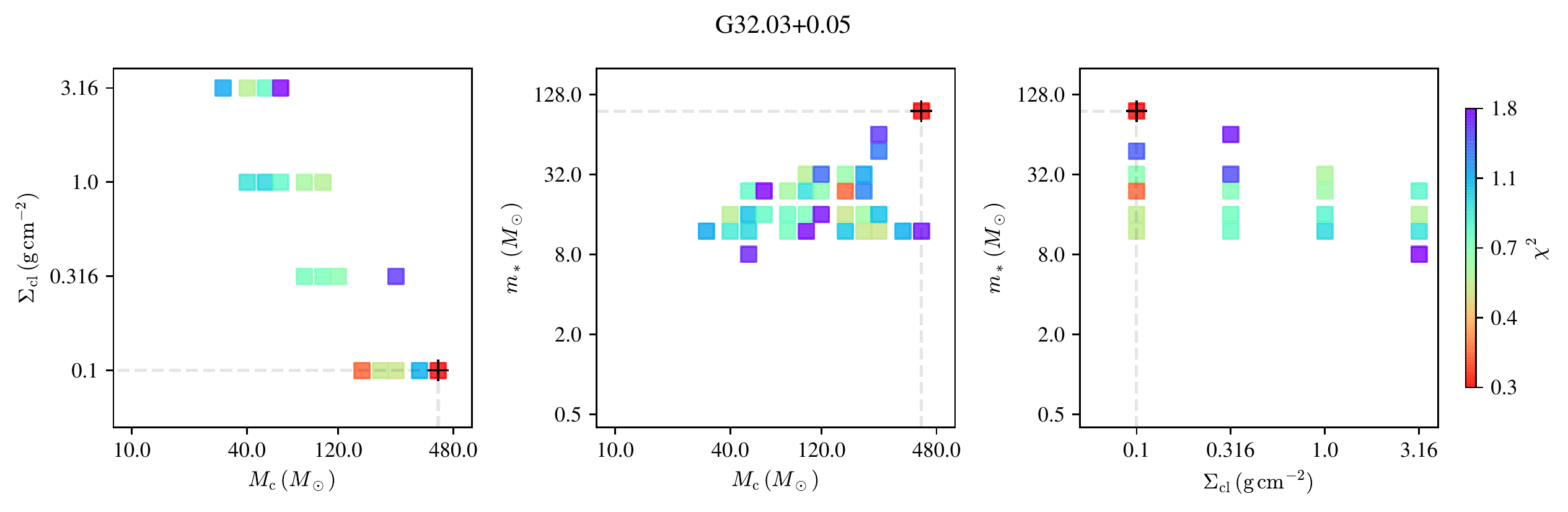}
\includegraphics[width=1.0\textwidth]{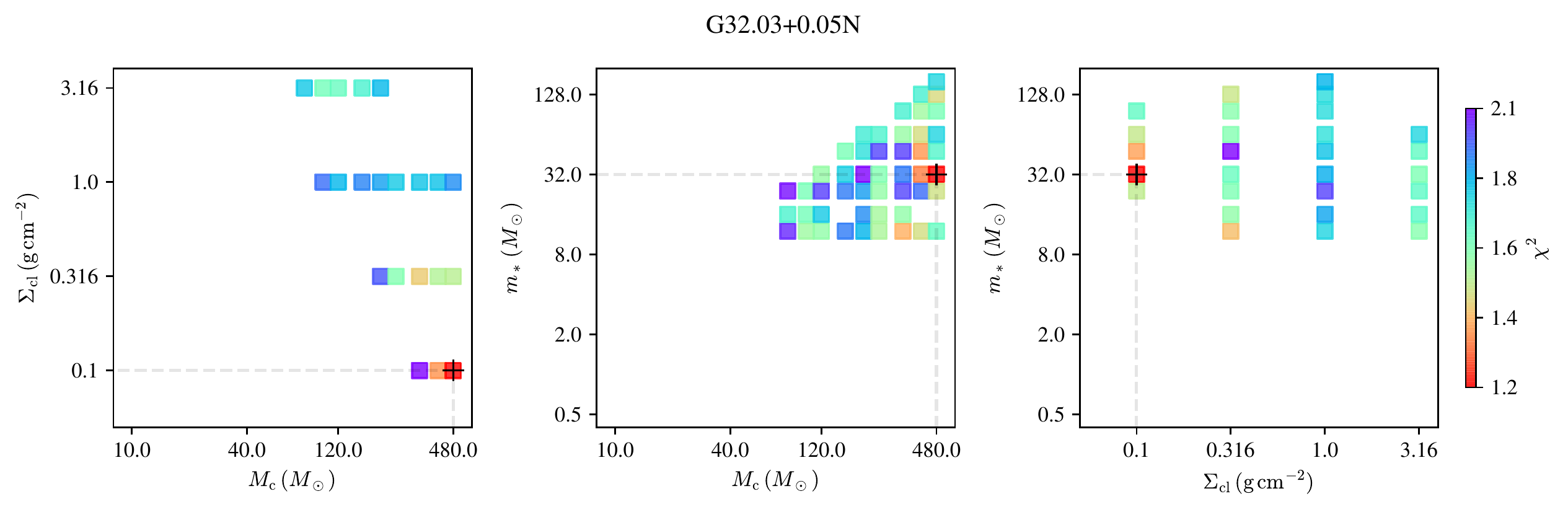}
\includegraphics[width=1.0\textwidth]{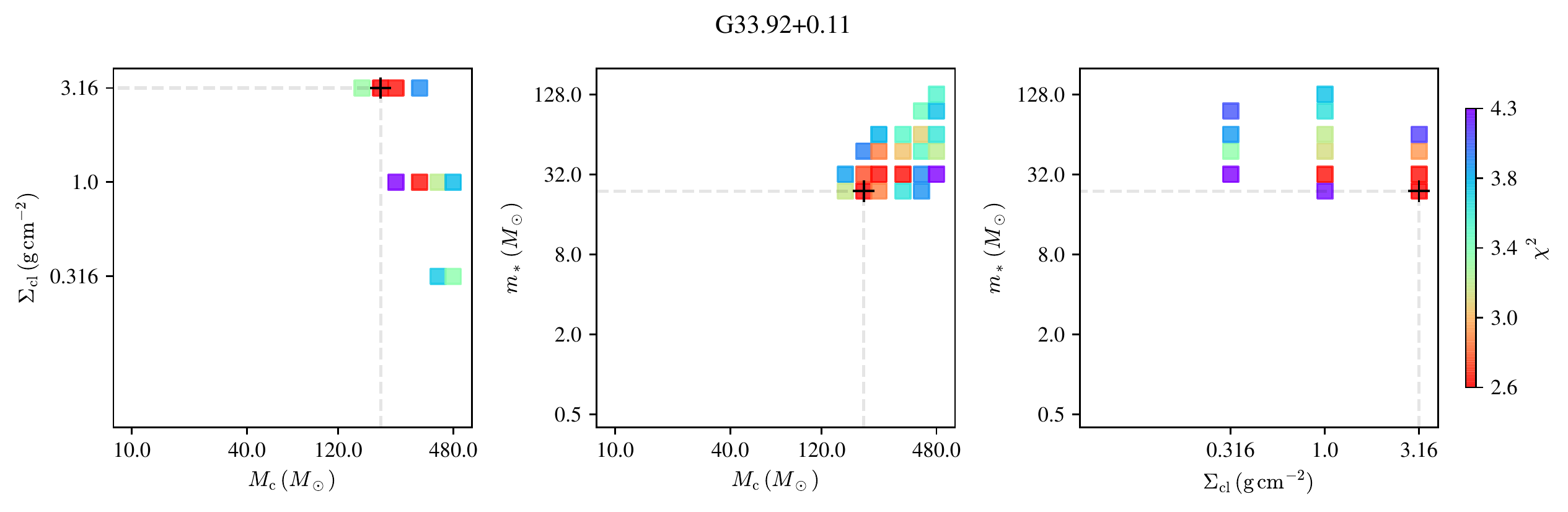}
\includegraphics[width=1.0\textwidth]{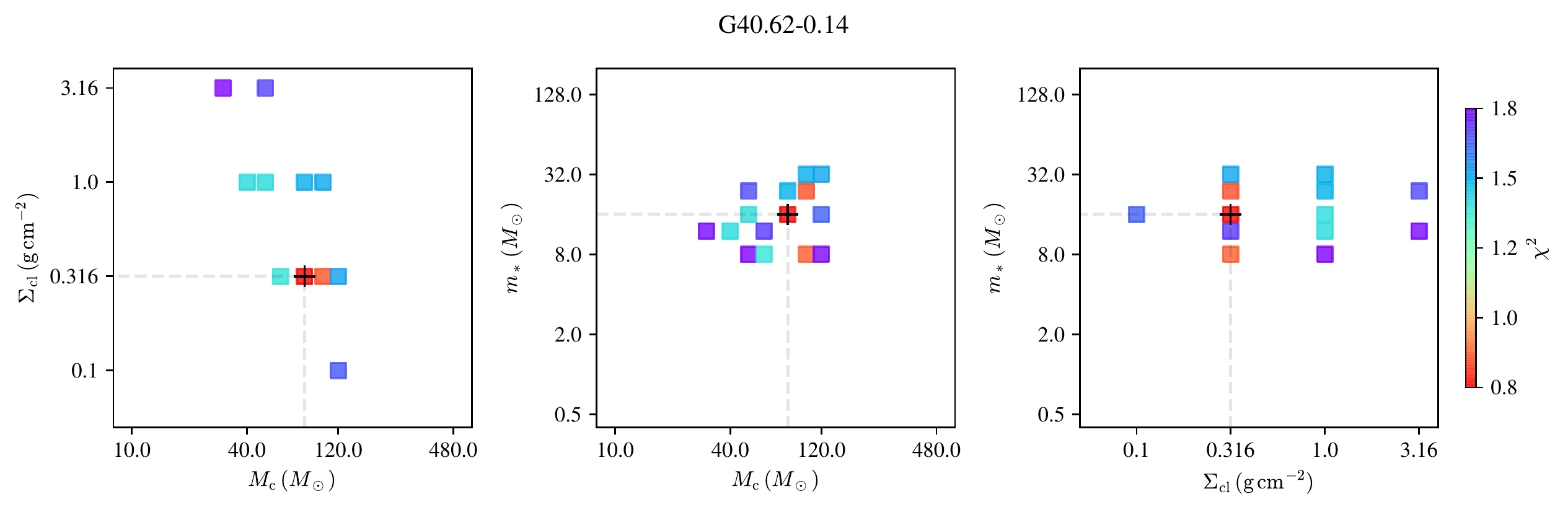}
\caption{(Continued.)}
\end{figure*}

\renewcommand{\thefigure}{\arabic{figure}}
\addtocounter{figure}{-1}
\begin{figure*}[!htb]
\includegraphics[width=1.0\textwidth]{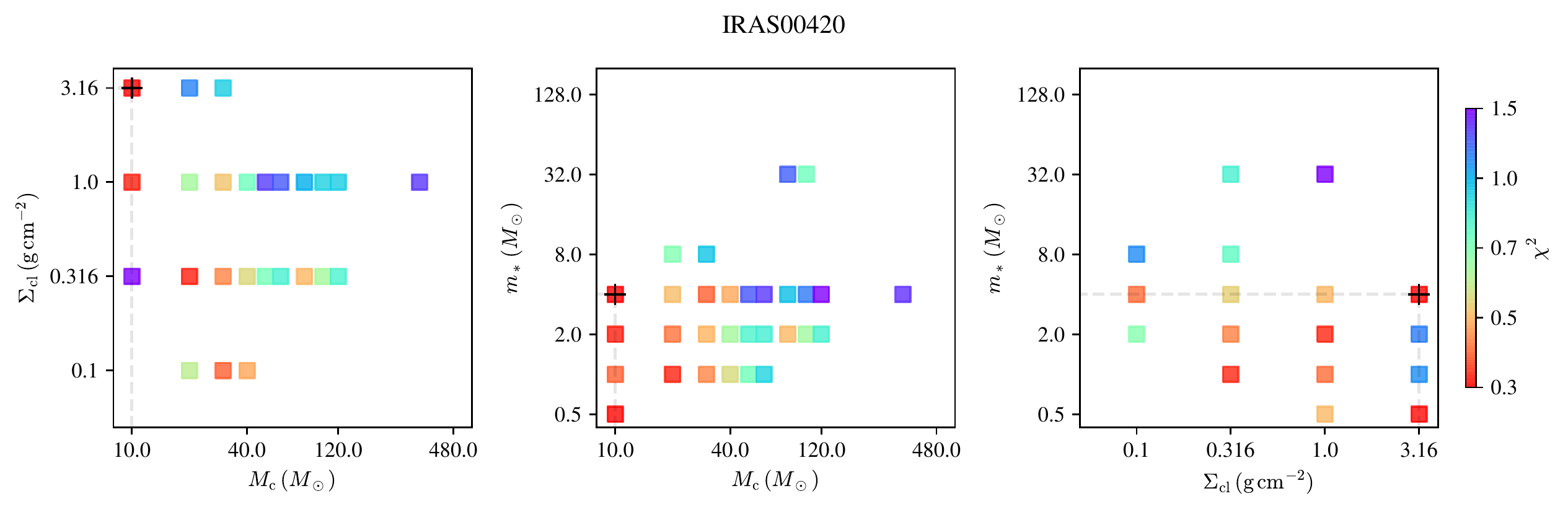}
\includegraphics[width=1.0\textwidth]{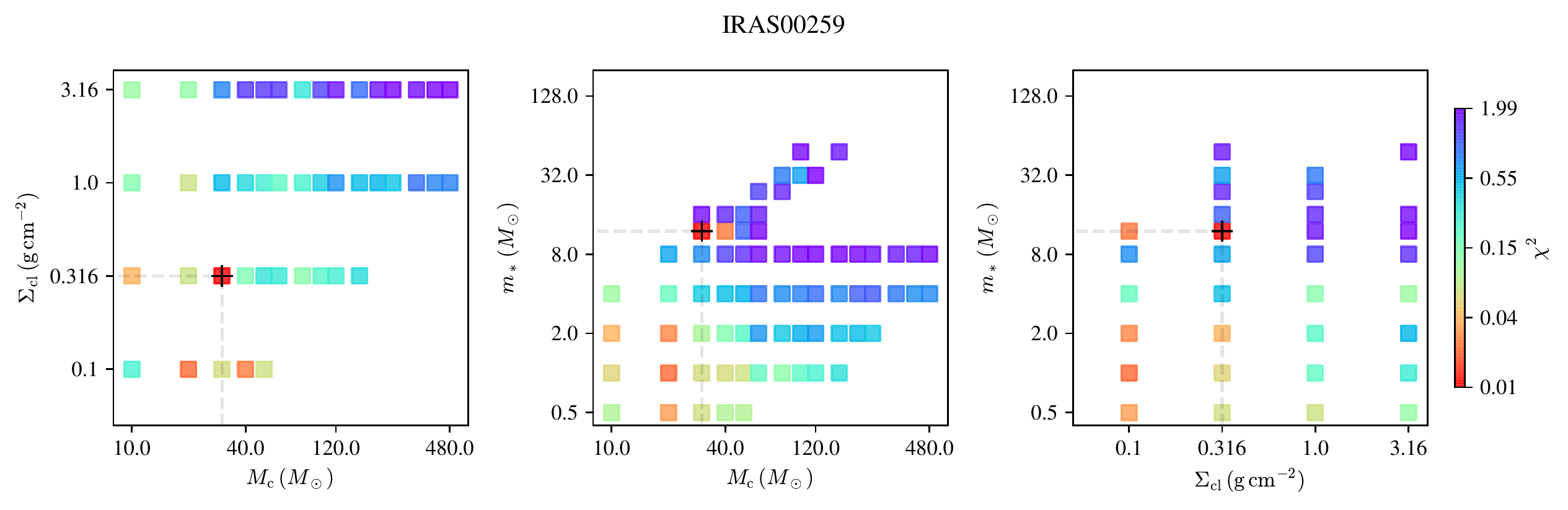}
\includegraphics[width=1.0\textwidth]{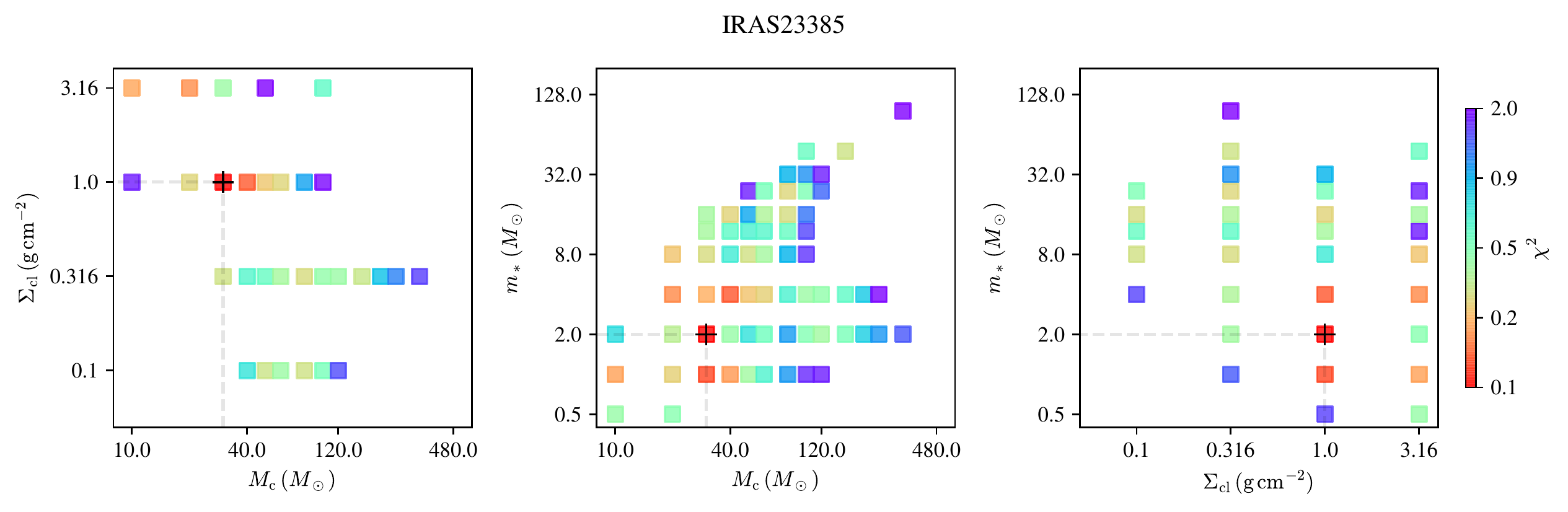}
\includegraphics[width=1.0\textwidth]{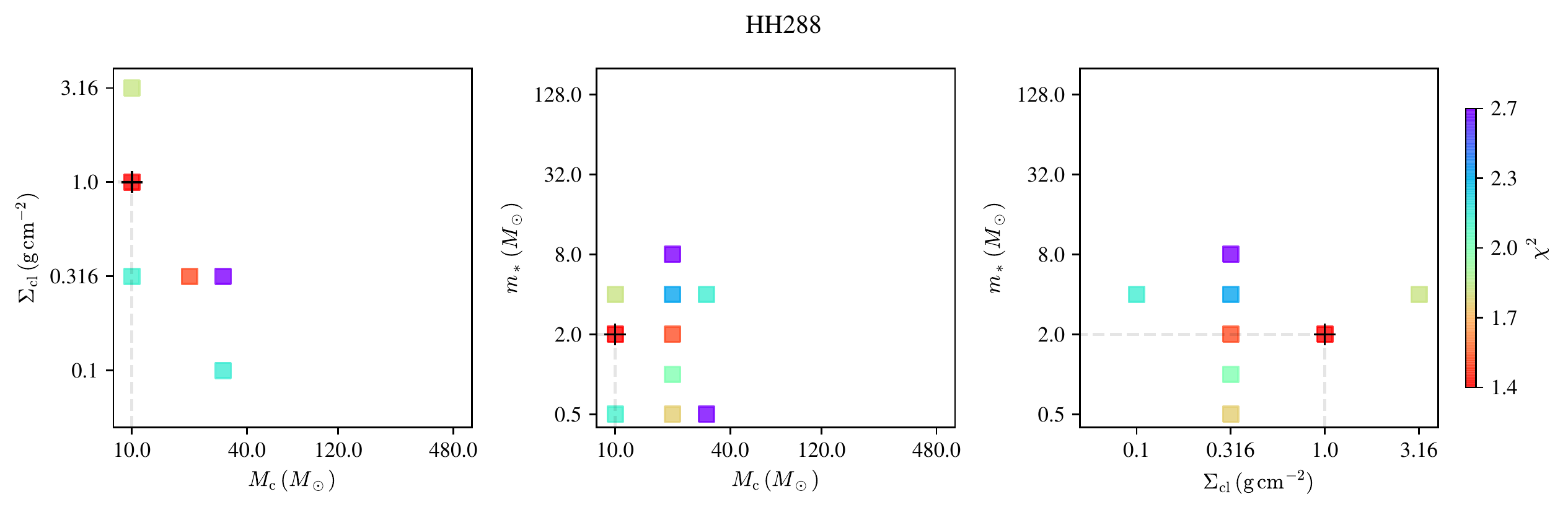}
\caption{(Continued.)}
\end{figure*}

\begin{longrotatetable}
\begin{deluxetable}{lcccccccccccc}
\tabletypesize{\small}
\tablecaption{Parameters of the Best Five Fitted Models and Average and Dispersion of Good Models\label{tab:best_models_soma_iv}} 
\tablewidth{18pt}
\tablehead{
\colhead{Source} &\colhead{$\chi^{2}$} & \colhead{$M_{\rm c}$} & \colhead{$\Sigma_{\rm cl}$} & \colhead{$R_{\rm core}$}  &\colhead{$m_{*}$} & \colhead{$\theta_{\rm view}$} &\colhead{$A_{V}$} & \colhead{$M_{\rm env}$} &\colhead{$\theta_{w,\rm esc}$} & \colhead{$\dot {M}_{\rm disk}$} & \colhead{$L_{\rm bol, iso}$} & \colhead{$L_{\rm bol}$} \\
\colhead{} & \colhead{} & \colhead{($M_\odot$)} & \colhead{(g $\rm cm^{-2}$)} & \colhead{(pc)} & \colhead{($M_{\odot}$)} & \colhead{(\arcdeg)} & \colhead{(mag)} & \colhead{($M_{\odot}$)} & \colhead{(deg)} &\colhead{($M_{\odot}$/yr)} & \colhead{($L_{\odot}$)} & \colhead{($L_{\odot}$)}
}
\startdata
AFGL2591 & 1.06 & 480 & 0.316 & 0.29 & 32 & 13 & 76 & 405.71 & 22 & 3.9$\times10^{-4}$ & 1.3$\times10^{6}$ & 2.0$\times10^{5}$ \\
$d=3.3$ kpc & 1.24 & 400 & 0.316 & 0.26 & 48 & 29 & 86 & 272.62 & 34 & 4.1$\times10^{-4}$ & 1.5$\times10^{6}$ & 4.1$\times10^{5}$ \\
$R_\mathrm{ap}=18.75\arcsec$ & 1.25 & 400 & 0.316 & 0.26 & 32 & 13 & 69 & 317.05 & 25 & 3.6$\times10^{-4}$ & 1.2$\times10^{6}$ & 2.0$\times10^{5}$ \\
$R_\mathrm{ap}=0.30$ pc & 1.41 & 400 & 0.316 & 0.26 & 64 & 39 & 91 & 222.98 & 42 & 4.3$\times10^{-4}$ & 1.4$\times10^{6}$ & 6.6$\times10^{5}$ \\
 & 1.52 & 160 & 3.160 & 0.05 & 24 & 22 & 90 & 114.57 & 23 & 1.4$\times10^{-3}$ & 1.3$\times10^{6}$ & 3.0$\times10^{5}$ \\
Average model & \#74 &  $313^{+159}_{-105}$ & $0.699^{+1.879}_{-0.509}$ & $0.16^{+0.21}_{-0.09}$ & $51^{+23}_{-16}$ & $38 \pm 16$ & $67 \pm 42$ & $178^{+110}_{-68}$ & $37 \pm 8$ & $6.9^{+9.3}_{-4.0}\times10^{-4}$ & $7.0^{+16.8}_{-4.9}\times10^{5}$ & $5.6^{+4.3}_{-2.4}\times10^{5}$\\
\hline
G25.40-0.14 & 2.24 & 480 & 3.160 & 0.09 & 24 & 22 & 0 & 440.54 & 12 & 2.0$\times10^{-3}$ & 2.9$\times10^{5}$ & 2.9$\times10^{5}$ \\
$d=5.7$ kpc & 2.27 & 480 & 1.000 & 0.16 & 64 & 48 & 0 & 324.63 & 32 & 1.2$\times10^{-3}$ & 2.8$\times10^{5}$ & 8.4$\times10^{5}$ \\
$R_\mathrm{ap}=48.50\arcsec$ & 2.28 & 400 & 3.160 & 0.08 & 24 & 22 & 1 & 361.65 & 13 & 1.9$\times10^{-3}$ & 3.0$\times10^{5}$ & 3.0$\times10^{5}$ \\
$R_\mathrm{ap}=1.34$ pc & 2.44 & 480 & 1.000 & 0.16 & 48 & 34 & 7 & 366.96 & 25 & 1.1$\times10^{-3}$ & 3.0$\times10^{5}$ & 5.4$\times10^{5}$ \\
 & 2.92 & 320 & 3.160 & 0.07 & 24 & 29 & 0 & 276.82 & 15 & 1.8$\times10^{-3}$ & 2.7$\times10^{5}$ & 3.1$\times10^{5}$ \\
Average model & \#40 &  $436^{+70}_{-61}$ & $1.678^{+1.318}_{-0.738}$ & $0.12^{+0.05}_{-0.03}$ & $42^{+30}_{-18}$ & $44 \pm 13$ & $7 \pm 17$ & $324^{+67}_{-55}$ & $24 \pm 11$ & $1.5^{+0.4}_{-0.3}\times10^{-3}$ & $2.7^{+0.6}_{-0.5}\times10^{5}$ & $5.4^{+4.6}_{-2.5}\times10^{5}$\\
\hline
G30.59-0.04 & 0.75 & 480 & 3.160 & 0.09 & 24 & 29 & 81 & 440.54 & 12 & 2.0$\times10^{-3}$ & 2.7$\times10^{5}$ & 2.9$\times10^{5}$ \\
$d=11.8$ kpc & 0.77 & 400 & 3.160 & 0.08 & 24 & 39 & 80 & 361.65 & 13 & 1.9$\times10^{-3}$ & 2.6$\times10^{5}$ & 3.0$\times10^{5}$ \\
$R_\mathrm{ap}=13.50\arcsec$ & 0.99 & 320 & 3.160 & 0.07 & 24 & 71 & 57 & 276.82 & 15 & 1.8$\times10^{-3}$ & 2.1$\times10^{5}$ & 3.1$\times10^{5}$ \\
$R_\mathrm{ap}=0.77$ pc & 1.07 & 480 & 1.000 & 0.16 & 24 & 34 & 3 & 433.43 & 15 & 8.2$\times10^{-4}$ & 1.7$\times10^{5}$ & 2.1$\times10^{5}$ \\
 & 1.08 & 480 & 1.000 & 0.16 & 32 & 39 & 36 & 414.30 & 19 & 9.3$\times10^{-4}$ & 2.0$\times10^{5}$ & 3.0$\times10^{5}$ \\
Average model & \#180 &  $409^{+96}_{-78}$ & $1.584^{+1.204}_{-0.684}$ & $0.12^{+0.05}_{-0.04}$ & $31^{+13}_{-9}$ & $63 \pm 18$ & $45 \pm 32$ & $338^{+91}_{-71}$ & $19 \pm 6$ & $1.2^{+0.5}_{-0.4}\times10^{-3}$ & $2.0^{+0.4}_{-0.4}\times10^{5}$ & $3.5^{+1.7}_{-1.1}\times10^{5}$\\
\hline
G32.03+0.05 & 0.28 & 400 & 0.100 & 0.47 & 96 & 86 & 103 & 45.80 & 76 & 8.3$\times10^{-5}$ & 3.9$\times10^{4}$ & 1.2$\times10^{6}$ \\
$d=5.5$ kpc & 0.36 & 160 & 0.100 & 0.29 & 24 & 51 & 71 & 86.57 & 45 & 8.5$\times10^{-5}$ & 2.8$\times10^{4}$ & 7.8$\times10^{4}$ \\
$R_\mathrm{ap}=22.75\arcsec$ & 0.56 & 240 & 0.100 & 0.36 & 12 & 29 & 21 & 210.90 & 19 & 8.5$\times10^{-5}$ & 1.7$\times10^{4}$ & 2.0$\times10^{4}$ \\
$R_\mathrm{ap}=0.61$ pc & 0.56 & 160 & 0.100 & 0.29 & 16 & 62 & 21 & 115.88 & 32 & 8.1$\times10^{-5}$ & 1.5$\times10^{4}$ & 3.3$\times10^{4}$ \\
 & 0.57 & 200 & 0.100 & 0.33 & 12 & 62 & 1 & 174.14 & 20 & 8.0$\times10^{-5}$ & 1.3$\times10^{4}$ & 2.0$\times10^{4}$ \\
Average model & \#398 &  $148^{+139}_{-72}$ & $0.222^{+0.475}_{-0.151}$ & $0.19^{+0.27}_{-0.11}$ & $20^{+15}_{-8}$ & $62 \pm 18$ & $52 \pm 60$ & $73^{+128}_{-46}$ & $39 \pm 17$ & $1.4^{+1.3}_{-0.7}\times10^{-4}$ & $2.6^{+5.3}_{-1.7}\times10^{4}$ & $6.7^{+13.2}_{-4.4}\times10^{4}$\\
\hline
G32.03+0.05N & 1.17 & 480 & 0.100 & 0.51 & 32 & 29 & 251 & 390.89 & 26 & 1.6$\times10^{-4}$ & 1.5$\times10^{5}$ & 1.6$\times10^{5}$ \\
$d=5.5$ kpc & 1.30 & 400 & 0.100 & 0.47 & 32 & 34 & 230 & 304.06 & 29 & 1.5$\times10^{-4}$ & 1.1$\times10^{5}$ & 1.6$\times10^{5}$ \\
$R_\mathrm{ap}=13.50\arcsec$ & 1.33 & 400 & 0.100 & 0.47 & 48 & 44 & 272 & 248.45 & 39 & 1.6$\times10^{-4}$ & 1.8$\times10^{5}$ & 3.6$\times10^{5}$ \\
$R_\mathrm{ap}=0.36$ pc & 1.37 & 320 & 0.316 & 0.23 & 12 & 22 & 78 & 293.02 & 13 & 2.2$\times10^{-4}$ & 3.6$\times10^{4}$ & 4.0$\times10^{4}$ \\
 & 1.45 & 480 & 0.316 & 0.29 & 128 & 22 & 367 & 91.15 & 67 & 3.8$\times10^{-4}$ & 4.9$\times10^{6}$ & 2.0$\times10^{6}$ \\
Average model & \#848 &  $245^{+187}_{-106}$ & $0.465^{+1.063}_{-0.324}$ & $0.17^{+0.22}_{-0.10}$ & $31^{+36}_{-17}$ & $60 \pm 20$ & $174 \pm 103$ & $137^{+128}_{-66}$ & $35 \pm 16$ & $3.7^{+4.1}_{-1.9}\times10^{-4}$ & $9.4^{+28.6}_{-7.1}\times10^{4}$ & $2.1^{+5.2}_{-1.5}\times10^{5}$\\
\hline
G33.92+0.11 & 2.60 & 200 & 3.160 & 0.06 & 24 & 29 & 46 & 155.51 & 20 & 1.5$\times10^{-3}$ & 2.6$\times10^{5}$ & 3.1$\times10^{5}$ \\
$d=7.1$ kpc & 2.62 & 240 & 3.160 & 0.06 & 32 & 29 & 92 & 175.32 & 23 & 1.9$\times10^{-3}$ & 4.5$\times10^{5}$ & 5.0$\times10^{5}$ \\
$R_\mathrm{ap}=25.25\arcsec$ & 2.62 & 320 & 1.000 & 0.13 & 32 & 29 & 17 & 252.35 & 24 & 8.2$\times10^{-4}$ & 2.0$\times10^{5}$ & 2.7$\times10^{5}$ \\
$R_\mathrm{ap}=0.87$ pc & 2.76 & 200 & 3.160 & 0.06 & 32 & 34 & 58 & 140.13 & 25 & 1.7$\times10^{-3}$ & 2.8$\times10^{5}$ & 4.6$\times10^{5}$ \\
 & 2.88 & 240 & 3.160 & 0.06 & 48 & 39 & 76 & 138.07 & 33 & 2.1$\times10^{-3}$ & 3.6$\times10^{5}$ & 7.5$\times10^{5}$ \\
Average model & \#154 &  $339^{+155}_{-106}$ & $0.934^{+1.639}_{-0.595}$ & $0.14^{+0.14}_{-0.07}$ & $52^{+32}_{-20}$ & $54 \pm 17$ & $21 \pm 26$ & $200^{+96}_{-65}$ & $35 \pm 11$ & $9.0^{+7.6}_{-4.1}\times10^{-4}$ & $2.0^{+0.7}_{-0.5}\times10^{5}$ & $6.3^{+5.2}_{-2.8}\times10^{5}$\\
\hline
G40.62-0.14 & 0.83 & 80 & 0.316 & 0.12 & 16 & 51 & 89 & 41.63 & 42 & 1.5$\times10^{-4}$ & 1.2$\times10^{4}$ & 4.2$\times10^{4}$ \\
$d=2.2$ kpc & 0.90 & 100 & 0.316 & 0.13 & 24 & 68 & 85 & 35.52 & 54 & 1.5$\times10^{-4}$ & 1.1$\times10^{4}$ & 8.8$\times10^{4}$ \\
$R_\mathrm{ap}=13.50\arcsec$ & 0.91 & 100 & 0.316 & 0.13 & 8 & 29 & 12 & 83.08 & 20 & 1.3$\times10^{-4}$ & 7.5$\times10^{3}$ & 1.0$\times10^{4}$ \\
$R_\mathrm{ap}=0.14$ pc & 1.35 & 60 & 0.316 & 0.10 & 8 & 55 & 0 & 43.41 & 28 & 1.1$\times10^{-4}$ & 5.8$\times10^{3}$ & 1.2$\times10^{4}$ \\
 & 1.37 & 50 & 1.000 & 0.05 & 16 & 62 & 60 & 16.19 & 48 & 2.8$\times10^{-4}$ & 9.2$\times10^{3}$ & 6.7$\times10^{4}$ \\
Average model & \#142 &  $64^{+34}_{-22}$ & $0.615^{+1.011}_{-0.382}$ & $0.08^{+0.07}_{-0.04}$ & $15^{+8}_{-5}$ & $69 \pm 14$ & $59 \pm 39$ & $24^{+27}_{-13}$ & $44 \pm 11$ & $2.1^{+1.9}_{-1.0}\times10^{-4}$ & $9.2^{+3.5}_{-2.5}\times10^{3}$ & $4.9^{+7.0}_{-2.9}\times10^{4}$\\
\hline
IRAS00259 & 0.01 & 30 & 0.316 & 0.07 & 12 & 39 & 169 & 0.76 & 81 & 2.2$\times10^{-5}$ & 2.2$\times10^{4}$ & 1.2$\times10^{4}$ \\
$d=2.5$ kpc & 0.02 & 20 & 0.100 & 0.10 & 1 & 29 & 49 & 17.28 & 20 & 1.3$\times10^{-5}$ & 1.2$\times10^{2}$ & 1.5$\times10^{2}$ \\
$R_\mathrm{ap}=7.00\arcsec$ & 0.03 & 40 & 0.100 & 0.15 & 12 & 13 & 162 & 2.10 & 82 & 9.5$\times10^{-6}$ & 2.1$\times10^{4}$ & 1.1$\times10^{4}$ \\
$R_\mathrm{ap}=0.08$ pc & 0.03 & 20 & 0.100 & 0.10 & 2 & 48 & 35 & 14.56 & 30 & 1.7$\times10^{-5}$ & 1.1$\times10^{2}$ & 1.9$\times10^{2}$ \\
 & 0.03 & 20 & 0.100 & 0.10 & 0 & 22 & 0 & 18.84 & 13 & 9.6$\times10^{-6}$ & 8.6$\times10^{1}$ & 9.0$\times10^{1}$ \\
Average model & \#2427 &  $45^{+57}_{-25}$ & $0.517^{+1.012}_{-0.342}$ & $0.07^{+0.07}_{-0.03}$ & $3^{+6}_{-2}$ & $55 \pm 22$ & $230 \pm 206$ & $25^{+65}_{-18}$ & $27 \pm 20$ & $8.6^{+15.0}_{-5.4}\times10^{-5}$ & $1.7^{+9.4}_{-1.5}\times10^{3}$ & $2.5^{+12.5}_{-2.0}\times10^{3}$\\
\hline
IRAS00420 & 0.33 & 10 & 3.160 & 0.01 & 4 & 51 & 160 & 1.65 & 56 & 1.9$\times10^{-4}$ & 1.9$\times10^{3}$ & 1.9$\times10^{3}$ \\
$d=2.2$ kpc & 0.33 & 10 & 3.160 & 0.01 & 0 & 13 & 170 & 9.05 & 14 & 1.1$\times10^{-4}$ & 6.8$\times10^{3}$ & 9.1$\times10^{2}$ \\
$R_\mathrm{ap}=7.00\arcsec$ & 0.35 & 10 & 1.000 & 0.02 & 2 & 22 & 153 & 5.33 & 39 & 7.5$\times10^{-5}$ & 3.2$\times10^{3}$ & 7.6$\times10^{2}$ \\
$R_\mathrm{ap}=0.07$ pc & 0.35 & 20 & 0.316 & 0.06 & 1 & 22 & 10 & 17.54 & 18 & 3.2$\times10^{-5}$ & 3.3$\times10^{2}$ & 3.9$\times10^{2}$ \\
 & 0.40 & 30 & 0.100 & 0.13 & 4 & 51 & 80 & 20.50 & 33 & 2.7$\times10^{-5}$ & 4.1$\times10^{2}$ & 7.7$\times10^{2}$ \\
Average model & \#429 &  $32^{+31}_{-16}$ & $0.365^{+0.619}_{-0.230}$ & $0.07^{+0.07}_{-0.03}$ & $3^{+3}_{-2}$ & $49 \pm 23$ & $103 \pm 92$ & $17^{+38}_{-12}$ & $32 \pm 18$ & $6.1^{+5.5}_{-2.9}\times10^{-5}$ & $1.2^{+3.2}_{-0.9}\times10^{3}$ & $1.5^{+3.5}_{-1.0}\times10^{3}$\\
\hline
IRAS23385 & 0.11 & 30 & 1.000 & 0.04 & 2 & 13 & 294 & 25.89 & 19 & 1.2$\times10^{-4}$ & 1.3$\times10^{4}$ & 1.7$\times10^{3}$ \\
$d=4.9$ kpc & 0.14 & 30 & 1.000 & 0.04 & 1 & 13 & 145 & 27.97 & 12 & 8.4$\times10^{-5}$ & 5.1$\times10^{3}$ & 1.0$\times10^{3}$ \\
$R_\mathrm{ap}=5.75\arcsec$ & 0.15 & 40 & 1.000 & 0.05 & 4 & 22 & 297 & 31.54 & 23 & 1.7$\times10^{-4}$ & 8.3$\times10^{3}$ & 2.2$\times10^{3}$ \\
$R_\mathrm{ap}=0.14$ pc & 0.18 & 20 & 3.160 & 0.02 & 4 & 13 & 347 & 11.60 & 34 & 3.1$\times10^{-4}$ & 1.9$\times10^{4}$ & 3.3$\times10^{3}$ \\
 & 0.20 & 40 & 1.000 & 0.05 & 1 & 13 & 95 & 38.55 & 10 & 9.1$\times10^{-5}$ & 3.6$\times10^{3}$ & 1.0$\times10^{3}$ \\
Average model & \#1692 &  $52^{+54}_{-26}$ & $0.517^{+1.067}_{-0.348}$ & $0.07^{+0.10}_{-0.04}$ & $5^{+10}_{-3}$ & $57 \pm 22$ & $265 \pm 255$ & $26^{+52}_{-17}$ & $32 \pm 20$ & $1.1^{+1.1}_{-0.5}\times10^{-4}$ & $3.5^{+16.9}_{-2.9}\times10^{3}$ & $5.6^{+27.4}_{-4.7}\times10^{3}$\\
\hline
HH288 & 1.44 & 10 & 1.000 & 0.02 & 2 & 39 & 183 & 5.33 & 39 & 7.5$\times10^{-5}$ & 1.0$\times10^{3}$ & 7.6$\times10^{2}$ \\
$d=2.0$ kpc & 1.53 & 20 & 0.316 & 0.06 & 2 & 29 & 149 & 15.05 & 27 & 4.2$\times10^{-5}$ & 7.3$\times10^{2}$ & 4.8$\times10^{2}$ \\
$R_\mathrm{ap}=7.50\arcsec$ & 1.74 & 20 & 0.316 & 0.06 & 0 & 13 & 86 & 19.17 & 11 & 2.3$\times10^{-5}$ & 8.3$\times10^{2}$ & 2.0$\times10^{2}$ \\
$R_\mathrm{ap}=0.07$ pc & 1.80 & 10 & 3.160 & 0.01 & 4 & 65 & 69 & 1.65 & 56 & 1.9$\times10^{-4}$ & 2.4$\times10^{2}$ & 1.9$\times10^{3}$ \\
 & 1.96 & 20 & 0.316 & 0.06 & 1 & 13 & 226 & 17.54 & 18 & 3.2$\times10^{-5}$ & 2.7$\times10^{3}$ & 3.9$\times10^{2}$ \\
Average model & \#92 &  $18^{+9}_{-6}$ & $0.449^{+1.046}_{-0.314}$ & $0.05^{+0.06}_{-0.03}$ & $3^{+3}_{-1}$ & $52 \pm 23$ & $148 \pm 99$ & $7^{+13}_{-5}$ & $40 \pm 15$ & $5.5^{+5.9}_{-2.8}\times10^{-5}$ & $6.6^{+14.6}_{-4.6}\times10^{2}$ & $1.1^{+1.7}_{-0.7}\times10^{3}$\\
\hline
\enddata
\tablecomments{
For each source the first five rows refer to the best five models taken from the 432 physical models, whereas the sixth row shows the average and dispersion of good model fits (see the text). The number next to the symbol \# represents the number of models considered in the average of the good models. Upper and lower scripts in the average models row refer to the upper and lower errors.}
\end{deluxetable}
\end{longrotatetable}

\subsection{MIR to FIR Variability Constraints from IRAS}\label{sect:iras_results}

Given the imaging presented above, and especially based on the 70$\mu$m images, AFGL~2591, G25.40-0.14, G30.59-0.04, G33.92+0.11, G40.62-0.14 and HH 288 appear to be single sources within large enough regions that can be reasonably well fit with IRAS Gaussian point spread function fitting. However, the following considerations still apply.

While G25.40-0.14 appears to be a single source at 12, 25, and 60\,$\mu$m within the IRAS square areas for Gaussian fitting based on its 70\,$\mu$m image, there is another source to its SE and at 100\,$\mu$m the two sources merge and become unresolved; thus, in this case, we cannot derive a reliable measurement at 100\,$\mu$m.

For G30.59-0.04, we find that it is moderately contaminated by a source to its east, especially at 100\,$\mu$m, which skews the Gaussian fitting. Also, the P.A. of the beam provided by the HIRES beam sample map does not align with the extension of the source. Thus, here we set P.A. as a free parameter and only fix the lengths of the major and minor axes. In addition, the 12\,$\mu$m emission of G30.59-0.04 has very poor S/N and makes it hard to define the source. 

We did not derive a valid flux from the IRAS images for G32.03+0.05 due to the presence of multiple sources. Within the IRAS square area, there are at least two strong sources revealed at 37.1\,$\mu$m and 70\,$\mu$m. In the IRAS 60\,$\mu$m and 100\,$\mu$m images, the main protostar lies at the lower edge of the emission, so clearly the 60\,$\mu$m and 100\,$\mu$m emission is dominated by other sources.

IRAS~00259+5625 and IRAS~00420+5530 do not have Herschel data. Their IRAS 12\,$\mu$m and 25\,$\mu$m emission have poor S/N and appear very extended. Their IRAS 60\,$\mu$m and 100\,$\mu$m emission exhibit a ``tail" structure extending along the major axis, indicating impact by other sources within the beam.

IRAS~23385 has a companion located 95\arcsec to the west, and has bright arc-shaped surrounding diffuse emission at 70\,$\mu$m. The 60\,$\mu$m IRAS image shows a \textit{tail} to the west of the main source, which should be from this companion source. Thus a valid flux could not be retrieved for this source.

The S/N of the 12\,$\mu$m emission of HH 288 is very poor, so that the source cannot be well characterized at this wavelength.

With the above caveats noted, we plot our estimated IRAS-derived fluxes of the sources in Figure~\ref{fig:sed_1D_results_soma_iv} (shown as red open squares). We note that because of the possibility of source contamination in the larger IRAS beam, these should generally be regarded as being upper limits on the source fluxes. Along with these data, we also show the expected fluxes in the IRAS bands from the average of the good models (shown as red open circles). We consider that a clear signature of variability would be when the model-fit SEDs, i.e., that are derived from the more recent SOFIA observations, and their prediction of IRAS fluxes exceeds the actual observed IRAS fluxes. We do not find any clear cases of this situation (i.e., red circles being higher than red squares) and so conclude there is no evidence for variability in this sample over the $\sim 40$\,yr time baseline between the IRAS and SOFIA observations.

\section{Discussion of Global SOMA Sample Results}\label{sect:discussion}

Here, we discuss the overall results from the SOMA survey, i.e., the sources of Papers I-III, which have been reanalyzed with the methods of this paper, and the 11 new sources we have presented here in Paper IV. In total, this is a combined sample of 40 high- and intermediate-mass protostars that have been analyzed uniformly. In addition, we will also make a comparison to protostars that have been identified in IRDCs by \citet{moser2020} and \citet{liu2021}: these SEDs were measured based on Spitzer and Herschel data and also fitted to the ZT-RT model grid.

\subsection{The SOMA Sample Space and the Evolutionary Sequence of Massive Star Formation}

\begin{figure*}[!htb]
\centering
\includegraphics[width=0.43\textwidth]{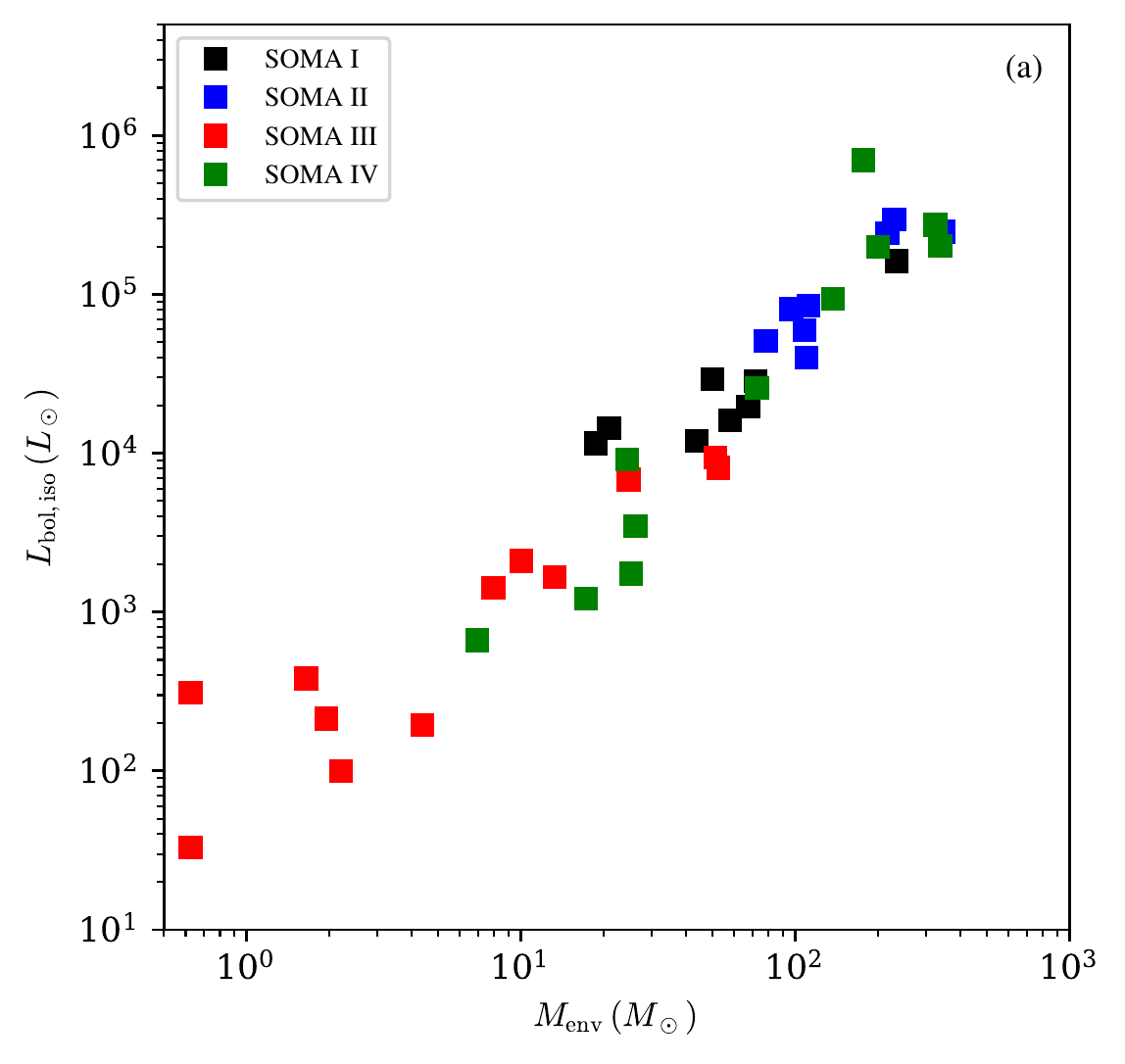}
\includegraphics[width=0.43\textwidth]{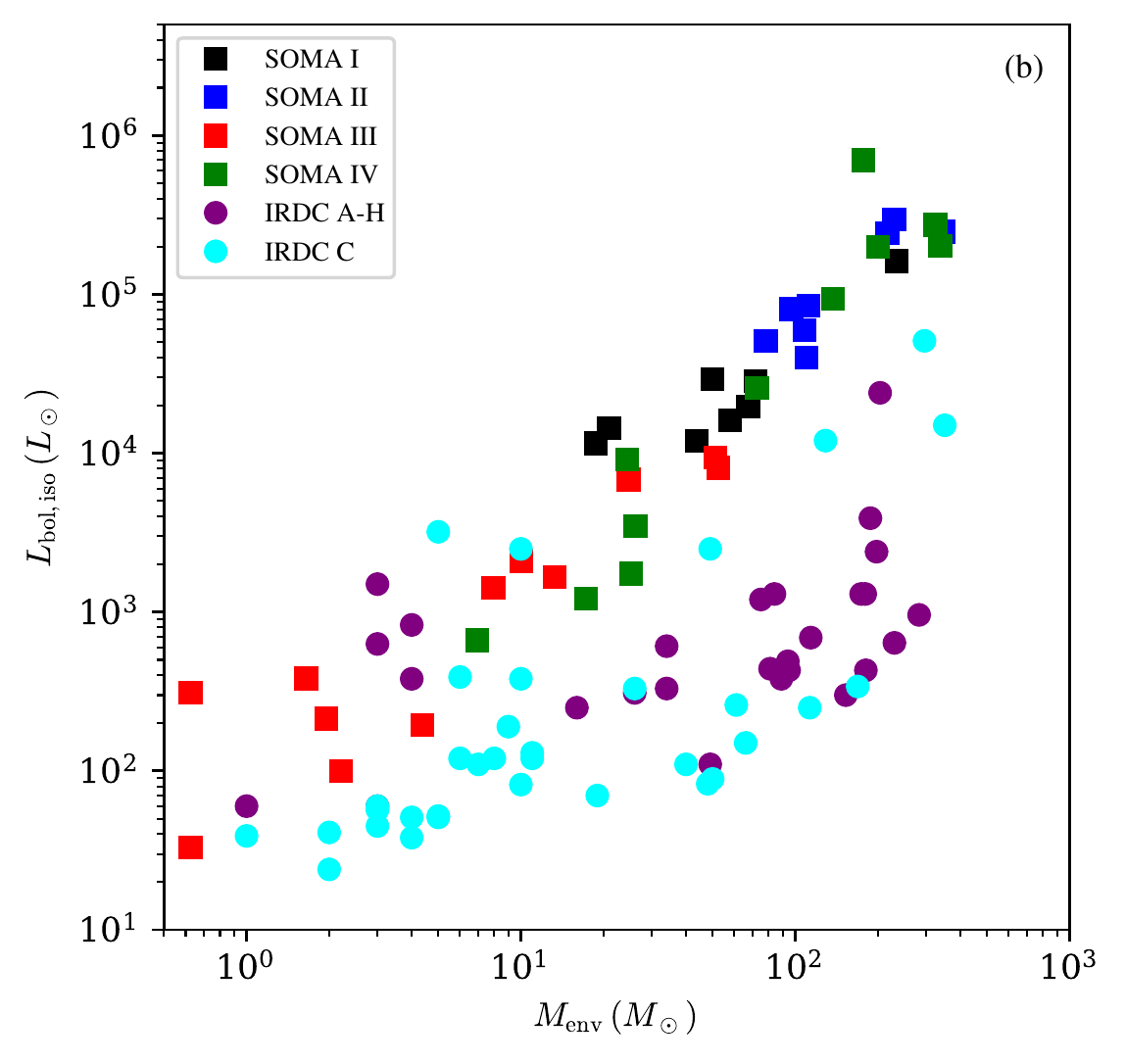}
\includegraphics[width=0.43\textwidth]{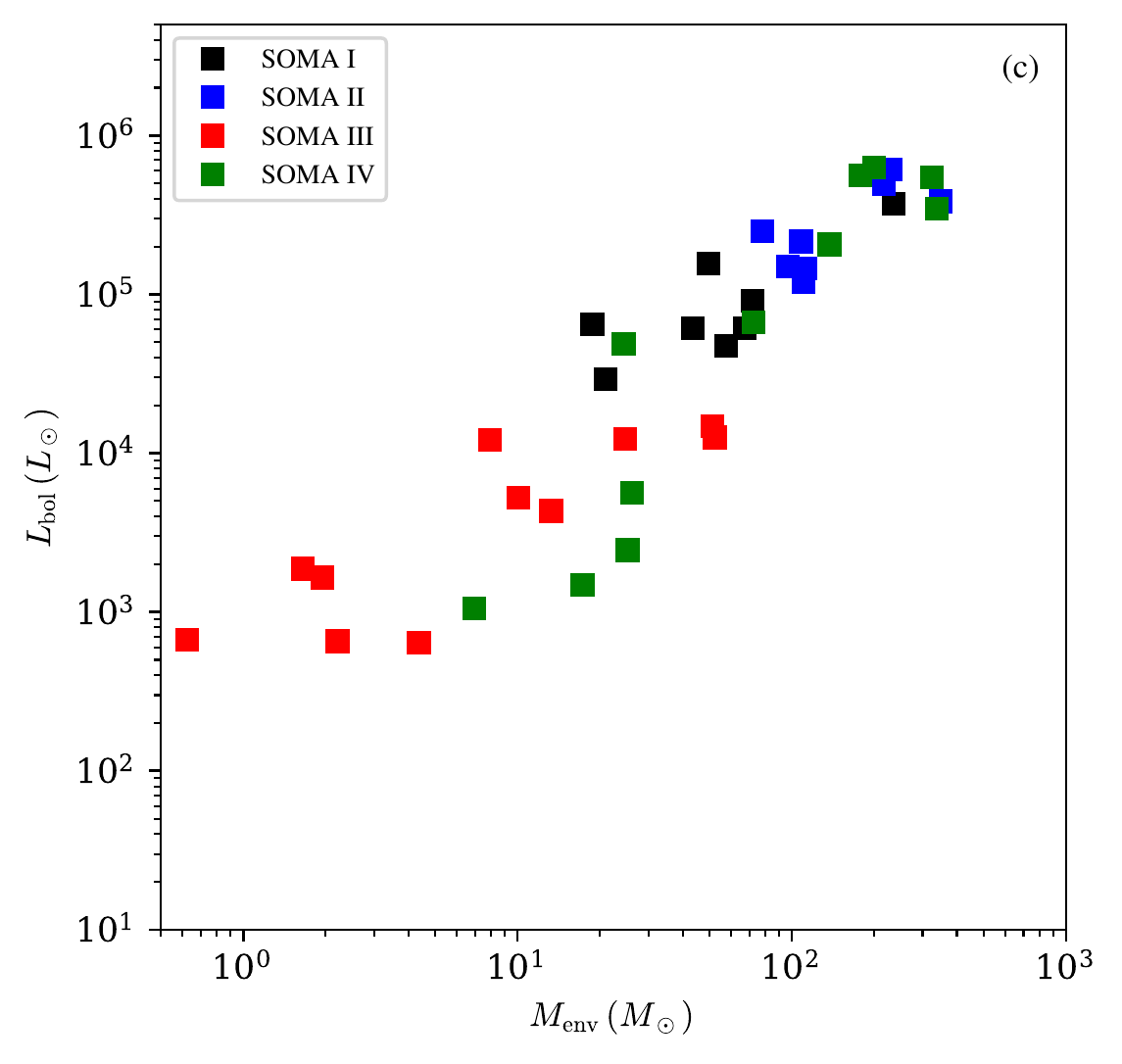}
\includegraphics[width=0.43\textwidth]{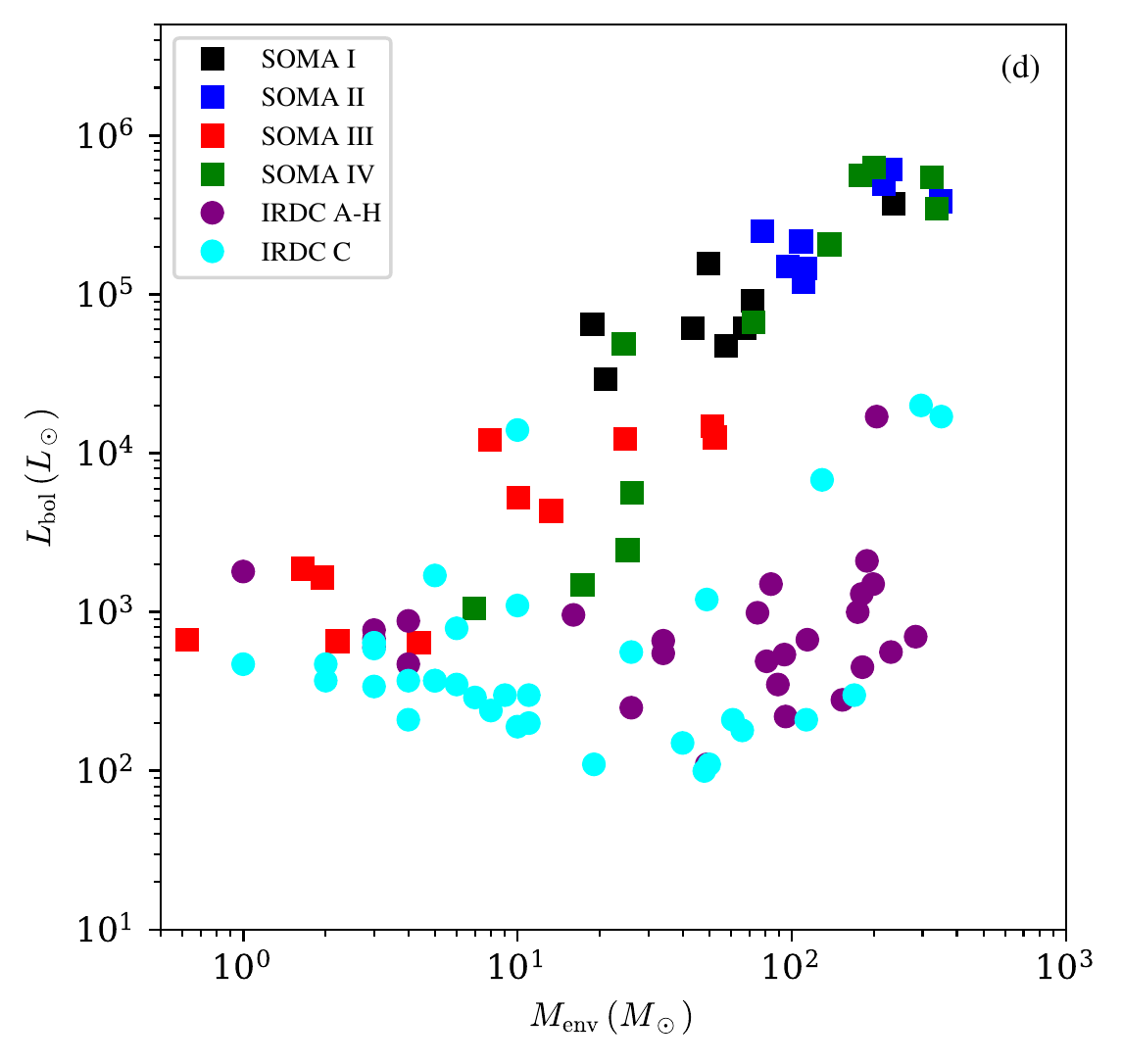}
\includegraphics[width=0.43\textwidth]{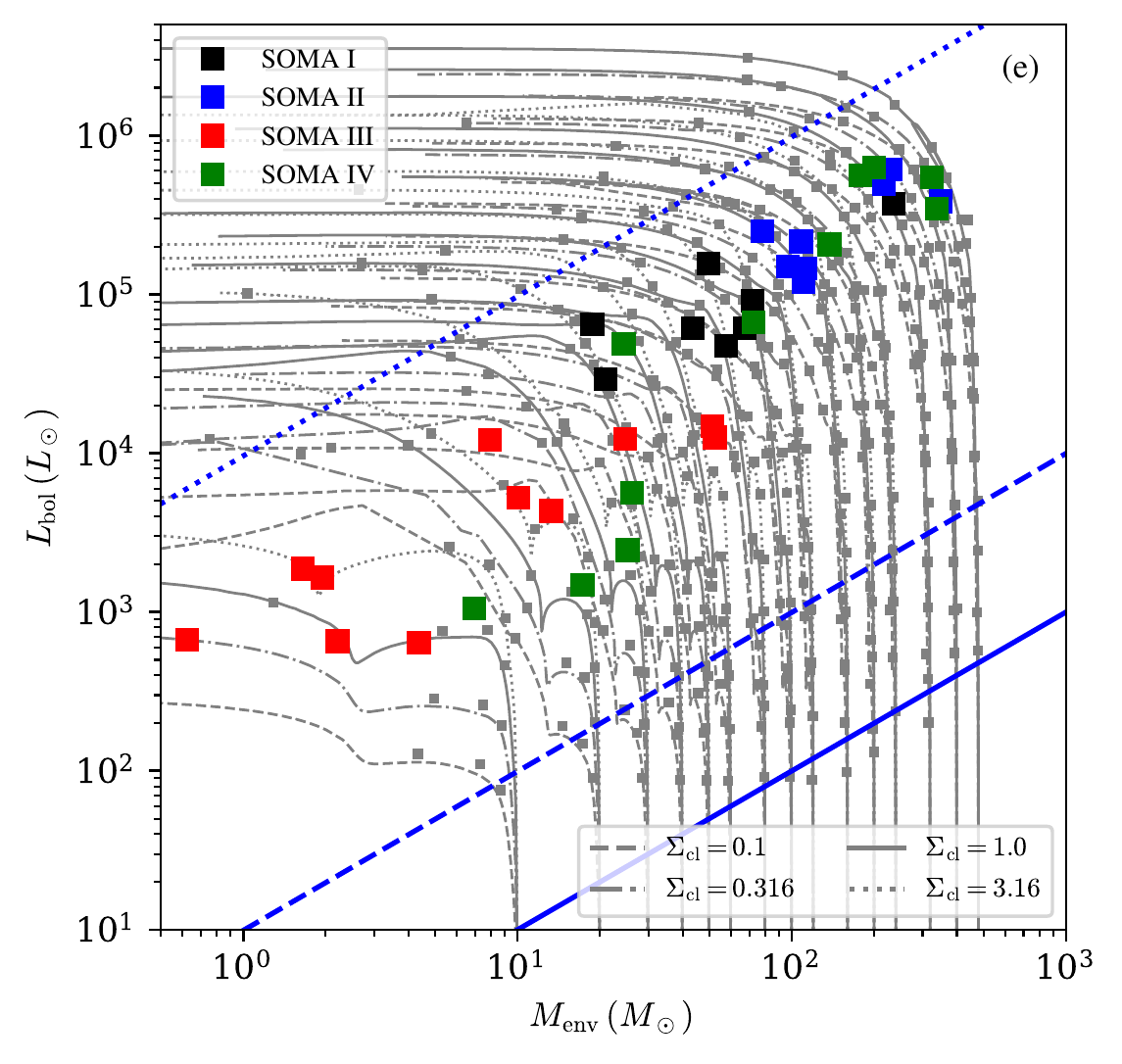}
\includegraphics[width=0.43\textwidth]{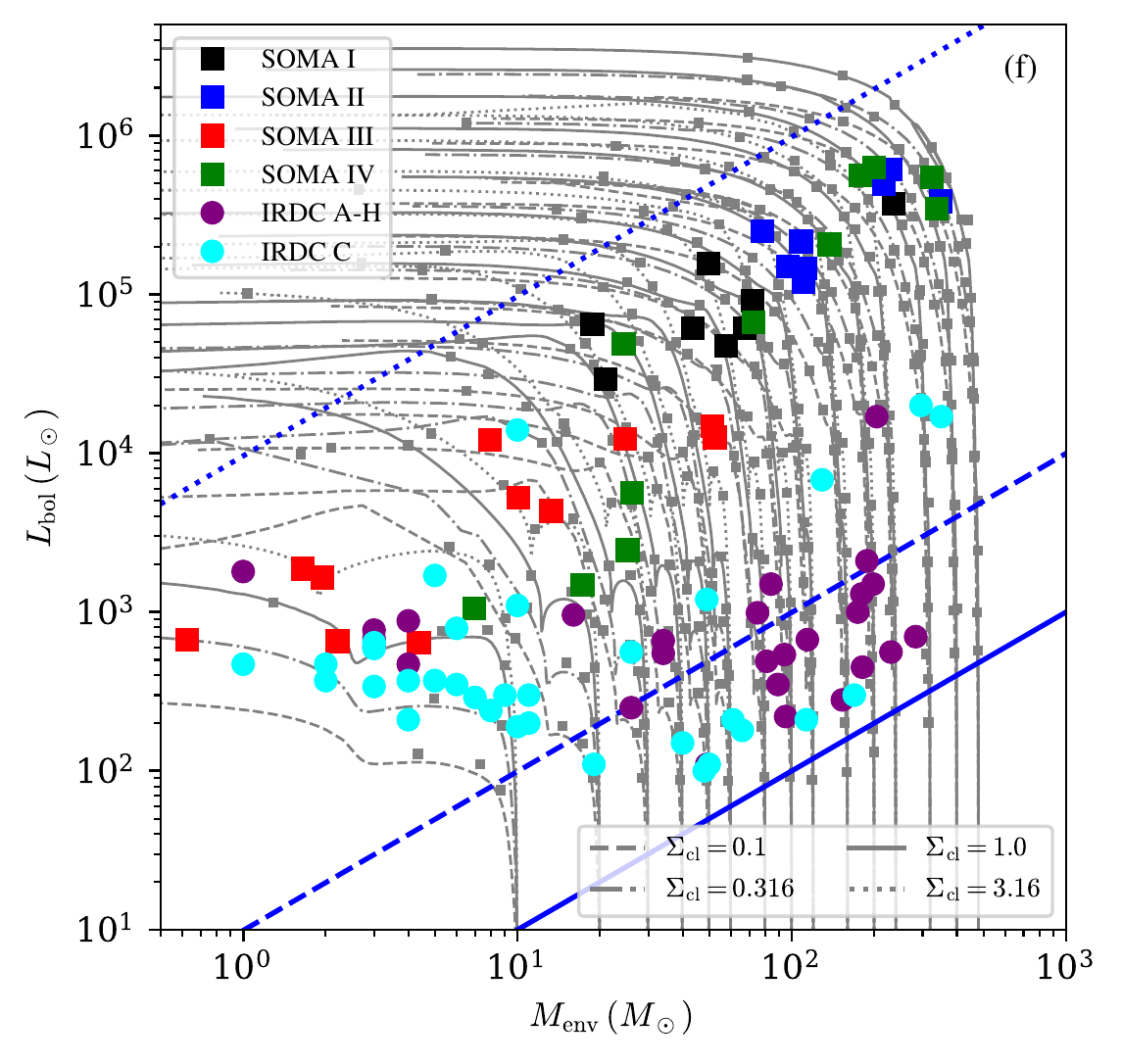}
\caption{{\it (a) Top left:} Isotropic bolometric luminosity ($L_\mathrm{bol,iso}$) versus mass of the envelope ($M_\mathrm{env}$). Data from SOMA Papers I-IV are shown, as indicated. {\it (b) Top right:} As (a), but now including protostars selected from IRDCs by \citet{liu2020} (IRDC A-H) and \citet{moser2020} (IRDC C). {\it (c) Middle left:} As (a), but now for bolometric luminosity ($L_\mathrm{bol}$) versus $M_\mathrm{env}$. {\it (d) Middle right:} As (c), but now also including IRDC protostars. 
{\it (e) Bottom left:} As (c), but now including the \citet{zhang2018} protostellar evolutionary tracks (gray lines and squares) for different initial core masses and clump mass surface densities (see legend). The three blue lines indicate $L_\mathrm{bol}/M_\mathrm{env}=1$ (solid line), 10 (dashed line) and $10^4\,L_\odot/M_\odot$ (dotted line).
{\it (f) Bottom right:} As (f), but now also including IRDC protostars. \label{fig:menv_lbol}}
\end{figure*}

Figure\,\ref{fig:menv_lbol} shows the isotropic bolometric luminosity ($L_\mathrm{bol,iso}$, top row) and true bolometric luminosity ($L_\mathrm{bol}$, middle and bottom rows) versus envelope mass ($M_\mathrm{env}$) of all the SOMA protostars. Note, these quantities are derived as averages of all good models that are fit to the SEDs. We see that the current sample of protostars presented in this paper span a range of luminosities from the upper end of the intermediate-mass sources presented in Paper III up to and beyond the highest end of the luminosity range probed previously in Paper II. Indeed, the most luminous source in terms of $L_\mathrm{bol,iso}$ is AFGL~2591 with $7\times 10^5\:L_\odot$. However, in terms of the ratio of luminosity to mass, we do not perceive any significant difference of the relatively isolated protostars of SOMA IV compared to the other sources. 
While there is an apparent tight correlation between $L_\mathrm{bol,iso}$ and $M_{\rm env}$, this becomes weaker when considering $L_\mathrm{bol}$ and $M_{\rm env}$. The theoretical protostellar evolutionary tracks (shown in the bottom row of Figure\,\ref{fig:menv_lbol}) cover a wide range in this parameter space, i.e., going from low-luminosity cores with high envelope masses to high-luminosity protostars with only small amounts of residual envelope mass. Including the IRDC sources of \citet{moser2020} and \citet{liu2021} gives more extensive coverage of the low-luminosity, high-mass end of these tracks.

The lack of sources at the high-luminosity low-mass end of the evolutionary sequence may reflect choices made in the selection of SOMA sources. Such sources would tend to be near the end of their formation phase and, for the most massive protostars, would be producing ionizing gas, i.e., likely appearing as UC HII regions. Furthermore, the lifetime of sources in this phase may be relatively short, which would reduce the number of such systems that may be observed or selected to be analyzed. Finally, with a focus on MIR to FIR emission, i.e., selecting sources that are relatively bright in these wave bands, may mean that the later protostellar evolutionary stages, perhaps with most luminosity emerging in the ultraviolet, optical, or NIR, are underrepresented.

\subsection{MIR-FIR SED Shape}

As discussed in SOMA Papers I-III, the shape of the MIR to FIR SED is expected to correlate with intrinsic protostellar properties, such as viewing angle with respect to the outflow axis and evolutionary stage. Essentially, the protostars appear relatively brighter in the MIR if we look down their outflow cavities or if we see them at later stages when their envelopes are warmer and have less internal extinction.

\begin{figure*}[!htb]
\centering
\includegraphics[width=0.43\textwidth]{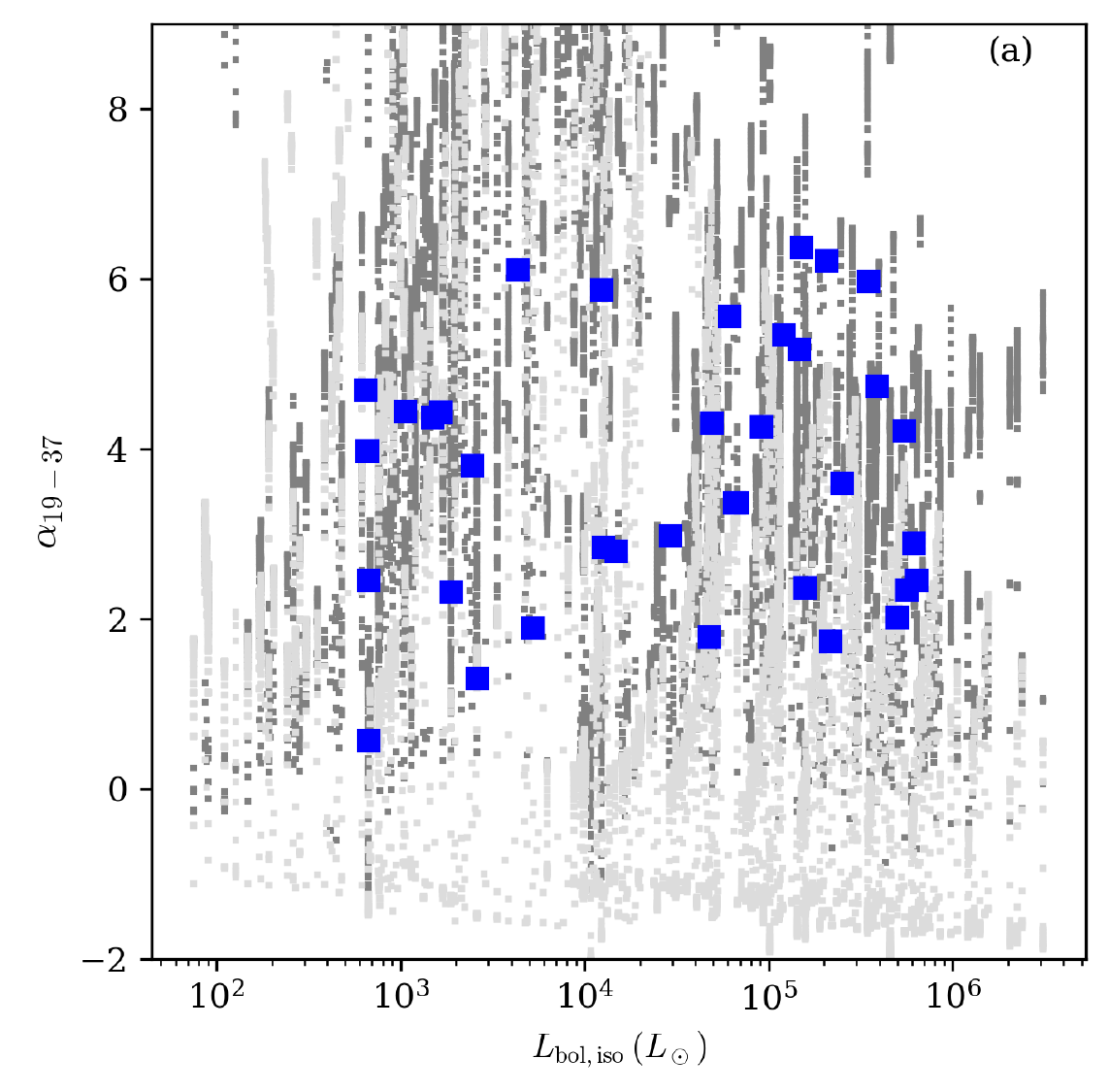}
\includegraphics[width=0.43\textwidth]{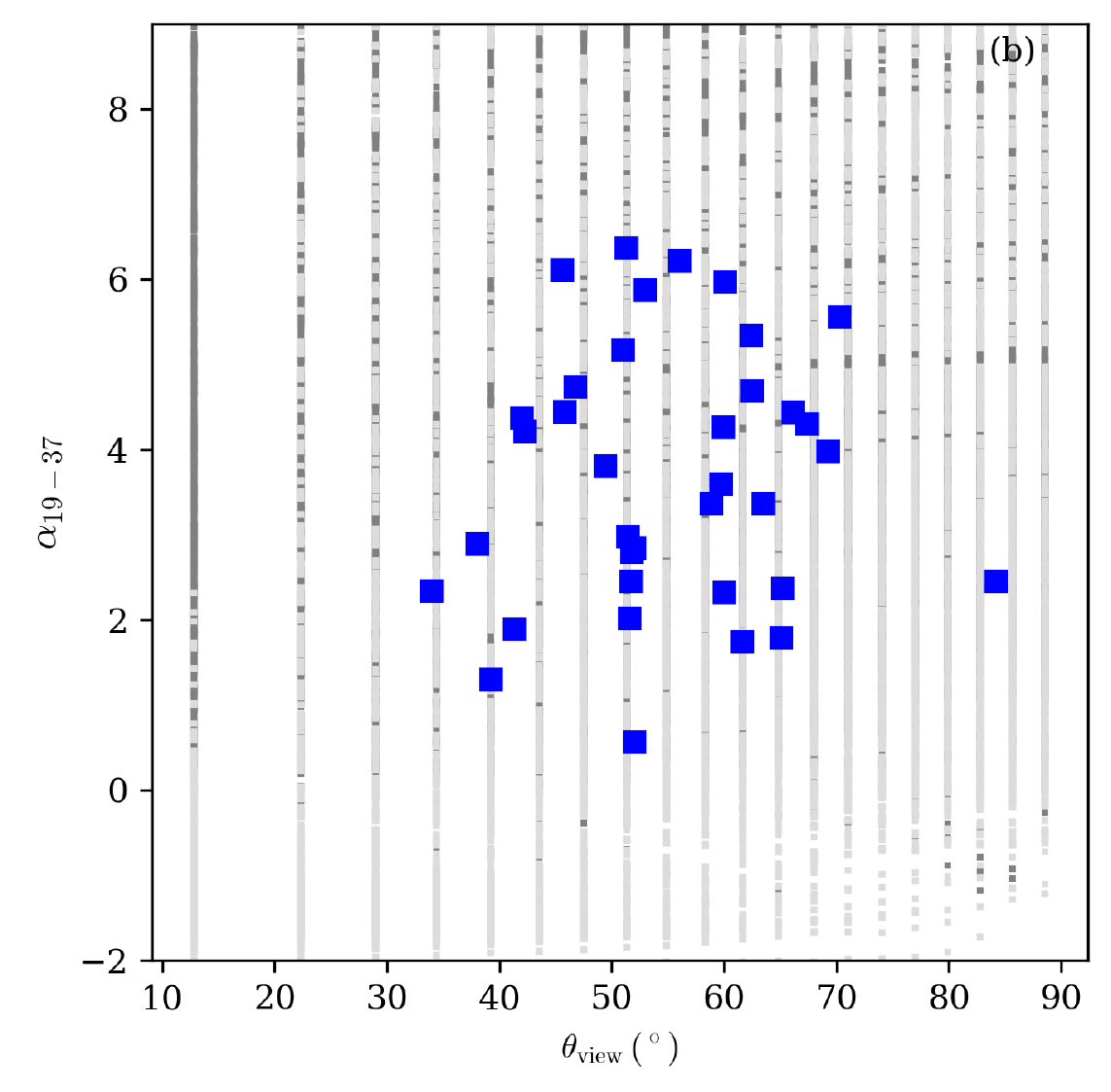}
\includegraphics[width=0.43\textwidth]{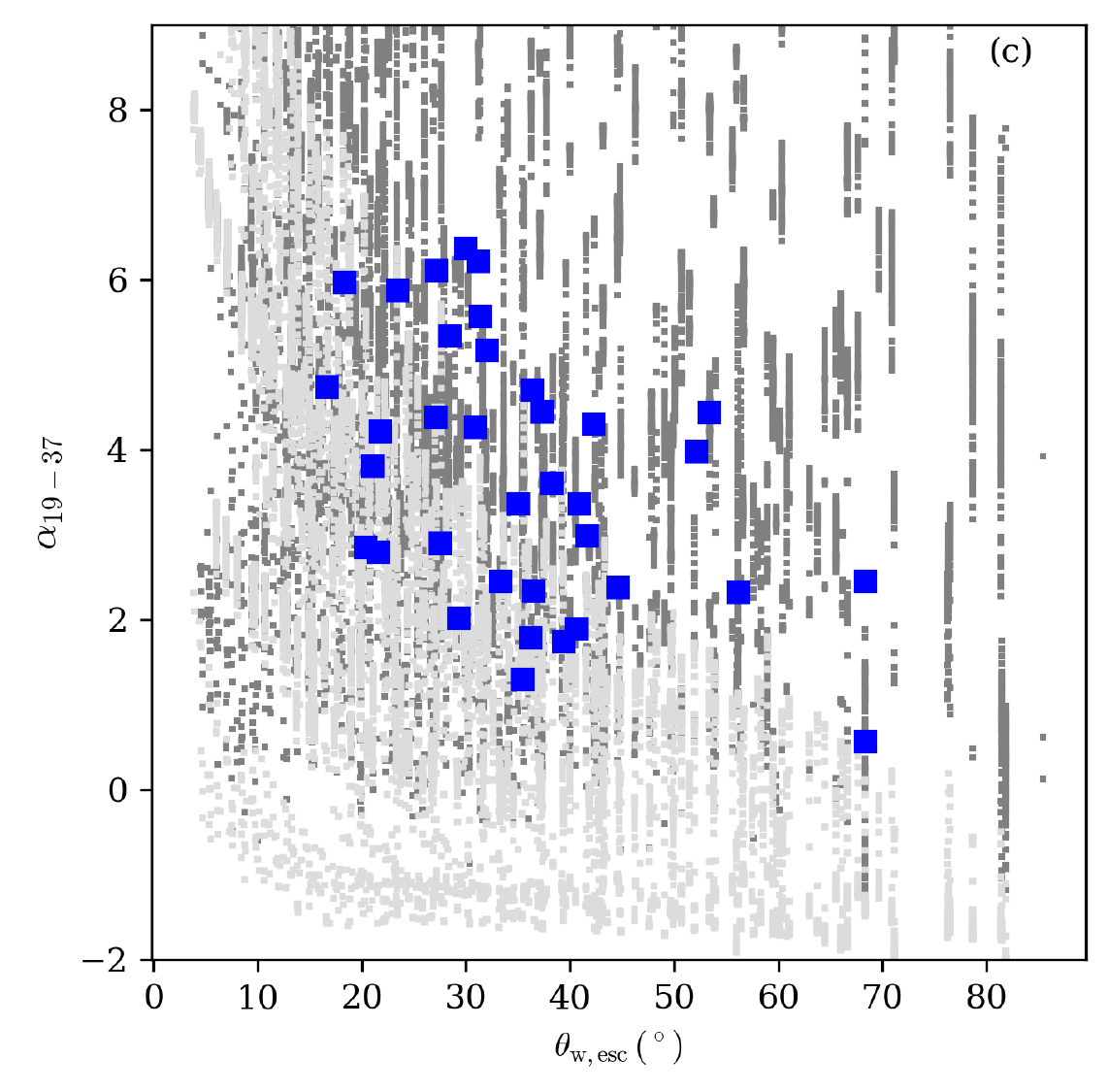}
\includegraphics[width=0.43\textwidth]{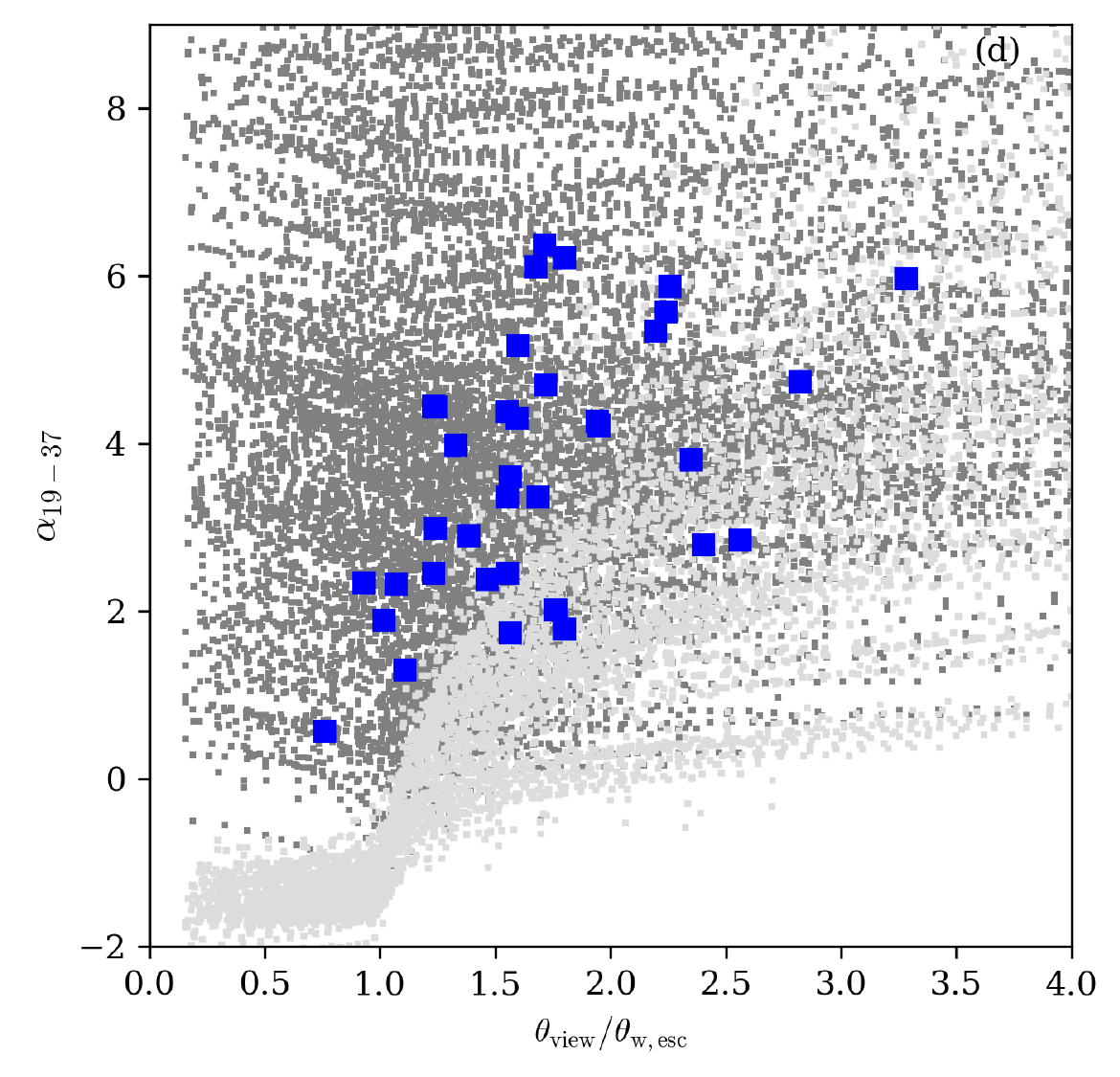}
\includegraphics[width=0.43\textwidth]{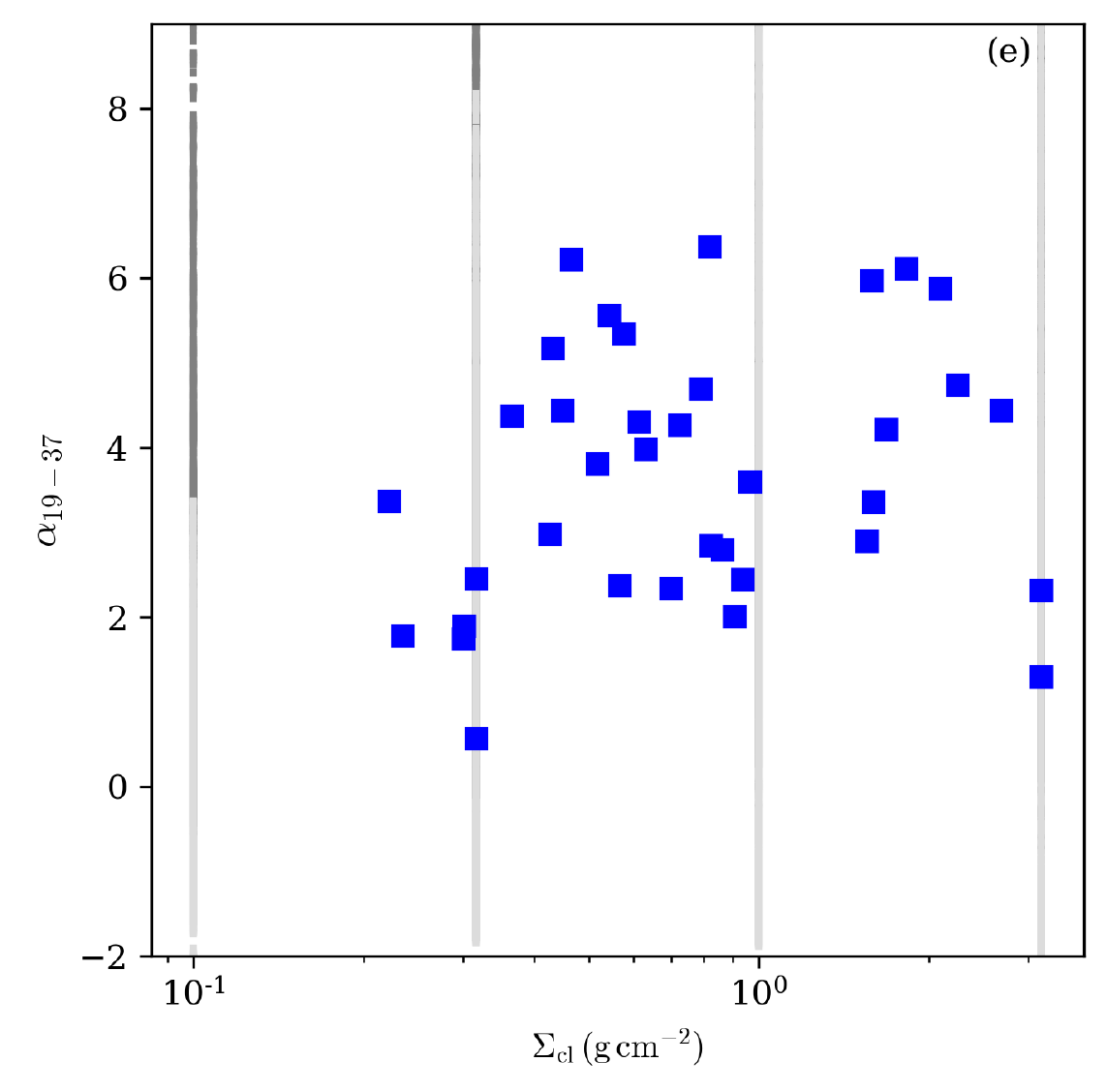}
\includegraphics[width=0.43\textwidth]{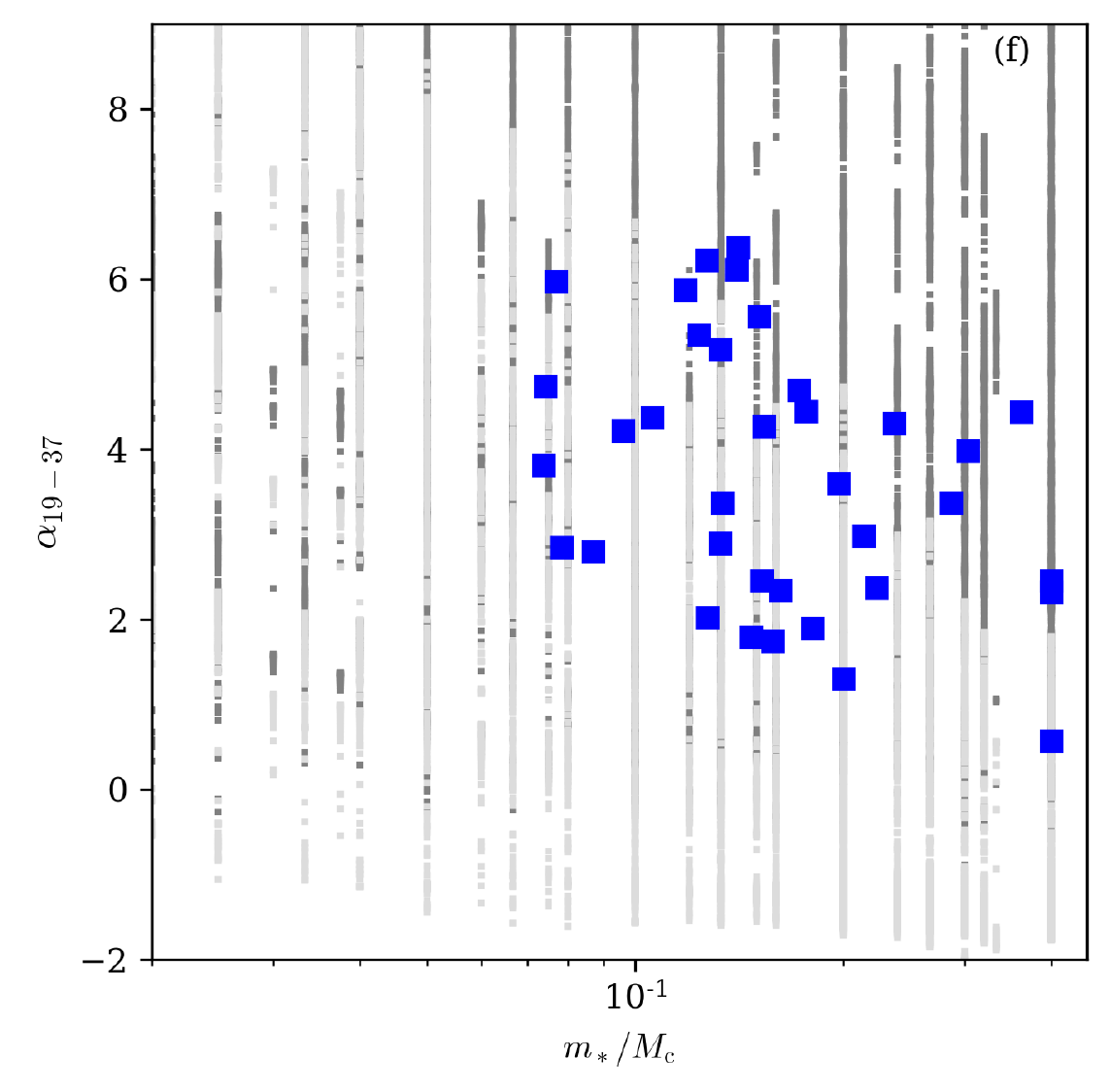}
\caption{Spectral index, $\alpha_{19-37}$ between 19\,$\mu$m and 37\,$\mu$m (see the text) vs. the geometric mean isotropic luminosity $L_\mathrm{bol,iso}$ (\textit{(a) top left}), the arithmetic mean inclination of viewing angle $\theta_\mathrm{view}$ (\textit{(b) top right}), the arithmetic mean opening angle $\theta_\mathrm{w,esc}$ (\textit{(c) middle left}), the arithmetic mean $\theta_\mathrm{view}/\theta_\mathrm{w,esc}$ (\textit{(d) middle right}), the geometric mean clump surface density $\Sigma_\mathrm{cl}$ (\textit{(e) bottom left}), and the geometric mean $m_*/M_\mathrm{c}$ (\textit{(f) bottom right}) returned by the good models. Small light gray points represent models without correction for foreground extinction whereas dark gray points include the correction for foreground extinction.
\label{fig:alpha_vs_properties}}
\end{figure*}

To explore the potential diagnostic power of the shape of the MIR to FIR SED, Figure \ref{fig:alpha_vs_properties} shows the values of the 19-37~$\rm \mu m$ spectral index,
\begin{equation}
\alpha_{19-37} = \frac{\log_{10}(\nu_{37\mu m}F_{\nu_{37\mu m}})-\log_{10}(\nu_{19\mu m}F_{\nu_{19\mu m}})}{\log_{10}(\lambda_{37\mu m})-\log_{10}(\lambda_{19\mu m})},
\end{equation}
of all the SOMA protostars versus various inferred properties of the systems, i.e., luminosity, inclination
of viewing angle, outflow cavity opening angle, and the ratio of inclination
of viewing angle to outflow cavity opening angle, $\Sigma_{\rm cl}$, and $m_{*}/M_{c}$. In this figure, the observed values of $\alpha_{19-37}$ of the SOMA protostars are shown with blue symbols, with the corresponding value of the protostellar property being the average of good models. For reference, we also show all the individual good models as light gray points for their unextincted spectral indices and as dark gray points after applying the best-fit extinction. Recall that it is the latter that is fit to the observed SEDs represented by the blue points.

Figure \ref{fig:alpha_vs_properties} shows that $\alpha_{19-37}$ has the strongest correlation with the ratio $\theta_{\rm view}/\theta_{\rm w,esc}$, but there are also related weaker correlations with $\theta_{\rm view}$ and $\theta_{\rm w,esc}$ individually, as well as with the evolutionary stage as parameterized via $m_*/M_c$. The data here allow one to gauge the uncertainty in estimating these intrinsic protostellar properties (as based on the average of good model fits to full SEDs), if only $\alpha_{19-37}$ is known. On the other hand, Figure \ref{fig:alpha_vs_properties} shows that there is limited apparent correlation of the value of $\alpha_{19-37}$ with $L_{\rm bol,iso}$ or $\Sigma_{\rm cl}$, i.e., these quantities are not well constrained by the MIR to FIR spectral slope.

Other important points that are illustrated in Figure \ref{fig:alpha_vs_properties} include the fact that the correction for foreground extinction can typically have a significant effect on $\alpha_{19-37}$ (e.g., compare the distribution of light and dark gray points). This means that in the ideal case observed value of $\alpha_{19-37}$ should be corrected for this extinction, which would require additional information, e.g., using colors based on filters at shorter wavelengths, or via some independent method, e.g., based on the ratio of $\rm H_2$ emission lines that may be detected in the outflow cavity \citep{fedriani2018,fedriani2019,fedriani2020,costa2022}. Another important caveat to note is that there are degeneracies in protostellar model properties for a given SED or value of $\alpha_{19-37}$, i.e., the dispersion in the light gray points is larger than that of the blue points. However, this is a problem that also affects protostellar properties derived from full SED fitting.

\subsection{The Environmental Dependence of Massive Star Formation}


Here, we examine if there are any dependencies of massive protostar properties on the mass surface density of the clump environment, $\Sigma_\mathrm{cl}$, in which it is forming. Higher values of $\Sigma_\mathrm{cl}$ imply higher pressures in a self-gravitating clump, which then lead to higher densities of prestellar cores in the turbulent core model of \citet{mckee2003}. This in turn leads to higher accretion rates, shorter formation times, and more efficient formation from a core of a given mass under the action of internal feedback processes from the protostar \citep{tanaka2017}. Furthermore, there have been proposed theoretical models that predict that the existence of massive prestellar cores requires certain conditions on $\Sigma_{\rm cl}$. In particular, \citet{krumholz2008} suggested that to prevent fragmentation of a massive prestellar core one requires $\Sigma_{\rm cl}\gtrsim 1\:{\rm g\:cm}^{-2}$ and the presence of a surrounding cluster of lower-mass protostars, which then have high enough accretion rates and high enough accretion luminosities to provide sufficient heating of the massive core so that the Jeans mass is raised to high-mass scales. On the other hand, \citet{butler2012} have discussed how the presence of moderate strength ($\sim 0.1\:$mG) magnetic fields within massive prestellar cores can prevent their fragmentation in cold conditions. In this case, massive prestellar cores and massive star formation could occur in environments with $\Sigma_{\rm cl}< 1\:{\rm g\:cm}^{-2}$.

Recall that the three main physical parameters that are derived in our SED fitting are the initial mass of the core ($M_c$), mass surface density of the clump ($\Sigma_\mathrm{cl}$) and current protostellar mass ($m_*$). We note that when considering these results, one should recall the caveats that these quantities are relatively indirect inferences from SED fitting, that there can be significant degeneracies (i.e., dispersion) in these properties among good-fitting SED models and that there could be systematic uncertainties, e.g., if the luminosity is overestimated because of the presence of multiple sources.

\begin{figure}[!htb]
\includegraphics[width=0.5\textwidth]{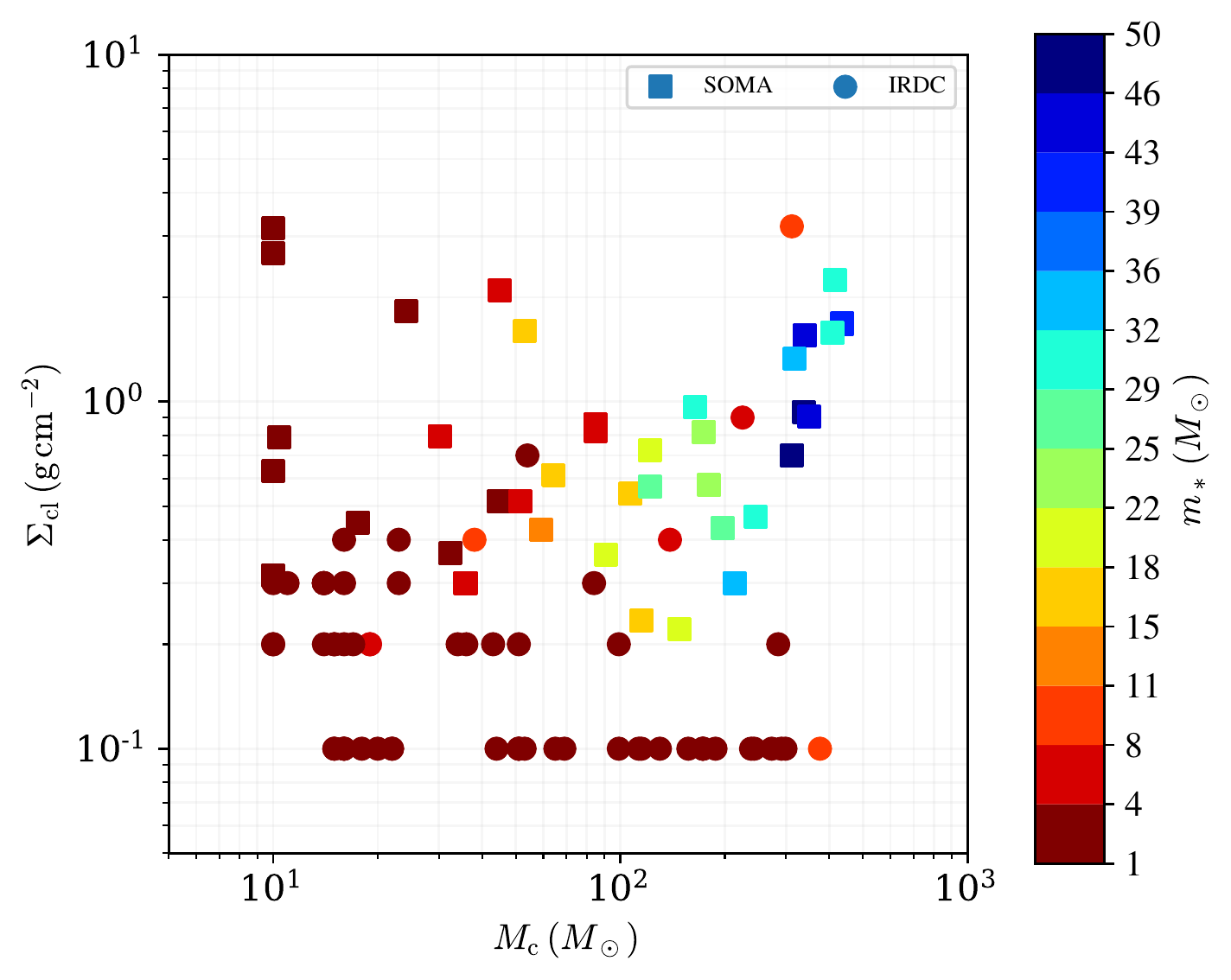}

\caption{Mass surface density of the clump environment ($\Sigma_\mathrm{cl}$) versus initial mass of the core ($M_\mathrm{c}$) for the full SOMA sample to date and the IRDC samples. Each data point is the average of good model fits. Each point is also color coded with the current mass of the protostar ($m_\mathrm{*}$).
\label{fig:mc_sigma_mstar}}
\end{figure}

Figure\,\ref{fig:mc_sigma_mstar} shows the distribution of the averaged good models (see \S\ref{sect:average_models}) as $M_c$ vs $\Sigma_\mathrm{cl}$ and color coded with $m_*$ for the 40 sources analyzed so far in the SOMA survey. We find that the analyzed sample contains protostars ranging in current mass from $\sim2$ to $\sim50\,M_\odot$ and spanning the full range in the ZT grid of $M_c$ and $\Sigma_\mathrm{cl}$ (see \S\ref{sect:ZT_model_grid}). 
Both low- and high-mass protostars are found to be forming from cores with high initial masses (and thus also generally high current envelope masses). One also notices a tendency for the most massive protostars to have higher values of $\Sigma_{\rm cl}$.


In Paper III, we explored the relationship between the mass surface density of the clump and the current mass of the protostar. In this earlier work, we found tentative evidence that the most massive protostars (i.e., $m_*\gtrsim25\,M_\odot$) in our sample require their cores to be in environments with $\Sigma_\mathrm{cl}\gtrsim1.0\,\mathrm{g\,cm^{-2}}$. With the addition of SOMA IV sources and the updated methods presented here, e.g., of aperture definition, that have been applied to all the SOMA sample, we now reexamine this result.

\begin{figure*}[!htb]
\includegraphics[width=0.5\textwidth]{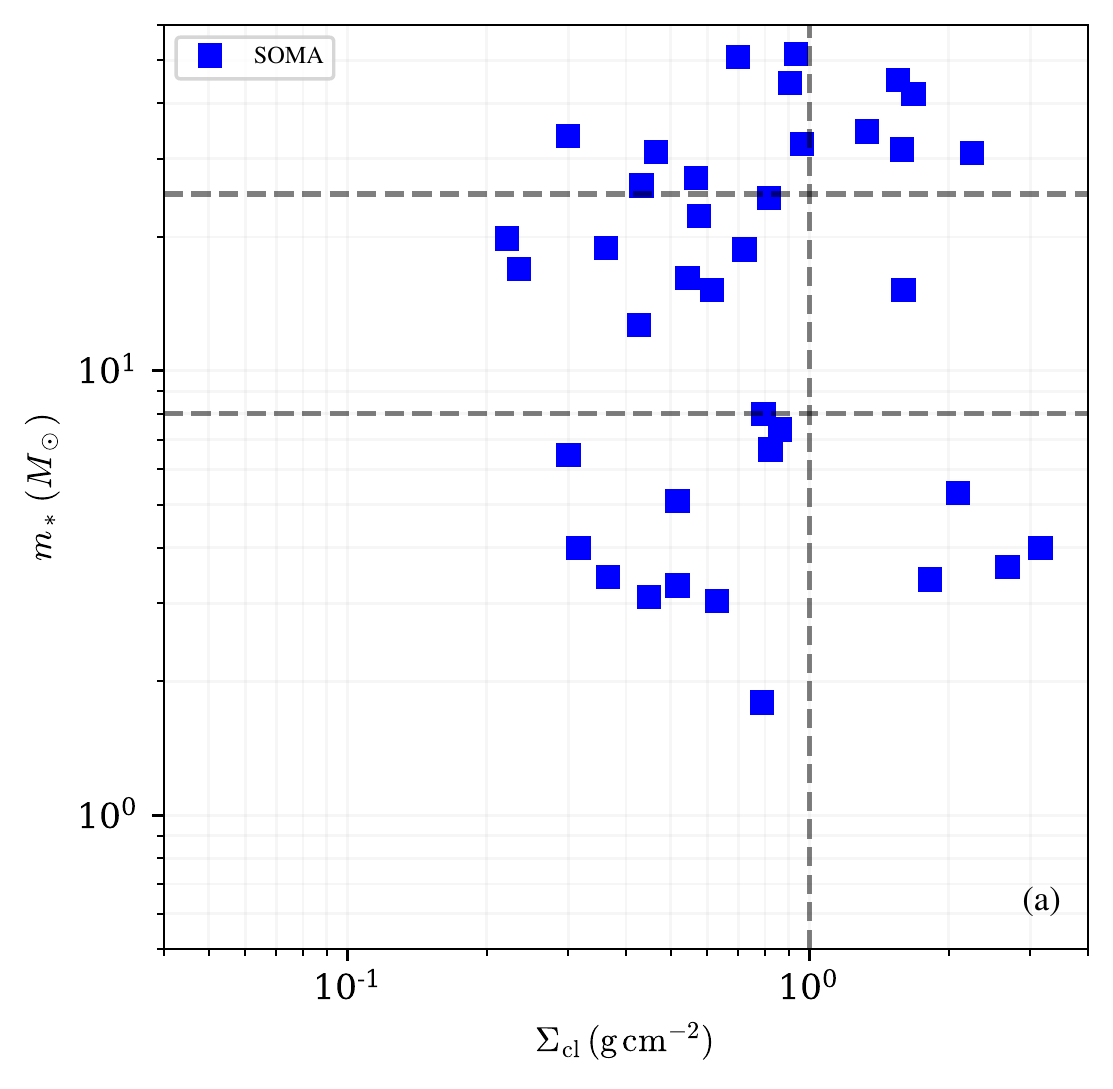}
\includegraphics[width=0.5\textwidth]{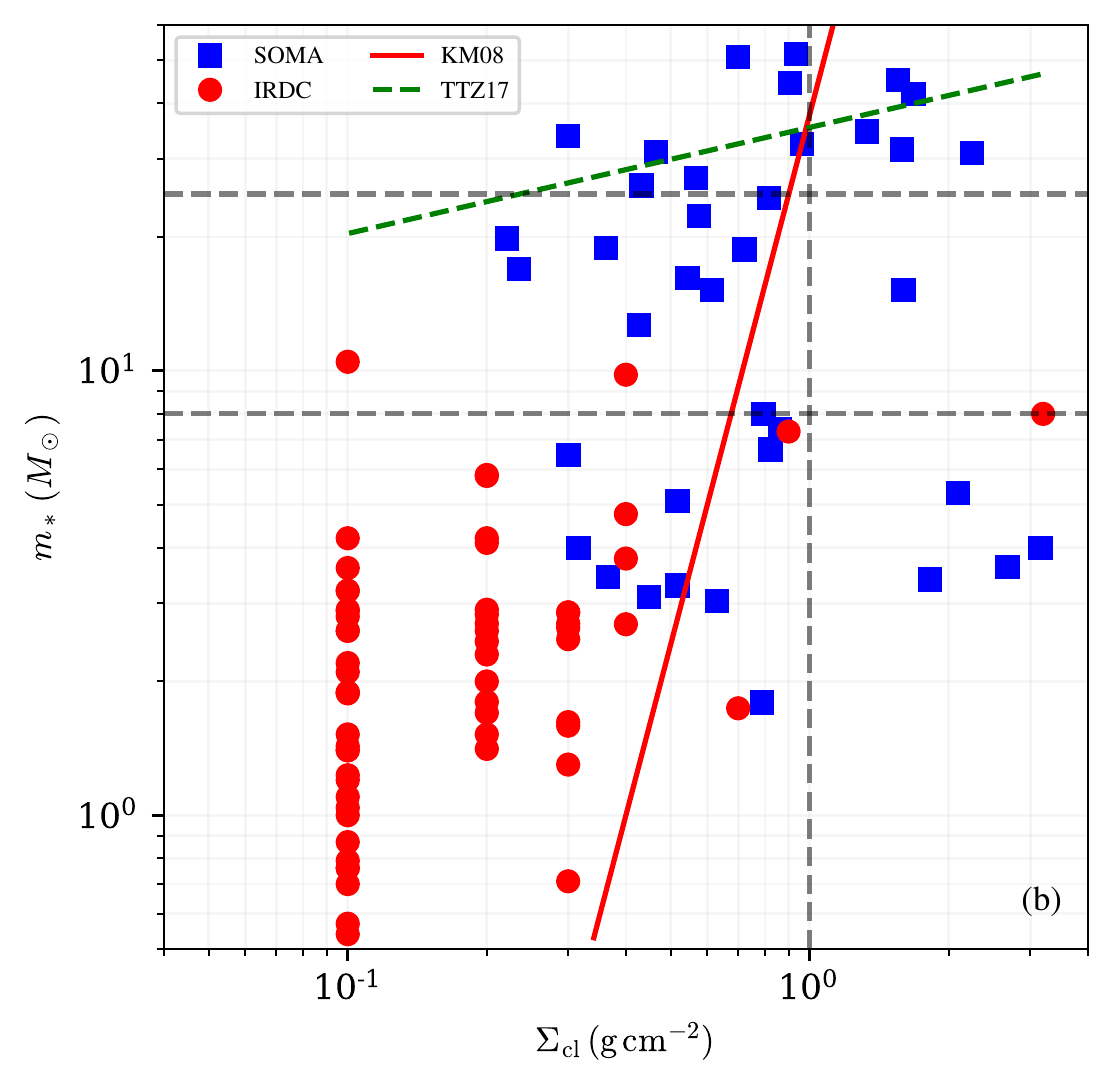}

\caption{{\it (a) Left panel:} Current protostellar mass ($m_\mathrm{*}$) versus clump environment mass surface density ($\Sigma_\mathrm{cl}$) for the 40 SOMA sources of Papers I to IV. Lines indicating reference values of $m_*=8$ and $25\:M_\odot$ and $\Sigma_{\rm cl}=1\:{\rm g\:cm}^{-2}$ (see the text) are highlighted.
{\it (b) Right panel:} As (a), but now including IRDC protostars (see the text). The red solid line shows the fiducial prediction of \citet{krumholz2008} (assuming their parameter values of $\delta=1$ and $T_b=10$\,K) for the minimum $\Sigma_{\rm cl}$ needed to form a star of given mass $m_*$. The green-dashed line shows the results for the final stellar mass formed from $100\:M_\odot$ prestellar cores as a function of $\Sigma_{\rm cl}$ \citep{tanaka2017}.
\label{fig:sigma_mstar}}
\end{figure*}

Figure\,\ref{fig:sigma_mstar}a shows the values of $m_*$ versus $\Sigma_\mathrm{cl}$ for the SOMA survey sample to date. One can see how the most massive protostars, i.e., with $m_*>25\,M_\odot$, tend to be concentrated in the higher $\Sigma_\mathrm{cl}$ region of parameter space. However, in contrast to the results presented in SOMA III, there are now some examples of such stars with $\Sigma_\mathrm{cl}$ in the range 0.3 to 1.0~$\rm g\:cm^{-2}$. Furthermore, as also found in SOMA III, there are numerous examples of \textit{high-mass} protostars, i.e., with $8\:M_\odot <m_*<25\:M_\odot$, that have $\Sigma_{\rm cl}$ spanning the full explored range from $\sim0.1$ to $3\:{\rm g\:cm}^{-2}$. Figure\,\ref{fig:sigma_mstar}b, which includes IRDC sources, also shows that lower-mass protostars, i.e., with $m_*<8\,M_\odot$, are also found across the full range of $\Sigma_\mathrm{cl}$.



Figure\,\ref{fig:sigma_mstar}b shows the fiducial condition on $\Sigma_{\rm cl}$ for massive star formation from \citet{krumholz2008} (KM08) (red solid line). The prediction is that massive protostars should only be found the right of this line, i.e., which defines a minimum $\Sigma_{\rm cl}$ for high-mass star formation. We see that the SOMA results are inconsistent with this prediction, i.e., there are numerous massive protostars that appear to be forming in conditions with $\Sigma_\mathrm{cl}\ll 1\,\mathrm{g\,cm^{-2}}$. We conclude that prevention of fragmentation of massive cores by $\sim$mG-strength magnetic fields \citep[e.g.,][]{butler2012} is more likely to be the condition needed for massive star formation. Indeed, such $B$-field strengths have been inferred to be present in some IRDCs \citep{pillai2015}, including in a massive prestellar core \citep{beuther2018a}, as well as in massive protostellar cores \citep{girart2009,zhangQ2014,beltran2019}.



While the KM08 relation does not appear to give a good description of the conditions needed to form massive stars, as noted, the data in Figure~\ref{fig:sigma_mstar} do suggest a trend that the most massive protostars tend to be in higher $\Sigma_{\rm cl}$ environments. The internal protostellar feedback model of \citet{tanaka2017} provides one way to help explain this result. Figure~\ref{fig:sigma_mstar}b shows the results of an example set of models from \citet{tanaka2017} (green-dashed line), which show the final value of $m_*$ that is expected to result from $M_c=100\:M_\odot$ prestellar cores under different $\Sigma_{\rm cl}$ conditions. As a result of internal feedback from the protostar on its infall envelope and disk, it is more difficult for the star to reach very high masses (i.e., $\gtrsim 20\:M_\odot$) when $\Sigma_{\rm cl}$ is low (i.e., $\sim 0.1\:{\rm g\:cm}^{-2}$). If the prestellar core mass function is universal under different $\Sigma_{\rm cl}$ conditions, e.g., with a maximum value of $\sim 200\:M_\odot$, then the upper envelope of the SOMA sample in the $m_*-\Sigma_{\rm cl}$ plane could be explained.


Further tests of the internal feedback model are needed. For example, the main source of protostellar feedback is that due to disk-wind outflows. A prediction is that the outflow cavity opening angle is larger at later evolutionary stages. Such a prediction can be tested by measuring the cavity opening angles based on the NIR to MIR morphology of the sources, i.e., with the shorter wavelength emission tending to emerge only from the outflow cavity. Analysis of high-resolution NIR imaging data of the SOMA sources or high-resolution CO outflow data is needed to help measure outflow cavity geometries in the sources.



\section{Conclusions}\label{sect:conclusions}

In the fourth paper of the SOMA series, we have analyzed 11 relatively isolated (as defined by SOFIA-FORCAST $37\:{\rm \mu m}$ morphology) massive protostars 
using observations from the MIR to FIR, including data from Spitzer, SOFIA, and Herschel telescopes. We have introduced new methods via the python package sedcreator, which helps with the construction of SEDs. It has a number of tools that allows one to measure fluxes and fit them to the model grid of \citet{zhang2018}. It also includes a tool to select in an algorithmic way the aperture size, important for a systematic measurement of the fluxes and subsequent SED fitting. Together with the protostars studied in SOMA Papers I-III, which we have also reanalyzed with the new methods, a total of 40 sources have been characterized in a uniform and systematic manner. Below, we summarize our main findings:

\begin{enumerate}
    \item The 11 \textit{isolated} protostars analyzed in this work span a wide range in bolometric luminosity, i.e., $\sim10^2-10^5\,L_\odot$.
    Fitting the SEDs with the RT models, we obtain protostellar masses ranging from $m_*\sim3-50\,M_\odot$, which are accreting at rates of $\dot{m}_\mathrm{disk}\sim10^{-5}-10^{-3}\,M_\odot\,\mathrm{yr^{-1}}$ from cores with initial masses $M_c\sim20-430\,M_\odot$ and in clump environments with mass surface densities ranging from $\Sigma_\mathrm{cl}\sim0.3-1.7\,\mathrm{g\,cm^{-2}}$. 
    \item We report the average results of all good fits to the SEDs and include information on the dispersion of the model parameters. This helps illustrate the range of degeneracies that are present in the SED-fitting method.
    \item The relatively isolated nature of the sources we have considered here allows us to search for MIR to FIR variability over an approximately 40\,yr time baseline by comparison with archival IRAS data. From this analysis, we do not find evidence of significant flux variability of the protostars.
    \item The distribution of SOMA and IRDC sources in the $L$ versus $M_{\rm env}$ plane shows that a large part of the theoretically expected evolutionary sequence is covered by these samples. However, these samples do not currently include sources at the very latest evolutionary stages, i.e., with relatively small envelope masses.
    \item The distribution of SOMA and IRDC sources in the $m_*$ versus $\Sigma_{\rm cl}$ plane places constraints on theoretical models for the conditions needed for massive star formation. The observed distribution is inconsistent with there being a sharp threshold minimum $\Sigma_{\rm cl}\sim 1\:{\rm g\:cm}^{-2}$ needed to form massive protostars as has been proposed by \citet{krumholz2008}. However, the upper end of the distribution appears to follow trends predicted by models of internal protostellar feedback that find higher star formation efficiency in higher $\Sigma_{\rm cl}$ conditions \citep{tanaka2017}.
    \item This sample of protostars, which appear relatively isolated in their MIR to FIR emission, is a further constraint on massive star formation theories. The images we present here can be used to test any predictions of such models, e.g., for $37\:{\rm \mu m}$ flux profiles. We anticipate that models of massive star formation that involve the presence of a surrounding cluster of lower-mass protostars may struggle to reproduce the morphologies seen in this sample.
    

    
    \item We released the open-source python package sedcreator to construct and fit SEDs. It includes several functions encapsulated into the two main classes SedFluxer and SedFitter that allow the user to measure fluxes on an image and/or fit the given fluxes to the \citet{zhang2018} model grid based on the TCA models.
   
\end{enumerate}

\begin{acknowledgments}
R.F. has received funding from the European Union's Horizon 2020 research and innovation programme under the Marie Sklodowska-Curie grant agreement No 101032092. J.C.T. acknowledges funding from NASA/USRA/SOFIA, from NSF grant AST1411527 and ERC project 788829 - MSTAR. K.E.I.T. acknowledges the support by JSPS KAKENHI Grant Numbers JP19K14760, JP19H05080, JP21H00058, and JP21H01145. Y.-L.Y acknowledges the support from the Virginia Initiative of Cosmic Origins Postdoctoral Fellowship. G.C. acknowledges funding from the Vetenskapsradet- Swedish Research Council (project code 2021-05589).
\end{acknowledgments}

\appendix 
\restartappendixnumbering

\section{SOMA I-III revisited}\label{sect:appendix_soma}

In order to achieve a systematic and uniform analysis with the methods developed in this paper, we reanalyzed all the protostars of SOMA Papers I-III with sedcreator, especially using the new automated algorithm to choose the aperture size. We then fit all the sources using the new methods detailed in the main text. In this appendix, we present the new fluxes, aperture sizes, and SED fit results. Tables \ref{tab:soma_all_fluxes} and \ref{tab:best_models_soma_ALL} present the revisited fluxes and best models, respectively, for SOMA I, II, and III sources.

\begin{longrotatetable}
\begin{deluxetable}{lccccccccccccccccc}
\tabletypesize{\tiny}
\tablecaption{Integrated Flux Densities for sources in SOMA I-III, as indicated \label{tab:soma_all_fluxes}}
\tablehead{
\colhead{Source} & \colhead{radius} & \colhead{$F_{3.6}$} & \colhead{$F_{4.5}$} & \colhead{$F_{5.8}$} & \colhead{$F_{7.7}$} & \colhead{$F_{8.0}$} & \colhead{$F_{11.1}$} & \colhead{$F_{19.1}$} & \colhead{$F_{24.4}$} & \colhead{$F_{31.5}$} & \colhead{$F_{37.1}$} &  \colhead{$F_{70}$} & \colhead{$F_{100}$} & \colhead{$F_{160}$} & \colhead{$F_{250}$}  & \colhead{$F_{350}$} & \colhead{$F_{500}$} \\
\colhead{} & \colhead{(\arcsec/pc)} & \colhead{(Jy)} & \colhead{(Jy)} & \colhead{(Jy)} & \colhead{(Jy)} & \colhead{(Jy)} & \colhead{(Jy)} & \colhead{(Jy)} & \colhead{(Jy)} & \colhead{(Jy)} & \colhead{(Jy)}  & \colhead{(Jy)} & \colhead{(Jy)} & \colhead{(Jy)} & \colhead{(Jy)}  & \colhead{(Jy)} & \colhead{(Jy)}
}
\startdata
\multicolumn{18}{c}{{\it SOMA I}}\\
\hline
\multirow{3}{*}{AFGL4029} & \multirow{2}{*}{12.75} & 0.84 & 1.03 & 4.81 & 12.88 &2.87 & $\cdots$ & 59.65 & $\cdots$ & 195.69 & 426.69 & 448.78 & 458.67 & 244.81 & 62.85 & 13.95 & 2.35 \\
& \multirow{2}{*}{0.12} & (0.92) & (1.13) & (5.26) & (12.76) & (3.56) & $\cdots$ & (60.33) & $\cdots$ & (200.14) & (434.95) & (489.45) & (533.01) & (339.53) & (118.64) & (40.23) & (10.59) \\
&  & [0.15] & [0.21] & [1.08] & [3.71] & [1.26] & $\cdots$ & [22.91] & $\cdots$ & [66.65] & [148.73] & [97.66] & [98.67] & [100.85] & [56.49] & [26.35] & [8.25] \\
\hline
\multirow{3}{*}{AFGL437} & \multirow{2}{*}{29.5} & 1.88 & 2.03 & 9.81 & 28.47 &14.36 & $\cdots$ & 266.45 & $\cdots$ & 718.34 & 865.59 & 1126.75 & $\cdots$ & 603.84 & 197.43 & 68.25 & 18.45 \\
& \multirow{2}{*}{0.29} & (1.98) & (2.10) & (10.61) & (28.47) & (15.87) & $\cdots$ & (266.45) & $\cdots$ & (718.34) & (865.59) & (1172.12) & $\cdots$ & (731.68) & (275.82) & (109.82) & (35.41) \\
&  & [0.27] & [0.29] & [1.44] & [4.03] & [2.16] & $\cdots$ & [37.68] & $\cdots$ & [101.59] & [122.41] & [160.60] & $\cdots$ & [153.74] & [83.21] & [42.68] & [17.15] \\
\hline
\multirow{3}{*}{IRAS07299} & \multirow{2}{*}{7.5} & 1.33 & 2.39 & 3.11 & 3.42 &2.97 & $\cdots$ & 73.75 & $\cdots$ & 443.77 & 678.91 & $\cdots$ & $\cdots$ & $\cdots$ & $\cdots$ & $\cdots$ & $\cdots$ \\
& \multirow{2}{*}{0.06} & (1.41) & (2.51) & (3.28) & (3.30) & (3.17) & $\cdots$ & (73.99) & $\cdots$ & (455.63) & (701.20) & $\cdots$ & $\cdots$ & $\cdots$ & $\cdots$ & $\cdots$ & $\cdots$ \\
&  & [0.19] & [0.34] & [0.44] & [0.50] & [0.42] & $\cdots$ & [10.44] & $\cdots$ & [62.95] & [96.46] & $\cdots$ & $\cdots$ & $\cdots$ & $\cdots$ & $\cdots$ & $\cdots$ \\
\hline
\multirow{3}{*}{G35.20-0.74} & \multirow{2}{*}{18.75} & 0.46 & 1.04 & 1.63 & $\cdots$ &$\cdots$ & $\cdots$ & 66.72 & $\cdots$ & 521.93 & 1109.92 & 2428.66 & $\cdots$ & 1899.93 & 656.47 & 212.66 & 41.06 \\
& \multirow{2}{*}{0.20} & (0.52) & (1.12) & (1.86) & $\cdots$ & $\cdots$ & $\cdots$ & (66.69) & $\cdots$ & (522.93) & (1110.19) & (2554.82) & $\cdots$ & (2211.92) & (858.19) & (318.12) & (84.94) \\
&  & [0.07] & [0.15] & [0.23] & $\cdots$ & $\cdots$ & $\cdots$ & [9.44] & $\cdots$ & [73.82] & [157.02] & [344.32] & $\cdots$ & [411.75] & [222.06] & [109.67] & [44.26] \\
\hline
\multirow{3}{*}{G45.47+0.05} & \multirow{2}{*}{15.0} & 0.07 & 0.20 & 0.10 & $\cdots$ &-0.11 & 0.17 & $\cdots$ & 39.28 & 137.63 & 201.95 & 1107.81 & $\cdots$ & 867.49 & 250.83 & 74.47 & 10.81 \\
& \multirow{2}{*}{0.61} & (0.13) & (0.28) & (0.64) & $\cdots$ & (1.41) & (0.17) & $\cdots$ & (33.17) & (137.63) & (201.95) & (1242.68) & $\cdots$ & (1124.52) & (378.40) & (130.56) & (32.27) \\
&  & [0.03] & [0.03] & [0.24] & $\cdots$ & [0.34] & [0.15] & $\cdots$ & [9.43] & [21.94] & [31.97] & [163.74] & $\cdots$ & [284.81] & [132.41] & [57.08] & [21.52] \\
\hline
\multirow{3}{*}{IRAS20126} & \multirow{2}{*}{12.75} & $\cdots$ & 1.52 & 1.95 & $\cdots$ &1.40 & 0.44 & $\cdots$ & 188.55 & 438.36 & 730.87 & 1569.11 & 1420.53 & 704.73 & 170.33 & 36.53 & 4.82 \\
& \multirow{2}{*}{0.10} & $\cdots$ & (1.58) & (2.28) & $\cdots$ & (2.07) & (-0.53) & $\cdots$ & (190.35) & (440.31) & (739.97) & (1637.86) & (1532.84) & (836.30) & (241.45) & (74.66) & (15.59) \\
&  & $\cdots$ & [0.22] & [0.28] & $\cdots$ & [0.22] & [0.29] & $\cdots$ & [26.68] & [62.01] & [103.39] & [223.21] & [230.15] & [165.06] & [75.09] & [38.47] & [10.79] \\
\hline
\multirow{3}{*}{CepA} & \multirow{2}{*}{36.25} & $\cdots$ & 12.59 & $\cdots$ & 8.62 &9.10 & $\cdots$ & 155.32 & $\cdots$ & 2684.87 & 6076.62 & 15565.73 & $\cdots$ & 8371.25 & $\cdots$ & 837.96 & 229.84 \\
& \multirow{2}{*}{0.12} & $\cdots$ & (13.06) & $\cdots$ & (8.62) & (11.04) & $\cdots$ & (155.32) & $\cdots$ & (2684.87) & (6076.62) & (16220.96) & $\cdots$ & (9388.70) & $\cdots$ & (1085.86) & (316.30) \\
&  & $\cdots$ & [2.17] & $\cdots$ & [1.80] & [1.45] & $\cdots$ & [27.90] & $\cdots$ & [385.09] & [864.25] & [2266.67] & $\cdots$ & [1561.01] & $\cdots$ & [274.77] & [92.36] \\
\hline
\multirow{3}{*}{NGC7538} & \multirow{2}{*}{13.75} & 0.80 & 2.21 & 7.83 & 63.05 &5.38 & $\cdots$ & 167.88 & $\cdots$ & 586.60 & 803.64 & 1407.51 & $\cdots$ & 624.73 & 142.91 & 33.90 & 3.69 \\
& \multirow{2}{*}{0.18} & (0.87) & (2.42) & (9.02) & (63.05) & (6.65) & $\cdots$ & (167.88) & $\cdots$ & (593.56) & (809.09) & (1495.68) & $\cdots$ & (804.30) & (245.16) & (79.40) & (19.80) \\
&  & [0.13] & [0.34] & [1.12] & [8.92] & [0.77] & $\cdots$ & [23.86] & $\cdots$ & [83.01] & [113.74] & [199.37] & $\cdots$ & [200.13] & [104.23] & [45.76] & [16.12] \\
\hline
\multicolumn{18}{c}{{\it SOMA II}}\\
\hline
\multirow{3}{*}{G45.12+0.13} & \multirow{2}{*}{47.0} & $\cdots$ & 7.30 & 40.61 & 92.62 &$\cdots$ & $\cdots$ & 1086.23 & $\cdots$ & 3042.48 & 4101.66 & 6542.52 & $\cdots$ & 3840.57 & $\cdots$ & 415.35 & 118.64 \\
& \multirow{2}{*}{1.69} & $\cdots$ & (7.79) & (44.50) & (92.62) & $\cdots$ & $\cdots$ & (1086.23) & $\cdots$ & (3042.48) & (4101.66) & (6857.41) & $\cdots$ & (4220.04) & $\cdots$ & (520.62) & (158.77) \\
&  & $\cdots$ & [1.11] & [5.99] & [16.42] & $\cdots$ & $\cdots$ & [174.86] & $\cdots$ & [476.23] & [635.76] & [959.77] & $\cdots$ & [662.57] & $\cdots$ & [120.55] & [43.50] \\
\hline
\multirow{3}{*}{G309.92+0.48} & \multirow{2}{*}{17.75} & 2.17 & $\cdots$ & 16.68 & 39.50 &23.00 & $\cdots$ & 365.21 & $\cdots$ & 1808.20 & 2466.17 & 3228.50 & $\cdots$ & 1669.32 & $\cdots$ & 130.49 & 23.16 \\
& \multirow{2}{*}{0.47} & (2.27) & $\cdots$ & (17.70) & (40.85) & (25.20) & $\cdots$ & (367.78) & $\cdots$ & (1826.86) & (2491.95) & (3353.09) & $\cdots$ & (1896.87) & $\cdots$ & (188.72) & (45.29) \\
&  & [0.31] & $\cdots$ & [2.38] & [5.65] & [3.36] & $\cdots$ & [51.73] & $\cdots$ & [255.73] & [348.85] & [456.64] & $\cdots$ & [327.89] & $\cdots$ & [61.09] & [22.37] \\
\hline
\multirow{3}{*}{G35.58-0.03} & \multirow{2}{*}{16.0} & 0.23 & 0.32 & 1.42 & 4.87 &3.36 & $\cdots$ & 21.53 & $\cdots$ & 265.07 & 490.17 & 1421.82 & $\cdots$ & 822.12 & 222.42 & 59.59 & 7.20 \\
& \multirow{2}{*}{0.79} & (0.36) & (0.44) & (2.21) & (4.89) & (5.83) & $\cdots$ & (23.27) & $\cdots$ & (266.56) & (499.77) & (1506.58) & $\cdots$ & (948.23) & (292.26) & (92.49) & (22.85) \\
&  & [0.05] & [0.05] & [0.29] & [1.00] & [0.47] & $\cdots$ & [3.32] & $\cdots$ & [37.78] & [69.77] & [202.44] & $\cdots$ & [171.53] & [76.60] & [33.96] & [15.68] \\
\hline
\multirow{3}{*}{IRAS16562} & \multirow{2}{*}{17.5} & 2.24 & $\cdots$ & 25.28 & 65.94 &18.90 & $\cdots$ & 237.75 & $\cdots$ & 1829.72 & 2554.34 & $\cdots$ & $\cdots$ & $\cdots$ & $\cdots$ & $\cdots$ & $\cdots$ \\
& \multirow{2}{*}{0.14} & (2.61) & $\cdots$ & (28.06) & (69.12) & (24.57) & $\cdots$ & (234.48) & $\cdots$ & (1856.85) & (2621.86) & $\cdots$ & $\cdots$ & $\cdots$ & $\cdots$ & $\cdots$ & $\cdots$ \\
&  & [0.34] & $\cdots$ & [3.70] & [9.84] & [3.34] & $\cdots$ & [34.17] & $\cdots$ & [260.96] & [364.44] & $\cdots$ & $\cdots$ & $\cdots$ & $\cdots$ & $\cdots$ & $\cdots$ \\
\hline
\multirow{3}{*}{G305.20+0.21} & \multirow{2}{*}{11.25} & $\cdots$ & $\cdots$ & $\cdots$ & 20.06 &$\cdots$ & $\cdots$ & 241.88 & $\cdots$ & 593.61 & 763.37 & 974.01 & $\cdots$ & 497.13 & $\cdots$ & 32.96 & 9.96 \\
& \multirow{2}{*}{0.22} & $\cdots$ & $\cdots$ & $\cdots$ & (21.01) & $\cdots$ & $\cdots$ & (241.87) & $\cdots$ & (614.25) & (797.89) & (1236.21) & $\cdots$ & (786.93) & $\cdots$ & (92.85) & (26.37) \\
&  & $\cdots$ & $\cdots$ & $\cdots$ & [4.48] & $\cdots$ & $\cdots$ & [44.24] & $\cdots$ & [96.84] & [119.98] & [254.26] & $\cdots$ & [298.21] & $\cdots$ & [60.07] & [16.47] \\
\hline
\multirow{3}{*}{G305.20+0.21A} & \multirow{2}{*}{10.0} & $\cdots$ & $\cdots$ & $\cdots$ & -0.15 &$\cdots$ & $\cdots$ & 2.33 & $\cdots$ & 92.97 & 154.87 & 944.78 & $\cdots$ & 814.09 & $\cdots$ & 54.49 & 5.30 \\
& \multirow{2}{*}{0.20} & $\cdots$ & $\cdots$ & $\cdots$ & (-0.30) & $\cdots$ & $\cdots$ & (0.64) & $\cdots$ & (100.43) & (164.48) & (1177.90) & $\cdots$ & (1262.40) & $\cdots$ & (134.98) & (29.66) \\
&  & $\cdots$ & $\cdots$ & $\cdots$ & [0.80] & $\cdots$ & $\cdots$ & [4.81] & $\cdots$ & [23.71] & [44.48] & [176.96] & $\cdots$ & [462.85] & $\cdots$ & [80.86] & [24.38] \\
\hline
\pagebreak
\multirow{3}{*}{G49.27-0.34} & \multirow{2}{*}{24.75} & 0.10 & 0.76 & 2.05 & 4.43 &$\cdots$ & $\cdots$ & 2.82 & $\cdots$ & 66.94 & 85.05 & 439.91 & $\cdots$ & 801.94 & 361.11 & 143.94 & 33.51 \\
& \multirow{2}{*}{0.67} & (0.24) & (0.91) & (3.33) & (4.43) & $\cdots$ & $\cdots$ & (2.82) & $\cdots$ & (66.94) & (85.05) & (547.65) & $\cdots$ & (1051.00) & (521.27) & (221.08) & (66.36) \\
&  & [0.02] & [0.11] & [0.32] & [1.34] & $\cdots$ & $\cdots$ & [0.68] & $\cdots$ & [9.69] & [12.17] & [62.75] & $\cdots$ & [273.66] & [168.11] & [79.77] & [33.19] \\
\hline
\multirow{3}{*}{G339.88-1.26} & \multirow{2}{*}{20.25} & $\cdots$ & $\cdots$ & $\cdots$ & 4.94 &$\cdots$ & $\cdots$ & 33.32 & $\cdots$ & 692.22 & 1125.47 & 3085.45 & $\cdots$ & 2021.52 & $\cdots$ & $\cdots$ & $\cdots$ \\
& \multirow{2}{*}{0.21} & $\cdots$ & $\cdots$ & $\cdots$ & (9.07) & $\cdots$ & $\cdots$ & (28.20) & $\cdots$ & (695.79) & (1145.93) & (3304.79) & $\cdots$ & (2261.84) & $\cdots$ & $\cdots$ & $\cdots$ \\
&  & $\cdots$ & $\cdots$ & $\cdots$ & [4.02] & $\cdots$ & $\cdots$ & [4.77] & $\cdots$ & [98.70] & [160.61] & [437.83] & $\cdots$ & [373.48] & $\cdots$ & $\cdots$ & $\cdots$ \\
\hline
\multicolumn{18}{c}{{\it SOMA III}}\\
\hline
\multirow{3}{*}{S235} & \multirow{2}{*}{7.0} & 0.28 & $\cdots$ & 1.94 & 5.61 &2.31 & $\cdots$ & 31.56 & $\cdots$ & 64.81 & 74.46 & $\cdots$ & $\cdots$ & $\cdots$ & $\cdots$ & $\cdots$ & $\cdots$ \\
& \multirow{2}{*}{0.06} & (0.30) & $\cdots$ & (2.05) & (5.67) & (2.43) & $\cdots$ & (32.20) & $\cdots$ & (66.47) & (77.11) & $\cdots$ & $\cdots$ & $\cdots$ & $\cdots$ & $\cdots$ & $\cdots$ \\
&  & [0.04] & $\cdots$ & [0.27] & [0.80] & [0.33] & $\cdots$ & [4.46] & $\cdots$ & [9.20] & [10.57] & $\cdots$ & $\cdots$ & $\cdots$ & $\cdots$ & $\cdots$ & $\cdots$ \\
\hline
\multirow{3}{*}{IRAS22198} & \multirow{2}{*}{6.25} & 0.01 & 0.03 & 0.09 & 0.19 &0.16 & $\cdots$ & 5.31 & $\cdots$ & 72.74 & 98.94 & $\cdots$ & 181.16 & 78.08 & 9.86 & 2.08 & $\cdots$ \\
& \multirow{2}{*}{0.02} & (0.01) & (0.04) & (0.11) & (0.32) & (0.20) & $\cdots$ & (5.95) & $\cdots$ & (75.10) & (102.04) & $\cdots$ & (216.74) & (119.06) & (33.07) & (10.21) & $\cdots$ \\
&  & $\cdots$ & [0.01] & [0.01] & [0.09] & [0.02] & $\cdots$ & [0.76] & $\cdots$ & [10.30] & [14.26] & $\cdots$ & [43.84] & [42.45] & [23.25] & [8.14] & $\cdots$ \\
\hline
\multirow{3}{*}{NGC2071} & \multirow{2}{*}{11.5} & $\cdots$ & $\cdots$ & $\cdots$ & 5.60 &$\cdots$ & $\cdots$ & 88.20 & $\cdots$ & 328.38 & 407.11 & 809.02 & 931.50 & 590.37 & $\cdots$ & 45.11 & 5.07 \\
& \multirow{2}{*}{0.02} & $\cdots$ & $\cdots$ & $\cdots$ & (5.60) & $\cdots$ & $\cdots$ & (88.20) & $\cdots$ & (335.19) & (408.69) & (854.79) & (1024.53) & (731.53) & $\cdots$ & (86.82) & (22.32) \\
&  & $\cdots$ & $\cdots$ & $\cdots$ & [0.81] & $\cdots$ & $\cdots$ & [12.63] & $\cdots$ & [47.73] & [59.44] & [126.76] & [161.27] & [164.00] & $\cdots$ & [42.19] & [17.27] \\
\hline
\multirow{3}{*}{CepE} & \multirow{2}{*}{8.5} & 0.02 & 0.07 & 0.12 & 0.18 &0.23 & $\cdots$ & 1.76 & $\cdots$ & 17.00 & 24.25 & 70.42 & $\cdots$ & 63.62 & 16.13 & 2.92 & 0.18 \\
& \multirow{2}{*}{0.03} & (0.02) & (0.07) & (0.14) & (0.03) & (0.27) & $\cdots$ & (1.89) & $\cdots$ & (17.77) & (24.86) & (73.77) & $\cdots$ & (72.78) & (24.58) & (8.07) & (1.90) \\
&  & $\cdots$ & [0.01] & [0.02] & [0.05] & [0.03] & $\cdots$ & [0.48] & $\cdots$ & [2.43] & [3.45] & [9.98] & $\cdots$ & [12.84] & [8.75] & [5.17] & [1.73] \\
\hline
\multirow{3}{*}{L1206$_\mathrm{A}$} & \multirow{2}{*}{6.5} & 0.07 & 0.18 & 0.22 & 0.26 &$\cdots$ & $\cdots$ & 1.85 & $\cdots$ & 60.30 & 104.16 & $\cdots$ & $\cdots$ & $\cdots$ & $\cdots$ & $\cdots$ & $\cdots$ \\
& \multirow{2}{*}{0.02} & (0.08) & (0.20) & (0.25) & (0.26) & $\cdots$ & $\cdots$ & (2.11) & $\cdots$ & (62.43) & (106.29) & $\cdots$ & $\cdots$ & $\cdots$ & $\cdots$ & $\cdots$ & $\cdots$ \\
&  & [0.01] & [0.03] & [0.04] & [0.09] & $\cdots$ & $\cdots$ & [0.40] & $\cdots$ & [8.54] & [14.82] & $\cdots$ & $\cdots$ & $\cdots$ & $\cdots$ & $\cdots$ & $\cdots$ \\
\hline
\multirow{3}{*}{L1206$_\mathrm{B}$} & \multirow{2}{*}{8.0} & 0.28 & 0.39 & 2.07 & 1.86 &$\cdots$ & $\cdots$ & 4.35 & $\cdots$ & 5.19 & 6.35 & $\cdots$ & $\cdots$ & $\cdots$ & $\cdots$ & $\cdots$ & $\cdots$ \\
& \multirow{2}{*}{0.03} & (0.30) & (0.41) & (2.13) & (1.88) & $\cdots$ & $\cdots$ & (4.05) & $\cdots$ & (3.87) & (5.11) & $\cdots$ & $\cdots$ & $\cdots$ & $\cdots$ & $\cdots$ & $\cdots$ \\
&  & [0.04] & [0.05] & [0.29] & [0.29] & $\cdots$ & $\cdots$ & [0.85] & $\cdots$ & [0.84] & [1.15] & $\cdots$ & $\cdots$ & $\cdots$ & $\cdots$ & $\cdots$ & $\cdots$ \\
\hline
\multirow{3}{*}{IRAS22172$_\mathrm{mir1}$} & \multirow{2}{*}{8.0} & 0.10 & 0.10 & 0.30 & 1.17 &0.72 & $\cdots$ & 0.93 & $\cdots$ & 3.28 & 6.12 & $\cdots$ & $\cdots$ & $\cdots$ & $\cdots$ & $\cdots$ & $\cdots$ \\
& \multirow{2}{*}{0.09} & (0.14) & (0.12) & (0.45) & (1.33) & (1.11) & $\cdots$ & (0.93) & $\cdots$ & (3.58) & (6.82) & $\cdots$ & $\cdots$ & $\cdots$ & $\cdots$ & $\cdots$ & $\cdots$ \\
&  & [0.10] & [0.16] & [0.40] & [0.57] & [0.42] & $\cdots$ & [1.70] & $\cdots$ & [3.01] & [3.49] & $\cdots$ & $\cdots$ & $\cdots$ & $\cdots$ & $\cdots$ & $\cdots$ \\
\hline
\multirow{3}{*}{IRAS22172$_\mathrm{mir2}$} & \multirow{2}{*}{26.75} & 0.62 & 0.74 & 2.73 & 5.41 &5.51 & $\cdots$ & 7.20 & $\cdots$ & 22.92 & 25.13 & $\cdots$ & $\cdots$ & $\cdots$ & $\cdots$ & $\cdots$ & $\cdots$ \\
& \multirow{2}{*}{0.31} & (0.68) & (0.79) & (3.05) & (5.41) & (6.26) & $\cdots$ & (7.20) & $\cdots$ & (22.92) & (25.13) & $\cdots$ & $\cdots$ & $\cdots$ & $\cdots$ & $\cdots$ & $\cdots$ \\
&  & [0.11] & [0.11] & [0.44] & [0.77] & [0.94] & $\cdots$ & [1.02] & $\cdots$ & [3.24] & [3.55] & $\cdots$ & $\cdots$ & $\cdots$ & $\cdots$ & $\cdots$ & $\cdots$ \\
\hline
\multirow{3}{*}{IRAS22172$_\mathrm{mir3}$} & \multirow{2}{*}{7.75} & 0.04 & 0.03 & 0.23 & 0.89 &0.62 & $\cdots$ & 0.90 & $\cdots$ & 4.26 & 5.71 & $\cdots$ & $\cdots$ & $\cdots$ & $\cdots$ & $\cdots$ & $\cdots$ \\
& \multirow{2}{*}{0.09} & (0.05) & (0.04) & (0.31) & (1.04) & (0.83) & $\cdots$ & (1.02) & $\cdots$ & (5.01) & (5.71) & $\cdots$ & $\cdots$ & $\cdots$ & $\cdots$ & $\cdots$ & $\cdots$ \\
&  & [0.04] & [0.07] & [0.21] & [0.29] & [0.37] & $\cdots$ & [1.11] & $\cdots$ & [1.77] & [2.17] & $\cdots$ & $\cdots$ & $\cdots$ & $\cdots$ & $\cdots$ & $\cdots$ \\
\hline
\multirow{3}{*}{IRAS21391$_\mathrm{bima2}$} & \multirow{2}{*}{7.0} & 0.01 & 0.06 & 0.11 & 0.24 &0.13 & $\cdots$ & 0.50 & $\cdots$ & 6.65 & 11.03 & $\cdots$ & $\cdots$ & $\cdots$ & $\cdots$ & $\cdots$ & $\cdots$ \\
& \multirow{2}{*}{0.03} & (0.02) & (0.08) & (0.15) & (0.34) & (0.19) & $\cdots$ & (0.53) & $\cdots$ & (6.78) & (11.52) & $\cdots$ & $\cdots$ & $\cdots$ & $\cdots$ & $\cdots$ & $\cdots$ \\
&  & [0.01] & [0.02] & [0.03] & [0.04] & [0.04] & $\cdots$ & [0.12] & $\cdots$ & [2.80] & [3.92] & $\cdots$ & $\cdots$ & $\cdots$ & $\cdots$ & $\cdots$ & $\cdots$ \\
\hline
\multirow{3}{*}{IRAS21391$_\mathrm{bima3}$} & \multirow{2}{*}{7.75} & 0.02 & 0.06 & 0.09 & 0.23 &0.11 & $\cdots$ & 0.27 & $\cdots$ & 8.66 & 12.94 & $\cdots$ & $\cdots$ & $\cdots$ & $\cdots$ & $\cdots$ & $\cdots$ \\
& \multirow{2}{*}{0.03} & (0.03) & (0.08) & (0.14) & (0.32) & (0.19) & $\cdots$ & (0.27) & $\cdots$ & (8.66) & (13.12) & $\cdots$ & $\cdots$ & $\cdots$ & $\cdots$ & $\cdots$ & $\cdots$ \\
&  & [0.01] & [0.03] & [0.04] & [0.10] & [0.06] & $\cdots$ & [0.28] & $\cdots$ & [2.58] & [4.21] & $\cdots$ & $\cdots$ & $\cdots$ & $\cdots$ & $\cdots$ & $\cdots$ \\
\hline
\multirow{3}{*}{IRAS21391$_\mathrm{mir48}$} & \multirow{2}{*}{7.75} & 0.05 & 0.08 & 0.14 & 0.23 &0.22 & $\cdots$ & 0.93 & $\cdots$ & 3.39 & 4.70 & $\cdots$ & $\cdots$ & $\cdots$ & $\cdots$ & $\cdots$ & $\cdots$ \\
& \multirow{2}{*}{0.03} & (0.05) & (0.09) & (0.17) & (0.29) & (0.26) & $\cdots$ & (0.93) & $\cdots$ & (3.14) & (4.04) & $\cdots$ & $\cdots$ & $\cdots$ & $\cdots$ & $\cdots$ & $\cdots$ \\
&  & [0.01] & [0.01] & [0.02] & [0.13] & [0.03] & $\cdots$ & [0.14] & $\cdots$ & [0.56] & [0.67] & $\cdots$ & $\cdots$ & $\cdots$ & $\cdots$ & $\cdots$ & $\cdots$ \\
\enddata
\tablecomments{
$F_{3.6}$, $F_{4.5}$, $F_{5.8}$, and $F_{8.0}$ refer to fluxes from Spitzer-IRAC at 3.6, 4.5, 5.8, and 8.0\,$\mu$m, respectively. $F_{7.7}$, $F_{11.1}$, $F_{19.1}$, $F_{24.4}$, $F_{31.5}$, and $F_{37.1}$ refer to fluxes from SOFIA-FORCAST at 7.7, 11.1, 19.1, 24.4, 31.5, and 37.5\,$\mu$m, respectively. $F_{70}$, $F_{100}$, $F_{160}$, $F_{250}$, $F_{350}$ and $F_{500}$ refer to fluxes from Herschel-PACS/SPIRE at 70, 100, 160, 250, 350, and 500\,$\mu$m, respectively. $\dagger$ No Herschel-PACS 70\,$\mu$m data is available and SOFIA-FORCAST 37\,$\mu$m was used to find the optimal aperture (Sect.\,\ref{sect:opt_rad}).}
\end{deluxetable}
\end{longrotatetable}

\begin{longrotatetable}
\begin{deluxetable}{lcccccccccccc}
\tabletypesize{\small}
\tablecaption{Parameters of the Five Best-Fitted Models and Average and Dispersion of good Models\label{tab:best_models_soma_ALL}} 
\tablewidth{18pt}
\tablehead{
\colhead{Source} &\colhead{$\chi^{2}$} & \colhead{$M_{\rm c}$} & \colhead{$\Sigma_{\rm cl}$} & \colhead{$R_{\rm core}$}  &\colhead{$m_{*}$} & \colhead{$\theta_{\rm view}$} &\colhead{$A_{V}$} & \colhead{$M_{\rm env}$} &\colhead{$\theta_{w,\rm esc}$} & \colhead{$\dot {M}_{\rm disk}$} & \colhead{$L_{\rm bol, iso}$} & \colhead{$L_{\rm bol}$} \\
\colhead{} & \colhead{} & \colhead{($M_\odot$)} & \colhead{(g $\rm cm^{-2}$)} & \colhead{(pc)} & \colhead{($M_{\odot}$)} & \colhead{(\arcdeg)} & \colhead{(mag)} & \colhead{($M_{\odot}$)} & \colhead{(deg)} &\colhead{($M_{\odot}$/yr)} & \colhead{($L_{\odot}$)} & \colhead{($L_{\odot}$)}
}
\startdata
\multicolumn{13}{c}{{\it SOMA I}}\\
\hline
AFGL4029 & 0.33 & 80 & 0.100 & 0.21 & 8 & 68 & 0 & 61.51 & 27 & 5.0$\times10^{-5}$ & 5.3$\times10^{3}$ & 9.7$\times10^{3}$ \\
$d=2.0$ kpc & 0.36 & 80 & 0.100 & 0.21 & 12 & 74 & 0 & 46.52 & 40 & 5.4$\times10^{-5}$ & 4.9$\times10^{3}$ & 1.6$\times10^{4}$ \\
$R_\mathrm{ap}=12.75\arcsec$ & 0.39 & 100 & 0.100 & 0.23 & 16 & 89 & 29 & 53.01 & 45 & 6.2$\times10^{-5}$ & 7.1$\times10^{3}$ & 3.0$\times10^{4}$ \\
$R_\mathrm{ap}=0.12$ pc & 0.40 & 40 & 0.316 & 0.08 & 8 & 44 & 0 & 22.29 & 36 & 9.2$\times10^{-5}$ & 5.0$\times10^{3}$ & 1.2$\times10^{4}$ \\
 & 0.42 & 50 & 0.316 & 0.09 & 12 & 58 & 0 & 21.56 & 46 & 1.0$\times10^{-4}$ & 4.8$\times10^{3}$ & 2.4$\times10^{4}$ \\
Average model & \#567 &  $59^{+36}_{-23}$ & $0.427^{+1.019}_{-0.301}$ & $0.09^{+0.10}_{-0.05}$ & $13^{+10}_{-6}$ & $57 \pm 22$ & $103 \pm 102$ & $21^{+32}_{-13}$ & $44 \pm 14$ & $1.4^{+1.8}_{-0.8}\times10^{-4}$ & $1.4^{+5.0}_{-1.1}\times10^{4}$ & $2.9^{+6.3}_{-2.0}\times10^{4}$\\
\hline
AFGL437 & 1.10 & 160 & 0.100 & 0.29 & 16 & 68 & 0 & 115.88 & 32 & 8.1$\times10^{-5}$ & 1.5$\times10^{4}$ & 3.3$\times10^{4}$ \\
$d=2.0$ kpc & 1.43 & 30 & 3.160 & 0.02 & 12 & 51 & 0 & 7.47 & 43 & 5.5$\times10^{-4}$ & 1.5$\times10^{4}$ & 4.9$\times10^{4}$ \\
$R_\mathrm{ap}=29.50\arcsec$ & 1.84 & 50 & 3.160 & 0.03 & 24 & 65 & 27 & 5.46 & 56 & 6.8$\times10^{-4}$ & 2.0$\times10^{4}$ & 1.9$\times10^{5}$ \\
$R_\mathrm{ap}=0.29$ pc & 1.97 & 200 & 0.100 & 0.33 & 12 & 39 & 0 & 174.14 & 20 & 8.0$\times10^{-5}$ & 1.5$\times10^{4}$ & 2.0$\times10^{4}$ \\
 & 1.97 & 160 & 0.100 & 0.29 & 24 & 86 & 26 & 86.57 & 45 & 8.5$\times10^{-5}$ & 1.9$\times10^{4}$ & 7.8$\times10^{4}$ \\
Average model & \#27 &  $115^{+99}_{-53}$ & $0.235^{+0.800}_{-0.181}$ & $0.16^{+0.30}_{-0.11}$ & $17^{+6}_{-4}$ & $67 \pm 16$ & $10 \pm 12$ & $58^{+130}_{-40}$ & $37 \pm 10$ & $1.3^{+1.8}_{-0.8}\times10^{-4}$ & $1.6^{+0.2}_{-0.2}\times10^{4}$ & $4.8^{+4.5}_{-2.3}\times10^{4}$\\
\hline
IRAS07299 & 0.26 & 40 & 1.000 & 0.05 & 12 & 51 & 37 & 15.56 & 42 & 2.5$\times10^{-4}$ & 1.1$\times10^{4}$ & 4.5$\times10^{4}$ \\
$d=1.7$ kpc & 0.31 & 80 & 0.316 & 0.12 & 16 & 68 & 14 & 41.63 & 42 & 1.5$\times10^{-4}$ & 9.7$\times10^{3}$ & 4.2$\times10^{4}$ \\
$R_\mathrm{ap}=7.50\arcsec$ & 0.33 & 50 & 1.000 & 0.05 & 16 & 65 & 7 & 16.19 & 48 & 2.8$\times10^{-4}$ & 8.6$\times10^{3}$ & 6.7$\times10^{4}$ \\
$R_\mathrm{ap}=0.06$ pc & 0.33 & 30 & 3.160 & 0.02 & 12 & 58 & 39 & 7.47 & 43 & 5.5$\times10^{-4}$ & 1.1$\times10^{4}$ & 4.9$\times10^{4}$ \\
 & 0.44 & 80 & 1.000 & 0.07 & 24 & 77 & 3 & 25.07 & 52 & 3.5$\times10^{-4}$ & 9.9$\times10^{3}$ & 1.2$\times10^{5}$ \\
Average model & \#106 &  $53^{+24}_{-17}$ & $1.594^{+2.033}_{-0.894}$ & $0.04^{+0.03}_{-0.02}$ & $15^{+7}_{-5}$ & $66 \pm 16$ & $28 \pm 30$ & $19^{+17}_{-9}$ & $42 \pm 10$ & $4.2^{+3.2}_{-1.8}\times10^{-4}$ & $1.2^{+0.4}_{-0.3}\times10^{4}$ & $6.5^{+6.6}_{-3.3}\times10^{4}$\\
\hline
G35.20-0.74 & 1.66 & 160 & 0.316 & 0.17 & 12 & 29 & 29 & 135.26 & 20 & 1.8$\times10^{-4}$ & 2.8$\times10^{4}$ & 3.8$\times10^{4}$ \\
$d=2.2$ kpc & 1.68 & 160 & 0.316 & 0.17 & 16 & 39 & 32 & 124.70 & 26 & 2.0$\times10^{-4}$ & 2.8$\times10^{4}$ & 5.0$\times10^{4}$ \\
$R_\mathrm{ap}=18.75\arcsec$ & 1.85 & 160 & 0.316 & 0.17 & 24 & 51 & 58 & 97.59 & 37 & 2.2$\times10^{-4}$ & 3.1$\times10^{4}$ & 9.9$\times10^{4}$ \\
$R_\mathrm{ap}=0.20$ pc & 1.93 & 80 & 3.160 & 0.04 & 12 & 39 & 26 & 58.05 & 22 & 8.4$\times10^{-4}$ & 2.8$\times10^{4}$ & 5.0$\times10^{4}$ \\
 & 1.97 & 200 & 0.316 & 0.19 & 16 & 29 & 64 & 162.47 & 22 & 2.2$\times10^{-4}$ & 3.7$\times10^{4}$ & 5.3$\times10^{4}$ \\
Average model & \#231 &  $122^{+67}_{-43}$ & $0.723^{+1.215}_{-0.453}$ & $0.10^{+0.10}_{-0.05}$ & $19^{+9}_{-6}$ & $62 \pm 17$ & $35 \pm 36$ & $72^{+64}_{-34}$ & $33 \pm 11$ & $3.5^{+3.3}_{-1.7}\times10^{-4}$ & $2.8^{+1.0}_{-0.7}\times10^{4}$ & $9.1^{+9.9}_{-4.7}\times10^{4}$\\
\hline
G45.47+0.05 & 0.96 & 320 & 3.160 & 0.07 & 24 & 22 & 178 & 276.82 & 15 & 1.8$\times10^{-3}$ & 3.3$\times10^{5}$ & 3.1$\times10^{5}$ \\
$d=8.4$ kpc & 0.97 & 240 & 3.160 & 0.06 & 24 & 29 & 156 & 194.50 & 18 & 1.6$\times10^{-3}$ & 2.6$\times10^{5}$ & 3.1$\times10^{5}$ \\
$R_\mathrm{ap}=15.00\arcsec$ & 1.04 & 240 & 3.160 & 0.06 & 32 & 34 & 182 & 175.32 & 23 & 1.9$\times10^{-3}$ & 3.2$\times10^{5}$ & 5.0$\times10^{5}$ \\
$R_\mathrm{ap}=0.61$ pc & 1.05 & 400 & 1.000 & 0.15 & 24 & 22 & 98 & 347.96 & 16 & 7.7$\times10^{-4}$ & 1.9$\times10^{5}$ & 2.0$\times10^{5}$ \\
 & 1.07 & 320 & 3.160 & 0.07 & 16 & 22 & 3 & 293.14 & 12 & 1.4$\times10^{-3}$ & 1.0$\times10^{5}$ & 1.1$\times10^{5}$ \\
Average model & \#330 &  $318^{+126}_{-90}$ & $1.331^{+1.636}_{-0.734}$ & $0.11^{+0.08}_{-0.05}$ & $35^{+18}_{-12}$ & $61 \pm 18$ & $88 \pm 47$ & $235^{+100}_{-70}$ & $25 \pm 8$ & $1.0^{+0.6}_{-0.4}\times10^{-3}$ & $1.6^{+0.6}_{-0.4}\times10^{5}$ & $3.7^{+2.8}_{-1.6}\times10^{5}$\\
\hline
IRAS20126 & 0.83 & 80 & 0.316 & 0.12 & 16 & 86 & 16 & 41.63 & 42 & 1.5$\times10^{-4}$ & 9.2$\times10^{3}$ & 4.2$\times10^{4}$ \\
$d=1.6$ kpc & 1.81 & 120 & 0.316 & 0.14 & 24 & 83 & 63 & 57.10 & 47 & 1.8$\times10^{-4}$ & 1.5$\times10^{4}$ & 9.3$\times10^{4}$ \\
$R_\mathrm{ap}=12.75\arcsec$ & 1.83 & 40 & 3.160 & 0.03 & 16 & 55 & 130 & 10.27 & 44 & 6.8$\times10^{-4}$ & 2.4$\times10^{4}$ & 1.1$\times10^{5}$ \\
$R_\mathrm{ap}=0.10$ pc & 2.11 & 40 & 3.160 & 0.03 & 12 & 44 & 115 & 17.69 & 35 & 6.5$\times10^{-4}$ & 2.1$\times10^{4}$ & 4.9$\times10^{4}$ \\
 & 2.17 & 100 & 0.316 & 0.13 & 16 & 58 & 55 & 61.06 & 36 & 1.6$\times10^{-4}$ & 1.4$\times10^{4}$ & 4.5$\times10^{4}$ \\
Average model & \#17 &  $91^{+31}_{-23}$ & $0.362^{+0.271}_{-0.155}$ & $0.12^{+0.06}_{-0.04}$ & $19^{+4}_{-4}$ & $77 \pm 9$ & $47 \pm 30$ & $44^{+22}_{-15}$ & $44 \pm 3$ & $1.7^{+0.8}_{-0.5}\times10^{-4}$ & $1.2^{+0.4}_{-0.3}\times10^{4}$ & $6.1^{+3.2}_{-2.1}\times10^{4}$\\
\hline
CepA & 1.04 & 120 & 0.316 & 0.14 & 12 & 55 & 69 & 93.48 & 24 & 1.6$\times10^{-4}$ & 2.0$\times10^{4}$ & 3.6$\times10^{4}$ \\
$d=0.7$ kpc & 1.34 & 160 & 0.316 & 0.17 & 24 & 68 & 112 & 97.59 & 37 & 2.2$\times10^{-4}$ & 2.8$\times10^{4}$ & 9.9$\times10^{4}$ \\
$R_\mathrm{ap}=36.25\arcsec$ & 1.61 & 160 & 0.316 & 0.17 & 32 & 89 & 119 & 72.00 & 48 & 2.2$\times10^{-4}$ & 2.5$\times10^{4}$ & 1.7$\times10^{5}$ \\
$R_\mathrm{ap}=0.12$ pc & 1.69 & 120 & 0.316 & 0.14 & 16 & 77 & 61 & 82.16 & 32 & 1.8$\times10^{-4}$ & 1.7$\times10^{4}$ & 4.6$\times10^{4}$ \\
 & 1.73 & 100 & 1.000 & 0.07 & 24 & 55 & 107 & 46.18 & 43 & 4.4$\times10^{-4}$ & 2.5$\times10^{4}$ & 1.3$\times10^{5}$ \\
Average model & \#106 &  $107^{+45}_{-31}$ & $0.544^{+0.626}_{-0.291}$ & $0.10^{+0.07}_{-0.04}$ & $16^{+6}_{-5}$ & $71 \pm 12$ & $66 \pm 31$ & $67^{+37}_{-24}$ & $32 \pm 8$ & $2.6^{+1.7}_{-1.0}\times10^{-4}$ & $2.0^{+0.5}_{-0.4}\times10^{4}$ & $6.1^{+4.5}_{-2.6}\times10^{4}$\\
\hline
NGC7538 & 0.55 & 50 & 3.160 & 0.03 & 16 & 44 & 61 & 20.52 & 37 & 7.7$\times10^{-4}$ & 4.7$\times10^{4}$ & 1.1$\times10^{5}$ \\
$d=2.6$ kpc & 0.57 & 200 & 0.100 & 0.33 & 24 & 89 & 21 & 128.41 & 37 & 9.9$\times10^{-5}$ & 2.7$\times10^{4}$ & 8.1$\times10^{4}$ \\
$R_\mathrm{ap}=13.75\arcsec$ & 0.60 & 60 & 3.160 & 0.03 & 12 & 34 & 25 & 37.85 & 27 & 7.6$\times10^{-4}$ & 3.3$\times10^{4}$ & 5.0$\times10^{4}$ \\
$R_\mathrm{ap}=0.18$ pc & 0.78 & 50 & 3.160 & 0.03 & 12 & 39 & 0 & 27.72 & 30 & 7.1$\times10^{-4}$ & 2.2$\times10^{4}$ & 5.1$\times10^{4}$ \\
 & 0.92 & 200 & 0.100 & 0.33 & 32 & 89 & 35 & 101.23 & 48 & 9.9$\times10^{-5}$ & 3.0$\times10^{4}$ & 1.5$\times10^{5}$ \\
Average model & \#55 &  $122^{+90}_{-52}$ & $0.568^{+1.573}_{-0.417}$ & $0.11^{+0.16}_{-0.06}$ & $27^{+14}_{-9}$ & $67 \pm 16$ & $23 \pm 19$ & $50^{+50}_{-25}$ & $46 \pm 9$ & $3.0^{+3.7}_{-1.7}\times10^{-4}$ & $2.9^{+0.5}_{-0.5}\times10^{4}$ & $1.6^{+1.5}_{-0.8}\times10^{5}$\\
\hline
\multicolumn{13}{c}{{\it SOMA II}}\\
\hline
G45.12+0.13 & 15.29 & 240 & 3.160 & 0.06 & 32 & 29 & 0 & 175.32 & 23 & 1.9$\times10^{-3}$ & 4.5$\times10^{5}$ & 5.0$\times10^{5}$ \\
$d=7.4$ kpc & 15.30 & 480 & 1.000 & 0.16 & 96 & 48 & 0 & 237.78 & 43 & 1.3$\times10^{-3}$ & 4.4$\times10^{5}$ & 1.6$\times10^{6}$ \\
$R_\mathrm{ap}=47.00\arcsec$ & 15.61 & 480 & 1.000 & 0.16 & 48 & 29 & 1 & 366.96 & 25 & 1.1$\times10^{-3}$ & 4.5$\times10^{5}$ & 5.4$\times10^{5}$ \\
$R_\mathrm{ap}=1.69$ pc & 17.32 & 480 & 1.000 & 0.16 & 64 & 39 & 0 & 324.63 & 32 & 1.2$\times10^{-3}$ & 3.6$\times10^{5}$ & 8.4$\times10^{5}$ \\
 & 17.95 & 400 & 1.000 & 0.15 & 64 & 39 & 4 & 246.43 & 36 & 1.1$\times10^{-3}$ & 4.2$\times10^{5}$ & 8.2$\times10^{5}$ \\
Average model & \#815 &  $350^{+150}_{-105}$ & $0.906^{+1.724}_{-0.594}$ & $0.14^{+0.14}_{-0.07}$ & $44^{+31}_{-18}$ & $57 \pm 21$ & $23 \pm 53$ & $217^{+138}_{-84}$ & $32 \pm 13$ & $8.3^{+8.6}_{-4.2}\times10^{-4}$ & $2.4^{+4.5}_{-1.6}\times10^{5}$ & $4.9^{+5.1}_{-2.5}\times10^{5}$\\
\hline
G309.92+0.48 & 1.66 & 480 & 1.000 & 0.16 & 96 & 51 & 20 & 237.78 & 43 & 1.3$\times10^{-3}$ & 3.3$\times10^{5}$ & 1.6$\times10^{6}$ \\
$d=5.5$ kpc & 1.70 & 240 & 3.160 & 0.06 & 32 & 39 & 6 & 175.32 & 23 & 1.9$\times10^{-3}$ & 2.7$\times10^{5}$ & 5.0$\times10^{5}$ \\
$R_\mathrm{ap}=17.75\arcsec$ & 1.88 & 240 & 3.160 & 0.06 & 24 & 29 & 1 & 194.50 & 18 & 1.6$\times10^{-3}$ & 2.6$\times10^{5}$ & 3.1$\times10^{5}$ \\
$R_\mathrm{ap}=0.47$ pc & 1.89 & 400 & 1.000 & 0.15 & 64 & 44 & 3 & 246.43 & 36 & 1.1$\times10^{-3}$ & 2.7$\times10^{5}$ & 8.2$\times10^{5}$ \\
 & 2.01 & 400 & 1.000 & 0.15 & 48 & 34 & 11 & 288.73 & 29 & 1.0$\times10^{-3}$ & 3.0$\times10^{5}$ & 5.3$\times10^{5}$ \\
Average model & \#21 &  $340^{+147}_{-102}$ & $1.550^{+1.798}_{-0.832}$ & $0.11^{+0.08}_{-0.05}$ & $45^{+31}_{-18}$ & $39 \pm 9$ & $14 \pm 18$ & $230^{+76}_{-57}$ & $29 \pm 9$ & $1.3^{+0.6}_{-0.4}\times10^{-3}$ & $3.0^{+0.7}_{-0.6}\times10^{5}$ & $6.1^{+4.7}_{-2.7}\times10^{5}$\\
\hline
G35.58-0.03 & 0.82 & 480 & 3.160 & 0.09 & 24 & 29 & 11 & 440.54 & 12 & 2.0$\times10^{-3}$ & 2.7$\times10^{5}$ & 2.9$\times10^{5}$ \\
$d=10.2$ kpc & 1.04 & 400 & 3.160 & 0.08 & 24 & 39 & 8 & 361.65 & 13 & 1.9$\times10^{-3}$ & 2.6$\times10^{5}$ & 3.0$\times10^{5}$ \\
$R_\mathrm{ap}=16.00\arcsec$ & 1.57 & 320 & 3.160 & 0.07 & 24 & 51 & 3 & 276.82 & 15 & 1.8$\times10^{-3}$ & 2.3$\times10^{5}$ & 3.1$\times10^{5}$ \\
$R_\mathrm{ap}=0.79$ pc & 1.70 & 480 & 1.000 & 0.16 & 48 & 48 & 7 & 366.96 & 25 & 1.1$\times10^{-3}$ & 2.4$\times10^{5}$ & 5.4$\times10^{5}$ \\
 & 1.93 & 480 & 1.000 & 0.16 & 64 & 55 & 24 & 324.63 & 32 & 1.2$\times10^{-3}$ & 2.6$\times10^{5}$ & 8.4$\times10^{5}$ \\
Average model & \#37 &  $417^{+73}_{-62}$ & $2.245^{+1.581}_{-0.928}$ & $0.10^{+0.04}_{-0.03}$ & $31^{+15}_{-10}$ & $48 \pm 12$ & $10 \pm 14$ & $349^{+59}_{-50}$ & $18 \pm 7$ & $1.6^{+0.4}_{-0.3}\times10^{-3}$ & $2.5^{+0.2}_{-0.2}\times10^{5}$ & $3.9^{+1.9}_{-1.3}\times10^{5}$\\
\hline
IRAS16562 & 0.38 & 200 & 0.316 & 0.19 & 32 & 48 & 67 & 114.70 & 41 & 2.6$\times10^{-4}$ & 5.5$\times10^{4}$ & 1.8$\times10^{5}$ \\
$d=1.7$ kpc & 0.41 & 200 & 0.316 & 0.19 & 24 & 39 & 48 & 139.72 & 32 & 2.5$\times10^{-4}$ & 4.8$\times10^{4}$ & 1.0$\times10^{5}$ \\
$R_\mathrm{ap}=17.50\arcsec$ & 0.45 & 100 & 3.160 & 0.04 & 16 & 29 & 119 & 68.61 & 23 & 1.1$\times10^{-3}$ & 1.0$\times10^{5}$ & 1.2$\times10^{5}$ \\
$R_\mathrm{ap}=0.14$ pc & 0.46 & 120 & 1.000 & 0.08 & 16 & 29 & 90 & 88.34 & 25 & 4.5$\times10^{-4}$ & 7.8$\times10^{4}$ & 9.7$\times10^{4}$ \\
 & 0.47 & 240 & 0.316 & 0.20 & 48 & 58 & 86 & 104.17 & 50 & 2.9$\times10^{-4}$ & 6.7$\times10^{4}$ & 3.8$\times10^{5}$ \\
Average model & \#279 &  $164^{+92}_{-59}$ & $0.963^{+1.686}_{-0.613}$ & $0.10^{+0.09}_{-0.05}$ & $32^{+24}_{-14}$ & $63 \pm 18$ & $44 \pm 39$ & $78^{+50}_{-30}$ & $40 \pm 12$ & $5.6^{+6.1}_{-2.9}\times10^{-4}$ & $5.1^{+2.5}_{-1.7}\times10^{4}$ & $2.5^{+3.9}_{-1.5}\times10^{5}$\\
\hline
G305.20+0.21 & 0.15 & 320 & 0.100 & 0.42 & 64 & 86 & 13 & 102.12 & 60 & 1.2$\times10^{-4}$ & 6.2$\times10^{4}$ & 6.0$\times10^{5}$ \\
$d=4.1$ kpc & 0.21 & 60 & 3.160 & 0.03 & 16 & 39 & 2 & 31.12 & 32 & 8.4$\times10^{-4}$ & 5.8$\times10^{4}$ & 1.1$\times10^{5}$ \\
$R_\mathrm{ap}=11.25\arcsec$ & 0.47 & 160 & 3.160 & 0.05 & 64 & 68 & 0 & 22.88 & 61 & 1.3$\times10^{-3}$ & 5.9$\times10^{4}$ & 8.6$\times10^{5}$ \\
$R_\mathrm{ap}=0.22$ pc & 0.48 & 100 & 3.160 & 0.04 & 32 & 48 & 48 & 36.92 & 42 & 1.2$\times10^{-3}$ & 1.1$\times10^{5}$ & 3.5$\times10^{5}$ \\
 & 0.53 & 80 & 3.160 & 0.04 & 24 & 44 & 45 & 35.12 & 37 & 1.1$\times10^{-3}$ & 1.0$\times10^{5}$ & 2.6$\times10^{5}$ \\
Average model & \#106 &  $213^{+142}_{-85}$ & $0.299^{+0.934}_{-0.227}$ & $0.20^{+0.31}_{-0.12}$ & $34^{+16}_{-11}$ & $64 \pm 16$ & $12 \pm 15$ & $108^{+124}_{-58}$ & $41 \pm 11$ & $2.5^{+3.7}_{-1.5}\times10^{-4}$ & $6.0^{+1.7}_{-1.3}\times10^{4}$ & $2.2^{+2.2}_{-1.1}\times10^{5}$\\
\hline
G305.20+0.21A & 0.20 & 320 & 0.316 & 0.23 & 12 & 22 & 98 & 293.02 & 13 & 2.2$\times10^{-4}$ & 3.6$\times10^{4}$ & 4.0$\times10^{4}$ \\
$d=4.1$ kpc & 0.21 & 320 & 0.316 & 0.23 & 16 & 13 & 237 & 283.06 & 16 & 2.5$\times10^{-4}$ & 3.6$\times10^{5}$ & 6.1$\times10^{4}$ \\
$R_\mathrm{ap}=10.00\arcsec$ & 0.22 & 100 & 3.160 & 0.04 & 12 & 48 & 58 & 76.73 & 20 & 9.4$\times10^{-4}$ & 2.9$\times10^{4}$ & 5.2$\times10^{4}$ \\
$R_\mathrm{ap}=0.20$ pc & 0.22 & 120 & 3.160 & 0.05 & 12 & 29 & 120 & 98.54 & 18 & 9.6$\times10^{-4}$ & 4.0$\times10^{4}$ & 5.2$\times10^{4}$ \\
 & 0.23 & 240 & 0.316 & 0.20 & 12 & 68 & 80 & 216.44 & 15 & 2.0$\times10^{-4}$ & 3.1$\times10^{4}$ & 4.1$\times10^{4}$ \\
Average model & \#1520 &  $174^{+130}_{-74}$ & $0.817^{+1.364}_{-0.511}$ & $0.11^{+0.11}_{-0.05}$ & $25^{+24}_{-12}$ & $57 \pm 22$ & $168 \pm 100$ & $97^{+101}_{-50}$ & $33 \pm 15$ & $4.6^{+4.5}_{-2.3}\times10^{-4}$ & $8.1^{+25.2}_{-6.1}\times10^{4}$ & $1.5^{+3.2}_{-1.0}\times10^{5}$\\
\hline
G49.27-0.34 & 1.73 & 480 & 0.100 & 0.51 & 24 & 22 & 222 & 417.72 & 21 & 1.4$\times10^{-4}$ & 2.0$\times10^{5}$ & 8.7$\times10^{4}$ \\
$d=5.5$ kpc & 1.90 & 400 & 0.100 & 0.47 & 24 & 29 & 192 & 331.01 & 24 & 1.3$\times10^{-4}$ & 7.0$\times10^{4}$ & 8.6$\times10^{4}$ \\
$R_\mathrm{ap}=24.75\arcsec$ & 1.94 & 320 & 0.100 & 0.42 & 32 & 29 & 257 & 228.39 & 34 & 1.3$\times10^{-4}$ & 5.3$\times10^{5}$ & 1.6$\times10^{5}$ \\
$R_\mathrm{ap}=0.67$ pc & 1.96 & 80 & 3.160 & 0.04 & 16 & 29 & 237 & 50.01 & 27 & 9.5$\times10^{-4}$ & 1.8$\times10^{5}$ & 1.1$\times10^{5}$ \\
 & 1.96 & 240 & 0.316 & 0.20 & 16 & 22 & 173 & 205.72 & 20 & 2.3$\times10^{-4}$ & 1.0$\times10^{5}$ & 5.5$\times10^{4}$ \\
Average model & \#972 &  $197^{+167}_{-90}$ & $0.432^{+1.018}_{-0.303}$ & $0.16^{+0.22}_{-0.09}$ & $26^{+24}_{-12}$ & $56 \pm 21$ & $163 \pm 94$ & $112^{+126}_{-59}$ & $35 \pm 14$ & $3.0^{+3.3}_{-1.6}\times10^{-4}$ & $8.5^{+27.1}_{-6.4}\times10^{4}$ & $1.5^{+2.9}_{-1.0}\times10^{5}$\\
\hline
G339.88-1.26 & 2.34 & 200 & 0.316 & 0.19 & 12 & 80 & 20 & 172.68 & 17 & 1.9$\times10^{-4}$ & 2.8$\times10^{4}$ & 4.0$\times10^{4}$ \\
$d=2.1$ kpc & 2.38 & 200 & 0.316 & 0.19 & 24 & 86 & 73 & 139.72 & 32 & 2.5$\times10^{-4}$ & 3.7$\times10^{4}$ & 1.0$\times10^{5}$ \\
$R_\mathrm{ap}=20.25\arcsec$ & 2.43 & 200 & 0.316 & 0.19 & 16 & 83 & 36 & 162.47 & 22 & 2.2$\times10^{-4}$ & 3.0$\times10^{4}$ & 5.3$\times10^{4}$ \\
$R_\mathrm{ap}=0.21$ pc & 2.50 & 240 & 0.316 & 0.20 & 48 & 71 & 134 & 104.17 & 50 & 2.9$\times10^{-4}$ & 5.5$\times10^{4}$ & 3.8$\times10^{5}$ \\
 & 2.53 & 240 & 0.316 & 0.20 & 12 & 86 & 19 & 216.44 & 15 & 2.0$\times10^{-4}$ & 3.1$\times10^{4}$ & 4.1$\times10^{4}$ \\
Average model & \#322 &  $180^{+108}_{-68}$ & $0.576^{+0.764}_{-0.328}$ & $0.13^{+0.11}_{-0.06}$ & $22^{+20}_{-11}$ & $65 \pm 16$ & $74 \pm 55$ & $110^{+88}_{-49}$ & $31 \pm 14$ & $3.5^{+2.9}_{-1.6}\times10^{-4}$ & $4.0^{+1.9}_{-1.3}\times10^{4}$ & $1.2^{+2.4}_{-0.8}\times10^{5}$\\
\hline
\multicolumn{13}{c}{{\it SOMA III}}\\
\hline
S235 & 1.09 & 10 & 3.160 & 0.01 & 2 & 39 & 0 & 5.65 & 35 & 1.8$\times10^{-4}$ & 1.4$\times10^{3}$ & 2.6$\times10^{3}$ \\
$d=1.8$ kpc & 2.97 & 50 & 0.316 & 0.09 & 16 & 80 & 0 & 7.84 & 68 & 7.1$\times10^{-5}$ & 1.4$\times10^{3}$ & 3.1$\times10^{4}$ \\
$R_\mathrm{ap}=7.00\arcsec$ & 3.55 & 20 & 3.160 & 0.02 & 4 & 39 & 0 & 11.60 & 34 & 3.1$\times10^{-4}$ & 1.6$\times10^{3}$ & 3.3$\times10^{3}$ \\
$R_\mathrm{ap}=0.06$ pc & 5.59 & 60 & 1.000 & 0.06 & 24 & 89 & 31 & 4.87 & 71 & 1.9$\times10^{-4}$ & 2.1$\times10^{3}$ & 9.3$\times10^{4}$ \\
 & 5.76 & 80 & 1.000 & 0.07 & 32 & 89 & 33 & 2.70 & 79 & 1.4$\times10^{-4}$ & 1.6$\times10^{3}$ & 1.6$\times10^{5}$ \\
Average model & \#5 &  $30^{+33}_{-16}$ & $0.794^{+2.008}_{-0.569}$ & $0.05^{+0.08}_{-0.03}$ & $8^{+13}_{-5}$ & $64 \pm 20$ & $2 \pm 4$ & $8^{+2}_{-2}$ & $54 \pm 16$ & $1.2^{+1.2}_{-0.6}\times10^{-4}$ & $1.4^{+0.1}_{-0.1}\times10^{3}$ & $1.2^{+3.2}_{-0.9}\times10^{4}$\\
\hline
IRAS22198 & 6.05 & 10 & 3.160 & 0.01 & 4 & 62 & 69 & 1.65 & 56 & 1.9$\times10^{-4}$ & 2.9$\times10^{2}$ & 1.9$\times10^{3}$ \\
$d=0.8$ kpc & 9.05 & 10 & 1.000 & 0.02 & 2 & 44 & 51 & 5.33 & 39 & 7.5$\times10^{-5}$ & 2.6$\times10^{2}$ & 7.6$\times10^{2}$ \\
$R_\mathrm{ap}=6.25\arcsec$ & 13.15 & 10 & 0.316 & 0.04 & 1 & 34 & 24 & 7.50 & 28 & 2.5$\times10^{-5}$ & 1.4$\times10^{2}$ & 2.6$\times10^{2}$ \\
$R_\mathrm{ap}=0.02$ pc & 13.23 & 10 & 0.316 & 0.04 & 0 & 22 & 38 & 8.75 & 18 & 1.9$\times10^{-5}$ & 1.7$\times10^{2}$ & 1.9$\times10^{2}$ \\
 & 13.33 & 10 & 1.000 & 0.02 & 4 & 74 & 47 & 1.29 & 59 & 7.7$\times10^{-5}$ & 1.3$\times10^{2}$ & 1.1$\times10^{3}$ \\
Average model & \#7 &  $10^{+0}_{-0}$ & $2.681^{+1.461}_{-0.945}$ & $0.01^{+0.00}_{-0.00}$ & $4^{+1}_{-1}$ & $67 \pm 13$ & $41 \pm 18$ & $2^{+1}_{-1}$ & $54 \pm 6$ & $1.7^{+0.7}_{-0.5}\times10^{-4}$ & $2.1^{+0.5}_{-0.4}\times10^{2}$ & $1.6^{+0.7}_{-0.5}\times10^{3}$\\
\hline
NGC2071 & 2.71 & 10 & 3.160 & 0.01 & 4 & 62 & 0 & 1.65 & 56 & 1.9$\times10^{-4}$ & 2.9$\times10^{2}$ & 1.9$\times10^{3}$ \\
$d=0.4$ kpc & 8.62 & 10 & 1.000 & 0.02 & 2 & 44 & 0 & 5.33 & 39 & 7.5$\times10^{-5}$ & 2.6$\times10^{2}$ & 7.6$\times10^{2}$ \\
$R_\mathrm{ap}=11.50\arcsec$ & 13.50 & 10 & 1.000 & 0.02 & 4 & 65 & 0 & 1.29 & 59 & 7.7$\times10^{-5}$ & 1.7$\times10^{2}$ & 1.1$\times10^{3}$ \\
$R_\mathrm{ap}=0.02$ pc & 17.96 & 10 & 0.316 & 0.04 & 0 & 22 & 0 & 8.75 & 18 & 1.9$\times10^{-5}$ & 1.7$\times10^{2}$ & 1.9$\times10^{2}$ \\
 & 19.31 & 10 & 1.000 & 0.02 & 1 & 29 & 26 & 7.78 & 25 & 6.0$\times10^{-5}$ & 5.6$\times10^{2}$ & 7.7$\times10^{2}$ \\
Average model & \#2 &  $10^{+0}_{-0}$ & $3.160^{+0.000}_{-0.000}$ & $0.01^{+0.00}_{-0.00}$ & $4^{+0}_{-0}$ & $60 \pm 2$ & $26 \pm 26$ & $2^{+0}_{-0}$ & $56 \pm 0$ & $1.9^{+0.0}_{-0.0}\times10^{-4}$ & $3.8^{+1.7}_{-1.2}\times10^{2}$ & $1.9^{+0.0}_{-0.0}\times10^{3}$\\
\hline
CepE & 2.09 & 10 & 1.000 & 0.02 & 4 & 89 & 49 & 1.29 & 59 & 7.7$\times10^{-5}$ & 1.1$\times10^{2}$ & 1.1$\times10^{3}$ \\
$d=0.7$ kpc & 2.35 & 10 & 0.316 & 0.04 & 2 & 55 & 4 & 4.96 & 43 & 3.0$\times10^{-5}$ & 7.2$\times10^{1}$ & 2.8$\times10^{2}$ \\
$R_\mathrm{ap}=8.50\arcsec$ & 6.28 & 10 & 0.316 & 0.04 & 1 & 34 & 53 & 7.50 & 28 & 2.5$\times10^{-5}$ & 1.4$\times10^{2}$ & 2.6$\times10^{2}$ \\
$R_\mathrm{ap}=0.03$ pc & 11.84 & 10 & 0.316 & 0.04 & 0 & 22 & 84 & 8.75 & 18 & 1.9$\times10^{-5}$ & 1.7$\times10^{2}$ & 1.9$\times10^{2}$ \\
 & 12.94 & 10 & 3.160 & 0.01 & 4 & 55 & 359 & 1.65 & 56 & 1.9$\times10^{-4}$ & 1.5$\times10^{3}$ & 1.9$\times10^{3}$ \\
Average model & \#15 &  $10^{+0}_{-0}$ & $0.631^{+0.501}_{-0.279}$ & $0.03^{+0.01}_{-0.01}$ & $3^{+1}_{-1}$ & $70 \pm 11$ & $47 \pm 40$ & $2^{+2}_{-1}$ & $53 \pm 8$ & $5.3^{+3.2}_{-2.0}\times10^{-5}$ & $10.0^{+4.1}_{-2.9}\times10^{1}$ & $6.6^{+6.7}_{-3.3}\times10^{2}$\\
\hline
L1206$_\mathrm{_A}$ & 0.75 & 40 & 1.000 & 0.05 & 4 & 39 & 0 & 31.54 & 23 & 1.7$\times10^{-4}$ & 1.2$\times10^{3}$ & 2.2$\times10^{3}$ \\
$d=0.8$ kpc & 0.81 & 30 & 1.000 & 0.04 & 2 & 29 & 17 & 25.89 & 19 & 1.2$\times10^{-4}$ & 1.2$\times10^{3}$ & 1.7$\times10^{3}$ \\
$R_\mathrm{ap}=6.50\arcsec$ & 0.93 & 40 & 1.000 & 0.05 & 2 & 22 & 67 & 35.79 & 16 & 1.3$\times10^{-4}$ & 1.6$\times10^{3}$ & 2.0$\times10^{3}$ \\
$R_\mathrm{ap}=0.02$ pc & 1.01 & 30 & 1.000 & 0.04 & 4 & 44 & 4 & 21.60 & 28 & 1.5$\times10^{-4}$ & 9.2$\times10^{2}$ & 2.0$\times10^{3}$ \\
 & 1.03 & 20 & 3.160 & 0.02 & 2 & 29 & 212 & 15.99 & 22 & 2.4$\times10^{-4}$ & 2.6$\times10^{3}$ & 3.9$\times10^{3}$ \\
Average model & \#46 &  $24^{+7}_{-5}$ & $1.823^{+1.436}_{-0.803}$ & $0.03^{+0.01}_{-0.01}$ & $3^{+3}_{-2}$ & $49 \pm 18$ & $110 \pm 84$ & $13^{+15}_{-7}$ & $30 \pm 13$ & $2.0^{+0.8}_{-0.6}\times10^{-4}$ & $1.7^{+0.6}_{-0.4}\times10^{3}$ & $4.3^{+4.0}_{-2.1}\times10^{3}$\\
\hline
L1206$_\mathrm{_B}$ & 2.18 & 10 & 0.316 & 0.04 & 4 & 77 & 32 & 0.63 & 68 & 2.4$\times10^{-5}$ & 4.9$\times10^{1}$ & 6.7$\times10^{2}$ \\
$d=0.8$ kpc & 4.60 & 30 & 1.000 & 0.04 & 0 & 13 & 134 & 29.10 & 8 & 6.0$\times10^{-5}$ & 1.0$\times10^{3}$ & 4.2$\times10^{2}$ \\
$R_\mathrm{ap}=8.00\arcsec$ & 5.67 & 10 & 0.316 & 0.04 & 2 & 44 & 186 & 4.96 & 43 & 3.0$\times10^{-5}$ & 3.3$\times10^{2}$ & 2.8$\times10^{2}$ \\
$R_\mathrm{ap}=0.03$ pc & 6.18 & 20 & 1.000 & 0.03 & 0 & 13 & 237 & 19.18 & 10 & 5.4$\times10^{-5}$ & 1.7$\times10^{3}$ & 4.5$\times10^{2}$ \\
 & 6.39 & 10 & 0.316 & 0.04 & 1 & 29 & 246 & 7.50 & 28 & 2.5$\times10^{-5}$ & 5.3$\times10^{2}$ & 2.6$\times10^{2}$ \\
Average model & \#20 &  $10^{+0}_{-0}$ & $0.316^{+0.000}_{-0.000}$ & $0.04^{+0.00}_{-0.00}$ & $4^{+0}_{-0}$ & $57 \pm 21$ & $101 \pm 57$ & $1^{+0}_{-0}$ & $68 \pm 0$ & $2.4^{+0.0}_{-0.0}\times10^{-5}$ & $3.1^{+11.4}_{-2.4}\times10^{2}$ & $6.7^{+0.0}_{-0.0}\times10^{2}$\\
\hline
IRAS22172$_\mathrm{_mir1}$ & 0.15 & 30 & 0.100 & 0.13 & 1 & 13 & 73 & 27.26 & 15 & 1.5$\times10^{-5}$ & 8.7$\times10^{2}$ & 1.7$\times10^{2}$ \\
$d=2.4$ kpc & 0.15 & 60 & 0.100 & 0.18 & 1 & 13 & 57 & 57.46 & 10 & 1.8$\times10^{-5}$ & 6.1$\times10^{2}$ & 2.0$\times10^{2}$ \\
$R_\mathrm{ap}=8.00\arcsec$ & 0.15 & 40 & 0.100 & 0.15 & 1 & 13 & 63 & 38.03 & 12 & 1.6$\times10^{-5}$ & 6.8$\times10^{2}$ & 1.7$\times10^{2}$ \\
$R_\mathrm{ap}=0.09$ pc & 0.16 & 30 & 0.100 & 0.13 & 2 & 22 & 67 & 24.61 & 23 & 2.0$\times10^{-5}$ & 8.0$\times10^{2}$ & 2.4$\times10^{2}$ \\
 & 0.16 & 20 & 0.100 & 0.10 & 4 & 48 & 47 & 9.85 & 43 & 2.1$\times10^{-5}$ & 3.4$\times10^{2}$ & 6.8$\times10^{2}$ \\
Average model & \#6240 &  $85^{+140}_{-53}$ & $0.822^{+1.625}_{-0.546}$ & $0.07^{+0.07}_{-0.03}$ & $7^{+20}_{-5}$ & $57 \pm 21$ & $425 \pm 286$ & $52^{+133}_{-38}$ & $26 \pm 18$ & $2.1^{+6.1}_{-1.6}\times10^{-4}$ & $8.1^{+92.8}_{-7.5}\times10^{3}$ & $1.3^{+15.2}_{-1.2}\times10^{4}$\\
\hline
IRAS22172$_\mathrm{_mir2}$ & 0.57 & 20 & 0.316 & 0.06 & 8 & 77 & 11 & 1.63 & 66 & 4.4$\times10^{-5}$ & 8.6$\times10^{2}$ & 9.9$\times10^{3}$ \\
$d=2.4$ kpc & 0.65 & 100 & 0.316 & 0.13 & 32 & 86 & 27 & 4.49 & 81 & 5.0$\times10^{-5}$ & 1.3$\times10^{3}$ & 1.4$\times10^{5}$ \\
$R_\mathrm{ap}=26.75\arcsec$ & 0.66 & 30 & 0.100 & 0.13 & 8 & 65 & 28 & 8.93 & 57 & 2.6$\times10^{-5}$ & 1.2$\times10^{3}$ & 6.3$\times10^{3}$ \\
$R_\mathrm{ap}=0.31$ pc & 0.86 & 80 & 0.100 & 0.21 & 4 & 22 & 0 & 70.84 & 18 & 3.7$\times10^{-5}$ & 9.0$\times10^{2}$ & 8.5$\times10^{2}$ \\
 & 0.86 & 60 & 0.100 & 0.18 & 4 & 22 & 37 & 51.25 & 21 & 3.4$\times10^{-5}$ & 2.1$\times10^{3}$ & 8.9$\times10^{2}$ \\
Average model & \#45 &  $36^{+45}_{-20}$ & $0.300^{+0.574}_{-0.197}$ & $0.08^{+0.10}_{-0.04}$ & $6^{+10}_{-4}$ & $52 \pm 30$ & $49 \pm 42$ & $10^{+23}_{-7}$ & $46 \pm 22$ & $5.5^{+4.9}_{-2.6}\times10^{-5}$ & $2.1^{+3.2}_{-1.3}\times10^{3}$ & $5.2^{+28.5}_{-4.4}\times10^{3}$\\
\hline
IRAS22172$_\mathrm{_mir3}$ & 0.18 & 10 & 1.000 & 0.02 & 4 & 62 & 31 & 1.29 & 59 & 7.7$\times10^{-5}$ & 2.4$\times10^{2}$ & 1.1$\times10^{3}$ \\
$d=2.4$ kpc & 0.19 & 40 & 0.100 & 0.15 & 2 & 22 & 48 & 35.65 & 19 & 2.2$\times10^{-5}$ & 3.9$\times10^{2}$ & 2.7$\times10^{2}$ \\
$R_\mathrm{ap}=7.75\arcsec$ & 0.21 & 60 & 0.100 & 0.18 & 2 & 13 & 134 & 55.43 & 15 & 2.5$\times10^{-5}$ & 1.7$\times10^{3}$ & 3.5$\times10^{2}$ \\
$R_\mathrm{ap}=0.09$ pc & 0.21 & 20 & 0.100 & 0.10 & 4 & 62 & 23 & 9.85 & 43 & 2.1$\times10^{-5}$ & 2.3$\times10^{2}$ & 6.8$\times10^{2}$ \\
 & 0.21 & 50 & 0.100 & 0.16 & 2 & 13 & 128 & 45.52 & 16 & 2.4$\times10^{-5}$ & 1.6$\times10^{3}$ & 3.1$\times10^{2}$ \\
Average model & \#5778 &  $85^{+142}_{-53}$ & $0.862^{+1.744}_{-0.577}$ & $0.07^{+0.06}_{-0.03}$ & $7^{+20}_{-5}$ & $57 \pm 21$ & $381 \pm 313$ & $51^{+133}_{-37}$ & $27 \pm 18$ & $2.3^{+6.6}_{-1.7}\times10^{-4}$ & $9.4^{+102.5}_{-8.6}\times10^{3}$ & $1.5^{+16.8}_{-1.4}\times10^{4}$\\
\hline
IRAS21391$_\mathrm{_bima2}$ & 0.19 & 10 & 0.316 & 0.04 & 2 & 48 & 103 & 4.96 & 43 & 3.0$\times10^{-5}$ & 9.0$\times10^{1}$ & 2.8$\times10^{2}$ \\
$d=0.8$ kpc & 0.55 & 10 & 0.316 & 0.04 & 0 & 22 & 117 & 8.75 & 18 & 1.9$\times10^{-5}$ & 1.7$\times10^{2}$ & 1.9$\times10^{2}$ \\
$R_\mathrm{ap}=7.00\arcsec$ & 0.55 & 10 & 1.000 & 0.02 & 4 & 62 & 204 & 1.29 & 59 & 7.7$\times10^{-5}$ & 2.4$\times10^{2}$ & 1.1$\times10^{3}$ \\
$R_\mathrm{ap}=0.03$ pc & 0.58 & 10 & 0.316 & 0.04 & 1 & 34 & 102 & 7.50 & 28 & 2.5$\times10^{-5}$ & 1.4$\times10^{2}$ & 2.6$\times10^{2}$ \\
 & 0.62 & 10 & 3.160 & 0.01 & 4 & 86 & 101 & 1.65 & 56 & 1.9$\times10^{-4}$ & 1.6$\times10^{2}$ & 1.9$\times10^{3}$ \\
Average model & \#121 &  $10^{+2}_{-2}$ & $0.788^{+1.275}_{-0.487}$ & $0.03^{+0.02}_{-0.01}$ & $2^{+2}_{-1}$ & $66 \pm 18$ & $96 \pm 97$ & $4^{+5}_{-2}$ & $39 \pm 14$ & $5.7^{+8.8}_{-3.5}\times10^{-5}$ & $2.0^{+3.6}_{-1.3}\times10^{2}$ & $6.4^{+10.4}_{-4.0}\times10^{2}$\\
\hline
IRAS21391$_\mathrm{_bima3}$ & 0.35 & 10 & 1.000 & 0.02 & 4 & 62 & 212 & 1.29 & 59 & 7.7$\times10^{-5}$ & 2.4$\times10^{2}$ & 1.1$\times10^{3}$ \\
$d=0.8$ kpc & 0.36 & 10 & 1.000 & 0.02 & 2 & 39 & 295 & 5.33 & 39 & 7.5$\times10^{-5}$ & 1.0$\times10^{3}$ & 7.6$\times10^{2}$ \\
$R_\mathrm{ap}=7.75\arcsec$ & 0.48 & 10 & 0.316 & 0.04 & 2 & 51 & 69 & 4.96 & 43 & 3.0$\times10^{-5}$ & 7.7$\times10^{1}$ & 2.8$\times10^{2}$ \\
$R_\mathrm{ap}=0.03$ pc & 0.49 & 10 & 0.316 & 0.04 & 0 & 22 & 120 & 8.75 & 18 & 1.9$\times10^{-5}$ & 1.7$\times10^{2}$ & 1.9$\times10^{2}$ \\
 & 0.51 & 10 & 0.316 & 0.04 & 1 & 34 & 106 & 7.50 & 28 & 2.5$\times10^{-5}$ & 1.4$\times10^{2}$ & 2.6$\times10^{2}$ \\
Average model & \#1809 &  $45^{+57}_{-25}$ & $2.091^{+1.845}_{-0.980}$ & $0.03^{+0.01}_{-0.01}$ & $5^{+13}_{-4}$ & $58 \pm 21$ & $489 \pm 306$ & $25^{+51}_{-17}$ & $28 \pm 16$ & $3.3^{+5.9}_{-2.1}\times10^{-4}$ & $6.8^{+47.9}_{-6.0}\times10^{3}$ & $1.2^{+8.7}_{-1.1}\times10^{4}$\\
\hline
IRAS21391$_\mathrm{mir48}$ & 0.92 & 10 & 0.316 & 0.04 & 4 & 89 & 48 & 0.63 & 68 & 2.4$\times10^{-5}$ & 2.9$\times10^{1}$ & 6.7$\times10^{2}$ \\
$d=0.8$ kpc & 8.25 & 10 & 0.316 & 0.04 & 2 & 44 & 235 & 4.96 & 43 & 3.0$\times10^{-5}$ & 3.3$\times10^{2}$ & 2.8$\times10^{2}$ \\
$R_\mathrm{ap}=7.75\arcsec$ & 9.79 & 10 & 0.316 & 0.04 & 1 & 29 & 299 & 7.50 & 28 & 2.5$\times10^{-5}$ & 5.3$\times10^{2}$ & 2.6$\times10^{2}$ \\
$R_\mathrm{ap}=0.03$ pc & 10.03 & 20 & 3.160 & 0.02 & 0 & 13 & 334 & 19.18 & 9 & 1.3$\times10^{-4}$ & 2.8$\times10^{3}$ & 8.6$\times10^{2}$ \\
 & 10.04 & 10 & 3.160 & 0.01 & 4 & 89 & 192 & 1.65 & 56 & 1.9$\times10^{-4}$ & 1.6$\times10^{2}$ & 1.9$\times10^{3}$ \\
Average model & \#4 &  $10^{+0}_{-0}$ & $0.316^{+0.000}_{-0.000}$ & $0.04^{+0.00}_{-0.00}$ & $4^{+0}_{-0}$ & $84 \pm 3$ & $60 \pm 11$ & $1^{+0}_{-0}$ & $68 \pm 0$ & $2.4^{+0.0}_{-0.0}\times10^{-5}$ & $3.3^{+0.5}_{-0.5}\times10^{1}$ & $6.7^{+0.0}_{-0.0}\times10^{2}$\\
\enddata
\tablecomments{
For each source the first five rows refer to the best five models taken from the 432 physical models, whereas the sixth row shows the average and dispersion of good model fits (see the text). The number next to the symbol \# represents the number of models considered in the average of the good models. Upper and lower scripts in the average models row refer to the upper and lower errors.
}
\end{deluxetable}
\end{longrotatetable}

\subsection{SOMA I New SED fit}

We revisited the measurement and error estimation for the SOMA I sources \citep{debuizer2017} as well as refit their SEDs. Figures\,\ref{fig:sed_1D_results_soma_i} and \ref{fig:sed_2D_results_soma_i} show the revisited results.

\begin{figure*}[!htb]
\includegraphics[width=0.5\textwidth]{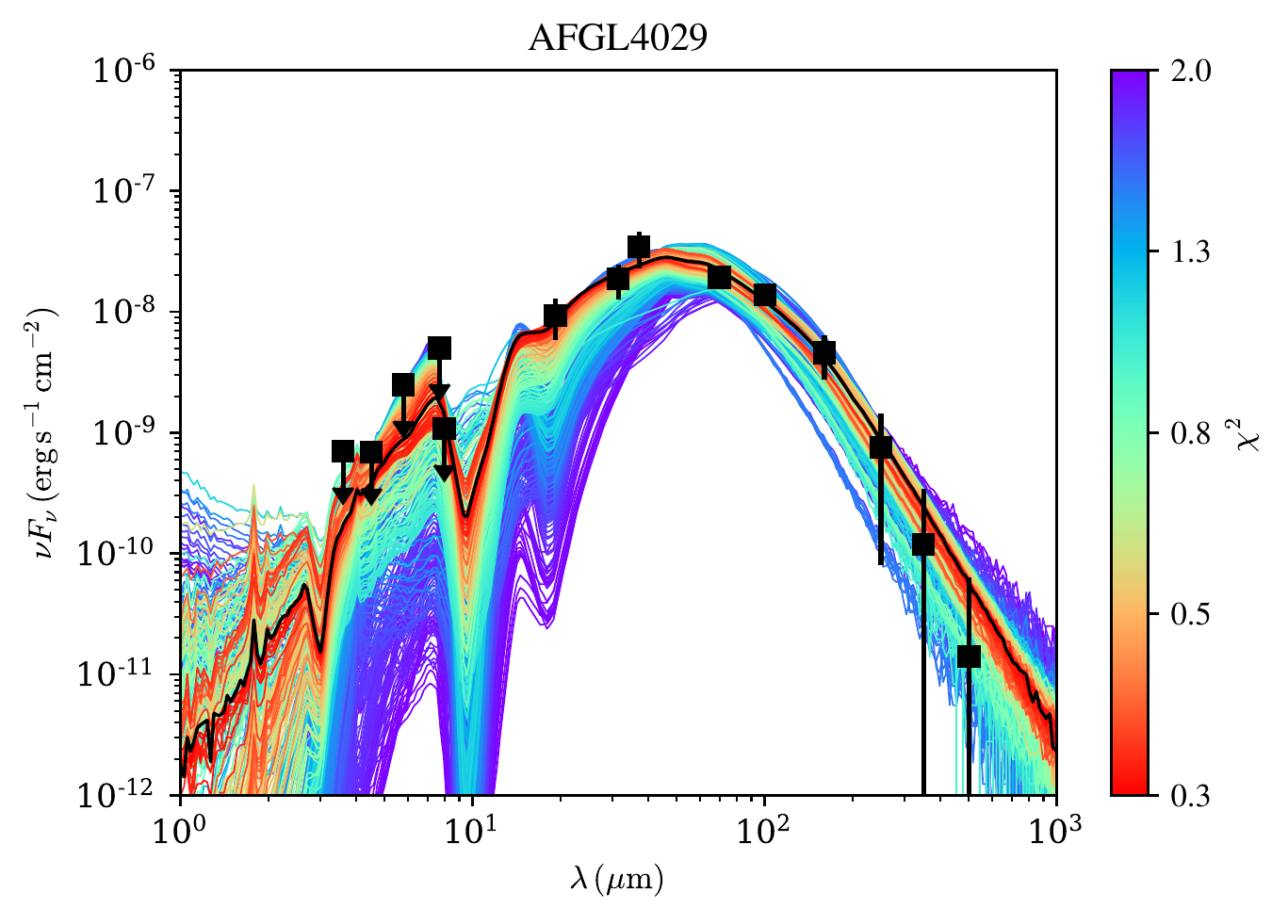}
\includegraphics[width=0.5\textwidth]{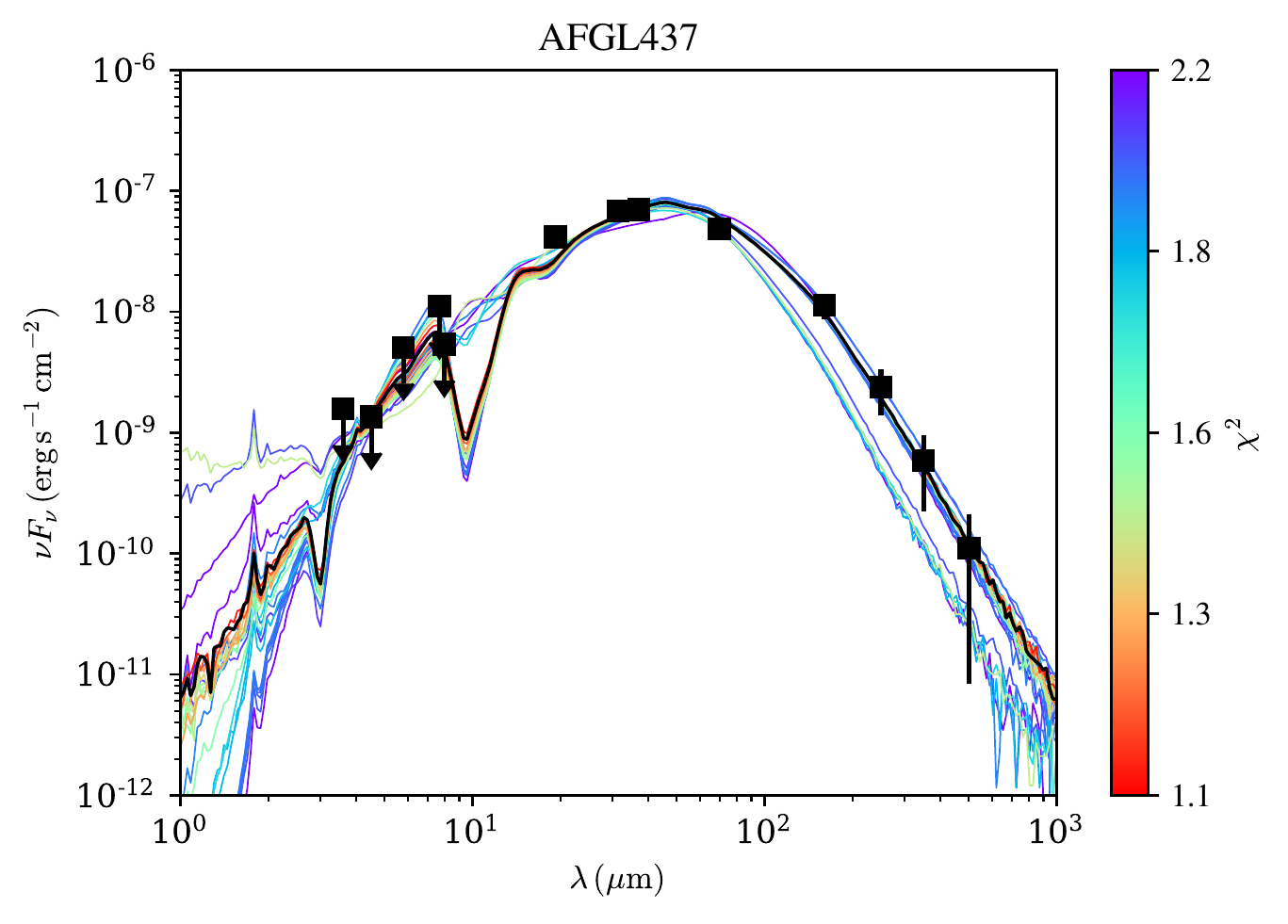}
\includegraphics[width=0.5\textwidth]{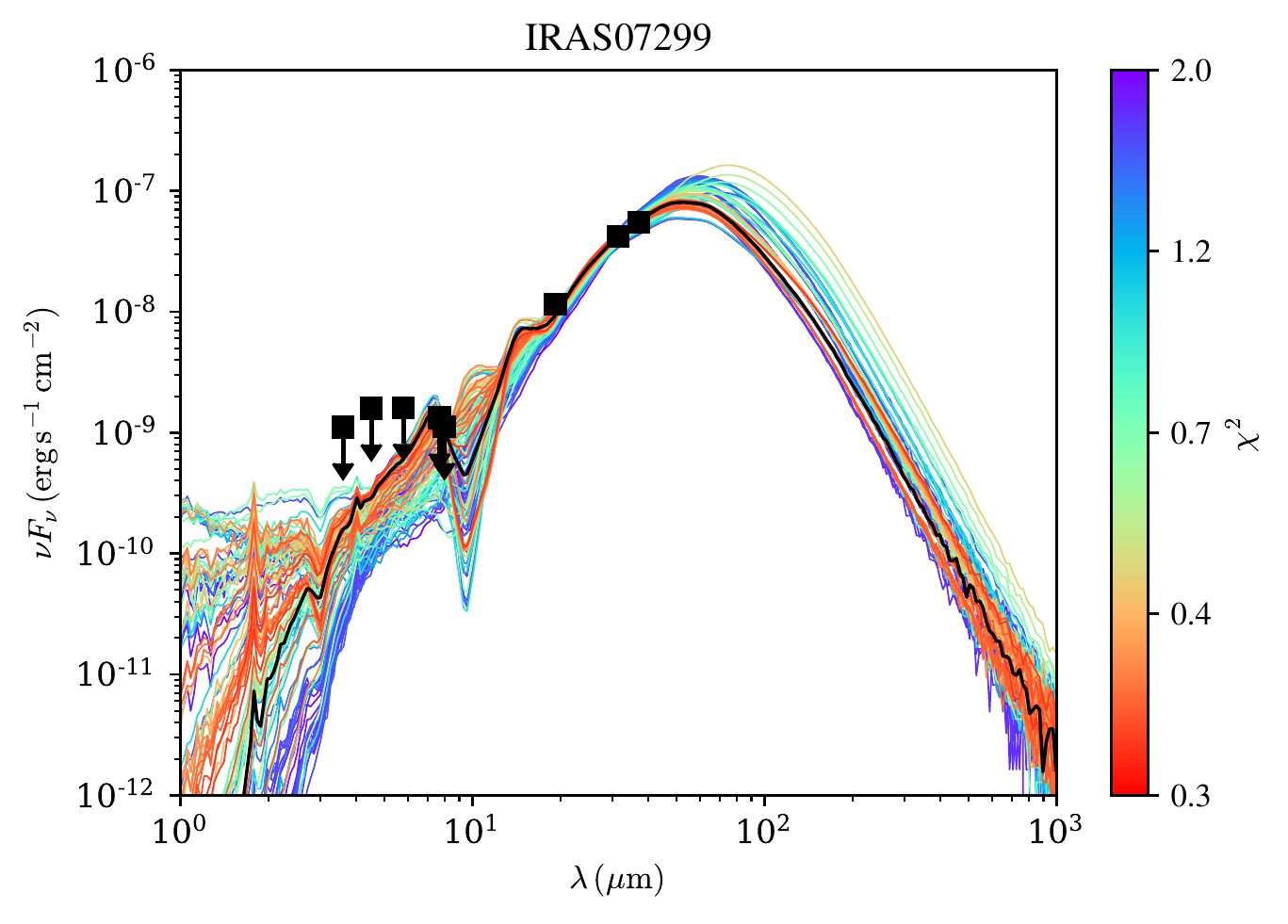}
\includegraphics[width=0.5\textwidth]{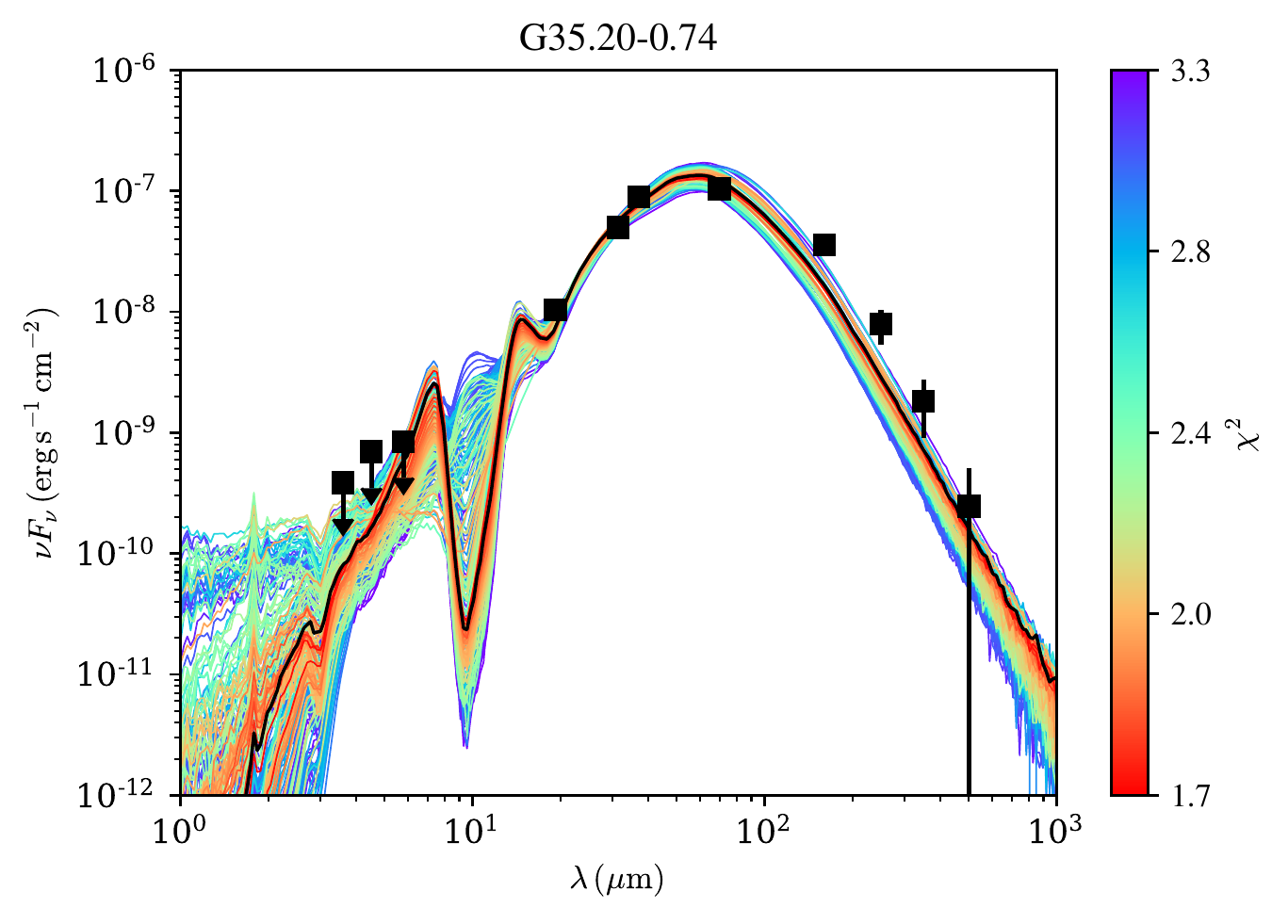}
\includegraphics[width=0.5\textwidth]{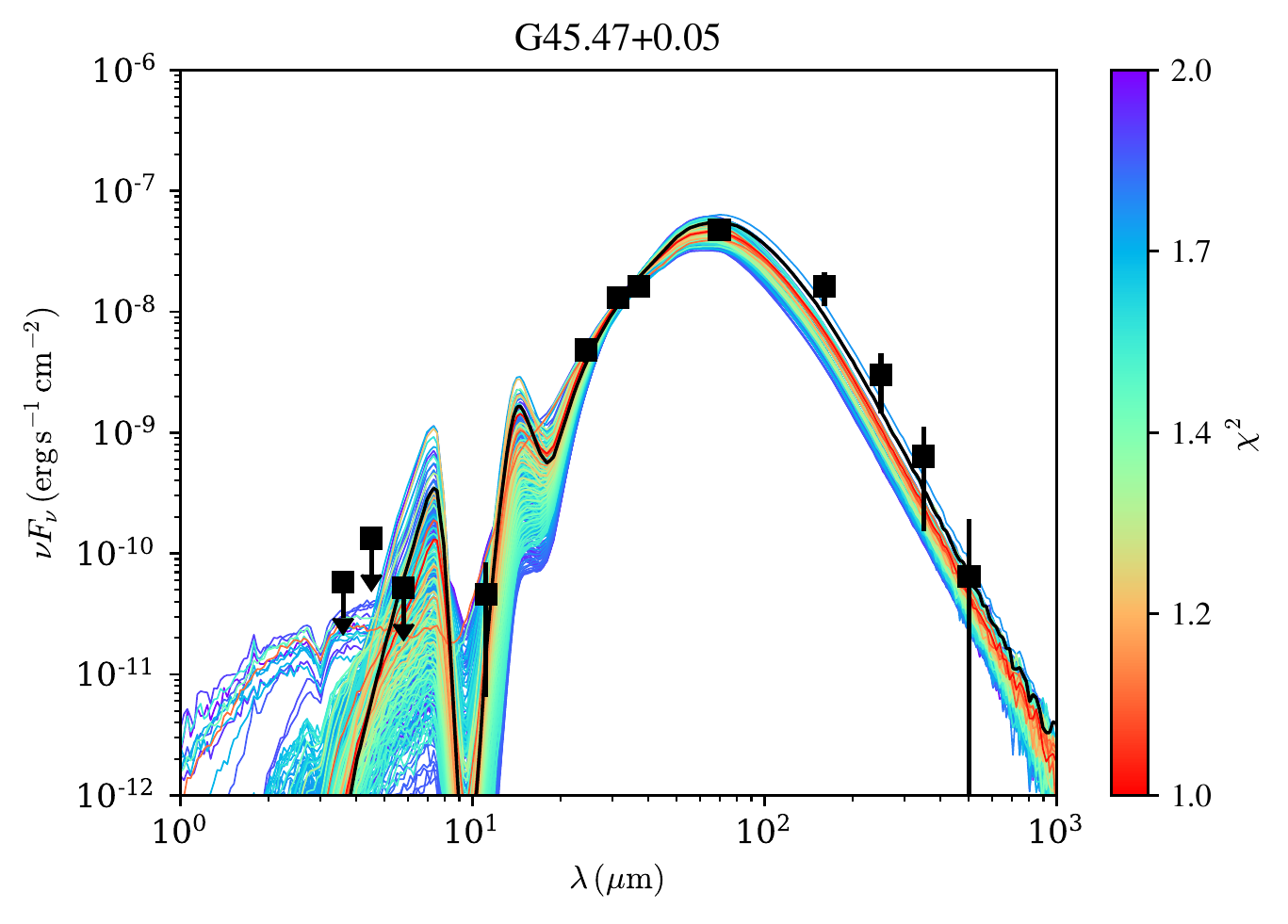}
\includegraphics[width=0.5\textwidth]{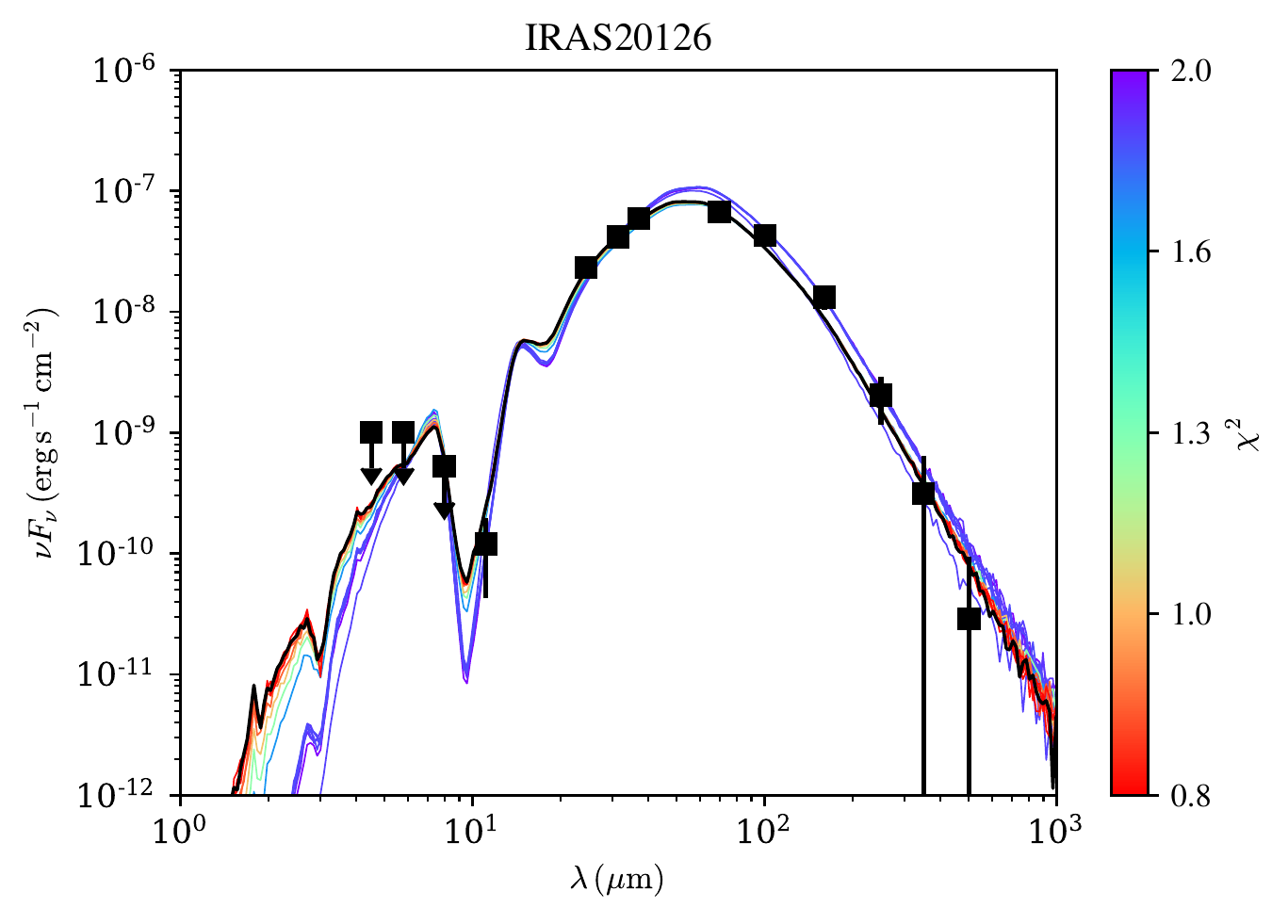}
\caption{SOMA I sources reanalyzed with sedcreator. Protostar model fitting to the fixed aperture, background-subtracted SED data using the ZT model grid. For each source (noted on top of each plot), the best fitting protostar model is shown with a black line, while all other good model fits (see the text) are shown with colored lines (red to blue with increasing $\chi^2$). Flux values are those from Table\,\ref{tab:soma_all_fluxes}. Note that the data at $\lesssim8\,{\rm \mu m}$ are treated as upper limits (see the text). The resulting model parameters are listed in Table\,\ref{tab:best_models_soma_ALL}.
\label{fig:sed_1D_results_soma_i}}
\end{figure*}

\renewcommand{\thefigure}{A\arabic{figure}}
\addtocounter{figure}{-1}
\begin{figure*}[!htb]
\includegraphics[width=0.5\textwidth]{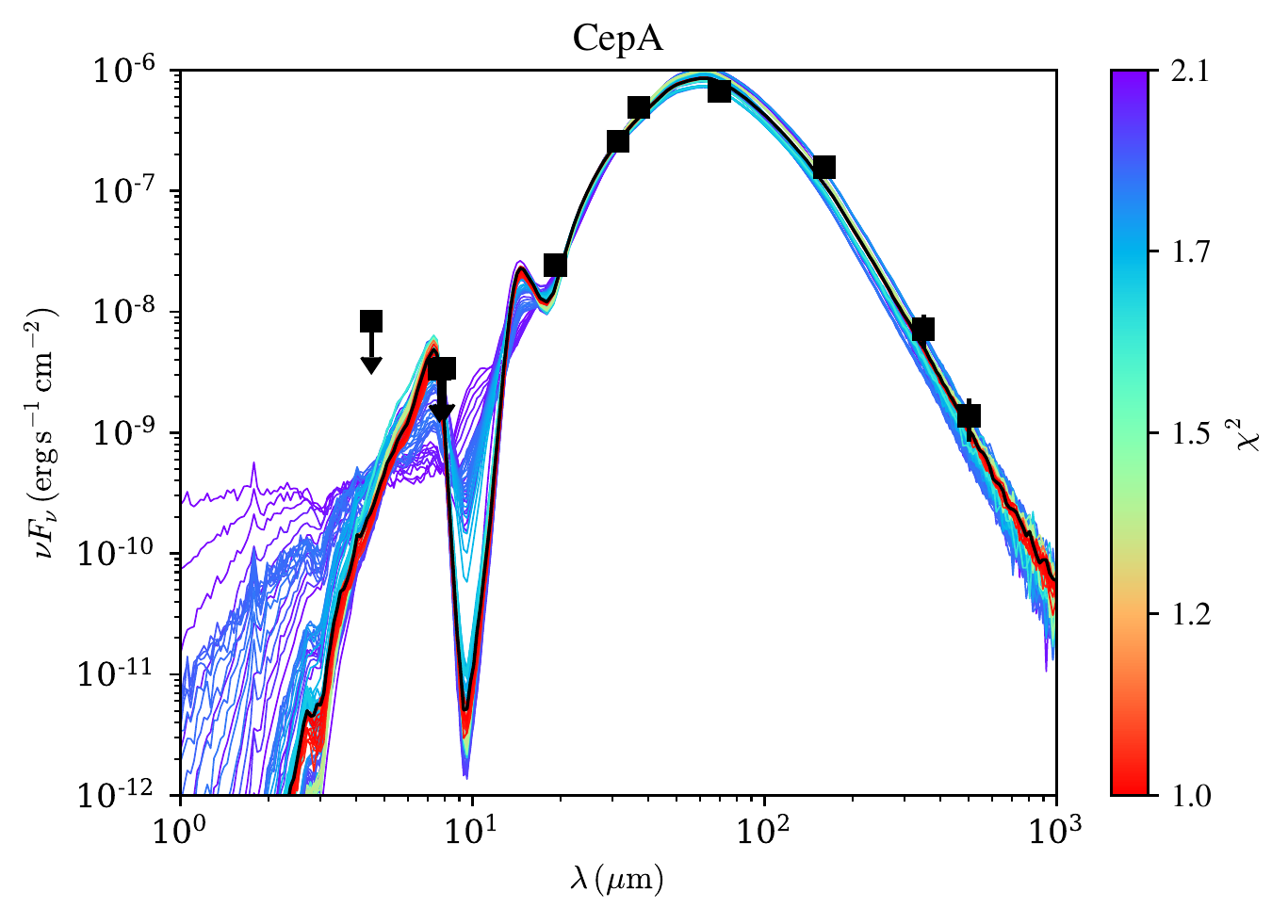}
\includegraphics[width=0.5\textwidth]{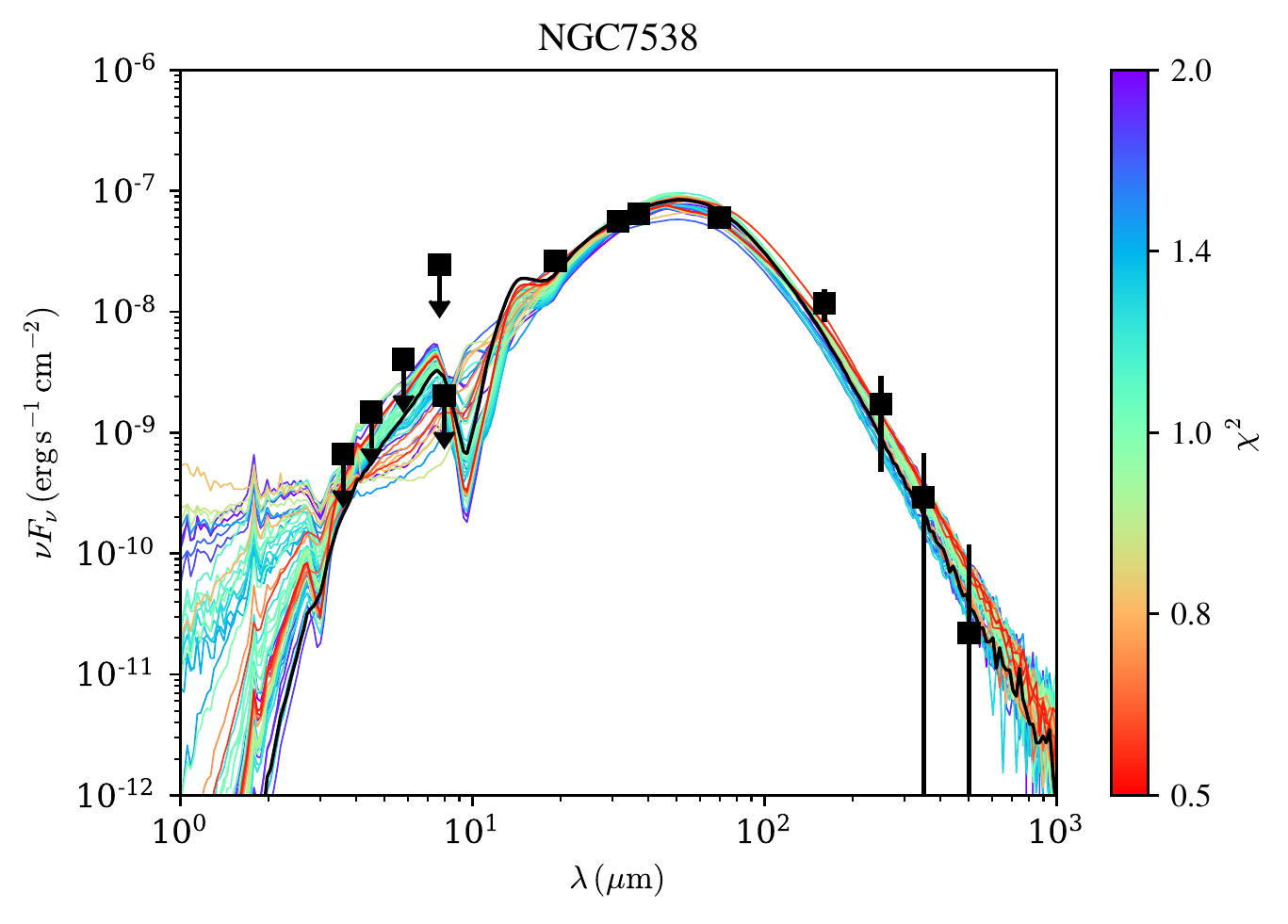}
\caption{(Continued.)}
\end{figure*}

\begin{figure*}[!htb]
\includegraphics[width=1.0\textwidth]{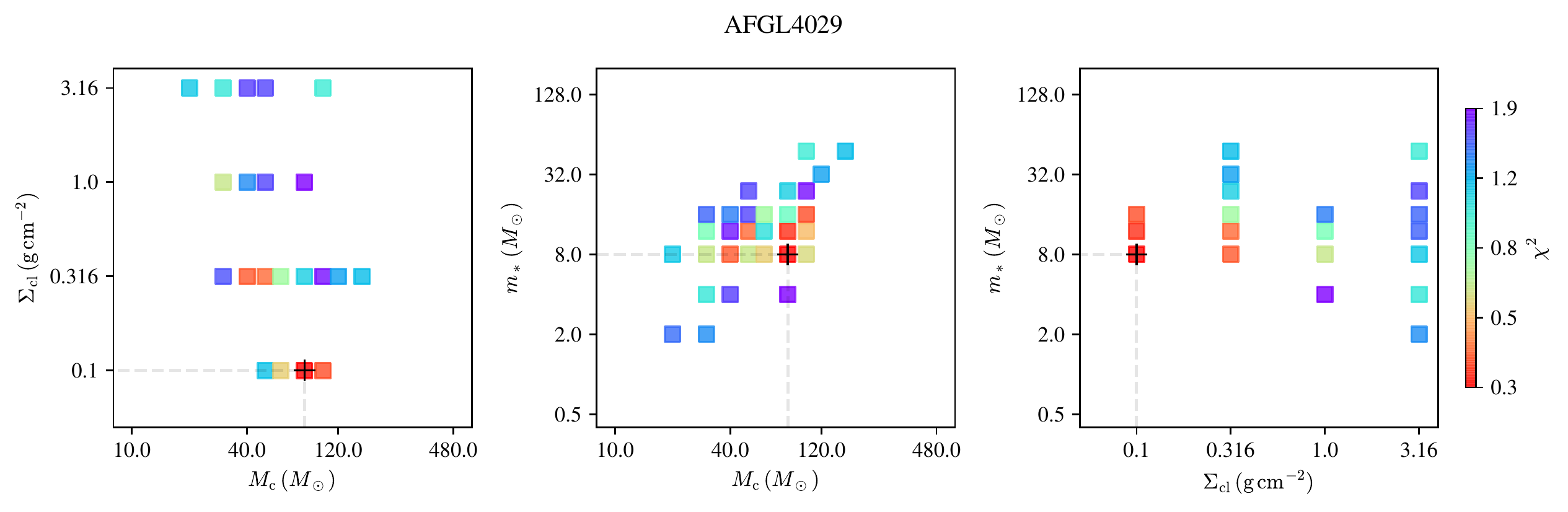}
\includegraphics[width=1.0\textwidth]{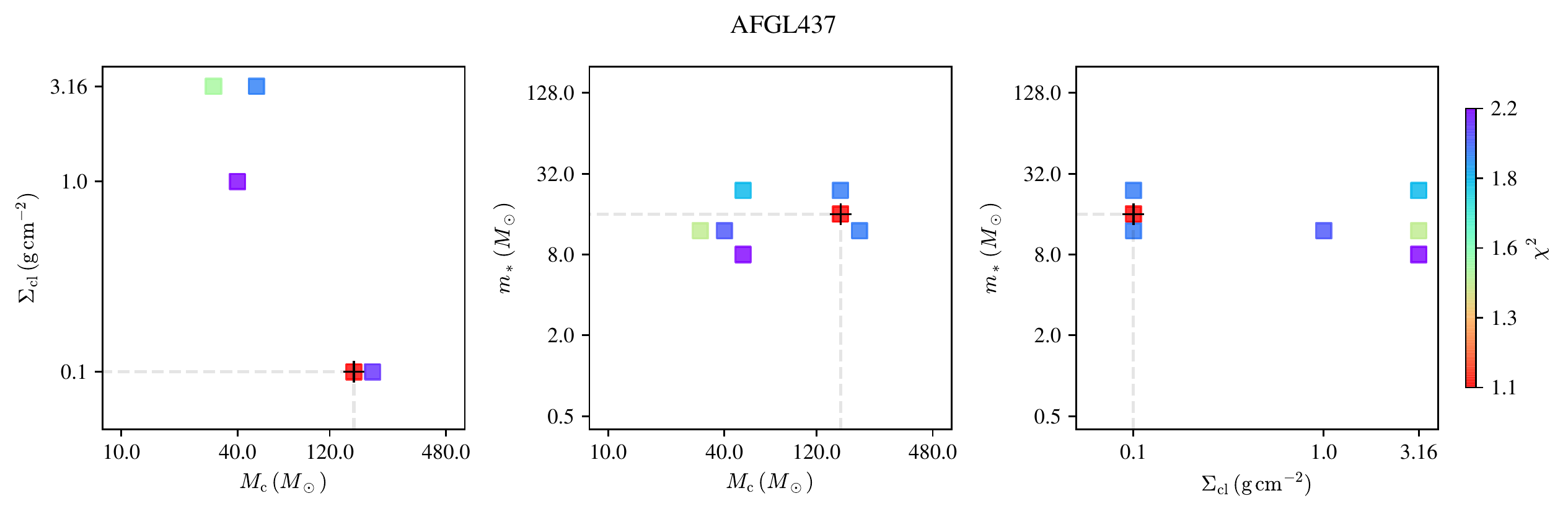}
\includegraphics[width=1.0\textwidth]{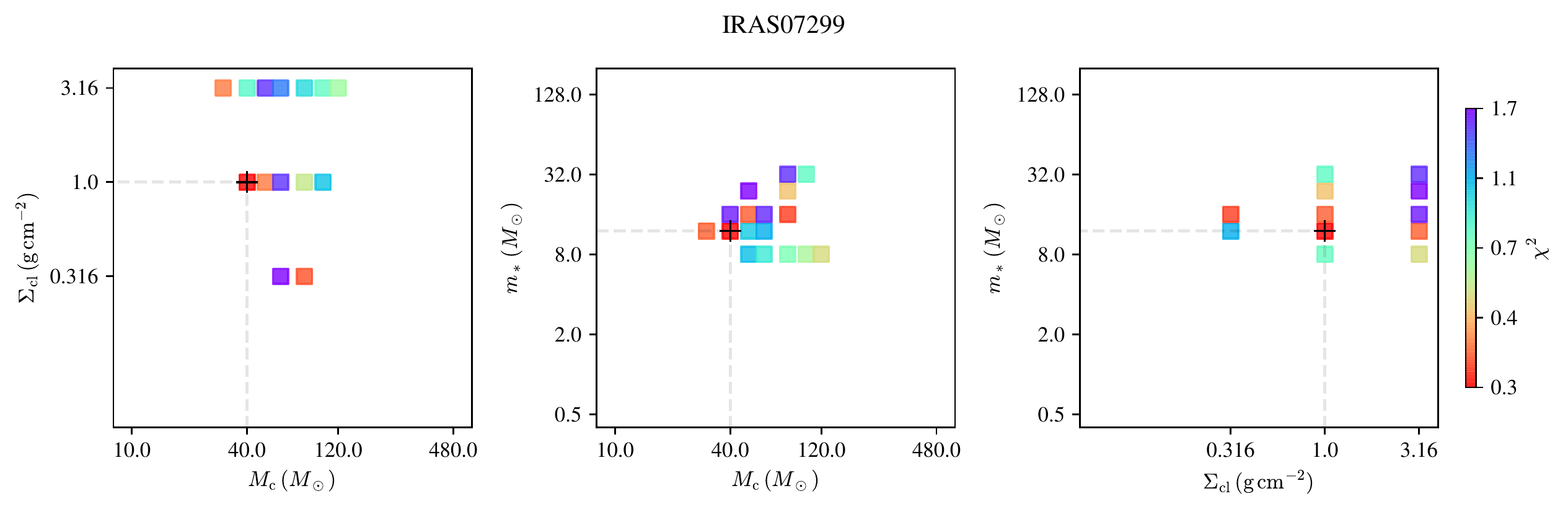}
\caption{Diagrams of $\chi^{2}$ distribution in $\Sigma_{\rm cl}$ - $M_{c}$ space (left), $m_{*}$ - $M_{\rm c}$ space (center) and $m_{*}$ - $\Sigma_{\rm  cl}$ space (right) for each source noted on top of each plot. The black cross is the best model.
\label{fig:sed_2D_results_soma_i}}
\end{figure*}

\renewcommand{\thefigure}{A\arabic{figure}}
\addtocounter{figure}{-1}
\begin{figure*}[!htb]
\includegraphics[width=1.0\textwidth]{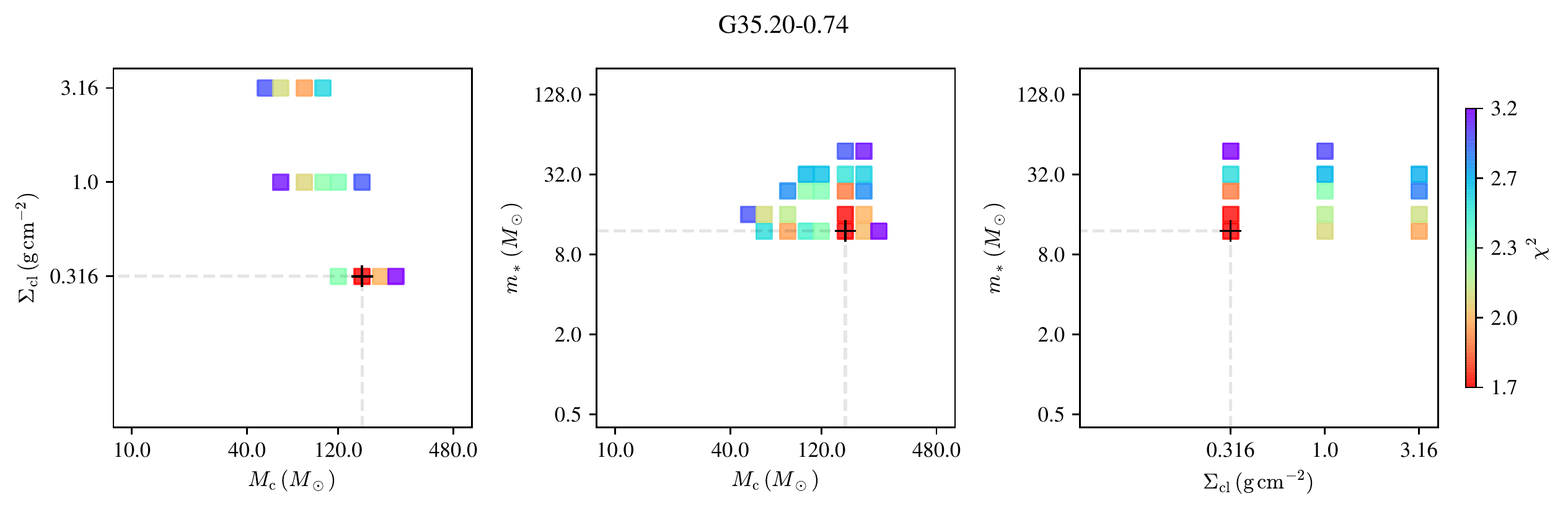}
\includegraphics[width=1.0\textwidth]{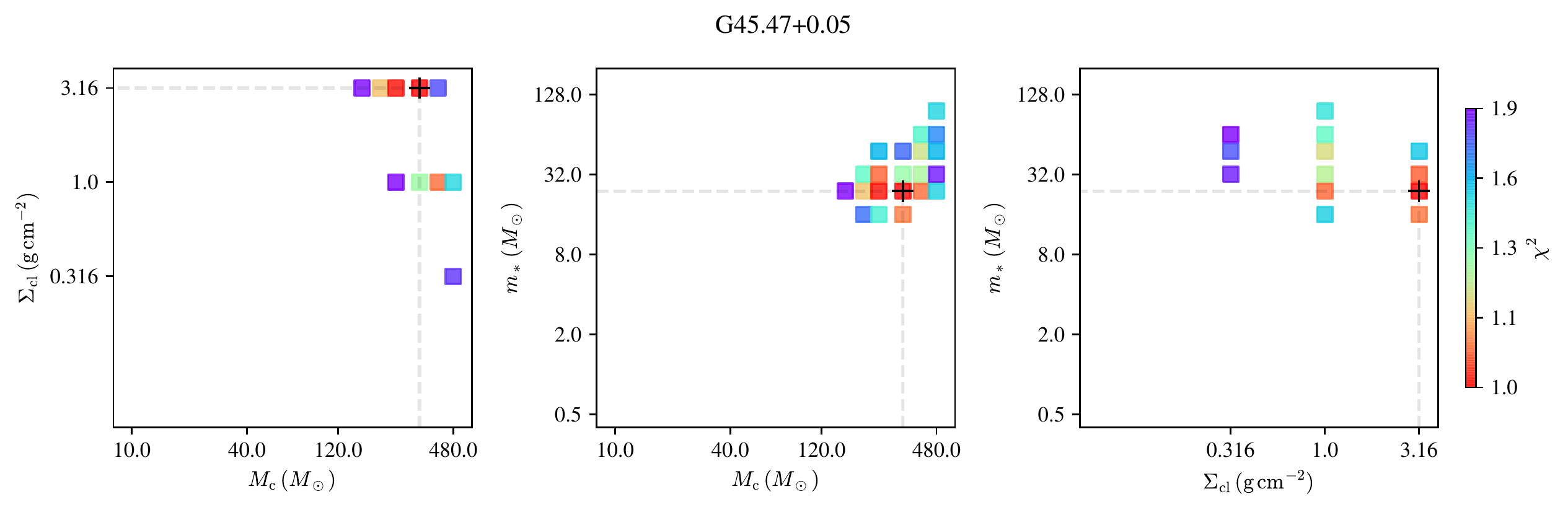}
\includegraphics[width=1.0\textwidth]{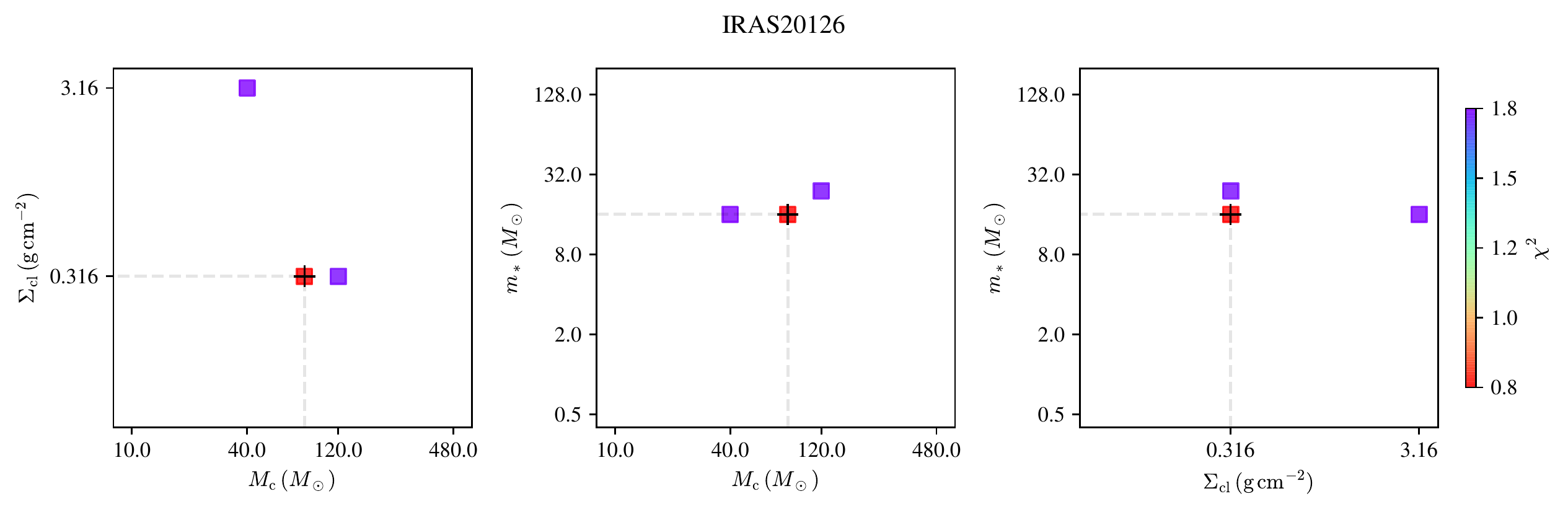}
\includegraphics[width=1.0\textwidth]{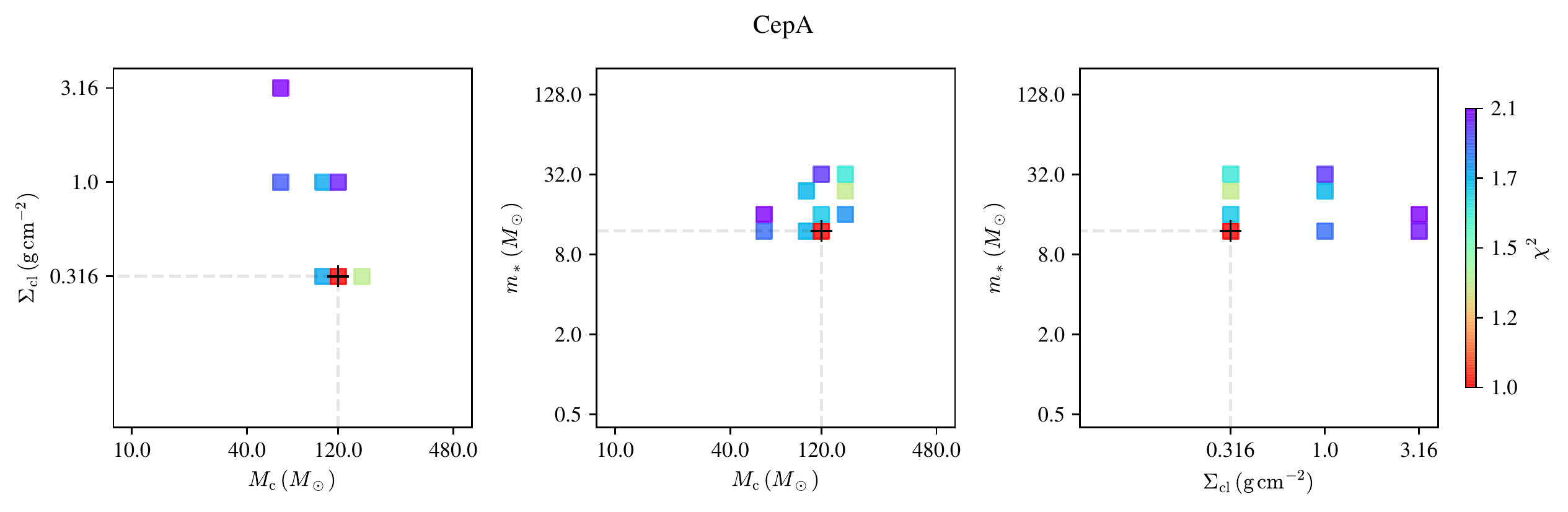}
\caption{(Continued.)}
\end{figure*}

\renewcommand{\thefigure}{A\arabic{figure}}
\addtocounter{figure}{-1}
\begin{figure*}[!htb]
\includegraphics[width=1.0\textwidth]{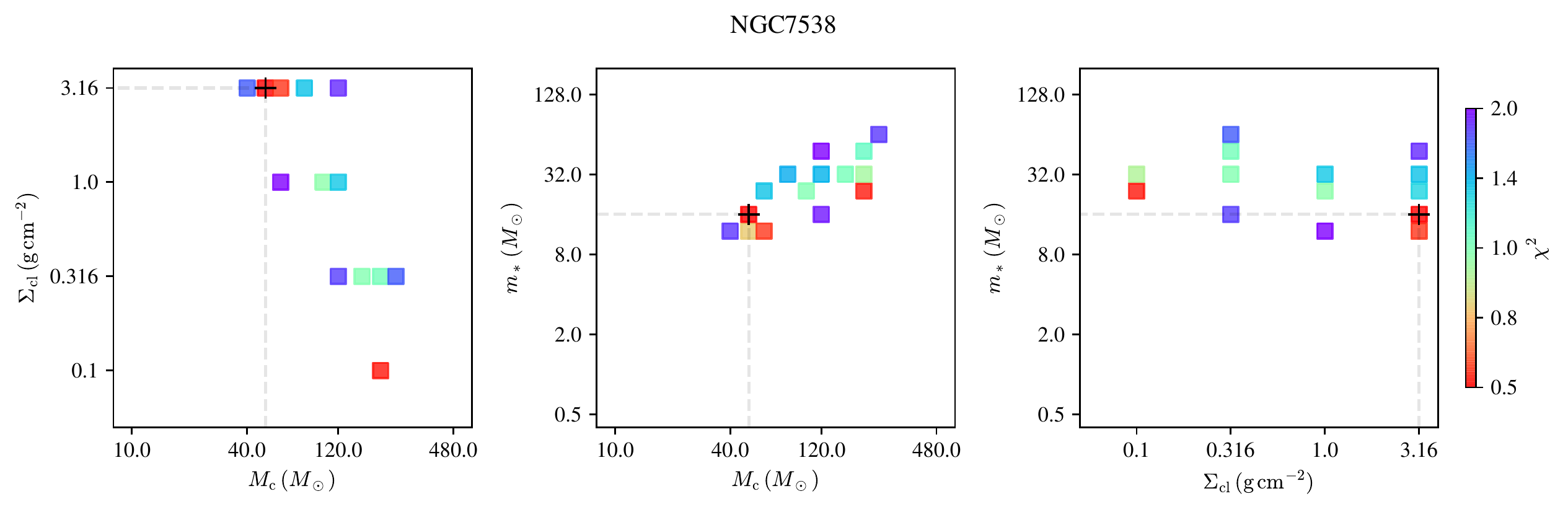}
\caption{(Continued.)}
\end{figure*}

\subsection{SOMA II New SED fit}

We revisited the measurement and error estimation for the SOMA II sources \citep{liu2019} as well as refit their SEDs. Figures\,\ref{fig:sed_1D_results_soma_ii} and \ref{fig:sed_2D_results_soma_ii} show the revisited results.

\begin{figure*}[!htb]
\includegraphics[width=0.5\textwidth]{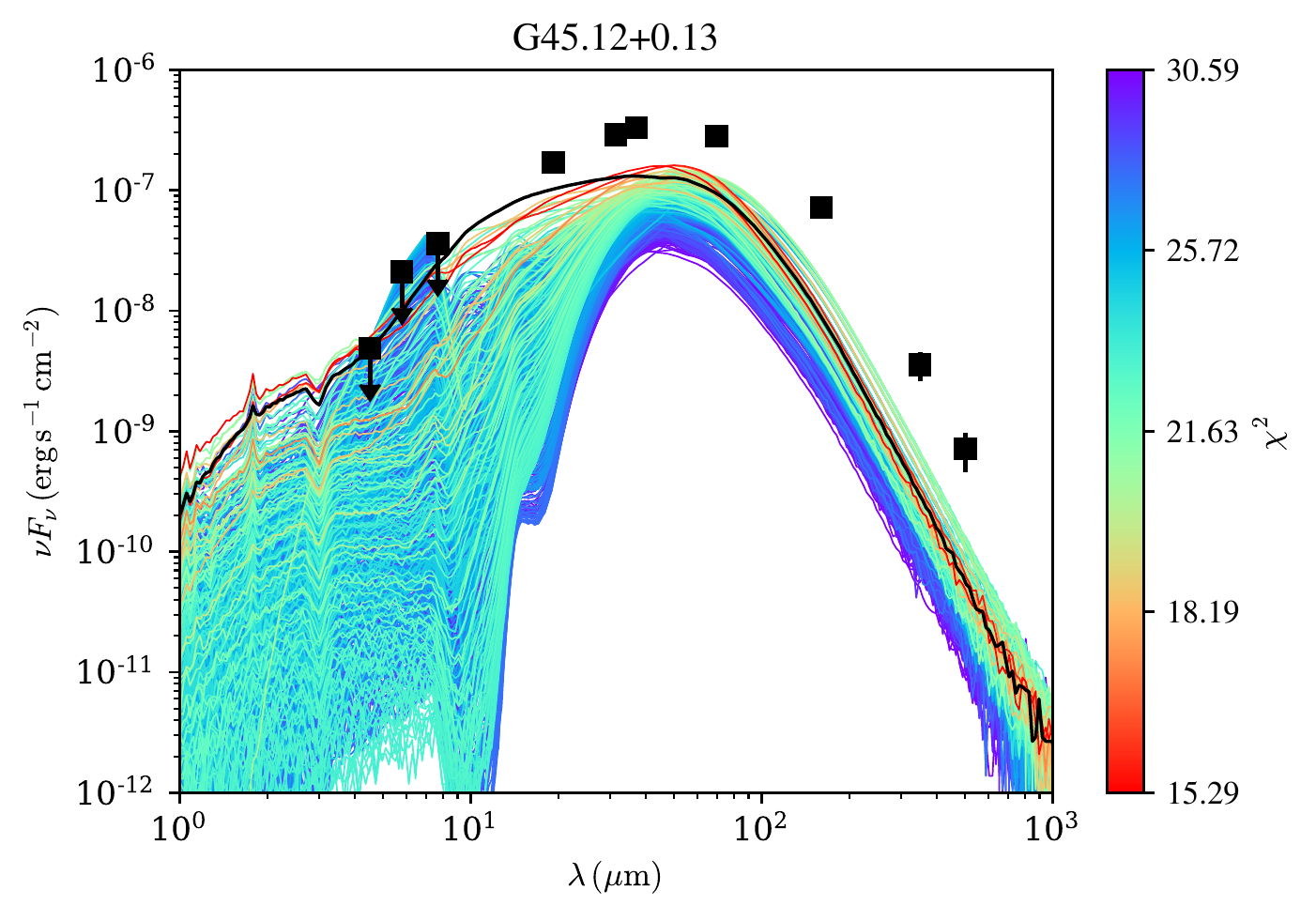}
\includegraphics[width=0.5\textwidth]{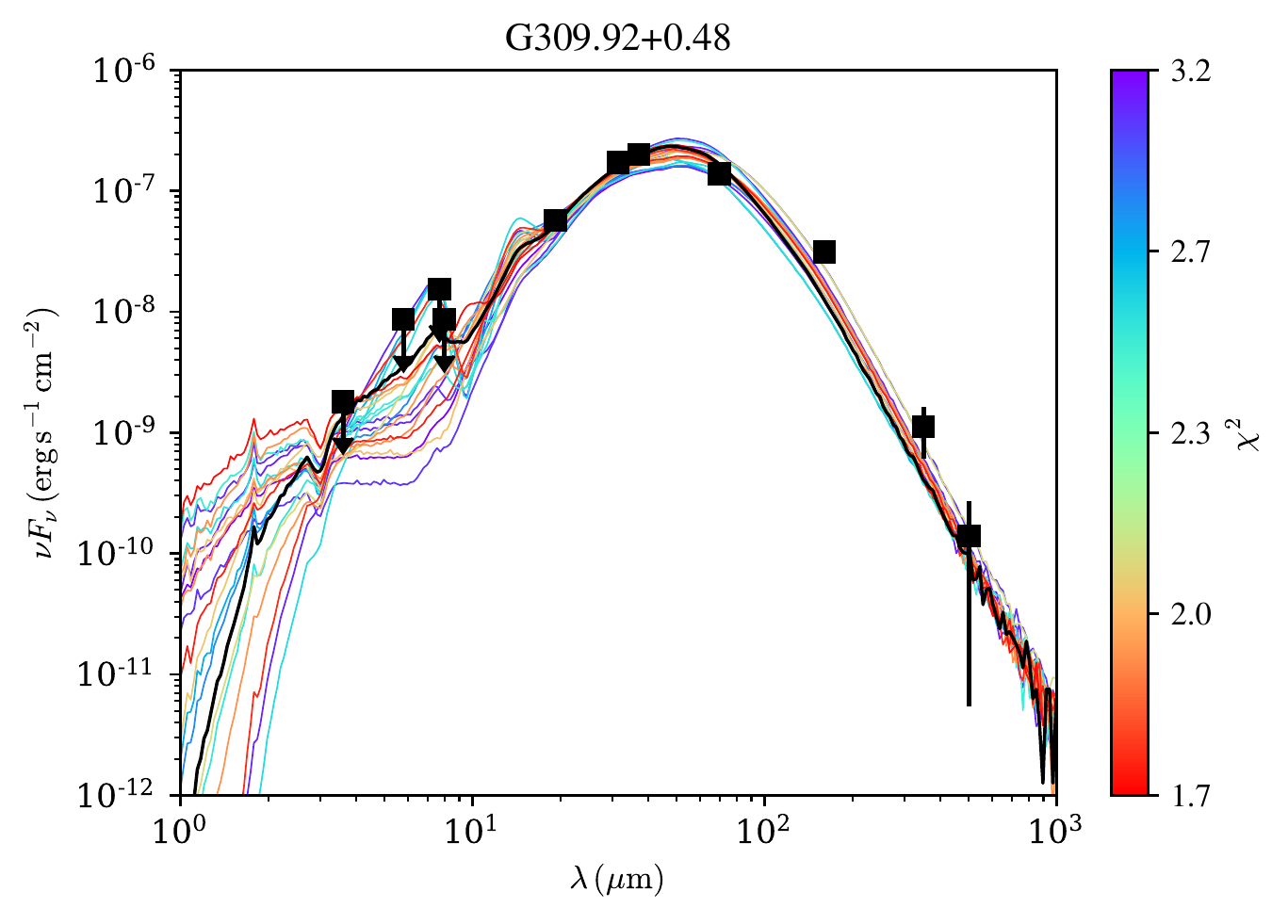}
\includegraphics[width=0.5\textwidth]{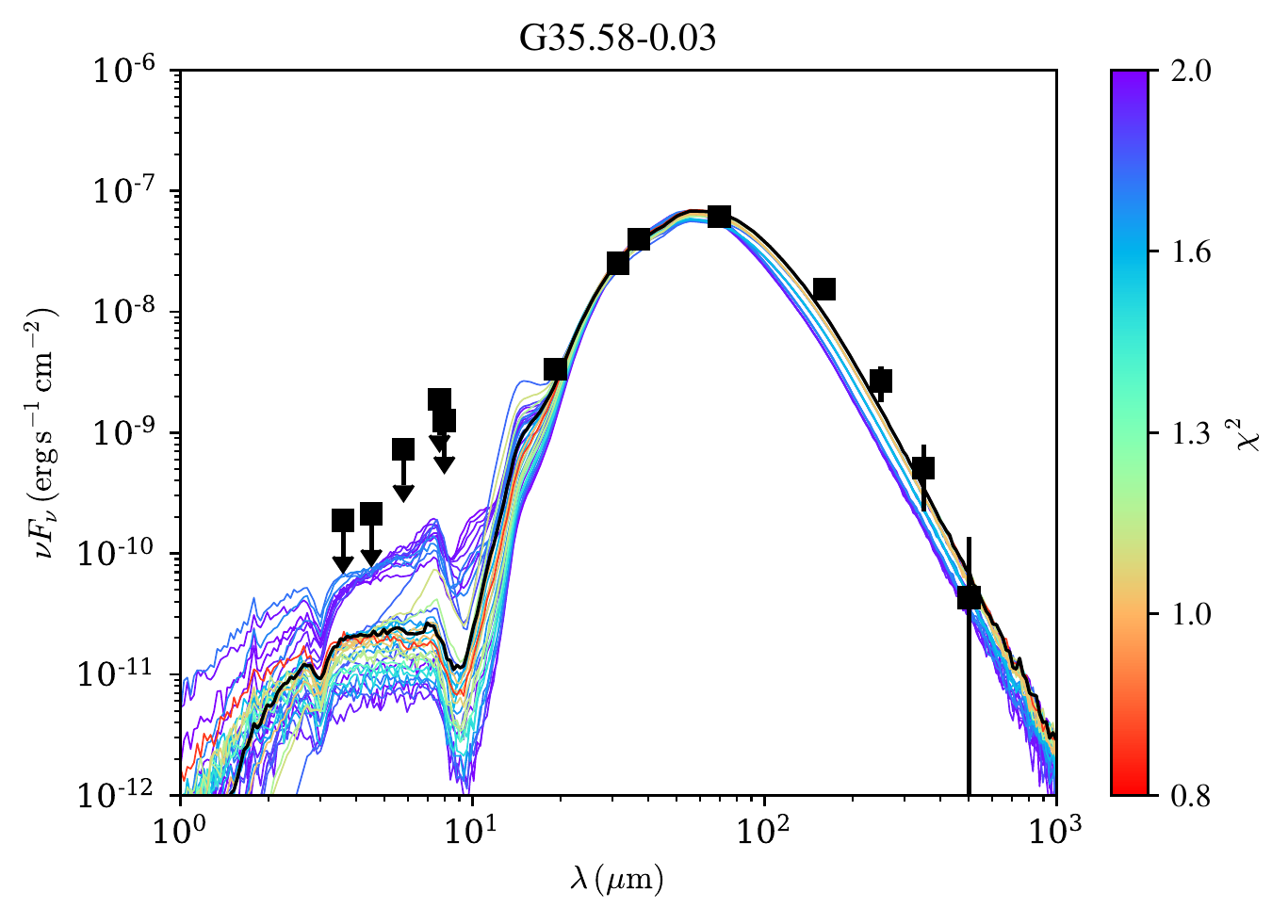}
\includegraphics[width=0.5\textwidth]{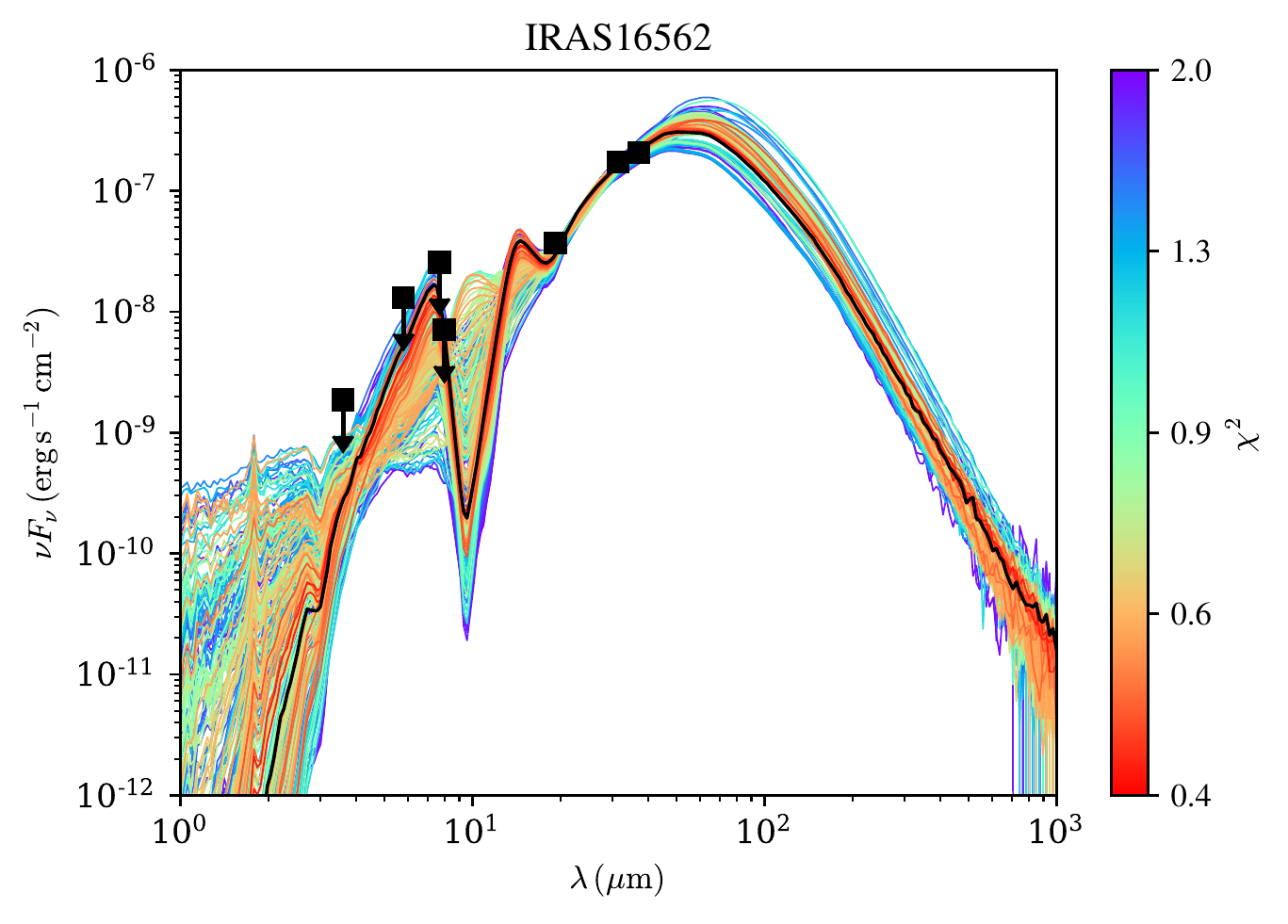}
\includegraphics[width=0.5\textwidth]{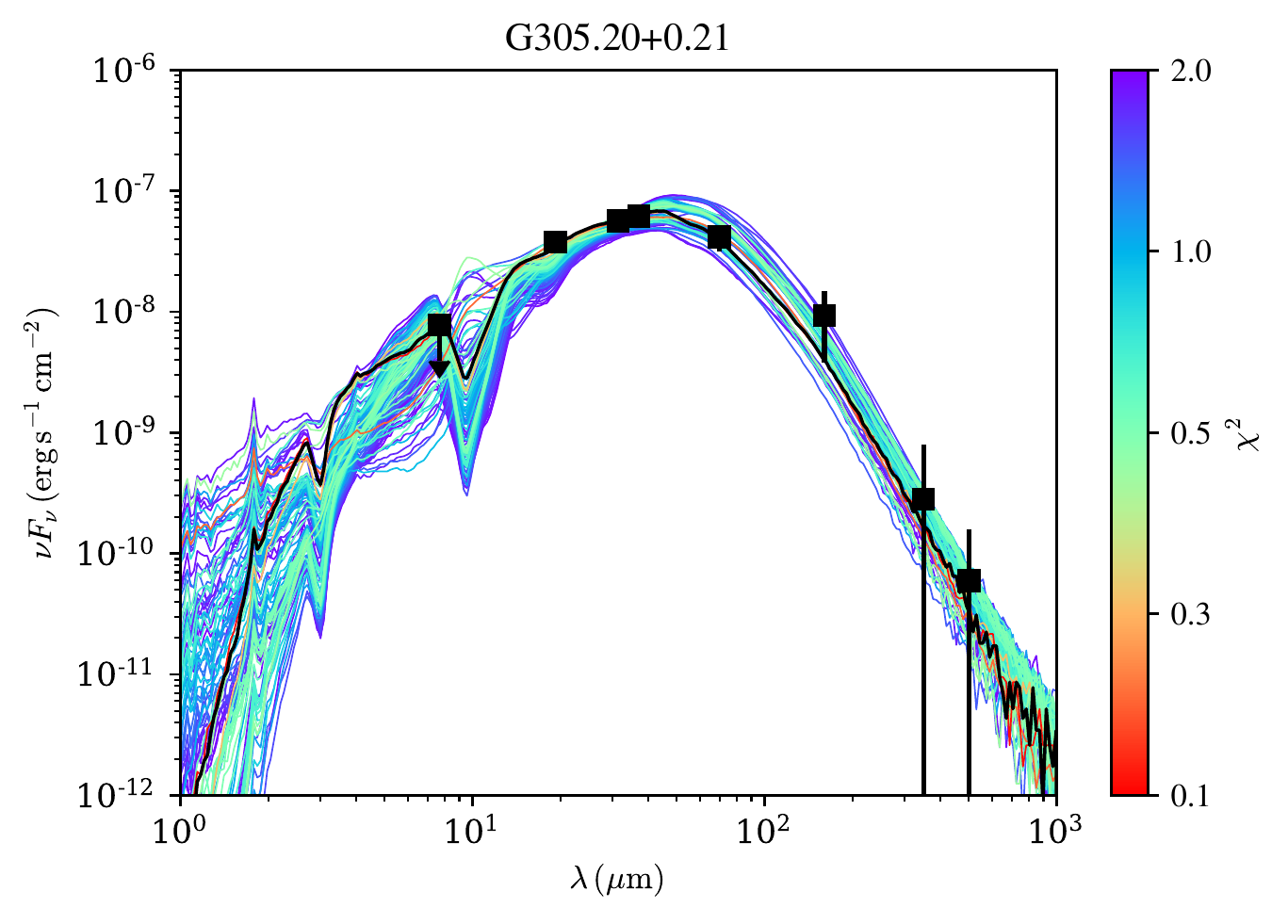}
\includegraphics[width=0.5\textwidth]{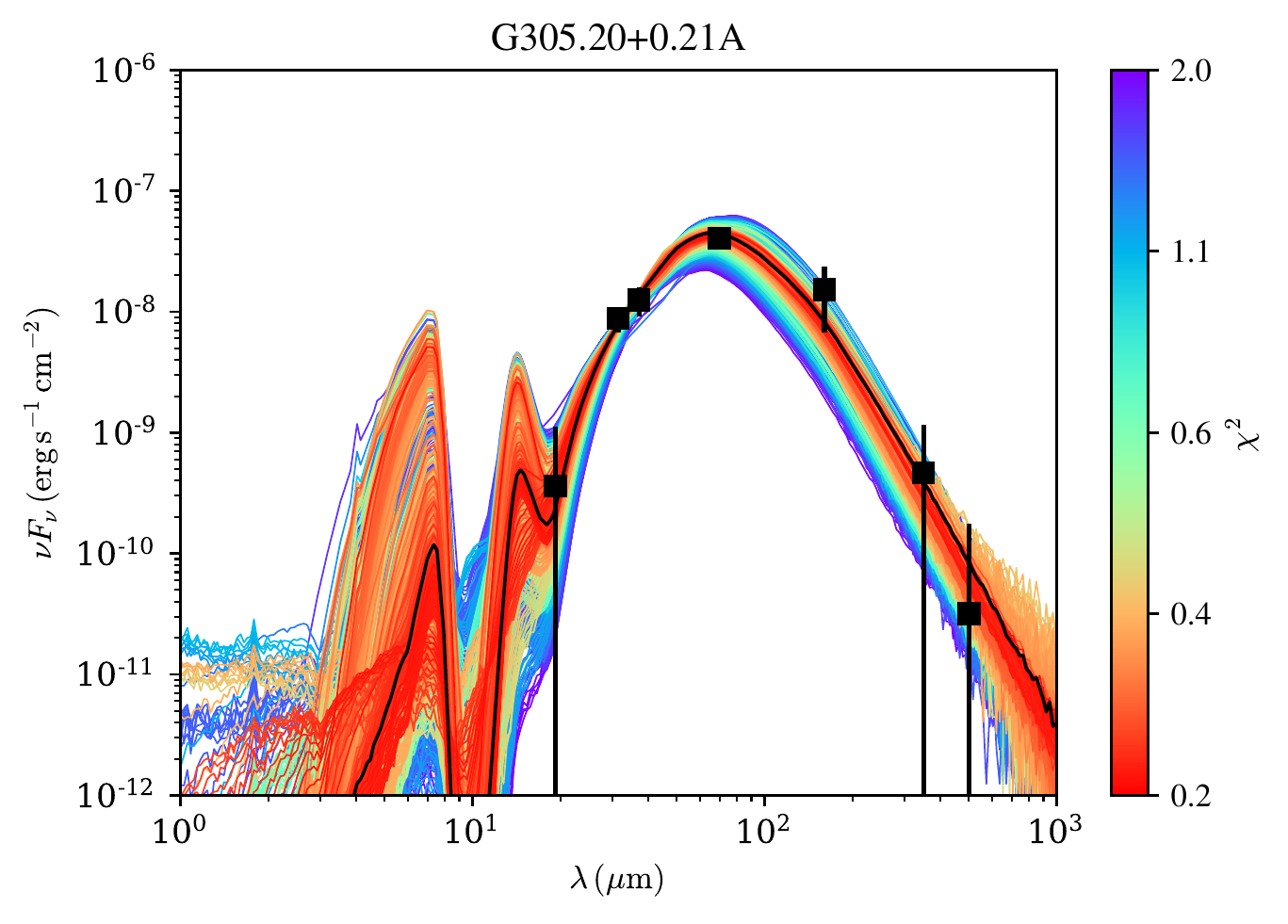}
\caption{SOMA II sources reanalyzed with sedcreator. Protostar model fitting to the fixed aperture, background-subtracted SED data using the ZT model grid. For each source (noted on top of each plot), the best fitting protostar model is shown with a black line, while all other good model fits (see the text) are shown with colored lines (red to blue with increasing $\chi^2$). Flux values are those from Table\,\ref{tab:soma_all_fluxes}. Note that the data at $\lesssim8\,{\rm \mu m}$ are treated as upper limits (see the text). The resulting model parameters are listed in Table\,\ref{tab:best_models_soma_ALL}.
\label{fig:sed_1D_results_soma_ii}}
\end{figure*}

\renewcommand{\thefigure}{A\arabic{figure}}
\addtocounter{figure}{-1}
\begin{figure*}[!htb]
\includegraphics[width=0.5\textwidth]{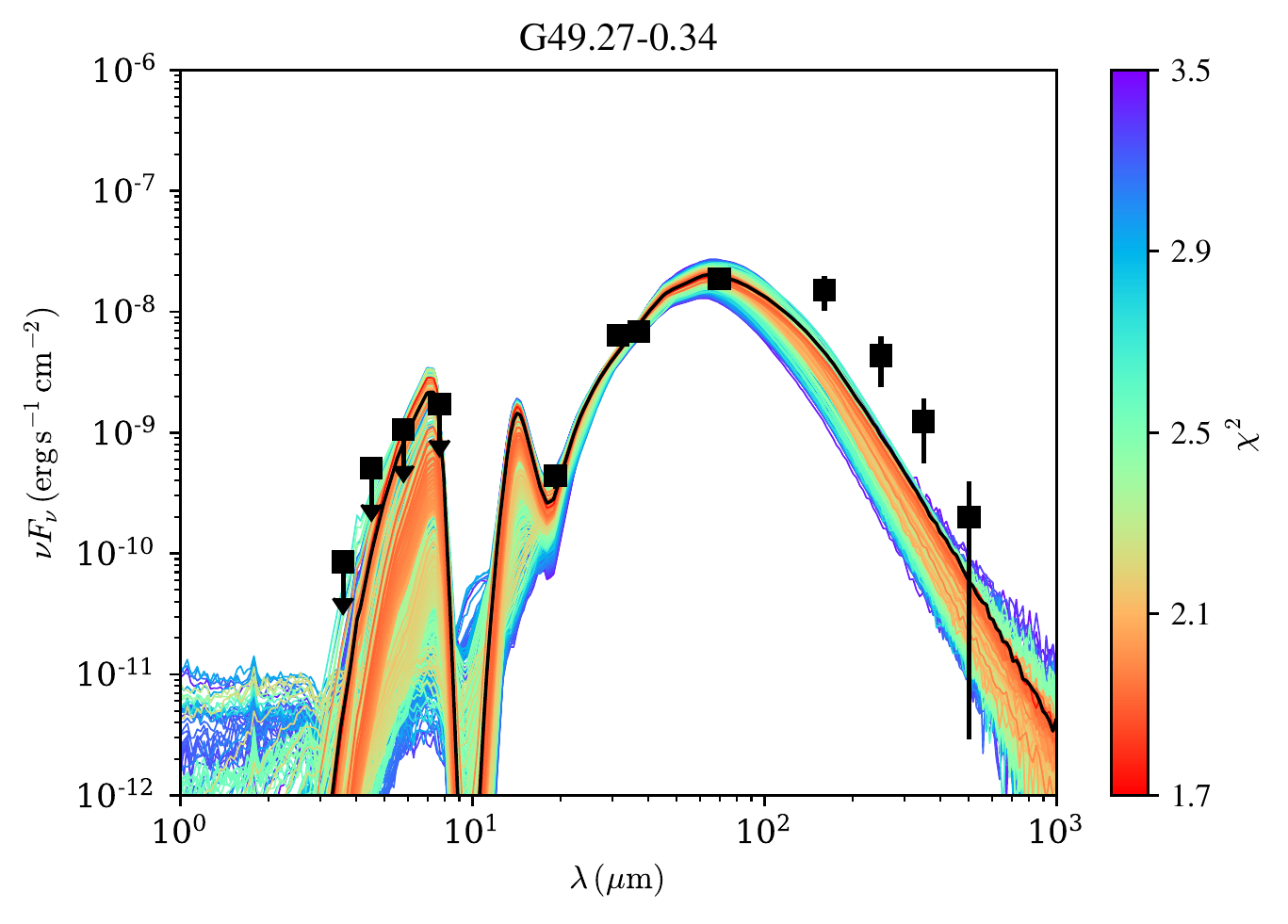}
\includegraphics[width=0.5\textwidth]{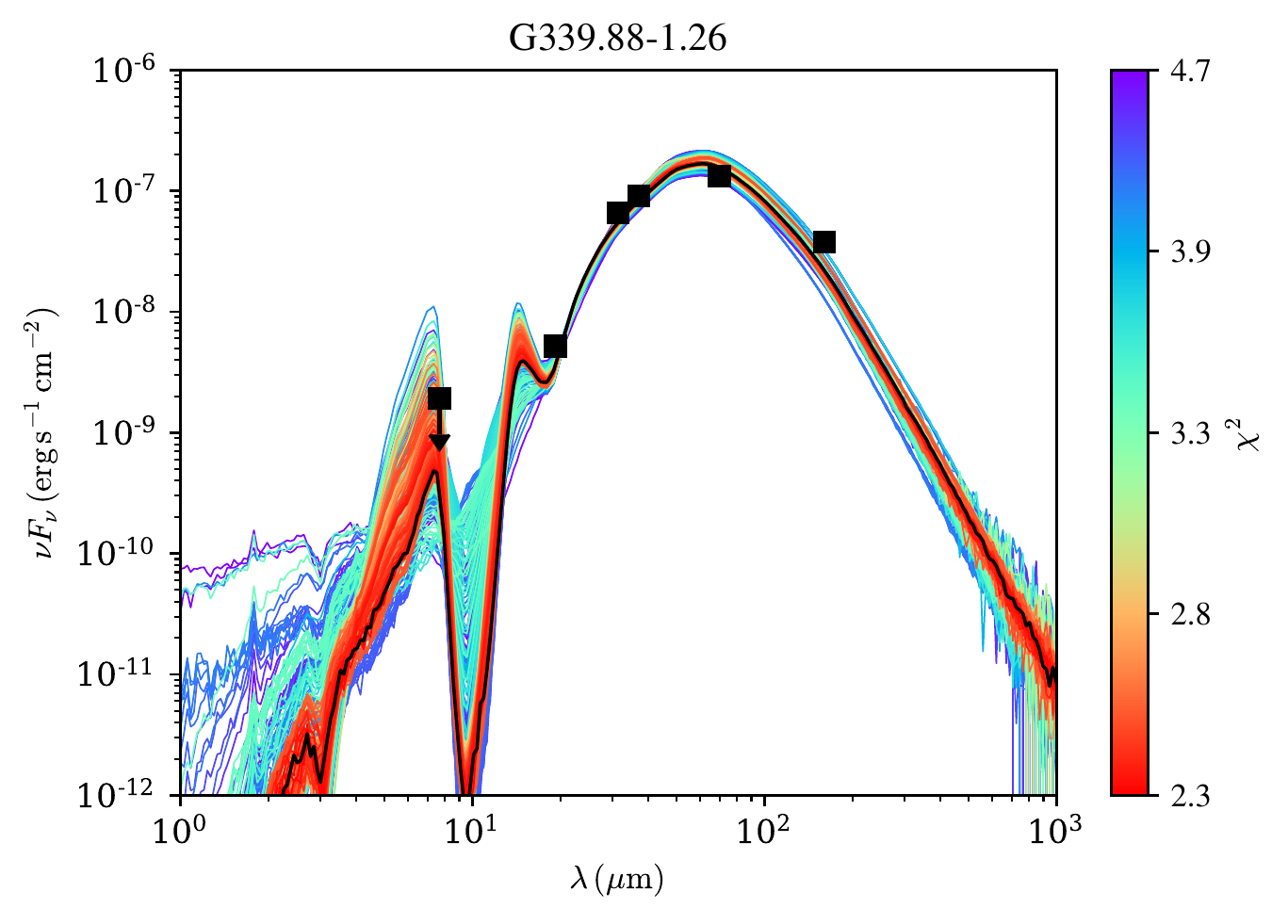}
\caption{(Continued.)}
\end{figure*}

\begin{figure*}[!htb]
\includegraphics[width=1.0\textwidth]{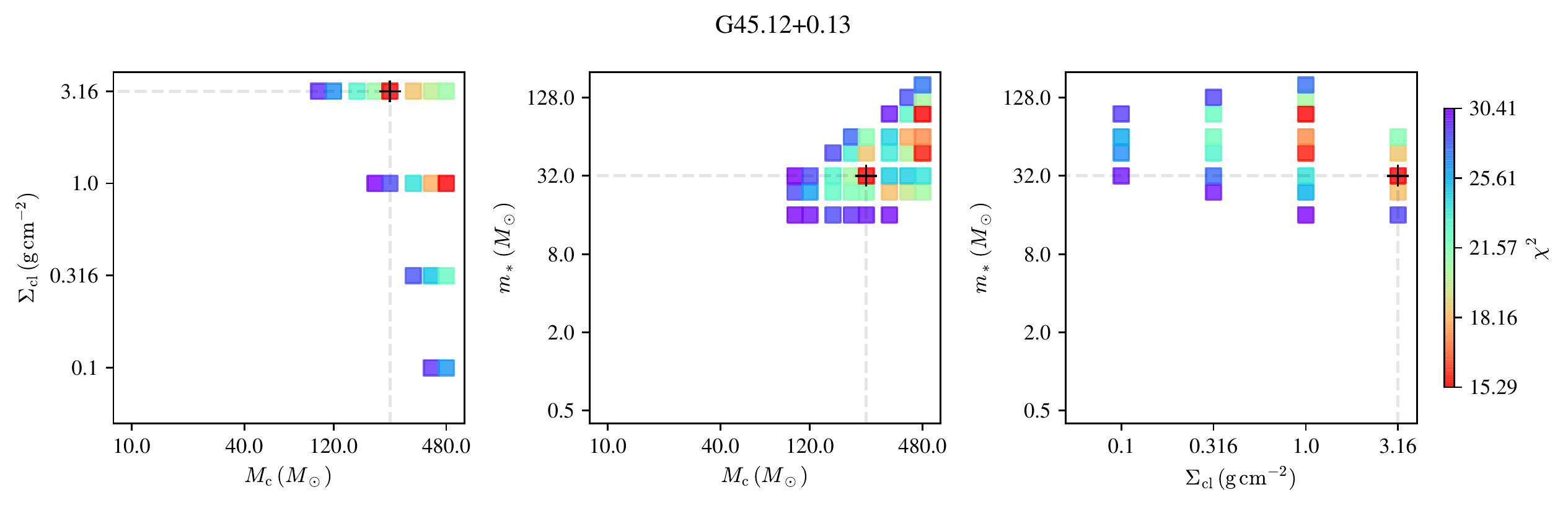}
\includegraphics[width=1.0\textwidth]{G30.59-0.04_2D_plot.pdf}
\includegraphics[width=1.0\textwidth]{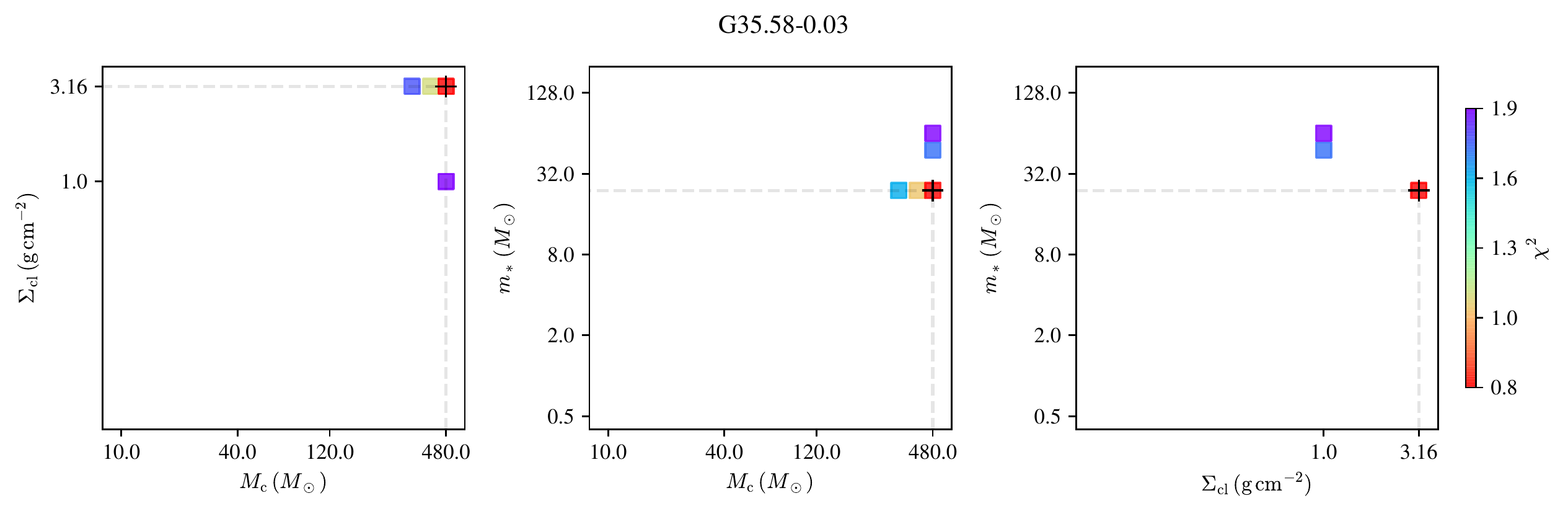}
\caption{Diagrams of $\chi^{2}$ distribution in $\Sigma_{\rm cl}$ - $M_{c}$ space (left), $m_{*}$ - $M_{\rm c}$ space (center) and $m_{*}$ - $\Sigma_{\rm  cl}$ space (right) for each source noted on top of each plot. The black cross is the best model.
\label{fig:sed_2D_results_soma_ii}}
\end{figure*}

\renewcommand{\thefigure}{A\arabic{figure}}
\addtocounter{figure}{-1}
\begin{figure*}[!htb]
\includegraphics[width=1.0\textwidth]{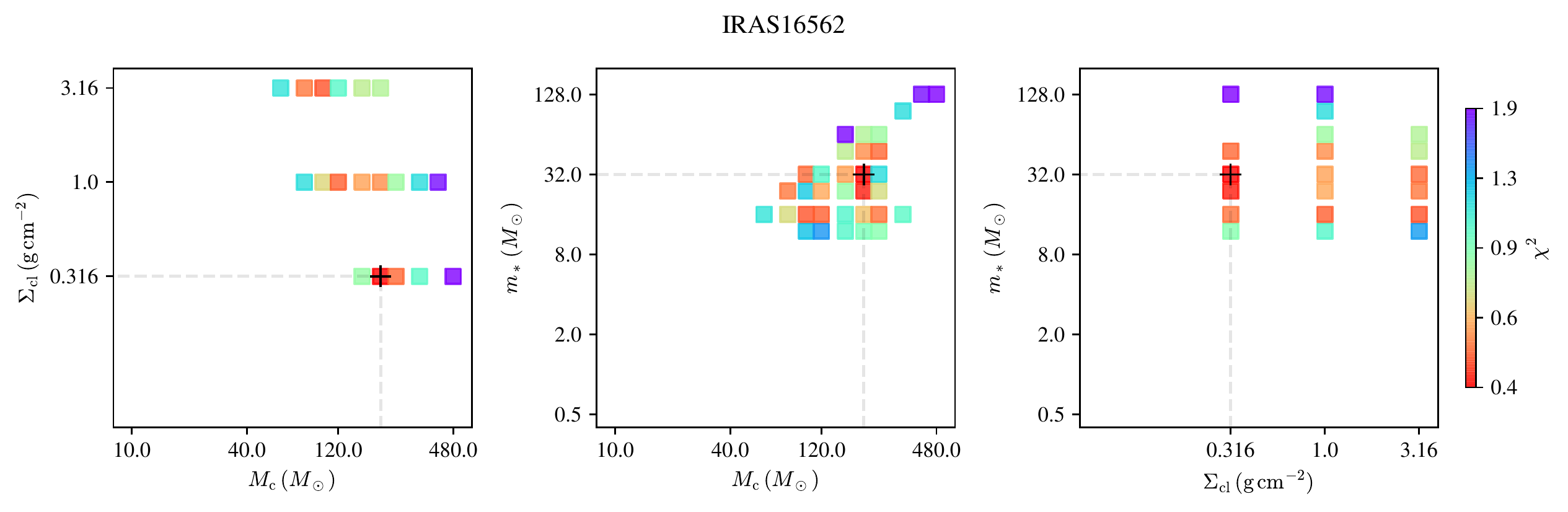}
\includegraphics[width=1.0\textwidth]{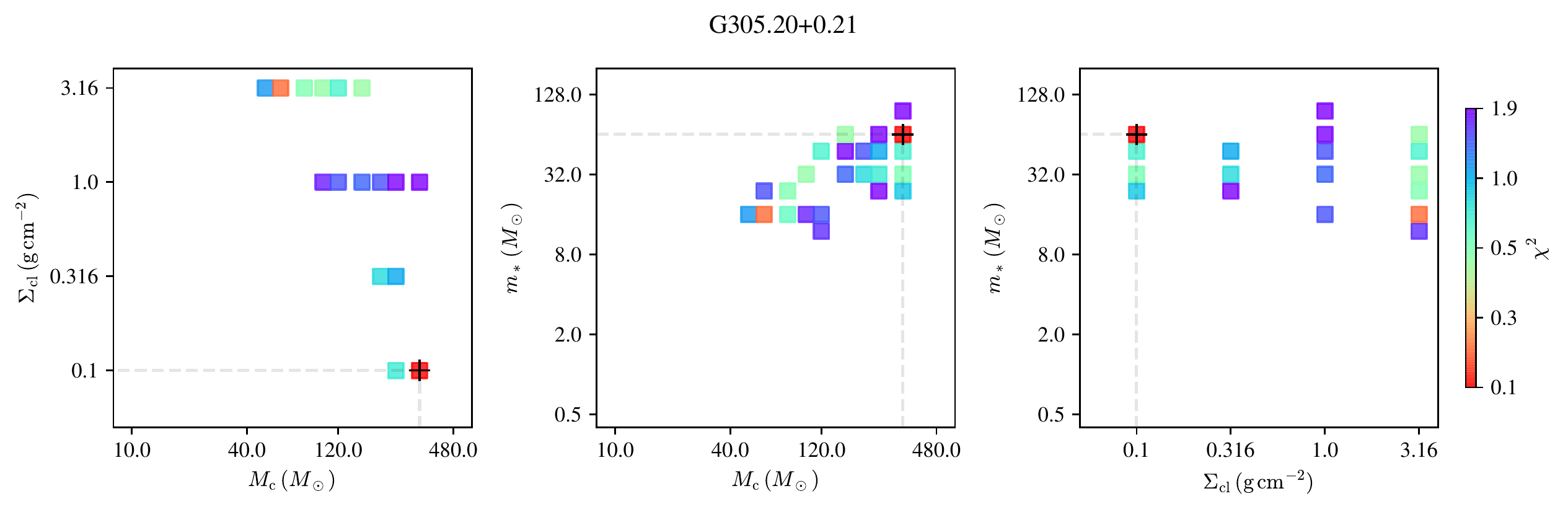}
\includegraphics[width=1.0\textwidth]{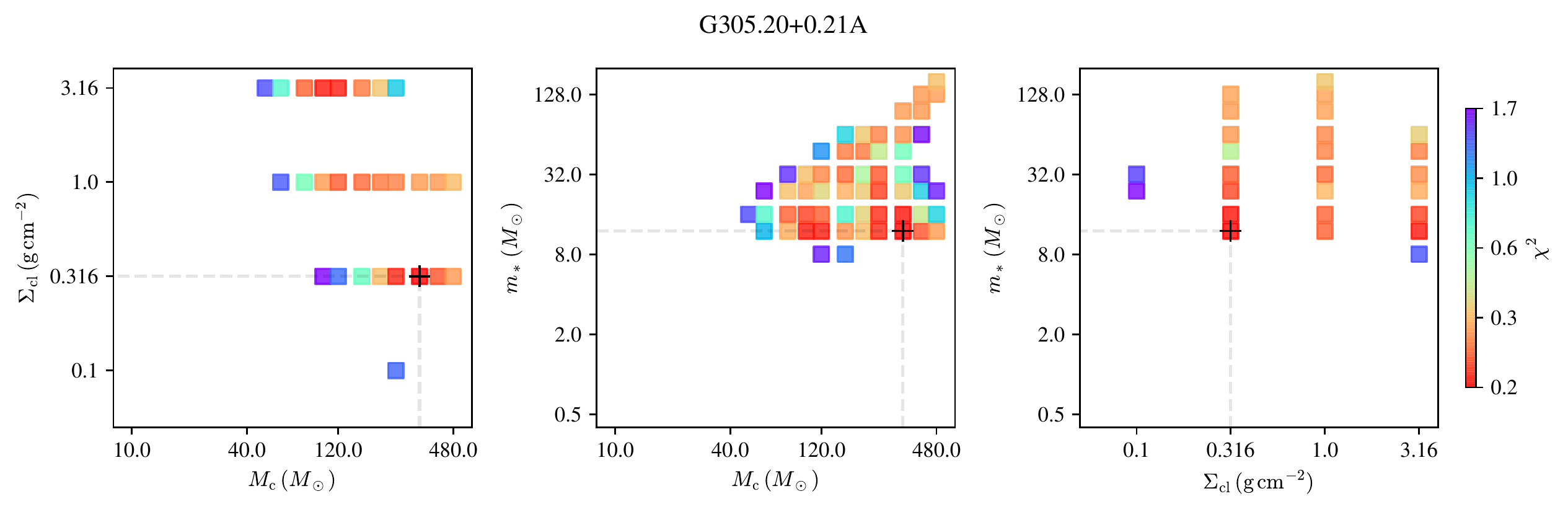}
\includegraphics[width=1.0\textwidth]{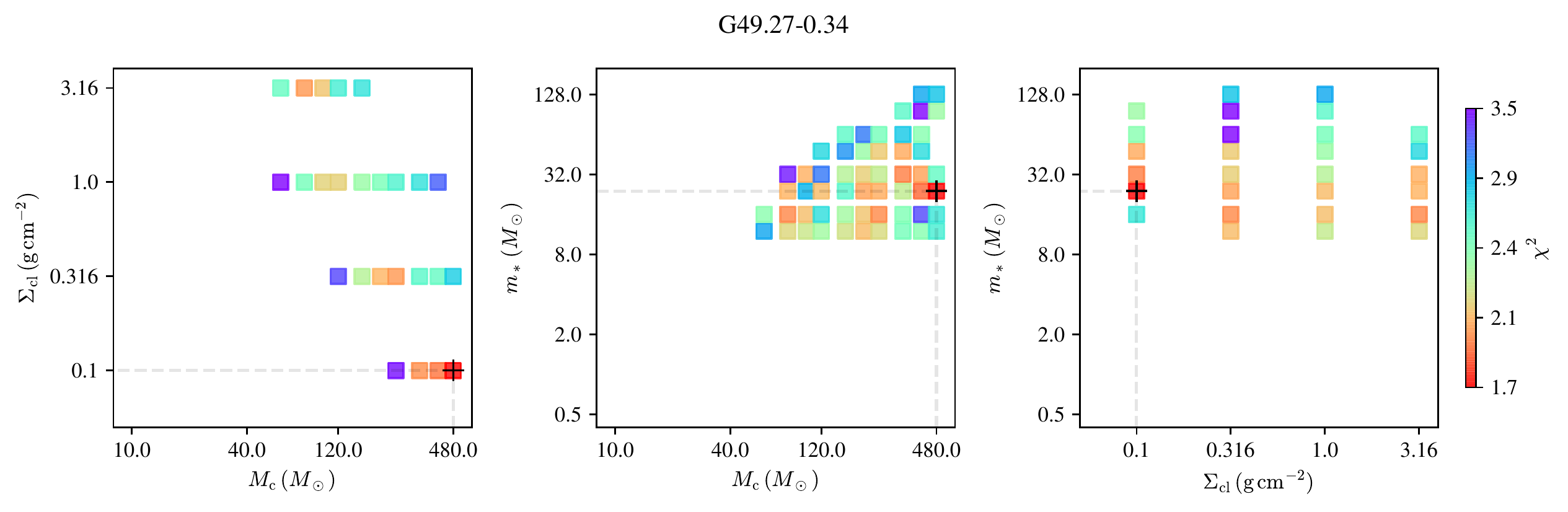}
\caption{(Continued.)}
\end{figure*}

\renewcommand{\thefigure}{A\arabic{figure}}
\addtocounter{figure}{-1}
\begin{figure*}[!htb]
\includegraphics[width=1.0\textwidth]{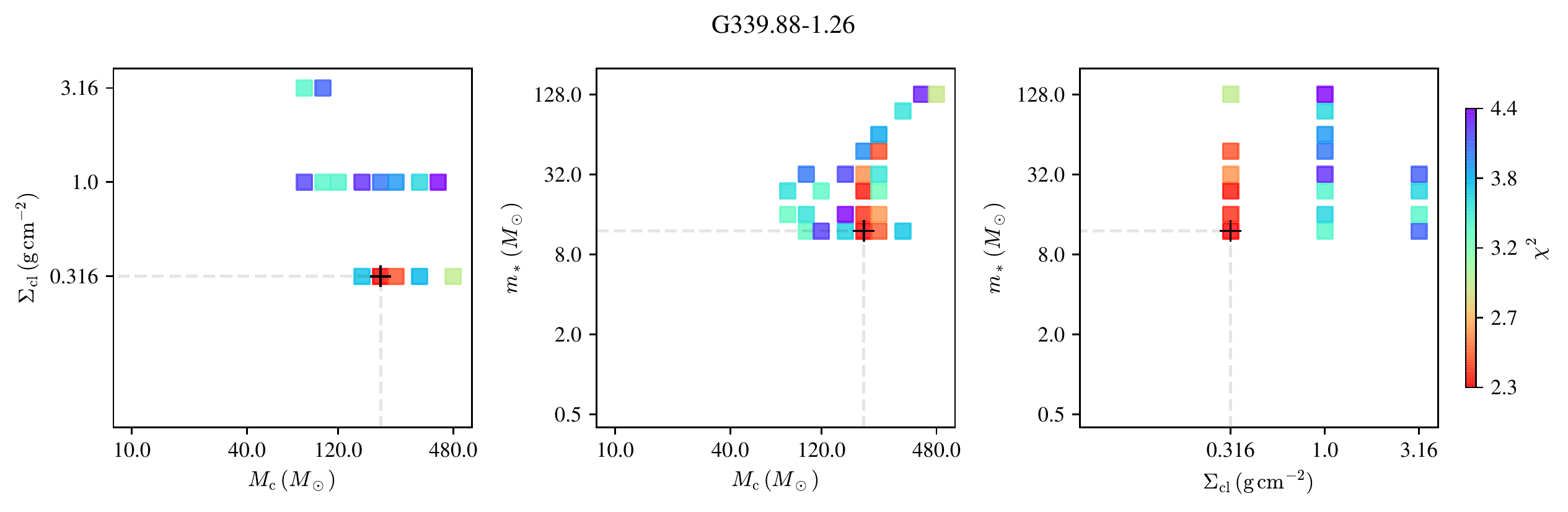}
\caption{(Continued.)}
\end{figure*}

\subsection{SOMA III New SED fit}

We revisited the measurement and error estimation for the SOMA I sources \citep{liu2020} as well as refit their SEDs. Figures\,\ref{fig:sed_1D_results_soma_iii} and \ref{fig:sed_2D_results_soma_iii} show the revisited results.

\begin{figure*}[!htb]
\includegraphics[width=0.5\textwidth]{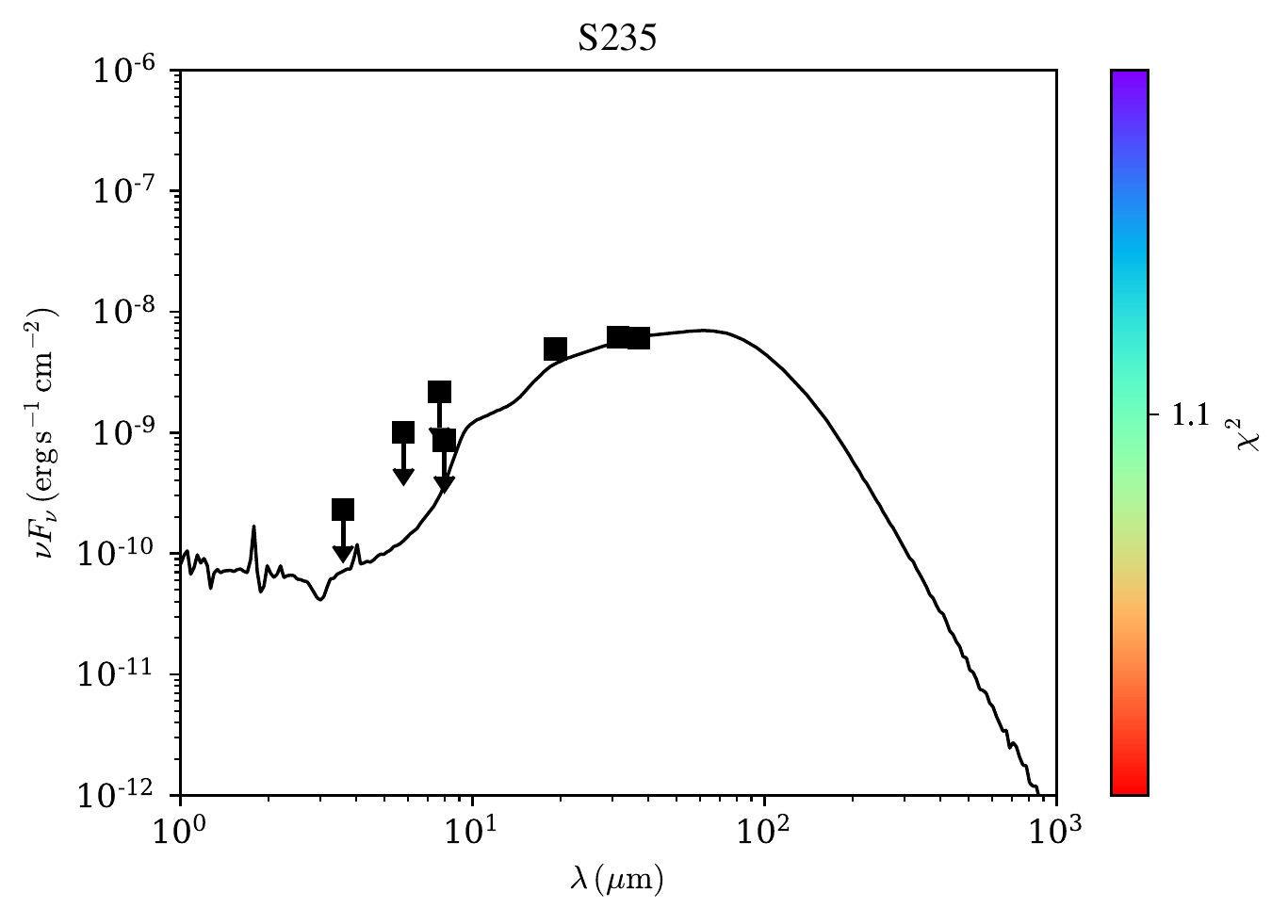}
\includegraphics[width=0.5\textwidth]{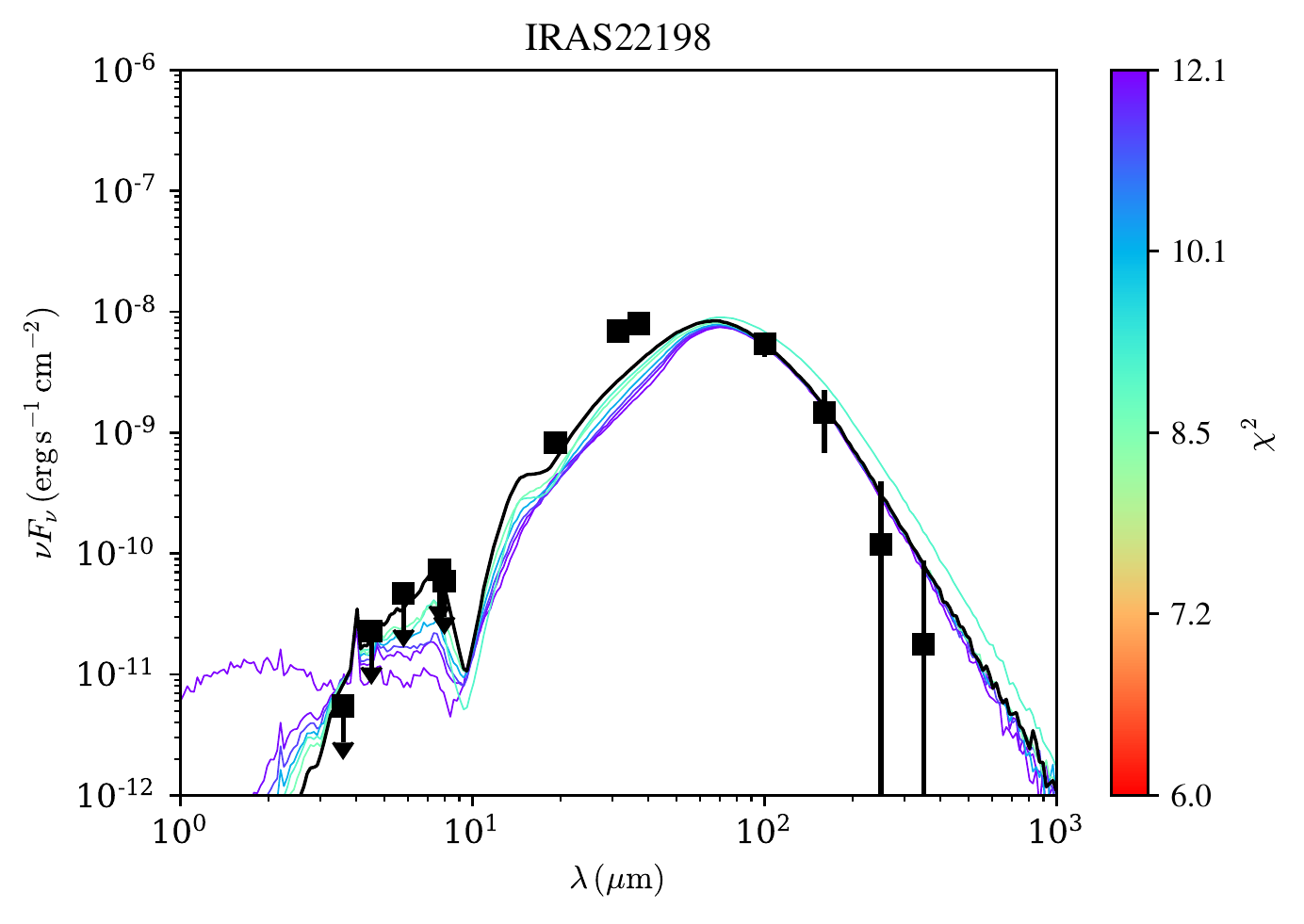}
\includegraphics[width=0.5\textwidth]{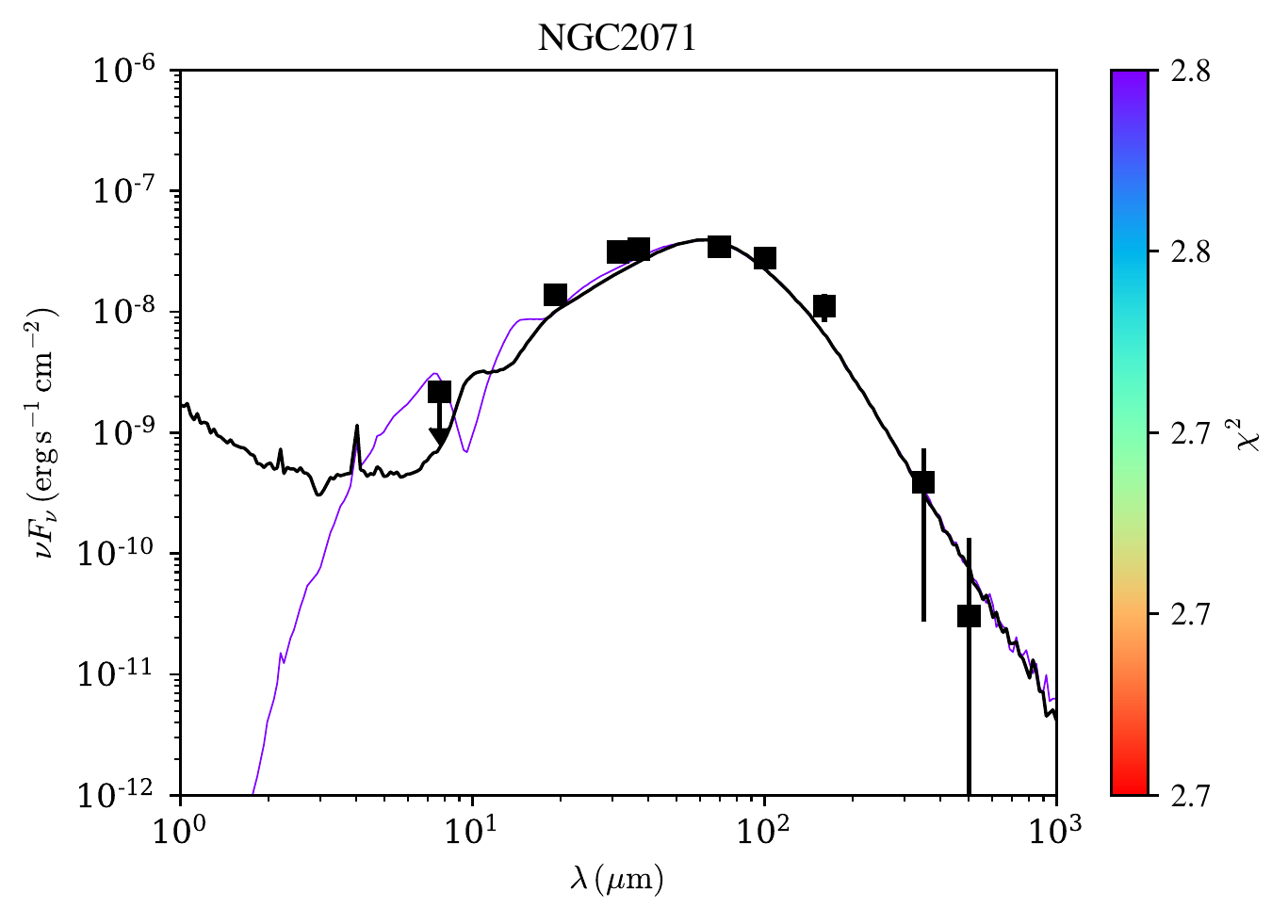}
\includegraphics[width=0.5\textwidth]{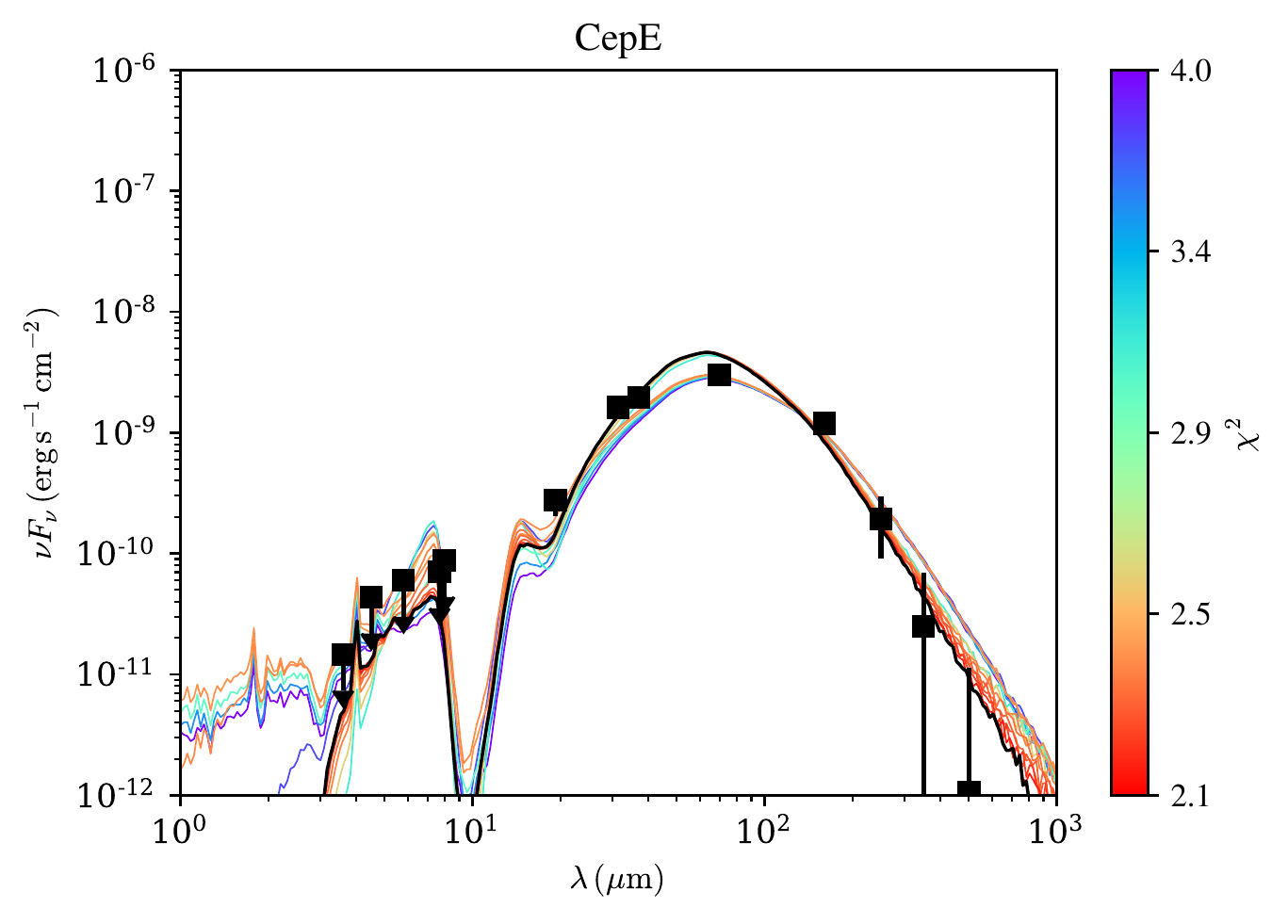}
\includegraphics[width=0.5\textwidth]{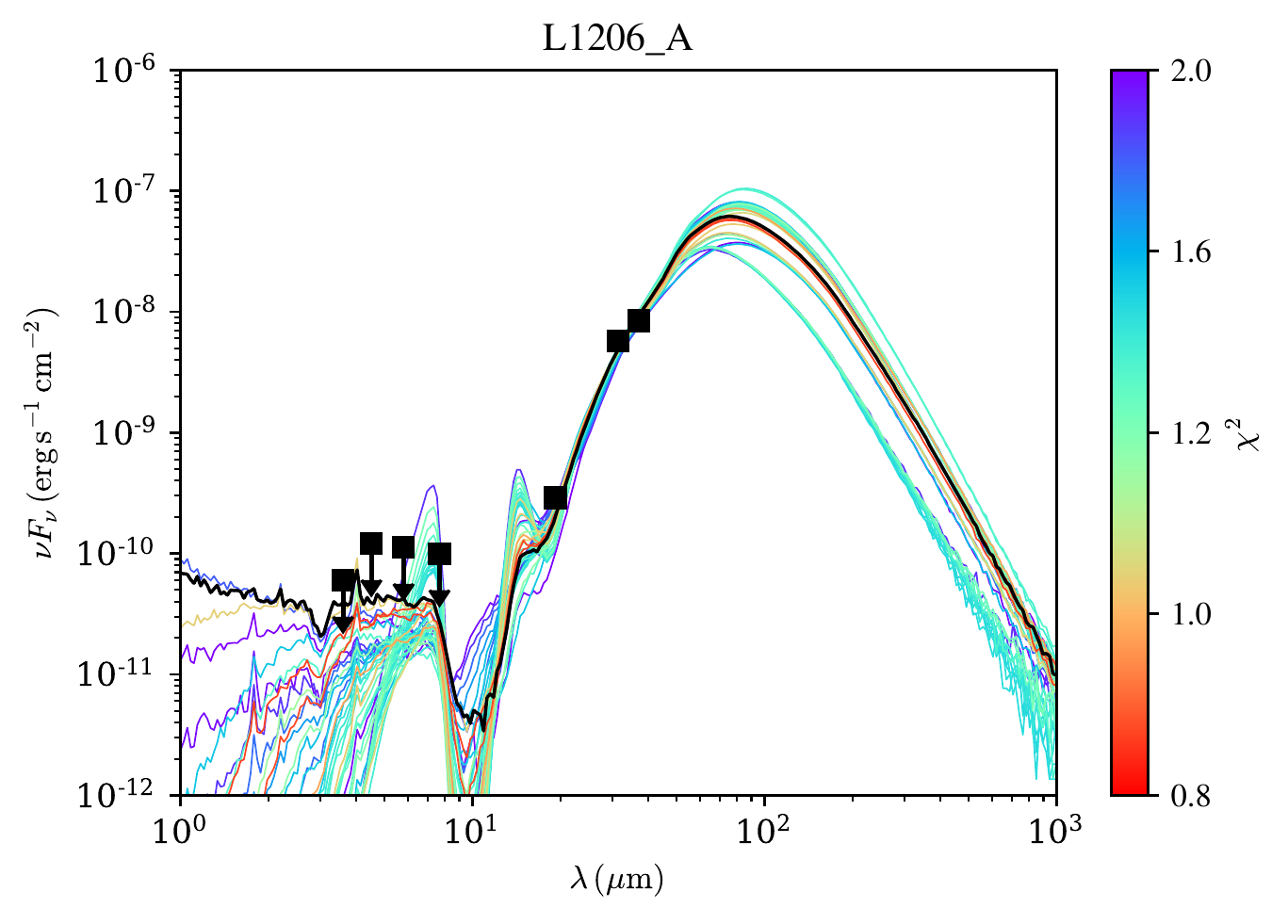}
\includegraphics[width=0.5\textwidth]{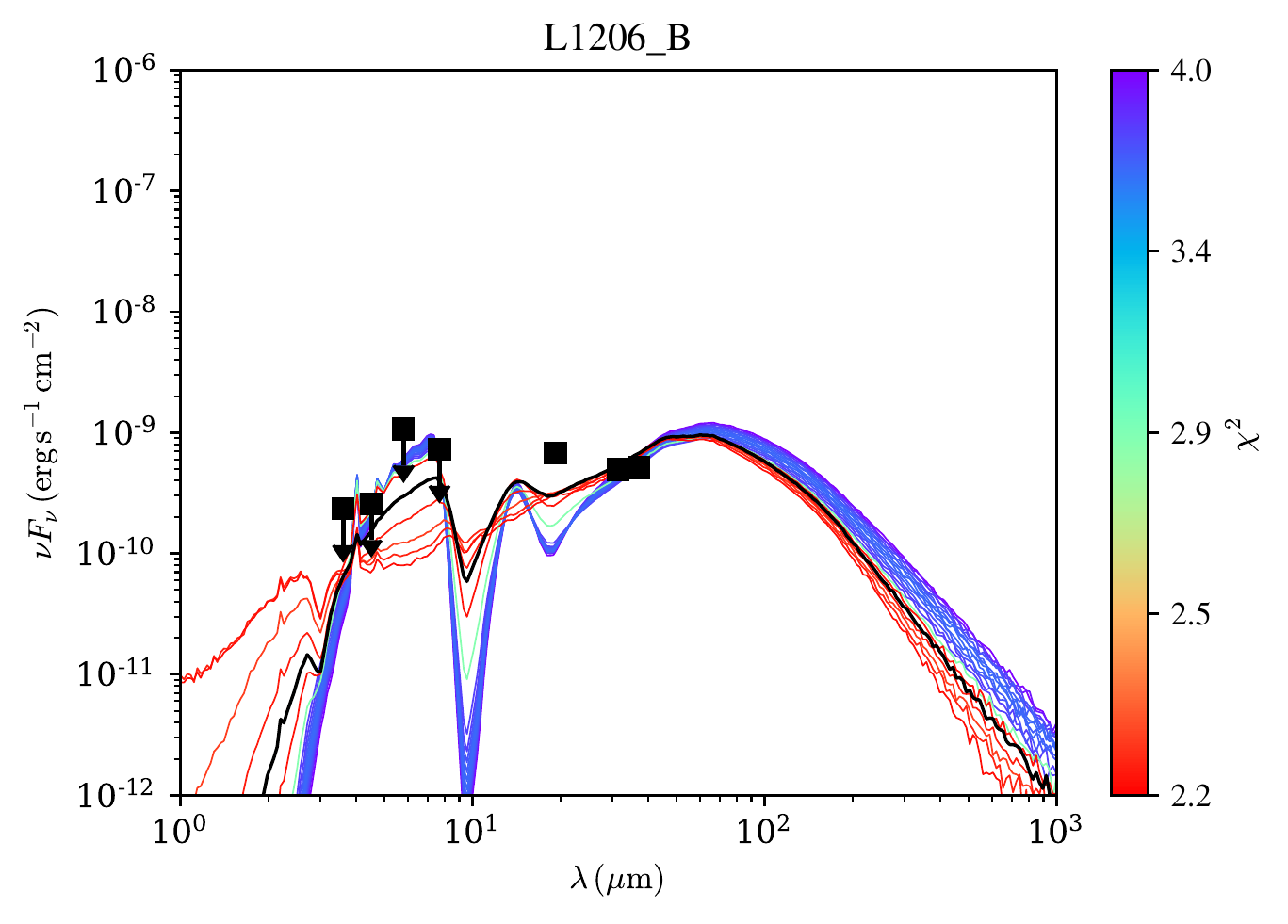}
\caption{SOMA III sources reanalyzed with sedcreator. Protostar model fitting to the fixed aperture, background-subtracted SED data using the ZT model grid. For each source (noted on top of each plot), the best fitting protostar model is shown with a black line, while all other good model fits (see the text) are shown with colored lines (red to blue with increasing $\chi^2$). Flux values are those from Table\,\ref{tab:soma_all_fluxes}. Note that the data at $\lesssim8\,{\rm \mu m}$ are treated as upper limits (see the text). The resulting model parameters are listed in Table\,\ref{tab:best_models_soma_ALL}.
\label{fig:sed_1D_results_soma_iii}}
\end{figure*}

\renewcommand{\thefigure}{A\arabic{figure}}
\addtocounter{figure}{-1}
\begin{figure*}[!htb]
\includegraphics[width=0.5\textwidth]{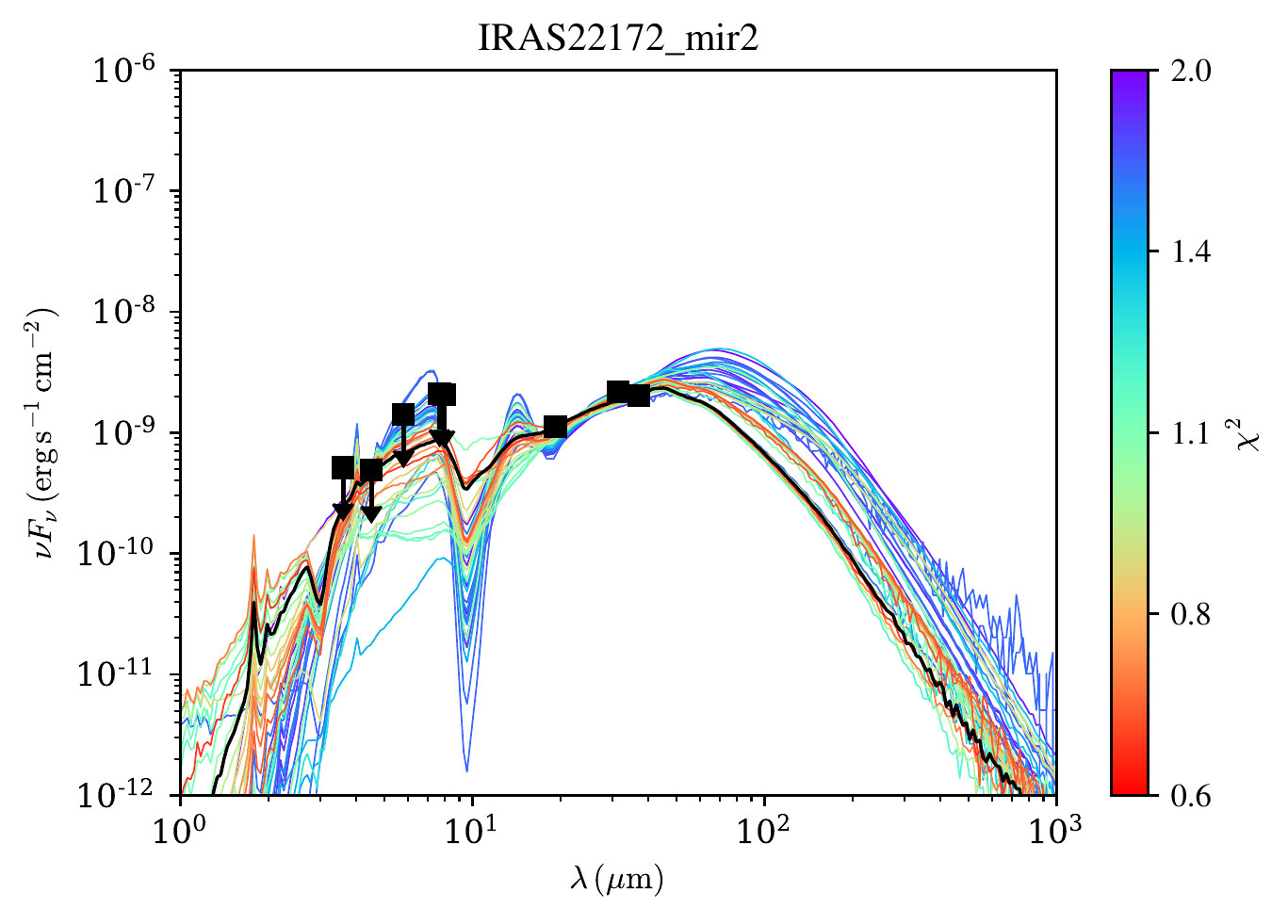}
\includegraphics[width=0.5\textwidth]{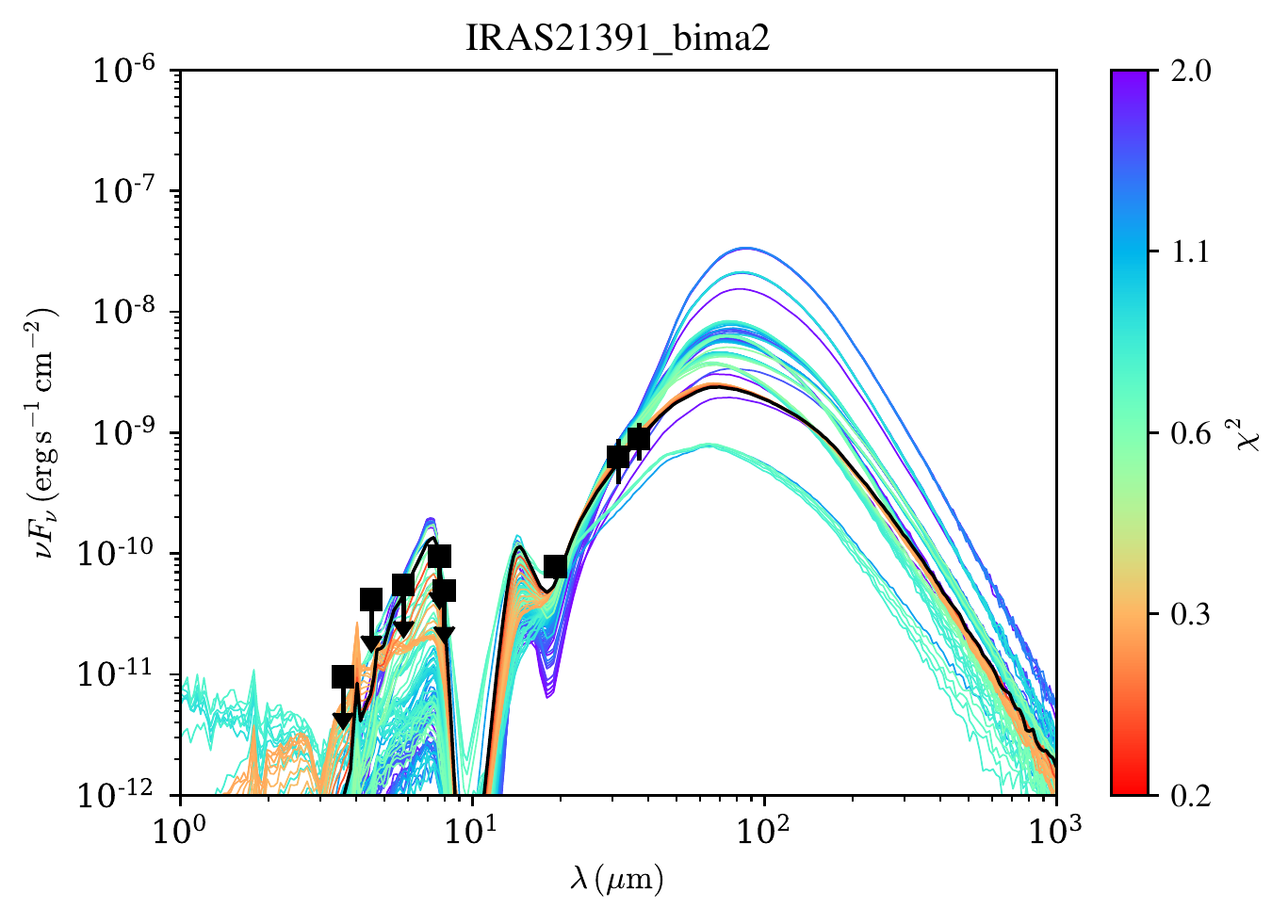}
\includegraphics[width=0.5\textwidth]{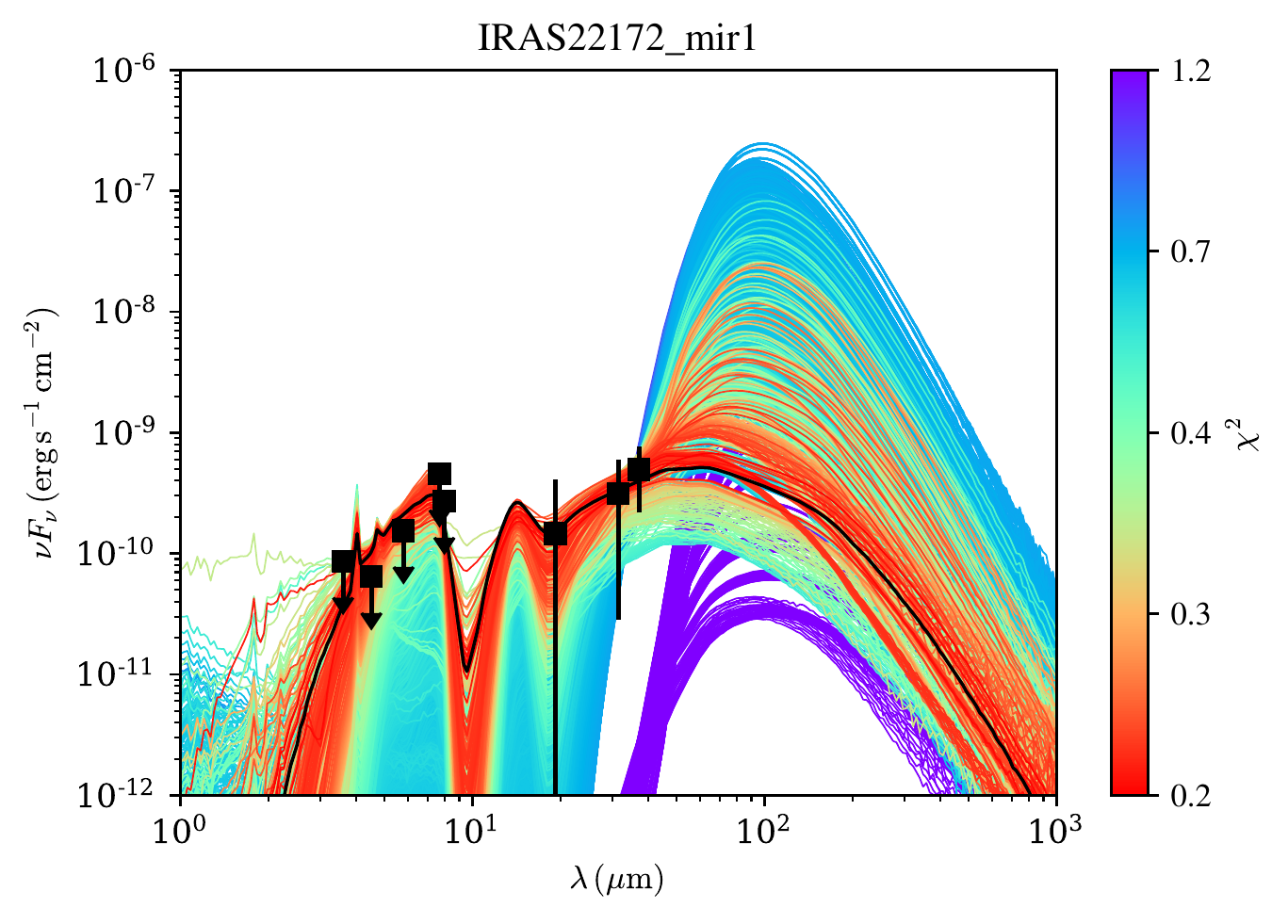}
\includegraphics[width=0.5\textwidth]{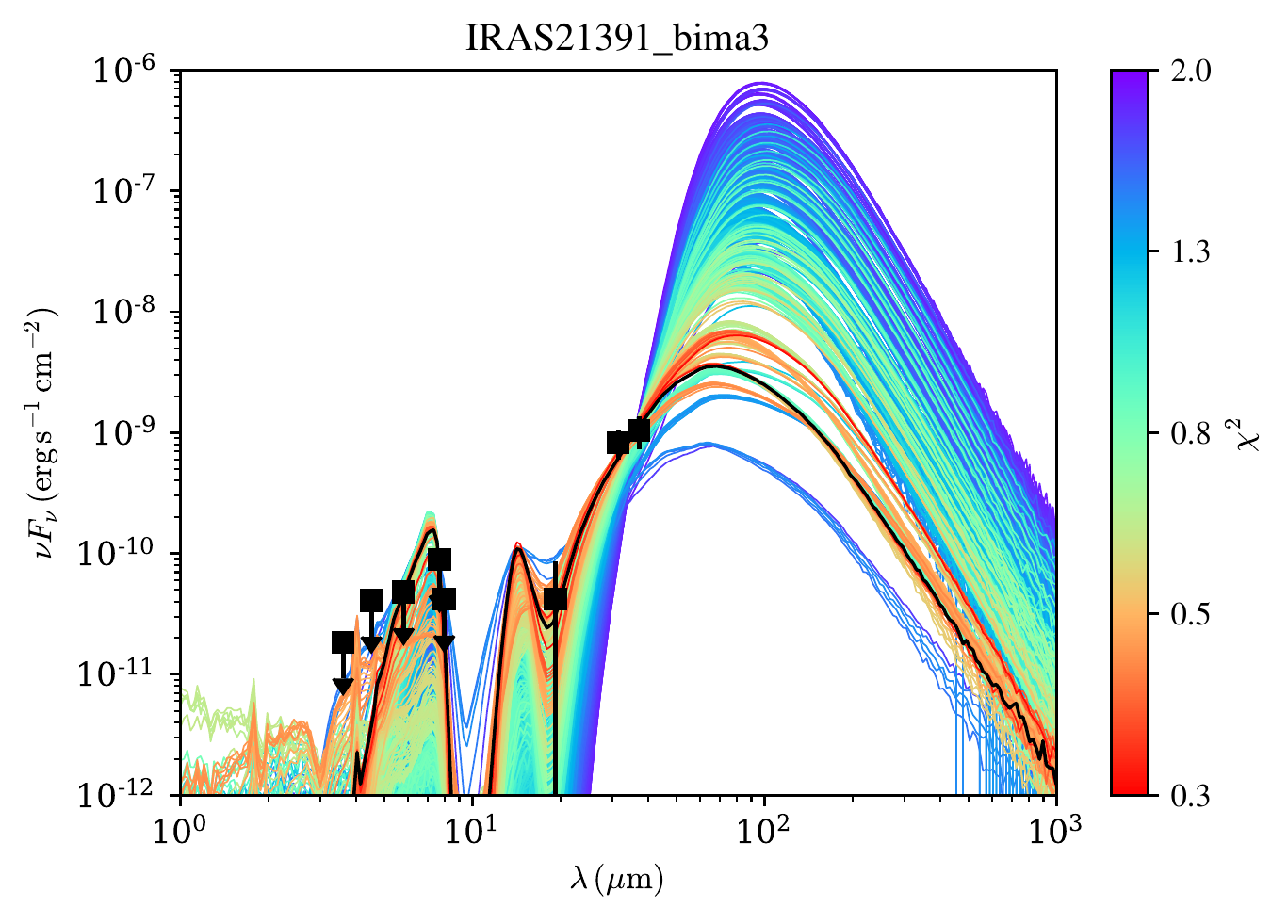}
\includegraphics[width=0.5\textwidth]{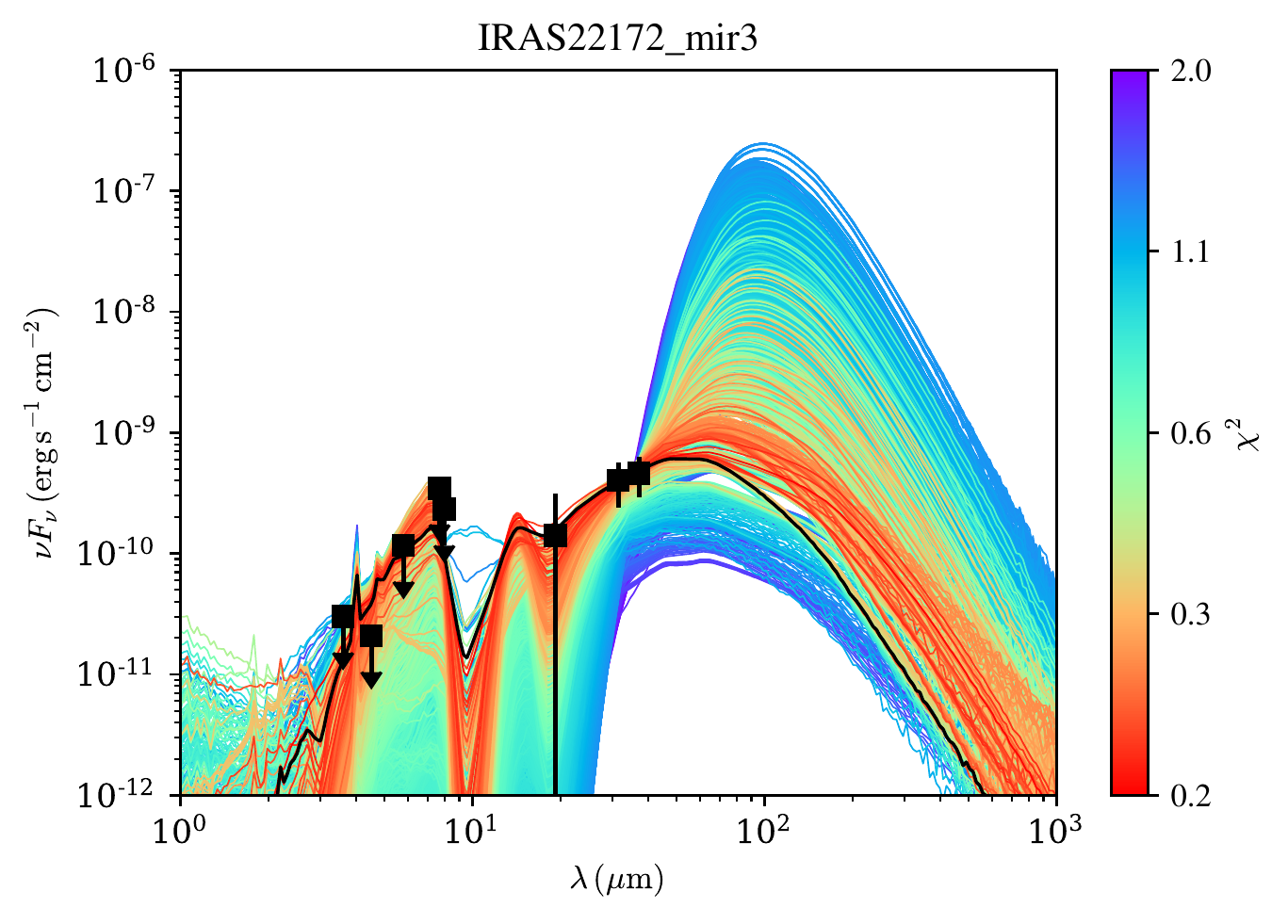}
\includegraphics[width=0.5\textwidth]{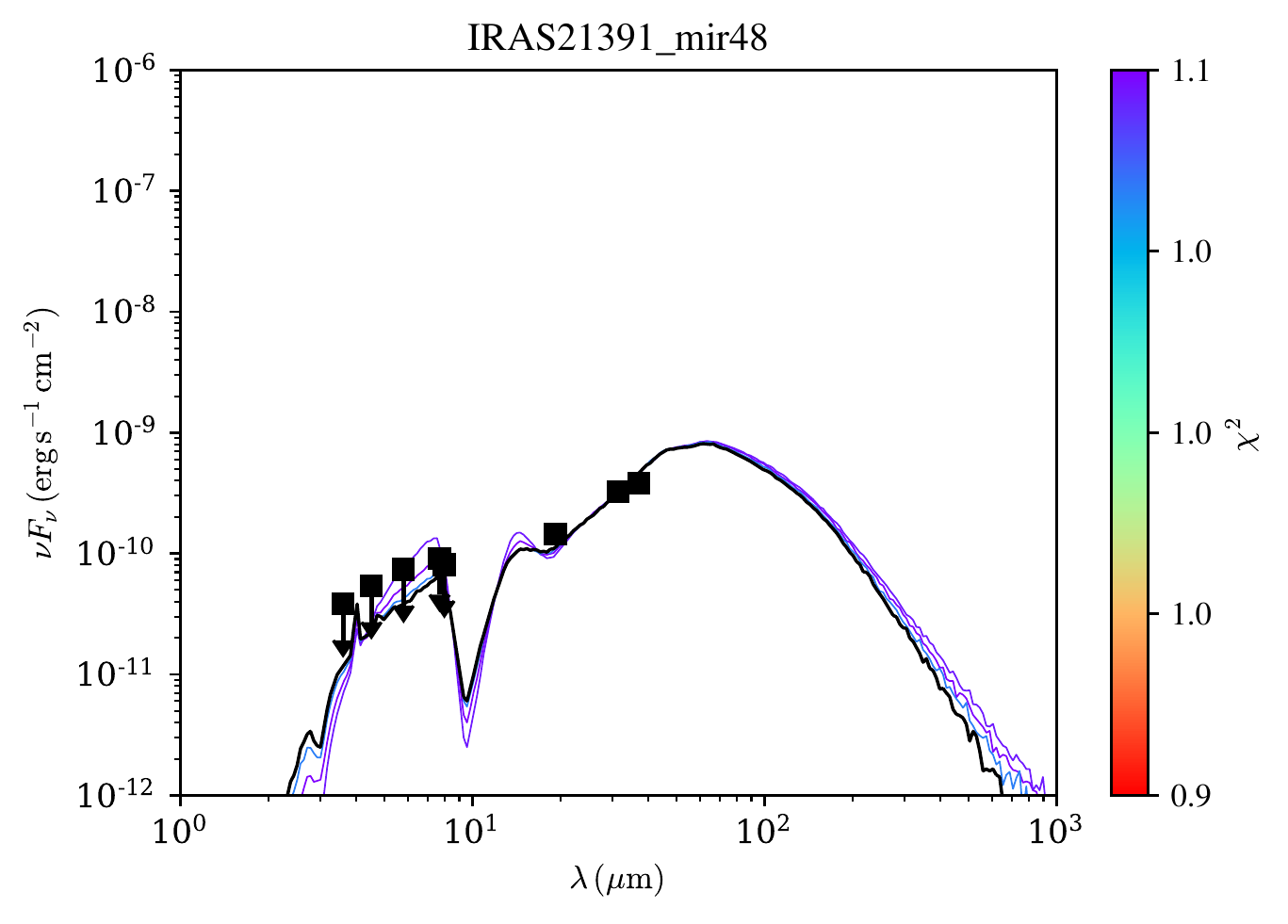}
\caption{(Continued.)}
\end{figure*}

\begin{figure*}[!htb]
\includegraphics[width=1.0\textwidth]{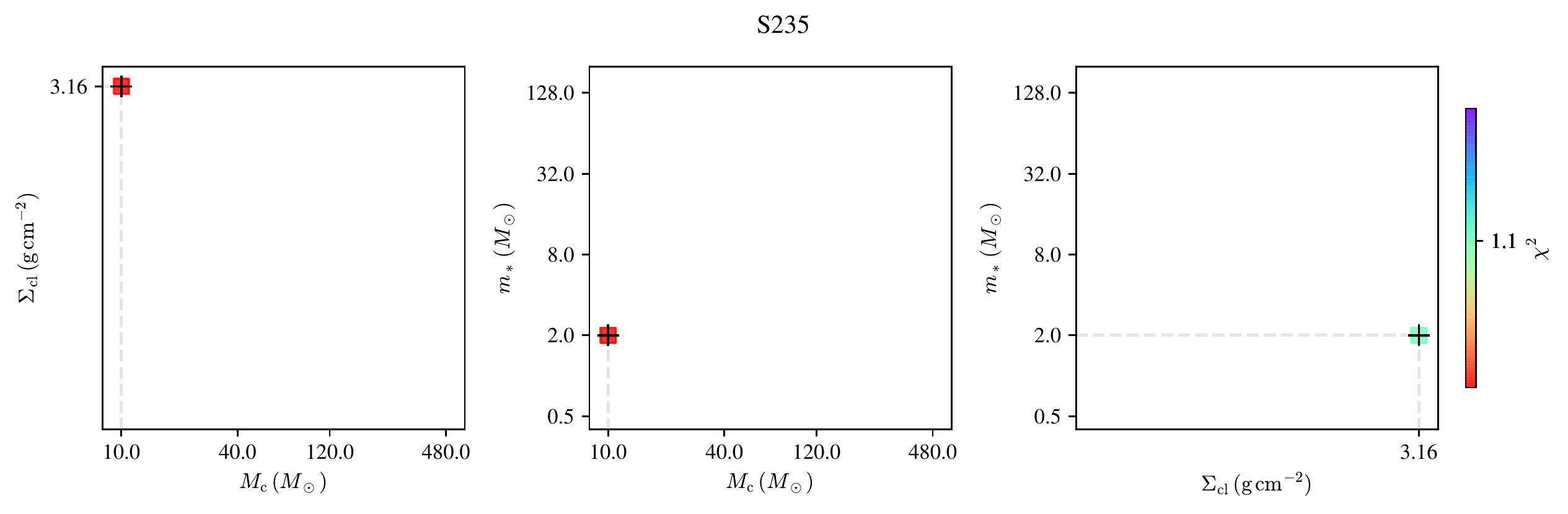}
\includegraphics[width=1.0\textwidth]{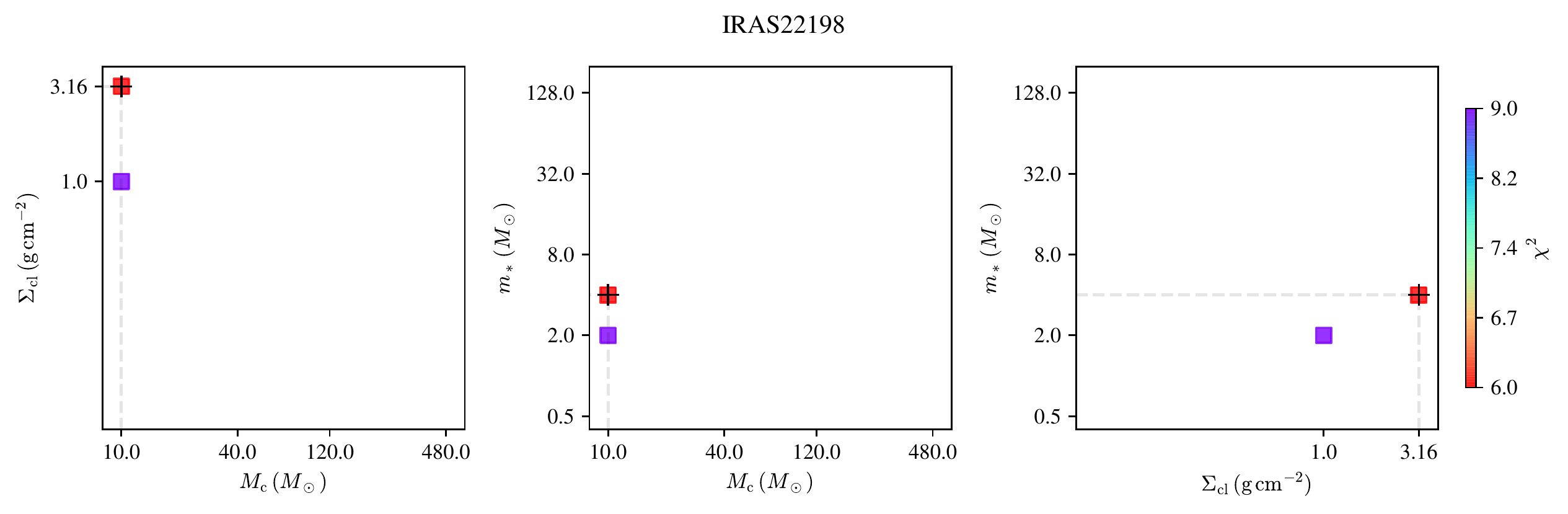}
\includegraphics[width=1.0\textwidth]{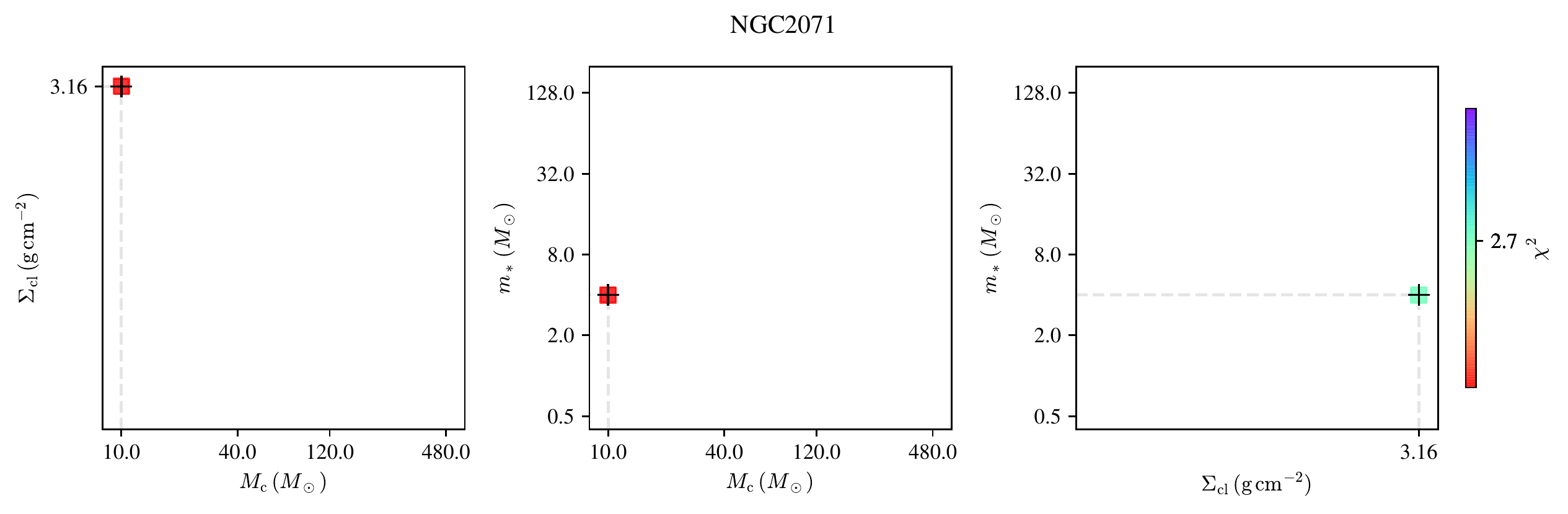}
\caption{Diagrams of $\chi^{2}$ distribution in $\Sigma_{\rm cl}$ - $M_{c}$ space (left), $m_{*}$ - $M_{\rm c}$ space (center) and $m_{*}$ - $\Sigma_{\rm  cl}$ space (right) for each source noted on top of each plot. The black cross is the best model.
\label{fig:sed_2D_results_soma_iii}}
\end{figure*}

\renewcommand{\thefigure}{A\arabic{figure}}
\addtocounter{figure}{-1}
\begin{figure*}[!htb]
\includegraphics[width=1.0\textwidth]{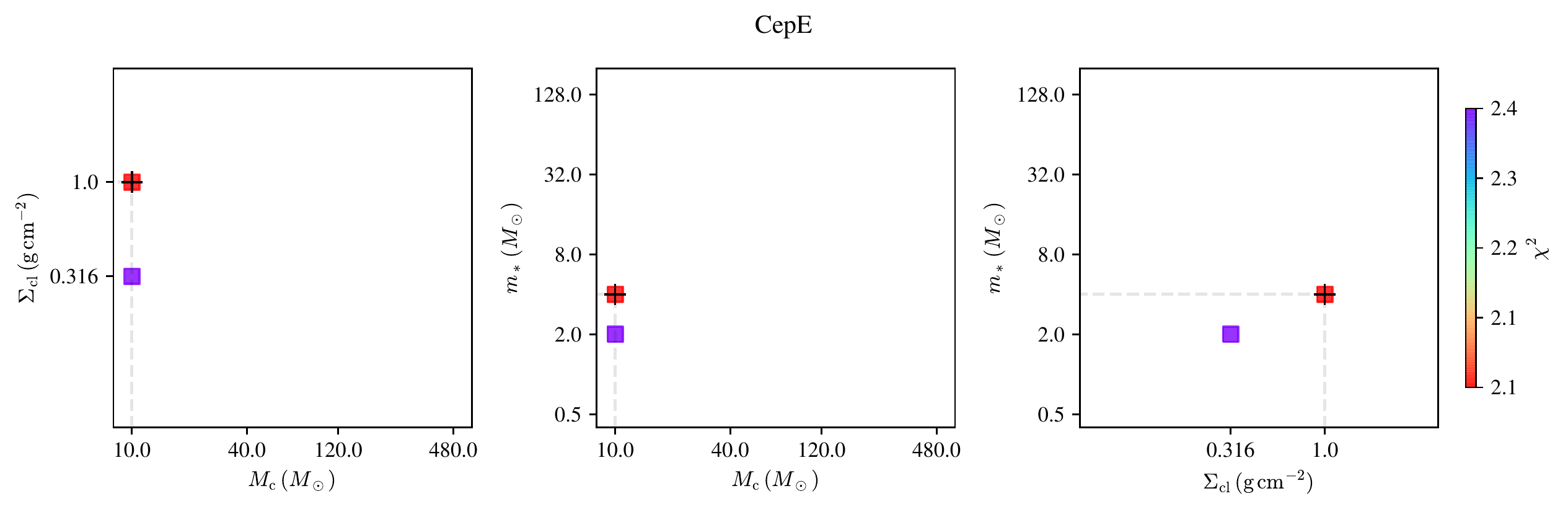}
\includegraphics[width=1.0\textwidth]{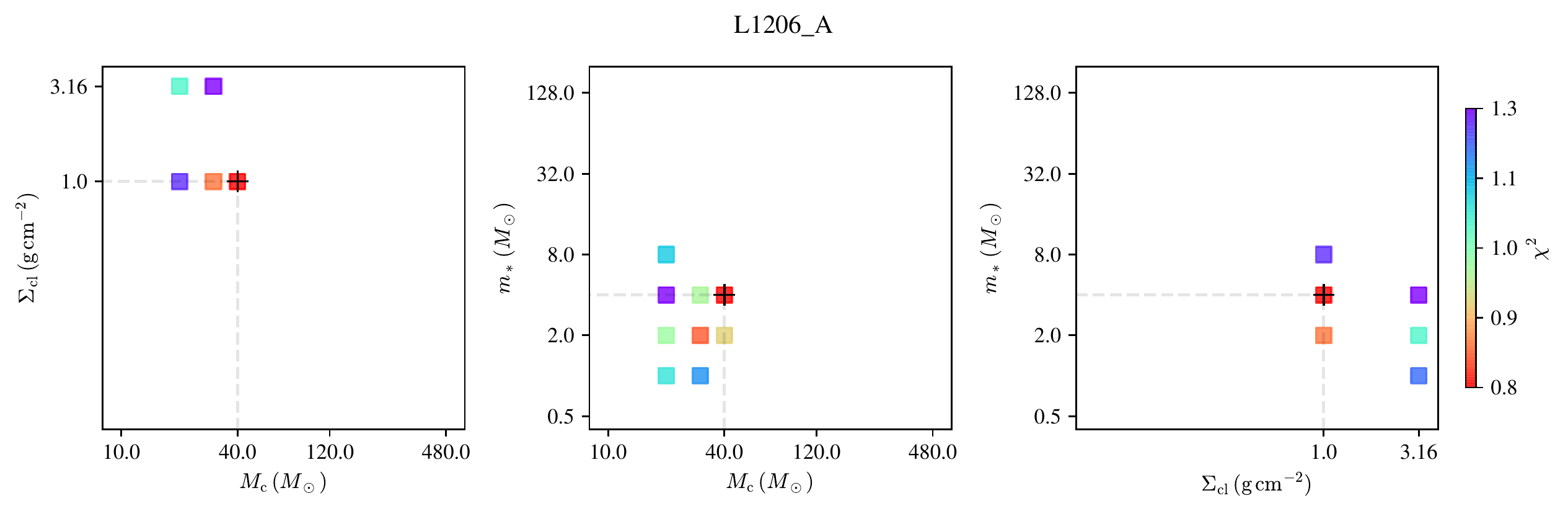}
\includegraphics[width=1.0\textwidth]{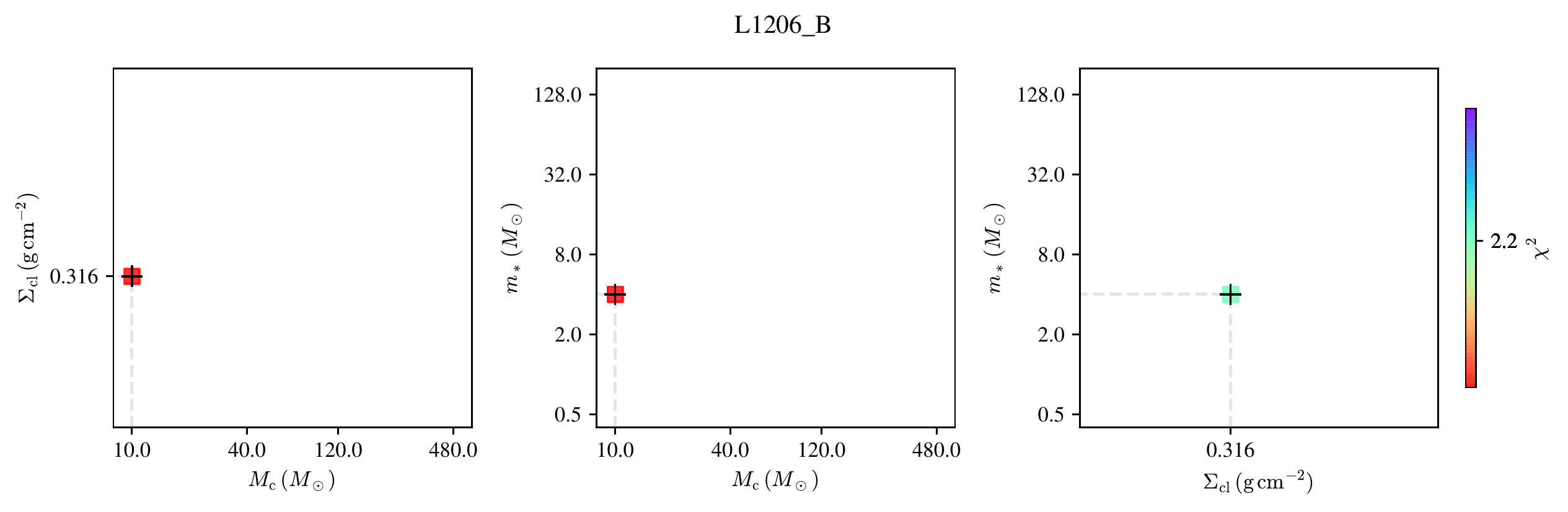}
\includegraphics[width=1.0\textwidth]{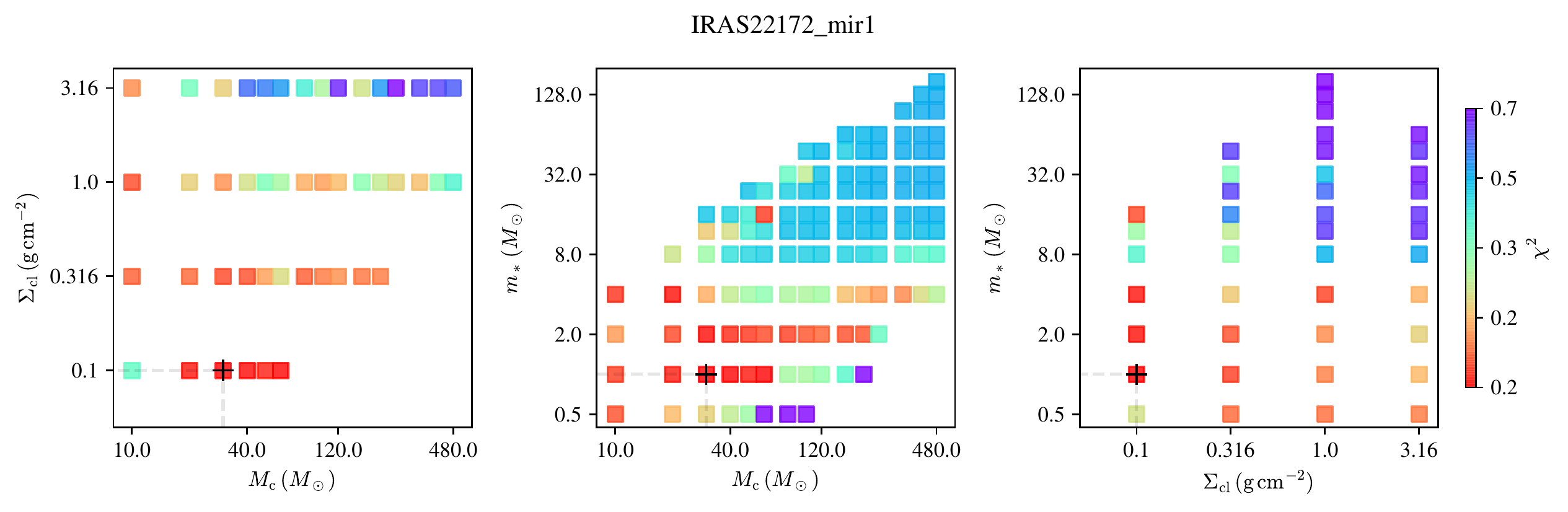}
\caption{(Continued.)}
\end{figure*}

\renewcommand{\thefigure}{A\arabic{figure}}
\addtocounter{figure}{-1}
\begin{figure*}[!htb]
\includegraphics[width=1.0\textwidth]{IRAS22172_mir1_2D_plot.pdf}
\includegraphics[width=1.0\textwidth]{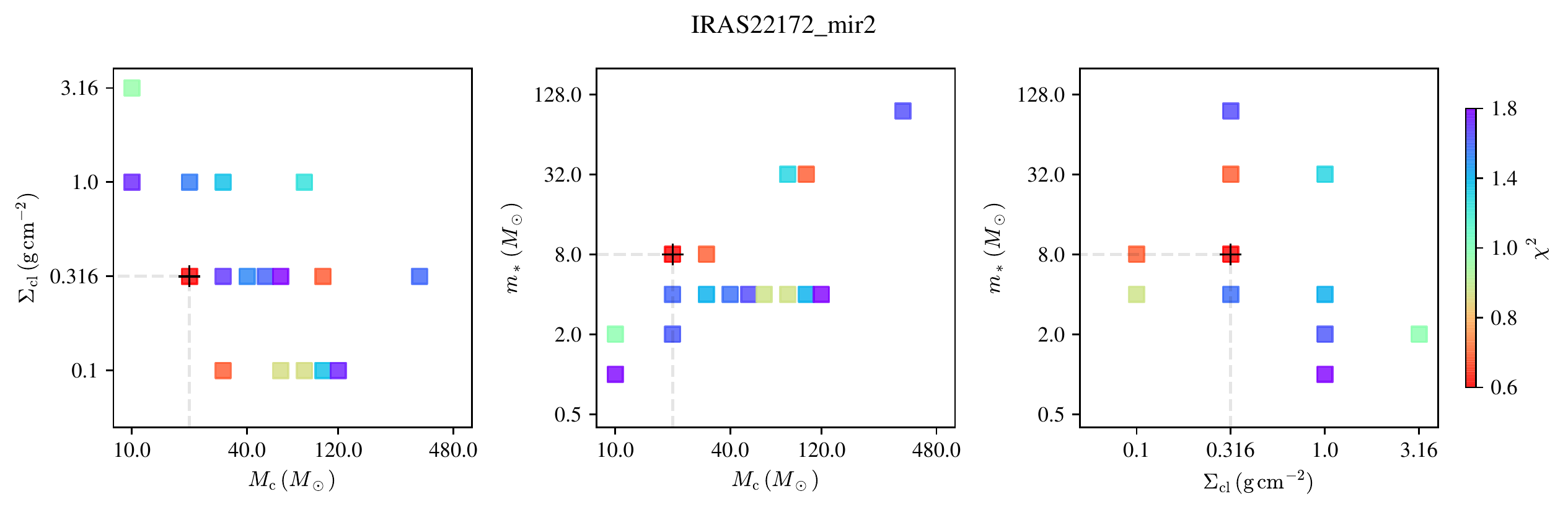}
\includegraphics[width=1.0\textwidth]{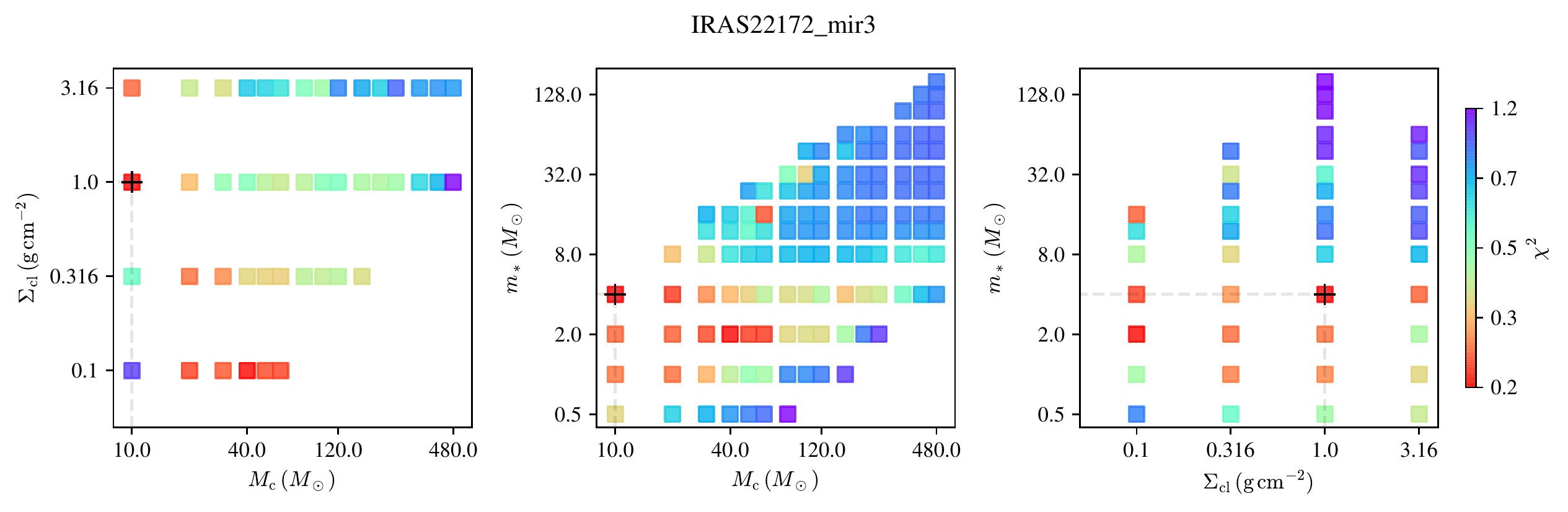}
\caption{(Continued.)}
\end{figure*}

\renewcommand{\thefigure}{A\arabic{figure}}
\addtocounter{figure}{-1}
\begin{figure*}[!htb]
\includegraphics[width=1.0\textwidth]{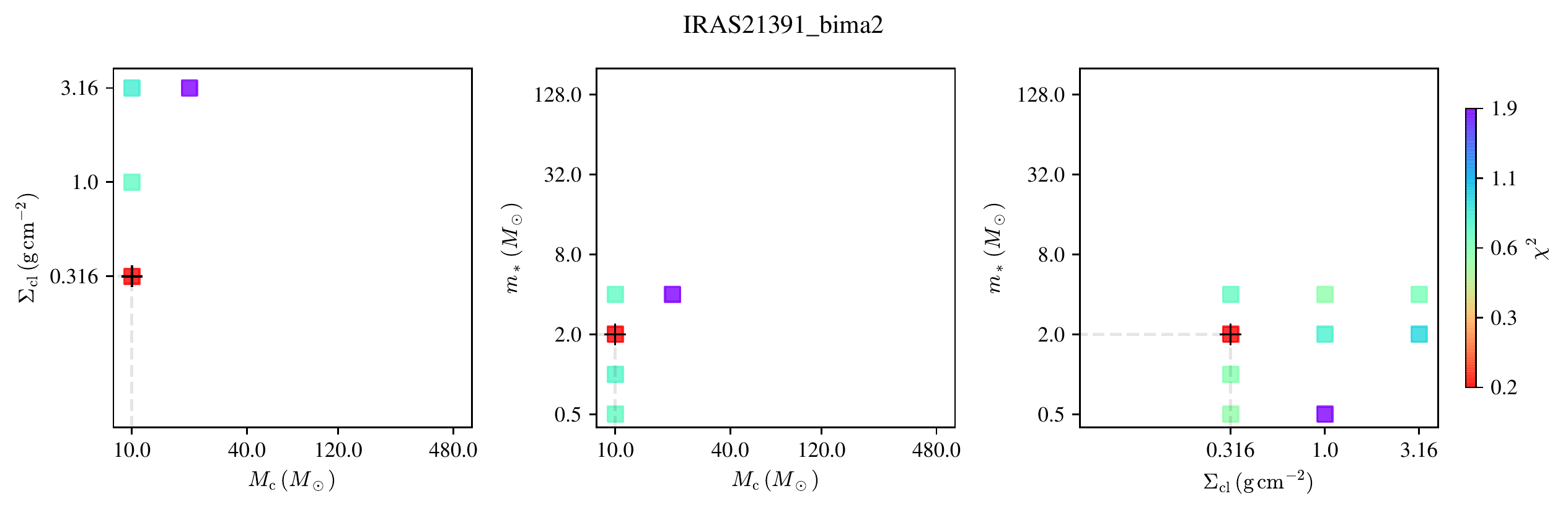}
\includegraphics[width=1.0\textwidth]{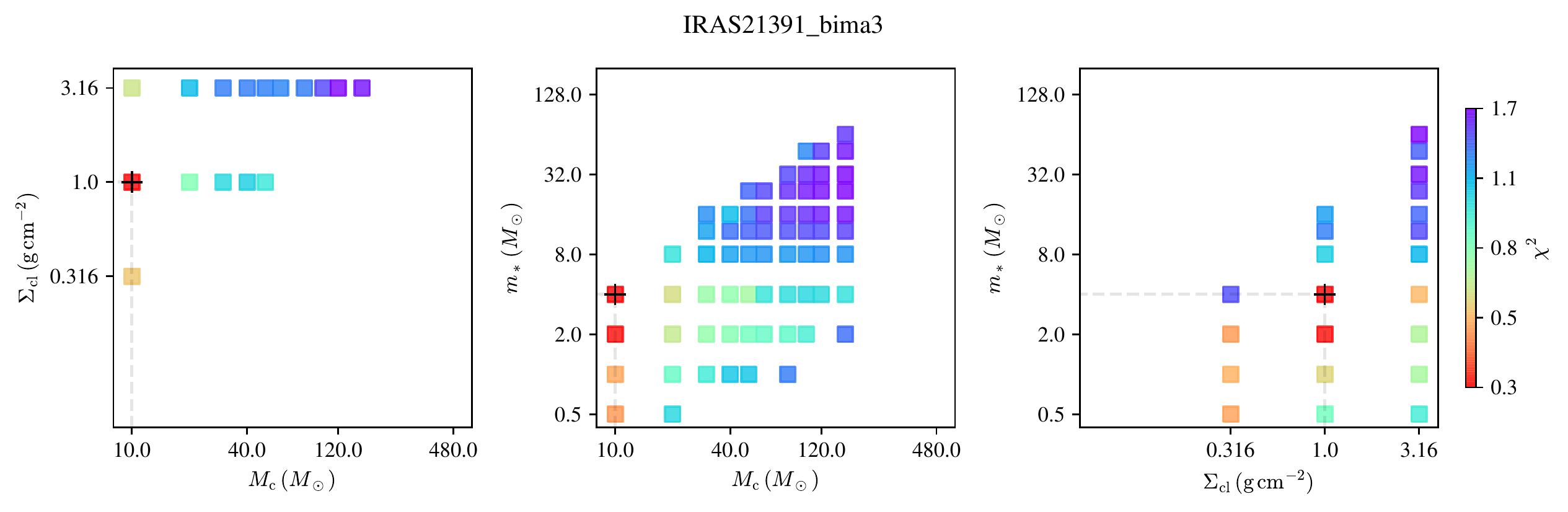}
\includegraphics[width=1.0\textwidth]{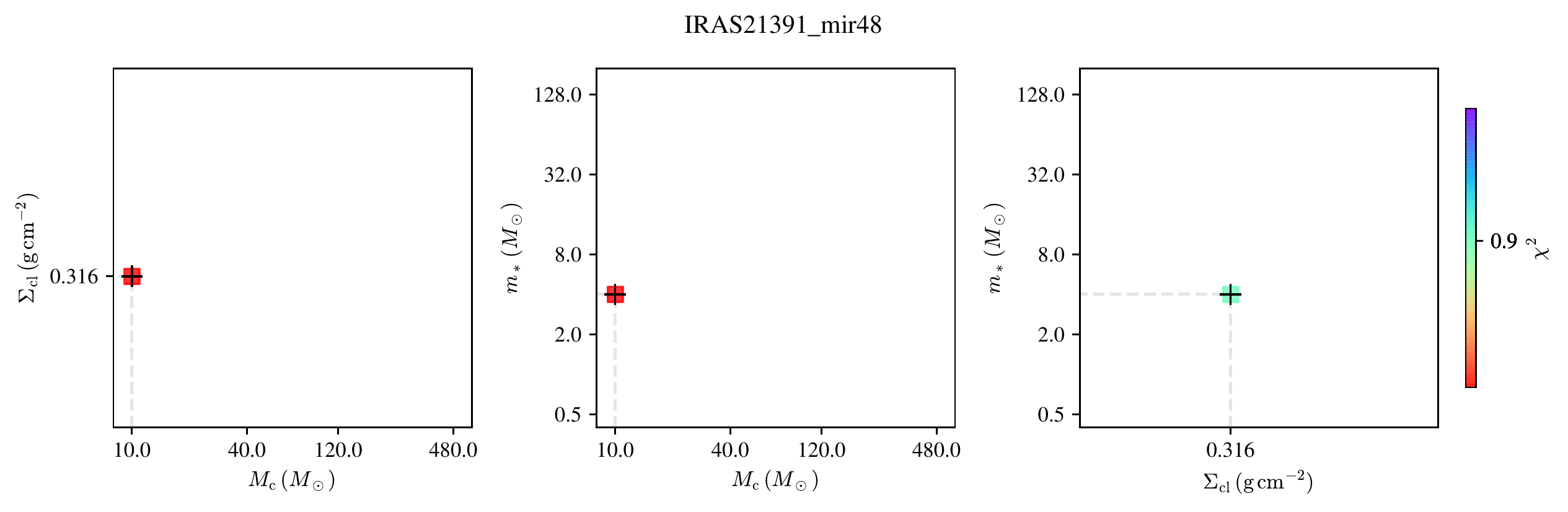}
\caption{(Continued.)}
\end{figure*}


\bibliography{phd_bibliography.bib}{}
\bibliographystyle{aasjournal}



\end{document}